%% file: thesis.tex
\documentclass[hidelinks, a4paper, 12pt, twoside]{report}
\usepackage{geometry}
\usepackage[T1]{fontenc}
\usepackage[utf8]{inputenc}
\usepackage[english]{babel} 
\usepackage{import}
\usepackage{tabularx}

\input{frontmatter/header}
\hypersetup {
pdfpagemode = {UseNone},
pdftitle = {Subsystem Symmetries and Fracton Models in Quantum Error Correction},
pdfauthor = {Giovanni Canossa},
pdflang = {en-GB}
}
\makeglossaries

\begin{document}
\nocite{Canossa23,Canossa24,Canossa25}

\input{frontmatter/acronyms}

%
\pagenumbering{roman}
\pagestyle{empty}
%
\include{frontmatter/first_title_page}
\afterpage{\blankpage}
\myemptypage
%
\include{frontmatter/second_title_page}
%
\include{frontmatter/begutachter.tex}
%
\include{frontmatter/abstract_german}

%
\include{frontmatter/abstract_english}

%
%
\include{frontmatter/list_of_publications}
%

\pagestyle{fancy}

\tableofcontents
\clearpage
\pagenumbering{arabic}
\setcounter{page}{1}
\include{mainmatter/introduction}

%
\include{mainmatter/theory}

%
\include{mainmatter/conclusion}

\clearpage
\phantomsection
\addcontentsline{toc}{chapter}{Bibliography}
\markboth{}{} 
\newrefcontext{nonmyarticles}
\printbibliography
\label{sec:bibliography} 

\appendix
\setcounter{section}{0} 
\renewcommand{\thesection}{\Alph{section}} 
\renewcommand{\thefigure}{\thesection.\arabic{figure}} 
\include{backmatter/appendices/appendices}


\listoffigures


%
\end{document}

%% file: frontmatter/header.tex
\usepackage[T1]{fontenc}
\usepackage[utf8]{inputenc}
\usepackage[english]{babel} 
\usepackage{lmodern}        

\usepackage{tocbibind} 

\usepackage{csquotes}
\usepackage{graphicx}
\usepackage{float}     
\usepackage[backend=biber,style=numeric-comp,sorting=none,defernumbers=true,maxnames=10]{biblatex}
\usepackage[unicode]{hyperref}

\usepackage{xcolor}
\usepackage{amsmath}
\usepackage[font=small,labelfont=bf]{caption}
\usepackage{fancyhdr}
\usepackage{titlesec}
\usepackage{pdfpages}
\usepackage{microtype}
\usepackage{placeins}
\usepackage{listings}
\usepackage{xpatch}
\usepackage[normalem]{ulem}
\usepackage{newtxtext,newtxmath}

\hypersetup {
pdfpagemode = {UseNone},
pdftitle = {Simulating quantum dissipative and vibrational environments. From single qubits to many-body physics},
pdfauthor = {Mattia Moroder},
pdflang = {en-US}
}

\usepackage{enumitem}
\newlist{enumerate_publications}{enumerate}{1}
\setlist[enumerate_publications,1]{
  label=Ref. [\arabic*],
  ref=Ref. [\arabic*],
  labelsep=1em
}

\usepackage{afterpage}
\newcommand\myemptypage{
    \null
    \thispagestyle{empty}
    \addtocounter{page}{-1}
    \newpage
    }

\DeclareBibliographyCategory{MyArticles}
\addtocategory{MyArticles}{Canossa23}
\addtocategory{MyArticles}{Canossa24}
\addtocategory{MyArticles}{Canossa25}
\nocite{Canossa23}
\nocite{Canossa24}
\nocite{Canossa25}

\AtEveryBibitem{%
  \ifcategory{MyArticles}{}{%
    \clearfield{doi}%
    \clearfield{url}%
  }%
}

\DeclareSourcemap{
  \maps[datatype=bibtex]{
    \map{
      \step[fieldset=month, null]
    }
  }
}
\DeclareRefcontext{myarticles}{labelprefix=A}

\DeclareFieldFormat{labelnumberwidth}{#1\hspace{10pt}}

\DeclareCiteCommand{\citenum}
  {\printtext[bibhyperref]{\printfield{labelprefix}}}
  {\printtext[bibhyperref]{\printfield{labelnumber}}}
  {}
  {}

\letbibmacro{ORIG-institution+location+date}{institution+location+date}
\renewbibmacro*{institution+location+date}
{\iffieldundef{url}
		{\usebibmacro{ORIG-institution+location+date}}
		{\href{\thefield{url}}{\usebibmacro{ORIG-institution+location+date}}}
}

\DeclareNameWrapperFormat{author}{\textsc{#1}}

\DeclareDelimFormat{finalnamedelim}{
\textup{
  \ifnumgreater{\value{liststop}}{2}{\finalandcomma}{}%
  \addspace\bibstring{and}\space
  }
}

\makeatletter

\abx@doentrytypes
\makeatother

\fancypagestyle{plain}{ %
  \fancyhf{} 
  
}

\def\headingStyle{\sffamily}

\titleformat*{\section}{\LARGE\headingStyle}
\titleformat*{\subsection}{\Large\headingStyle}
\titleformat*{\subsubsection}{\large\headingStyle}
\titleformat{\chapter}[display]
{\huge\headingStyle}{\chaptertitlename\ \thechapter}{20pt}{\Huge\headingStyle}

\usepackage[toc]{glossaries} 
\usepackage{hyphenat}
\usepackage[capitalise]{cleveref}
\usepackage{physics}
\usepackage{nicefrac}
\usepackage{braket}
\usepackage{booktabs}
\usepackage{tikz}
\usepackage{tikzpagenodes}
\usepackage{subfig}
\usepackage{algpseudocode}
\usepackage{bm}
\usepackage{amsfonts}
\usepackage{mathtools}		        
\usepackage{blkarray} 
\usepackage{mathrsfs} 
\usepackage[version=4]{mhchem} 

\newboolean{buildtikzpics}
\setboolean{buildtikzpics}{false}
\newif\ifrebuildtikz
\newif\ifChangeMode
\ChangeModetrue
\ChangeModefalse
\ifthenelse{\boolean{buildtikzpics}}
{
	\rebuildtikztrue
	\usetikzlibrary{external}
	\tikzexternalize[optimize=false,prefix=figures/autogen/]%
}
{
	\rebuildtikzfalse
}

\definecolor{mygreen}{RGB}{0,128,0}
\definecolor{mypurple}{RGB}{128,0,128}

\DeclareRobustCommand{\rchi}{{\mathpalette\irchi\relax}}
\newcommand{\irchi}[2]{\raisebox{\depth}{$#1\chi$}} 

\usepackage{afterpage}  
\newcommand\blankpage{%
    \null
    \thispagestyle{empty}%
    \addtocounter{page}{-1}%
    \newpage}

\usepackage{geometry}

%% file: frontmatter/acronyms.tex
\newacronym{1D}{1D}{one-dimensional}
\newacronym{2D}{2D}{two-dimensional}
\newacronym{3D}{3D}{three-dimensional}
\newacronym{KW}{KW}{Kramers-Wannier}
\newacronym{KW}{KW}{Kramers-Wannier}
\newacronym{CSS}{CSS}{Calderbank-Shor-Stead (code)}
\newacronym{SM}{SM}{Statistical-Mechanical (mapping)}
\newacronym{PBC}{PBC}{Periodic Boundary Conditions}
\newacronym{FCC}{FCC}{Face-Centered Cubic}
\newacronym{FSS}{FSS}{Finite-size scaling}
\newacronym{RG}{RG}{Renormalization Group}
\newacronym{GSD}{GSD}{Ground State Degeneracy}
\newacronym{TIM}{TIM}{Tetrahedral Ising Model}
\newacronym{FIM}{FIM}{Fractal Ising Model}
\newacronym{MCMC}{MCMC}{Markov Chain Monte Carlo}
\newacronym{PT}{PT}{Parallel Tempering}
\newacronym{MUCA}{MUCA}{Multicanonical}
\newacronym{LGT}{LGT}{Lattice Gauge Theory}
\newacronym{SSPT}{SSPT}{Subsystem Symmetry Protected Topological (phase)}
\newacronym{FTQC}{FTQC}{Fault-Tolerant Quantum Computation}
\newacronym{LDPC}{LDPC}{Low-Density Parity-Check}
\newacronym{qLDPC}{qLDPC}{Quantum Low-Density Parity-Check}
\newacronym{i.i.d.}{i.i.d.}{independent and identically distributed (noise)}
\newacronym{RBI}{RBI}{Random-Bond Ising (model)}

%% file: frontmatter/first_title_page.tex
\newcommand{\HRule}{\rule{\linewidth}{1mm}}

\begin{titlepage}
    \centering
    \HRule \\[1.5cm]
    {\Huge \bfseries \textsc{Subsystem Symmetries and Fracton Models in Quantum Error Correction} }\\[1.5cm]
    {\Large Giovanni Canossa}\\[1.5cm]
    \HRule \\[5.5cm]
        
    \includegraphics[width=0.4\textwidth]{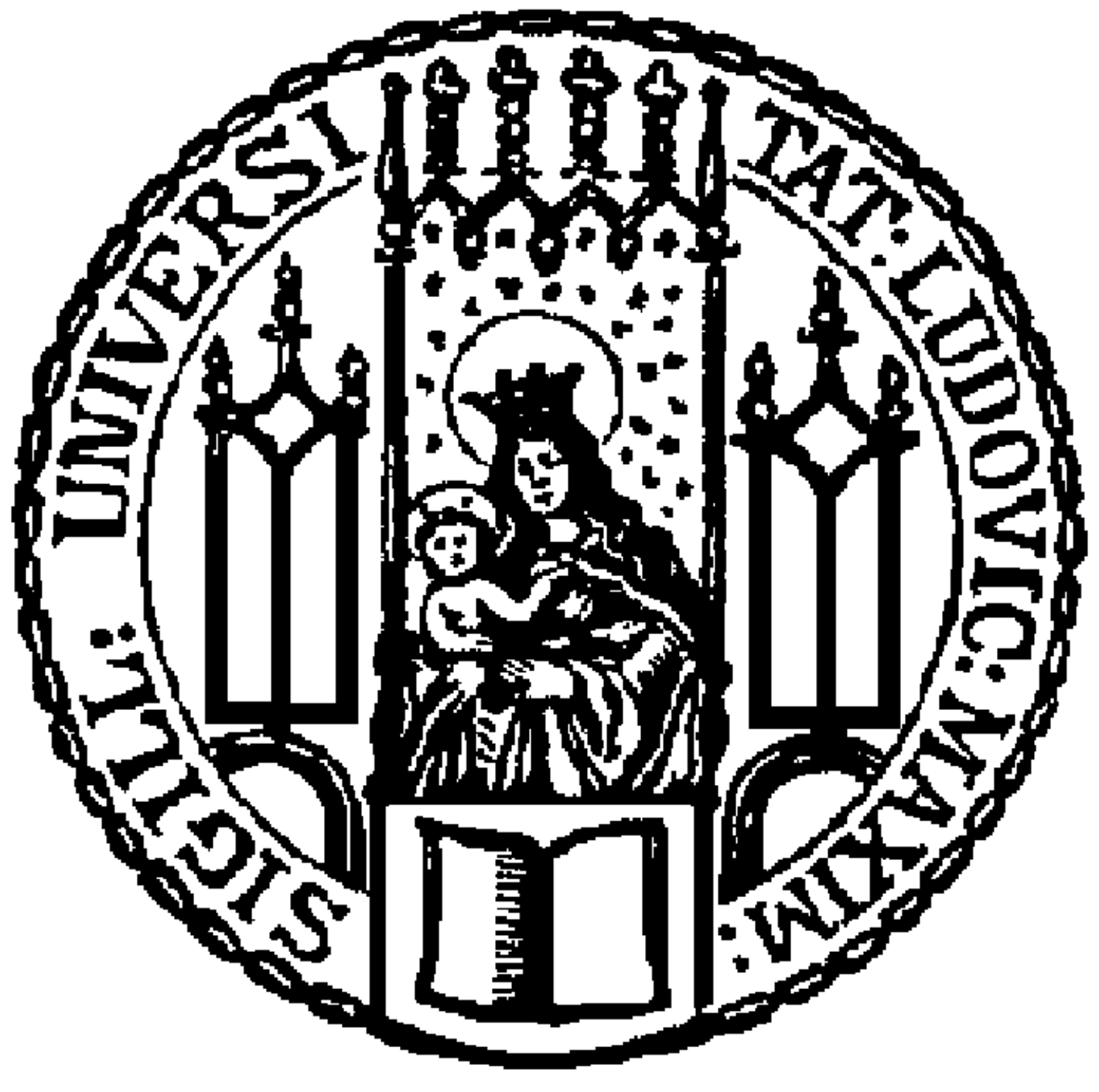}\\[1cm] 
    {\large München 2026}

\end{titlepage}
\restoregeometry 

%% file: frontmatter/second_title_page.tex
\begin{titlepage}
    \centering
    \HRule \\[1.5cm]
    {\Huge \bfseries \textsc{Subsystem Symmetries and Fracton Models in Quantum Error Correction} }\\[1.5cm]
    {\Large Giovanni Canossa}\\[1.5cm]
    \HRule \\[3cm]
        
    \large Dissertation \\
    \large an der Fakultät für Physik \\
    \large der Ludwig-Maximilians-Universität \\
    \large München \\[2cm]

    \large vorgelegt von \\
    \large Giovanni Canossa \\
    \large aus Mantova, Italien \\ [2cm]

    \large München, den 29. April 2026

\end{titlepage}

%% file: frontmatter/begutachter.tex
\newpage

\vspace*{\fill}
Erstgutachter: Prof. Dr. Lode Pollet \\
\vskip 0.1cm
Zweitgutachter: Prof. Dr. Annabelle Bohrdt \\
\vskip 0.1cm
\indent Tag der mündlichen Prüfung: 22. Juni 2026 

%% file: frontmatter/abstract_german.tex
\phantomsection
\addcontentsline{toc}{section}{\textbf{Zusammenfassung}}
\section*{Zusammenfassung}
Die Entwicklung neuer Quantenfehlerkorrekturcodes und das Verständnis ihrer Fehlertoleranz zählen zu den zentralen Herausforderungen bei der Realisierung robuster und effizienter Quantenspeicher. 
Unter den zahlreichen bislang untersuchten Quantencodes nehmen topologische Codes wie der torische Code und der Farbcode eine führende Stellung ein, da sie sowohl günstige 
Fehlerkorrektureigenschaften als auch tiefe Verbindungen zu Phasen der Materie in der Vielteilchenphysik aufweisen. Diese Codes stellen besondere Beispiele einer breiteren 
Klasse von Quantenfehlerkorrekturcodes dar, die mit gut etablierten Modellen der klassischen Spinphysik in Beziehung gesetzt werden können.
In dieser Dissertation untersuchen wir das Zusammenspiel zwischen klassischen Ising-Modellen und Quantenfehlerkorrektur, wobei wir uns auf den Zusammenhang zwischen Subsystemsymmetrien 
in klassischen Theorien, fraktonischer topologischer Ordnung und deren Bedeutung für Quantenspeicher konzentrieren. Zunächst betrachten wir zwei dreidimensionale klassische selbstduale 
Ising-Modelle mit Subsystemsymmetrien, das tetraedrische Ising-Modell und das fraktale Ising-Modell, und analysieren ihr thermisches Verhalten im Zusammenhang mit dem Brechen von 
Subsystemsymmetrien. Insbesondere untersuchen wir ihre stark erstordentlichen Phasenübergänge, die subextensive Grundzustandsentartung sowie Finite-Size-Effekte, die durch die Geometrie 
der zugrunde liegenden Symmetrien geprägt sind, und identifizieren diese als zentrale Merkmale der Phänomenologie des Brechens von Subsystemsymmetrien.
Anschließend zeigen wir, wie das Eichen von Subsystemsymmetrien zu fraktonischen Phasen mit eingeschränkten Anregungen und nichttrivialer Struktur logischer Operatoren führt. 
Der letzte Teil der Dissertation behandelt die Fehlerkorrekturleistung dieser fraktonischen Codes sowie der zugehörigen CSS-Codes mittels einer statistisch-mechanischen Abbildung 
zwischen Dekodierung und ungeordneten Ising-Modellen. In diesem Rahmen bestimmen wir die optimale Code-Capacity-Schwelle des Checkerboard Codes zu $0.107(3)$, einem Wert, 
der die theoretische Grenze nahezu sättigt und die höchste optimale Fehlerschwelle unter den bekannten dreidimensionalen Codes darstellt. Allgemeiner liefert dieses Ergebnis 
weitere Unterstützung für die verallgemeinerte Entropierelation für wechselseitig duale klassische Spinmodelle und erweitert deren Gültigkeit über standardmäßige topologische 
Codes hinaus auf fraktonische Modelle, was darauf hindeutet, dass auch Haahs Code eine Code-Capacity-Schwelle nahe der theoretischen Grenze besitzen sollte.
In ihrer Gesamtheit weisen diese Ergebnisse fraktonische Codes als hochgradig robuste Kandidaten für Quantenspeicher aus und verdeutlichen die Leistungsfähigkeit des 
statistisch-mechanischen Zugangs zusammen mit seinen Dualitätsvorhersagen für die Analyse und Konstruktion robuster Quantenfehlerkorrekturcodes.
\newpage

%% file: frontmatter/abstract_english.tex
\phantomsection
\addcontentsline{toc}{section}{\textbf{Abstract}}
\section*{Abstract}
Constructing new quantum codes and understanding their error resilience are central challenges in the development of robust and efficient quantum memories. 
Among the many quantum codes studied so far, topological codes such as the toric code and the color code have played a leading role because of their favorable 
error-correcting properties and their deep connections to phases of matter in many-body physics. These codes are particular examples of a broader class of 
quantum error-correcting codes that can be related to well-established models in classical spin physics.
In this thesis, we further explore the interplay between classical Ising models and quantum error correction by focusing on the relation between 
subsystem symmetries in classical theories, fracton topological order, and their significance for quantum memories. We first study two three-dimensional 
classical self-dual Ising models with subsystem symmetries, the Tetrahedral Ising model and the Fractal Ising model, and examine their thermal behavior 
in connection with subsystem symmetry breaking. In particular, we analyze their strongly first-order phase transitions, subextensive ground-state degeneracy, 
and finite-size effects shaped by the geometry of the underlying symmetries, identifying these as central features of the subsystem symmetry-breaking phenomenology.
We then establish how gauging subsystem symmetries gives rise to fracton phases with constrained excitations and nontrivial logical-operator structure. 
The final part of the thesis addresses the error-correcting performance of these fracton codes, and of the associated CSS codes, through a statistical-mechanical 
mapping between decoding and disordered Ising models. Within this framework, we determine the optimal code-capacity threshold of the Checkerboard code to be $0.107(3)$, 
a value that nearly saturates the theoretical limit, representing the highest optimal error threshold among known three-dimensional codes. More broadly, this result 
provides further support for the generalized entropy relation for mutually dual classical spin models, extending its relevance beyond standard topological codes 
to fracton models, therefore indicating that Haah’s code should likewise possess a code-capacity threshold close to the theoretical limit.
Taken together, these findings establish fracton codes as highly resilient candidates for quantum memories and demonstrate the power of the statistical-mechanical 
framework, together with its duality predictions, for the analysis and construction of robust quantum-error-correcting codes.

%% file: frontmatter/list_of_publications.tex
\phantomsection
\addcontentsline{toc}{section}{\textbf{Publications}}
\section*{Publications}

\noindent The following list contains the publications produced during my doctoral studies.
\bigskip
\begin{minipage}{\textwidth}
    \vspace{3em}
    \defbibheading{none}{}
    \setlength{\bibitemsep}{1.7em}
    \printbibliography[resetnumbers,category=MyArticles,title={Publications},heading=none]
\end{minipage}
\noindent My individual contributions to these projects are summarized below.
\begin{enumerate_publications}
    \item I implemented and benchmarked Parallel Tempering Monte Carlo methods for the study of different $3D$ compass models, investigating the nature of the phase transitions and their symmetry-breaking patterns
            for different coupling regimes.
    \item I implemented and benchmarked Multicanonical Monte Carlo methods for the study of the Tetrahedral and Fractal Ising model, performed the calculations, data analysis and discussion of the results.
    \item I investigated the Tetrahedral Ising model in the presence of quenched disorder, performed large-scale disorder simulations and data analysis, established the connection to the corresponding Checkerboard model, 
          and formulated the generalized Kramers–Wannier duality-based fixed-point condition as an argument for threshold bound saturation in certain families of CSS codes, extending the analysis to Haah's cubic code.
\end{enumerate_publications}
\noindent The contents of this thesis are centered primarily on the second and third publications.

%% file: mainmatter/introduction.tex
\chapter{Introduction}
\label{chap:introduction}

The relation between classical spin physics, topological phases of matter, and quantum computation runs deeper than one might initially expect. 
Each of these fields addresses a wide range of phenomena, and most of their conceptual frameworks and practical motivations appear largely independent. Nevertheless, a precise and nontrivial 
connection links them, a connection that can be traced back, as is often the case in condensed matter physics, to the Ising model.

Since its introduction in 1925, the Ising model has served as a cornerstone of modern statistical mechanics~\cite{Ising25}. Onsager’s exact solution of the two-dimensional model provided the first rigorous demonstration 
of a phase transition in a system with short-range interactions~\cite{Onsager44}. Its critical behavior later became a paradigmatic setting for the development of scaling theory and Wilson’s renormalization group, establishing 
the model as a testing ground for ideas such as universality, spontaneous symmetry breaking, and critical phenomena~\cite{Kadanoff66,Wilson71,Goldenfeld92,Cardy96}. 

Among these conceptual advances that emerged from the study of the Ising model, one was particularly fundamental for the formulation and development of the field of lattice gauge theory as it is known today. 
In 1941, Kramers and Wannier proposed the idea that the effective low-temperature description of the Ising model can be exactly mapped to the high-temperature expansion of another Ising model defined on the dual graph of the lattice, 
with an inverted effective coupling~\cite{Kramers41a,Kramers41b}. This relation, now formally known as the Kramers-Wannier (KW) duality, implies that the strongly ordered regime of one model corresponds to the disordered regime of its dual, and vice versa.
Beyond providing an estimate of the critical temperature, this duality revealed a deeper structural insight: the fundamental degrees of freedom and excitation structure
of the Ising model admit an equivalent description in terms of the domain wall configurations separating regions of opposite magnetization in its dual model, with the ordered phase of one model corresponding to the proliferation 
of domain walls in the disordered phase of the other. This formulation introduces the notion that fundamental excitations of a theory are not the microscopic variables themselves, but rather extended objects emerging from them.

The scope of this construction was significantly expanded by Wegner in 1971~\cite{Wegner71}, who developed a general framework for duality in arbitrary Ising-type models with multi-spin interactions. Within this framework, 
the natural degrees of freedom of the dual model describe the boundaries of frustrated regions in the original spin configuration. As boundaries, these variables cannot be specified independently: they must 
satisfy local closure conditions imposed by the connectivity of the underlying lattice. This local constraint structure distinguishes the dual theory from an ordinary spin system, as it implies the existence of redundancies 
under independent local transformations of the dual variables, redundancies that are nontrivially tied to the symmetry structure of the original model. 
Wegner’s generalized duality construction was shown to yield a type of theory that we currently refer to as $\mathbb{Z}_2$ lattice gauge theory, in which local gauge constraints emerge as a structural consequence of 
the original model's global symmetry~\cite{Kogut79,Savit80}. In the years that followed, lattice gauge theory developed into a broad and independent field. By the 1980s, gauge-theoretic ideas such as confinement, 
deconfinement, and the role of extended excitations had become part of the conceptual toolkit of both high-energy and condensed matter theory.

While these developments were unfolding, the field of quantum computation was starting to take shape. Beginning in the 1980s, it was recognized that quantum mechanical systems could process information in ways fundamentally 
inaccessible to classical devices~\cite{Feynman82,Deutsch85}. This insight led to the formulation of the quantum circuit model and to landmark algorithms such as Shor’s factoring algorithm in 1994~\cite{Shor97}, 
which demonstrated an exponential speedup over the best known classical methods. However, the quantum coherence that enables such computational advantages is also what renders quantum information inherently fragile, as even weak couplings 
to the environment are sufficient to induce decoherence, rapidly degrading stored or processed states~\cite{Zurek91}. By the mid-1990s, it had become clear that scalable quantum computation would require active protection against local noise. 
This realization led to the development of quantum error correction in 1995–1996, beginning with Shor’s nine-qubit code and followed by the Calderbank–Shor–Steane (CSS) constructions~\cite{Calderbank96}, which provided systematic methods for 
encoding quantum information in a fault-tolerant manner.

At the intersection of these evolving fields one can find the toric code. Introduced by Kitaev~\cite{Kitaev03} as an exemplary local, exactly solvable quantum spin Hamiltonian, the model represents a quantum realization of a 
$\mathbb{Z}_2$ lattice gauge theory in $2+1$ dimensions with a distinctive organizing principle: rather than having its low-energy physics captured by spontaneous symmetry breaking or any local order parameter, its ordering
is strictly non-local and depends entirely on the topology of the supporting manifold. This property follows directly from the local gauge constraints that define the ground-state subspace and makes the toric code one of 
the first exactly solvable lattice spin Hamiltonians characterized by topological order~\cite{Wen17}.

The gauge constraints that generate this non-local structure also provide the point of contact with the classical duality framework: from this perspective, the toric code can be obtained from an Ising-type model
through a procedure of \textit{global symmetry gauging} that parallels Wegner's generalized duality construction, in which the original globally symmetric model is replaced by one in which local constraints 
and corresponding gauge redundancy appear, with the additional promotion of the resulting variables to quantum operators~\cite{Fradkin78}. The same defect constraints that underlie the classical Ising model become the property that governs
the excitation structure of the resulting quantum model, and hence the detectability of errors under local noise~\cite{Dennis02}. This connection highlights the relation between the Ising model, the quantum toric code, 
and its performance in quantum error correction, with the stability of the toric code as a quantum memory being controlled by whether typical error processes remain confined to topologically trivial 
defect configurations, or proliferate into system-spanning events that connect distinct topological sectors.

These developments naturally raise a broader question: if the structure of symmetry and defects in classical spin models controls both the excitation content of the associated gauge theory and the logical 
failure mechanisms of the corresponding code, what new behavior might arise from \emph{different} symmetry-constraint structures? This question helped motivate the exploration of \emph{fracton phases of matter},
in which locality and constraints conspire to produce fundamentally altered excitation dynamics. This class of exactly solvable models was first inspired by the Chamon model in 2005, in which local 
commuting constraints imply that point-like excitations cannot be moved independently by local operators~\cite{Chamon05}. The resulting \emph{fractionalized mobility} introduced various dynamical consequences, 
including slow relaxation reminiscent of glassy behavior, a property that could potentially provide an alternative route towards robustness against quantum noise~\cite{Haah11}.

This line of inquiry naturally converged on the exploration of $\mathbb{Z}_2$ lattice gauge theories derived from Ising-type models with \emph{subsystem symmetries}, i.e. models characterized by
a subextensive number of symmetries acting on rigid lower-dimensional subsets of the entire spin system~\cite{Haah11,Vijai15,Vijai16}. Applying to subsystem-symmetric spin models a \textit{subsystem symmetry gauging} procedure 
analogous to the global symmetry gauging of the 2D Ising model, one obtains a new class of $\mathbb{Z}_2$ fracton models in the form of lattice gauge theories whose elementary excitations are fundamentally immobile.
An additional element of novelty to these fracton phases derived from subsystem symmetry gauging is that their ground-state degeneracy is determined not only by the topology of the global system, 
but also by geometric features of the lattice embedding and by the structure of the subsystem symmetries themselves, placing them outside the scope of conventional topological quantum field theory 
descriptions~\cite{Ma17,Shirley18,Slagle18,Shirley19,Nandkishore19}.

Beyond their inherent relevance as novel phases of matter with unconventional topological order, these models carry an element of novelty both in the classical and quantum computation settings.
The geometric structure of subsystem symmetries and the nature of the phase transitions associated with subsystem-symmetry breaking are still
not fully understood in terms of a single unifying mechanism, and the corresponding transitions have so far resisted a standard classification in terms of thermal universality classes 
due to the absence of conventional critical behavior in the models studied so far~\cite{Nussinov15,Canossa23}. 
In addition, the restricted mobility of elementary defects implies that local noise has a reduced ability to reorganize into extended, topologically nontrivial error processes in the context of quantum computation.
This suggests an alternative route to robustness of the topologically ordered phase against quantum fluctuations and, in the error-correction setting, to suppressing logical failure mechanisms~\cite{Haah11,Vijai16,Song22}. 
These considerations motivate the study of fracton phases as candidate quantum memories and quantum error-correcting codes, which may be viewed as stabilizer realizations of subsystem 
lattice gauge theories and as natural generalizations of the toric-code paradigm from ordinary gauge structure to subsystem gauge structure.

This thesis focuses on the connection between subsystem-symmetric phases of matter, the emergent fracton phases that arise upon gauging them, and the extent to which the 
resulting constraint structure can be leveraged for quantum error correction. This is investigated concretely through the analysis of two examples of subsystem-symmetric classical Ising models, namely
the Tetrahedral and Fractal Ising models, and of the fracton models derived from these classical models via gauging, respectively the Checkerboard model and Haah's cubic code.
Our first aim is therefore to investigate the occurrence of subsystem-symmetry breaking in these \emph{classical} Ising-type models, with the objective of bringing further coherence 
to the understanding of subsystem symmetry breaking, which has so far often evaded the standard Landau picture. To address this systematically, we analyze two representative classical subsystem-symmetric models, 
characterize their symmetry structure, and determine whether, and in what form, subsystem-symmetry breaking occurs at finite temperature, with the aim of identifying which aspects of subsystem 
symmetry control the existence and nature of phase transitions. 

This classical analysis provides the foundation for the later parts of the thesis, where these same symmetry and defect structures are promoted, via gauging, to local constraints in the corresponding 
fracton gauge theories and ultimately determine the syndrome and logical-failure structure of the associated quantum codes. In the next step, we discuss how the gauging procedure leads to the 
formulation of fracton gauge theories with generalized Gauss-law constraints, clarifying the role played by the geometry of the subsystem symmetries, 
the way in which local constraints encode subsystem charge conservation, and why these ingredients lead to excitation sectors with fractionalized mobility.

Having established the classical models and their relation to fracton models, we then turn to a distinct but complementary question: how does the statistical-mechanical structure of the 
classical models manifest in the concrete performance of the corresponding quantum codes? This involves formulating the decoding problem for the representative fracton stabilizer models 
and identifying the relevant noise models and syndrome structure. In this procedure, we will analyze in detail the extent to which the properties of the classical model can be leveraged 
to gain information about the performance of these fracton codes, and we will place this relation within a more general framework in the form of a classical-to-quantum statistical mechanical (SM) mapping.

The thesis is organized as follows. Chapter~\ref{chap:subsystem_symmetries} introduces subsystem symmetries in classical and quantum spin models, with emphasis on their 
geometric character and on the associated notions of subsystem charges and defects. Chapter~\ref{chap:fracton_models} both provides a more accurate characterization of 
fracton phases and develops the gauging procedure for subsystem symmetries, which is then used to construct the corresponding $\mathbb{Z}_2$ lattice gauge theories with
fracton topological order. Chapter~\ref{chap:fractons_in_QEC} then turns to the perspective of quantum computation: fracton stabilizer 
Hamiltonians are interpreted as codes, their syndrome structure is related to the underlying defect constraints, and the implications for decoding and memory stability 
under local noise are analyzed. Throughout the thesis, we aim to construct a narrative in line with the Ising-to-toric code thread developed above: subsystem symmetries 
provide the analogue of the original global symmetry, subsystem gauging replaces ordinary gauging, and fracton codes supply a new setting in which the behavior of 
quantum information is governed by the interplay between constraints, defect dynamics, and the proliferation of topologically nontrivial error processes.

Beyond these objectives, a higher-level purpose of this analysis is to show that the classical language of symmetry and defects does not merely provide intuition for 
topological quantum codes, but can potentially be leveraged in more general Ising-type settings as a tool for generating novel quantum codes and understanding the behavior in the
presence of errors, thereby highlighting the relation between the three themes developed above.

%% file: mainmatter/theory.tex
\part{Classical Spin Models with subsystem symmetries}

\input{mainmatter/theory/subsystem_symmetries.tex}
\part{Fracton models in Quantum Computation}

\input{mainmatter/theory/fracton_models.tex}
%
\input{mainmatter/theory/fractons_in_QEC.tex}

%% file: mainmatter/theory/subsystem_symmetries.tex
\newcommand{\mean}[1]{\left\langle #1 \right\rangle}

\chapter{Subsystem Symmetries in Classical Ising Models}
\label{chap:subsystem_symmetries}

\section{Introductory remarks}
Phase transitions constitute one of the central themes of condensed matter and statistical physics. Since the formulation of Landau theory, 
it has been widely understood that systems whose Hamiltonians possess a given symmetry may undergo a spontaneous breaking of that symmetry 
when the model reaches certain parameter regimes, be it temperature or interaction strength. 
Within this framework, the emergence of an ordered phase is accompanied by the spontaneous selection of one ordered state among a set 
of symmetry-related ground states. 
Subsequent investigations into theories endowed with local gauge symmetries revealed a fundamental limitation of the Landau paradigm, 
namely that local gauge symmetries cannot be spontaneously broken~\cite{Elitzur75}. 
As a result, systems characterized by a form of gauge-invariance evade the symmetry-breaking framework of the Landau description
based on local order parameters, which prompted the formulation of different paradigms that go beyond the notion of a local order 
parameter~\cite{Wegner71}. 

Following Anderson’s seminal proposal of the resonating-valence-bond picture to describe the non-trivial types of antiferromagnetic
order that characterize the insulating phases of strongly correlated antiferromagnetic models~\cite{Anderson73}, 
the attention shifted toward systems affected by magnetic frustration, in which competing interactions 
in the Hamiltonian favor distinct, equally likely ordering patterns that cannot be simultaneously satisfied, a feature that can 
lead to an extensive degenerate energy spectrum.
This intrinsic competition enables the realization of ground states with highly nontrivial structure, exotic low-energy excitations, 
and unconventional dynamical behavior, giving rise to the exceptionally broad and vibrant research area of frustrated magnetism
~\cite{Lacroix11}.
One of the hallmark properties of frustrated spin models is the presence of a macroscopically degenerate ground-state manifold. 
In many instances, this degeneracy enhances quantum or thermal fluctuations, suppressing conventional long-range order and 
favoring disordered phases such as spin liquids, in which no symmetry-breaking order parameter can be defined~\cite{Khomskii03}.

In classical settings, frustration frequently leads to accidental degeneracies on a macroscopic scale 
that are not protected by the symmetry group of the Hamiltonian. In such cases, it is often possible to continuously 
interpolate between distinct ground states through local transformations unrelated to any underlying symmetry. 
Well-known examples include classical spin-ice models on the pyrochlore lattice~\cite{Harris97,Moessner98,Bramwell01,Castelnovo08}, 
the Heisenberg antiferromagnet on the kagome lattice~\cite{Hastings00,Yan11}, and the triangular-lattice antiferromagnet with 
Ising anisotropy~\cite{Collins97}. These systems are also characterized by a proliferation of low-energy modes, 
whose accessibility may depend sensitively on the specific ground-state configuration of the system.
Despite the absence of an explicit energetic preference, a unique ordered state may still be selected via the mechanism known as 
\emph{order-by-disorder}. In this process, entropic contributions stabilize those configurations that maximize the density 
of low-energy modes, favoring ordered states that have access to the greatest number of low-energy excitations~\cite{Villain80}.
This phenomenon is also responsible for the presence of symmetry breaking in models whose highly degenerate spectrum is associated with the
presence of multiple symmetries; in such cases, the entropic part of the free energy plays a decisive role~\cite{Nussinov15}.

Over the past two decades, another class of models characterized by \textit{subdimensional symmetries} (or subsystem symmetries) emerged:
following their original formulation in classical spin compass models~\cite{Nussinov05,Nussinov15}, 
these are conventionally understood as lying between the notion of global and local symmetry, with the model Hamiltonian
being invariant under symmetry transformations involving all degrees of freedom contained in a support 
that lies on an extensively large submanifold of the system. 
Due to the intrinsic dependence of critical phenomena with respect to both the dimensionality of the system~\cite{Goldenfeld92}, 
the nature of its microscopic degrees of freedom and the type of symmetry of the Hamiltonian, 
changing the codimension of the symmetry support was shown to give rise to a variety of phases of matter with unique
degeneracy and ordering structures, both in the classical and quantum settings~\cite{Nussinov15,Nussinov09}. 
In classical spin systems, an ordering behavior commonly observed is the subdivision of the lattice in 
several subregions characterized by similar ordering behaviors, with each region determined by the subsystem symmetry acting upon it.
Since such symmetries operate on subsets of the whole system, one can separate the lattice into a set of submanifolds, 
each with its own ordering structure and contribution to the overall degeneracy of the energy spectrum, giving rise 
to a symmetry-driven subextensive ground-state degeneracy. In some cases, subsystem symmetries are accompanied by 
geometric frustration, showing that certain compass models can exhibit both subextensive ground-state degeneracy 
and subsystem ordering achieved through the order-by-disorder mechanism characteristic of frustrated models, while that 
same ordering can also be described by a generalized symmetry-breaking paradigm through the framework of dimensional reduction, 
thereby avoiding both the Landau paradigm and deconfinement scenarios~\cite{Nussinov15,Canossa23}.
Keeping the focus on models with subsystem symmetries, in the remainder of this chapter we 
will concentrate on lattice models with Ising degrees of freedom and ferromagnetic local 
interactions. Including the paradigmatic nearest-neighbor Ising model on D-dimensional cubic 
lattices, Ising-type interactions define an extensive, variegated class of models.
Among these, many fall within the paradigm of frustrated magnetism~\cite{Binder86,Collins97,Diep20},
or admit a direct connection to topologically ordered models, which can be derived from the former through a generalized Kramers-Wannier 
duality by recasting their behavior in the high-temperature regime in a gauge-theoretical formulation~\cite{Kogut79,Kitaev03,Castelnovo07}, 
a feature that will be discussed in more detail in Sec.~\ref{sec:gauging}.

Subsystem-symmetric Ising models have also attracted considerable attention due to the peculiar topological features they 
can give rise to in quantum spin systems, most prominently in 3D spin models characterized by 2D subsystem symmetry~\cite{Vijai16,You18,Shirley19,Stephen20}.
Despite the breadth of this research field, the body of work on \emph{classical} subsystem-symmetric Ising models is comparatively slim. 
This is partly due to the fact that such interactions are uncommon in real materials.
Furthermore, all classical 3D compass models and Ising models with subsystem symmetries studied so far exhibit non-standard 
temperature-driven first-order phase transitions~\cite{Lipowski97,Johnston17,Canossa23,Canossa24}. 
The absence of scale invariance at these transitions complicates their theoretical and numerical 
analysis and places them outside the standard RG framework associated with critical phenomena. In the absence of a formal theorem, it remains an open question 
whether this empirical pattern reflects a general principle or can be overturned by future observations.
To further characterize the properties of Ising models with subsystem symmetries, we therefore focus our attention on the
investigation of the 3D Tetrahedral Ising model and the Fractal Ising model. As will be shown in Sec.~\ref{sec:gauging} and
Sec.~\ref{sec:SMmapping}, these models are closely related to the construction of topologically ordered models with constrained excitation 
mobility and their performance as quantum memories in the context of quantum error correction, which will be the focal point of the latter part of this thesis.
Together with the Plaquette Ising model already discussed in Ref.~\cite{Johnston17}, these models provide a paradigm that allows us to characterize
classical ordering phenomena in subsystem-symmetric Ising systems. All these models exhibit strong first-order transitions with non-standard finite-size scaling 
behavior and the emergence of long-range order within each submanifold upon subsystem symmetry breaking, without any physical local order parameter being uniquely 
defined across the entire lattice.

This chapter of the thesis is organized as follows. We begin in Sec.~\ref{sec:1.1} with a brief overview of thermally driven phase transitions in classical statistical 
physics, with special attention to the finite-size scaling effects which will be used to predict the thermodynamic properties of the studied models and 
describe their qualitative changes in the presence of subsystem symmetries. We then move on to discuss the Tetrahedral Ising model and the Fractal Ising 
model in Sec.~\ref{sec:subsym_models}, with particular emphasis on the characterization of their symmetry properties and their expected ordering behavior. In Sec.~\ref{sec:NumericalMethods}
and Sec.~\ref{sec:DCDA} we review the Monte Carlo techniques used to simulate and analyze the two models, with special attention paid to strategies for mitigating metastability effects 
associated with first-order transitions. The chapter is concludes with a presentation of the numerical results and a final discussion in Sec.~\ref{sec:nodis_results}.

\section{Phase transitions and Symmetries in classical spin systems}

\subsection{Thermal properties}\label{sec:1.1}
Before introducing the specific models studied in this thesis, we briefly review the statistical-mechanics foundations underlying their analysis.
Consider a system of $N$ classical Ising spins $S_i=\pm1$ defined on a D-dimensional lattice and governed by a Hamiltonian $H(\mathbf S)$, where
$\mathbf S=\{S_1,\dots,S_N\}$ denotes a specific microscopic configuration living in the space of all possible configurations $\mathcal S$.
At thermal equilibrium, fixed temperature $T$ and with inverse temperature $\beta = 1/(k_B T)$, the thermodynamic properties of the system are encoded in the canonical partition function
\begin{equation}
Z(\beta) \coloneqq \sum_{\mathbf S\in\mathcal S} e^{-\beta H(\mathbf S)}.
\label{zeta}
\end{equation}
The free energy of the system at equilibrium can be derived from the partition function as
\begin{equation}
F\left(\beta\right) = -k_B T\, \log\, Z.
\end{equation}
As it encapsulates the equilibrium properties of the system, from this quantity it is possible to derive all other thermodynamic expectation values at 
equilibrium~\cite{Pathria96}.

For systems with local, short-range interactions, the free energy is an extensive quantity, 
meaning that it can scale at most proportionally to the volume of the system.
Subleading finite-size corrections depend on geometry, boundary conditions, and dimensionality, and may acquire singular behavior near criticality.
This suggests the possibility of splitting the contributions to the free energy into bulk effects ($\propto V$), surface effects ($\propto A$), 
and additional finite-size effects scaling at most as $\mathcal O(L^{D-2})$.
For a finite system of linear size $L$ and volume $V\sim L^D$, it is convenient to decompose the free energy as
\begin{equation}
F = V f_b + A f_s + \mathcal O(L^{D-2}),
\end{equation}
where $f_b$ is the bulk free energy density and $f_s$ is the surface free energy density.
In the thermodynamic limit ($N,V\rightarrow\infty$), the finite-size behavior and the surface effects 
become negligible compared to the contributions in the bulk, meaning that all thermodynamic properties of 
the system are captured by the bulk free energy density~\cite{Binder87}.

Given an observable $O(\mathbf{S})$, its thermal expectation value is defined as
\begin{equation}
\left<O\right>  =\frac{1}{Z} \displaystyle \sum_{\mathbf{S}\in\mathcal S}O \left(\mathbf{S}\right) e^{-\beta H(\mathbf{{S}})} 
                =  \displaystyle \sum_{\mathbf{S}\in\mathcal S}O \left(\mathbf{S}\right) P(\mathbf{S}),
\label{average}
\end{equation}
where $P(\mathbf S)= e^{-\beta  H(\mathbf S)}/Z$ is the Boltzmann probability distribution.

Other relevant thermodynamic quantities that can be derived from the partition function are the mean internal energy, 
the heat capacity and the susceptibilities of a given observable:
\begin{flalign}
&\left<E \right> = \frac{1}{Z} \displaystyle \sum_{\mathbf{S}\in\mathcal S} H(\mathbf{{S}})  e^{-\beta H(\mathbf{{S}})} = -\frac{\partial \log Z}{\partial \beta}, \\
&\,\,C_V = \frac{1}{V}\frac{\partial \langle E \rangle}{\partial T} = \frac{\beta^2}{V}\left(\langle E^2 \rangle - \langle E \rangle ^2\right), \label{Cv}\\
&\,\,\rchi_{O} = \frac{\beta}{V}\left(\langle O^2 \rangle - \langle O \rangle^2\right). \label{Susc}
\end{flalign}
To characterize spatial correlations of a given quantity $O$, one introduces the connected two-point correlation function $G(r)$:
\begin{equation}
G(r) =  \displaystyle \sum_{\substack{\mathbf{r}_i,\mathbf{r}_j\\|\mathbf{r}_i-\mathbf{r}_j|=r}} 
        \left(\,\langle  O (\mathbf{r}_i) O (\mathbf{r}_j)\rangle - \langle O (\mathbf{r}_i)\rangle\langle O (\mathbf{r}_j)\rangle \,\right)
\end{equation}
The first term denotes the similarity of the observable at different points in space, also referred to as the disconnected correlation function. 
The second term is used to get rid of any sort of statistical dependence between the two values not caused by thermal fluctuations.
In phases without long-range order, the connected correlation function typically decays exponentially,
$G(r)\sim e^{-r/\xi}$, defining the correlation length $\xi$.
The behavior of $\xi$ as a function of temperature provides a key diagnostic in the presence of a phase transition at a temperature $T_c$:
\begin{itemize}
\item for $T \gg T_c$, correlations are short-ranged and both the connected and disconnected correlation functions flatten out;
\item at $T\approx T_c$, $\xi$ either diverges for second-order phase transitions or converges to some finite value for first-order phase transitions in the thermodynamic limit;
\item in an ordered phase at $T \ll T_c$, the connected correlation function decays exponentially with distance with a finite correlation length $\xi$, 
        while the disconnected correlation function approaches a nonzero constant.
\end{itemize}

\subsection{Symmetries and symmetry breaking}

The symmetry properties of a Hamiltonian play a central role in determining the structure of its equilibrium phases.
For a system whose Hamiltonian is invariant under a global symmetry group, its spectrum and the set of physical states can be organized into 
representations of that group, with the possibility to map the different state representations into representations of that group, with the different state 
representations being related to one another by symmetry transformations belonging to the group.
While this symmetry is reflected by the thermal equilibrium distribution at high temperature, 
at sufficiently low temperatures, the system may spontaneously select one of these symmetry-related sectors: when this process occurs,
the symmetry possessed by the Hamiltonian is no longer reflected by the equilibrium state of the system.
This phenomenon is known as spontaneous symmetry breaking and underlies most order-disorder phase transitions.

Temperature-driven spontaneous symmetry breaking corresponds to a qualitative change in the equilibrium ordering of the system.
In the thermodynamic limit, this is accompanied by non-analyticities in the temperature dependence of the free energy.
As a consequence, at least one derivative of the free energy exhibits a singularity at the transition point $T_c$. 
These non-analyticities can be captured by the order parameter $O$, a thermal quantity which reflects how the microscopic components 
of the system arrange themselves following the symmetry breaking and acquires a non-zero expectation value in the 
symmetry-broken phase, thus providing a quantitative measure of the long-range order that accompanies the transition. 
Identifying an appropriate order parameter is strictly dependent on the model and its symmetry group, 
and may be nontrivial in the presence of competing interactions, frustration, or complex ground-state manifolds.

The existence of non-analyticities in the free energy implies that at least one of its derivatives 
presents a singularity at $T_c$. It is then possible to classify phase transitions in two categories:
\begin{itemize}
\item first-order, discontinuous phase transitions, characterized by a discontinuity in the first derivative of the free energy with respect to temperature (i.e., the total energy of the system);
\item higher-order transitions, which have continuous internal energy but discontinuous second or higher derivative of the free energy.
\end{itemize}
In a thermal first-order transition the system absorbs/ejects a finite amount of energy in latent heat before going above/below the transition temperature $T_c$.
Rather than fluctuating around a well-defined thermal state, near the transition there is a coexistence of ordered and disordered phases~\cite{Binder87}. 
As a result, the thermodynamic properties of the system will be given by their values in the two different phases and $T_c$ will be the temperature at which 
the two phases have equal weight, meaning that a given point in the system has the same probability of being in the ordered phase as in the disordered phase.
This property can be diagnosed by studying the energetic probability distribution,
\begin{equation}
P(e)=\langle\delta(e-e')\rangle,
\end{equation}
at the transition temperature, which will show two peaks centered around the mean energy of the ordered and disordered phases respectively,
reflecting the coexistence of the two phases.
Due to phase coexistence, microscopic details remain relevant even in the thermodynamic limit~\cite{Binder87}.
The system decomposes into distinct domains separated by interfaces, whose fluctuations introduce a finite characteristic length scale.
The correlation length $\xi$ remains finite at $T_c$, reaching a maximum value at the transition point, determined by interfacial properties.
The presence of a finite latent heat implies a discontinuity in the internal energy: in the thermodynamic limit, this manifests as a $\delta$-like 
singularity in the specific heat $C_V$, while susceptibilities associated with the order parameter exhibit analogous singular behavior.
The absence of a diverging correlation length implies that first-order transitions are not governed by universal long-wavelength physics, as most
of the fluctuations occur at a scale set by $\xi$.

Another consequence of finite latent heat, and a defining feature of first-order transitions, is the occurrence of metastability and the subsequent
presence of a hysteresis loop. The existence of free-energy barriers associated with interfacial tension allows the system to persist in a metastable state 
beyond $T_c$ due to the additional energy cost required to break down its domain walls. This implies that, upon cooling from $T\gg T_c$, the system may 
remain in the disordered phase below the transition temperature (supercooling), while upon heating it may remain ordered above $T_c$ (superheating). 
Consequently, the observed transition temperature depends on the thermal history of the system.
In numerical simulations, superheating and supercooling represent a challenge for an accurate estimation of the true transition point, which lies in 
the middle of the hysteresis loop~\cite{Goldenfeld92}.


The critical point of a second-order transition is, in a sense, more interesting to investigate, as it showcases richer physics than phase coexistence. 
At the critical temperature $T_c$, thermal fluctuations become relevant on all length scales, and correlations extend over macroscopic distances.
The connected correlation length $\xi$ diverges in the thermodynamic limit, signaling the absence of a characteristic length scale, 
making microscopic details negligible~\cite{Goldenfeld92}.

Near the critical point, the system enters a unique critical state that is independent of whether it is approached from the ordered 
or disordered phase. Unlike first-order transitions, continuous transitions do not exhibit phase coexistence or hysteresis: 
instead, the system displays scale invariance and universal behavior governed by long-wavelength fluctuations.

In the vicinity of the critical point $T_c$, the divergence of the correlation length in the thermodynamic limit is described by the scaling law:
\begin{equation}
\xi = \xi_{0_{\pm}} \left|1-\frac{T}{T_c}\right|^{-\nu} + .\,.\,.\,, \nonumber
\label{corr-nu}
\end{equation}
where $\xi_{0_{\pm}}$ are nonuniversal amplitudes corresponding to the ordered ($-$) and disordered ($+$) phases, 
and $\nu$ is a universal critical exponent.
Other thermodynamic quantities exhibit analogous power-law singularities near criticality, such as the specific heat, 
the order parameter, and its susceptibility:
\begin{align}
C &= C_{\text{bg}} + C_0 \left|1 - \frac{T}{T_c}\right|^{-\alpha} + \cdots, \nonumber \\
O &= O_0 \left(1 - \frac{T}{T_c}\right)^{\beta} + \cdots, \nonumber \\
\chi &= \chi_0 \left|1 - \frac{T}{T_c}\right|^{-\gamma} + \cdots, \nonumber
\label{scalinglaws}
\end{align}
where $\alpha$, $\beta$ and $\gamma$ are other critical exponents associated with the symmetry breaking.

A central consequence of the divergence of $\xi$ is that microscopic details become irrelevant near the critical point.
Instead, the critical behavior is determined solely by macroscopic features such as spatial dimensionality, 
symmetry properties of the Hamiltonian and the order parameter, and the interaction range.
This is also referred to as the universality principle, which allows us to classify continuous phase transitions in universality classes, 
determined by the macroscopic properties of the system, such as its spatial dimensionality and symmetries of the Hamiltonian, 
and characterized by the same critical exponents~\cite{Goldenfeld92}.

\subsection{Finite-size scaling}\label{sec:finite_size_scaling}
The singularities associated with continuous phase transitions arise in the thermodynamic limit.
For any finite system, the free energy is an analytic function of the control parameters, and critical singularities are therefore 
rounded and shifted over a finite region of parameter space~\cite{Binder87}.
Finite-size scaling (FSS) theory provides a systematic framework to extract thermodynamic-limit properties from simulations or 
experiments performed on finite systems~\cite{Fernandez09,Park10}.

The physical origin of finite-size scaling can be understood in terms of the correlation length $\xi$.
In an infinite system, $\xi$ diverges at the critical temperature $T_c$, whereas in a finite system of linear size $L$ correlations cannot extend beyond $L$.
As a result, sufficiently close to criticality one has $\xi \sim L$, which implies
\begin{equation}
\left|1-\frac{T}{T_c}\right| \sim \xi^{-1/\nu} \sim L^{-1/\nu},
\end{equation}
where $\nu$ is the correlation-length critical exponent.
This relation controls both the rounding and the finite-size shift of critical singularities.

The physical origin behind it can be understood in terms of the correlation length $\xi$.
While $\xi$ diverges at the critical temperature $T_c$ in an infinite system, in a finite system of linear size $L$ correlations 
cannot extend beyond the system size, meaning that the system can only achieve a maximum correlation length $\xi \simeq L$.
It then follows that 
\begin{equation}
\left|1-\frac{T}{T_c}\right|\propto\xi^{-\frac{1}{\nu}} \longrightarrow L^{-\frac{1}{\nu}}.
\end{equation}
where $\nu$ is the correlation length critical exponent introduced previously.

One then obtains the following finite-size scaling laws:
\begin{flalign}
C &= C_{\text{background}}+b\,\, L^{\alpha/\nu}+\,.\,.\,.\,\nonumber\\
O &= c \,\,L^{-\beta/\nu}+\,.\,.\,.\, \nonumber\\
\rchi &= d\,\, L^{\gamma/\nu}+\,.\,.\,.\,
\end{flalign}
These equations are valid in close proximity to $T_c$ so long as the scaling variable
\begin{equation}
x = \frac{T_c - T}{T_c} \,L^{1/\nu}
\end{equation}
remains at a fixed value~\cite{Gausterer92}.

Although no true phase transition can be defined in finite systems, it is possible to identify the pseudo-critical point $T_c(L)$ at 
a given lattice size $L$ as the temperature at which the susceptibility, the specific heat and the correlation length reach their maximum.
This is not to be confused with the true critical point $T_c \equiv T_c(L=\infty)$.
The requirement of fixed $x$ implies that the position of $T_c(L)$ is located at a particular value of $x_{\text{max}}$ and follows a specific 
finite-size scaling behavior given by:
\begin{flalign}
&T_c(L) = T_c (\infty)\left(1 + x_{\text{max}} L^{-1/\nu} + \,.\,.\,.\right)\longrightarrow \nonumber\\
&T_c(L) - T_c(\infty) \propto L^{-1/\nu}
\end{flalign}
Given the fact that $T_c(L)-T_c(\infty) \propto \beta_c(\infty) - \beta_c(L)$, this can be rewritten as
\begin{equation}
\beta_c(L) = \beta_c (\infty) - a\,\, L^{-1/\nu}
\label{2ndorderscaling}
\end{equation}
If one relaxes the requirement that $x$ must remain fixed, it is still possible to write the scaling laws as
\begin{equation}
C_V (T,L) = L^{\alpha/\nu} f(x) + \,.\,.\,.\,, \qquad \rchi(T,L) = L^{\gamma/\nu} g(x) + \,.\,.\,.\,,
\end{equation}
where $f(x)$ and $g(x)$ are universal scaling functions. Since, for a given $L$, $T_c(L)$ is identified by a value of $x_{\text{max}}$,
the maxima of the susceptibility curve and the heat capacity curve for different lattice sizes will behave as:
\begin{flalign}
&C_{\text{max}} (L) = f(x_{\text{max}}) L^{\alpha/\nu}\\
&\rchi_{\text{max}} (L) = g(x_{\text{max}}) L^{\gamma/\nu}
\end{flalign}
It is then possible to determine the critical exponents $\nu$,$\alpha$ and  $\gamma$ and the true critical point 
$T_c(\infty)$ by analyzing the behavior of $\beta_c(L)$, $C_{\text{max}}(L)$, and $\rchi_{\text{max}}(L)$.

Finite-size effects at first-order phase transitions exhibit qualitatively different behavior.
While in the thermodynamic limit these are characterized by $\delta$-like singularities in the energy density distribution $P(e)$ 
associated with a finite latent heat, these singularities become more rounded and are shifted in energy at finite system sizes.
Since bulk thermodynamic observables are dominated by fluctuations associated with the latent heat, while phase coexistence is controlled 
by the interfacial free-energy cost, finite-size effects at first-order transitions typically scale either with the system volume $L^D$
or with the interfacial area $L^{D-1}$~\cite{Binder87,Challa86,Borgs90}.

Treating the phase transition as a sharp jump between phases, one can say that, around $T_c$, the system spends a fraction $W_o$ of its 
time in the ordered phase and the remaining fraction $W_d = 1 - W_o$ in the disordered phase.
The ordered and disordered phases in the thermodynamic limit have energy densities $\hat{e}_o$ and $\hat{e}_d$. 
This means that, up to energy fluctuations, the energy density moment of order $n$ near the transition for finite lattice sizes becomes
\begin{equation}
\left<e^n\right> = W_{\text{o}} \hat{e}^{n}_{\text{o}} + (1-W_{\text{o}}) \hat{e}^{n}_{\text{d}}.
\end{equation}
The specific heat can then be written as
\begin{flalign}
C_V(\beta,L) &= \beta^2L^D\left(\left<e^2\right>-\left<e\right>^2\right) = .\,.\,. 
              = \beta^2 L^D W_{\text{o}} (1-W_{\text{o}}) (\hat{e}_{\text{o}} - \hat{e}_{\text{d}})^2 \nonumber\\
&=\beta^2 L^D W_{\text{o}} (1-W_{\text{o}}) \Delta \hat{e}^2.
\end{flalign}
This is referred to as the \textit{volume scaling law}. 
The probability of being in an ordered/disordered state is given by
\begin{equation}
p_{\text{o}} \propto e^{-\beta L^D \hat{f}_{\text{o}}},\qquad p_{\text{d}} \propto e^{-\beta L^D \hat{f}_{\text{d}}}.
\end{equation}
Since spontaneous symmetry breaking gives rise to a set of $q$ distinct Gibbs states with identical free energy contributions and likelihood,
the total probability weight of the ordered phase is proportional to $q p_o$~\cite{Janke14}. One can then write
\begin{equation}
\frac{W_{\text{o}}}{W_{\text{d}}} \simeq    \frac{q \,e^{-\beta L^D \hat{f}_{\text{o}}}}{e^{-\beta L^D \hat{f}_{\text{d}}}} 
                                            \quad\longrightarrow\quad \log \frac{W_o}{W_d} \simeq \log\, q + 
                                            L^D \beta \left(\hat{f}_{\text{d}} - \hat{f}_{\text{o}}\right).
\end{equation}
The specific heat maximum is obtained when $W_o = W_d$, which occurs at a size-dependent temperature $T(L)\neq T_c(\infty)$. 
Expanding around $\beta^{\infty}$, it is possible to obtain the finite-size scaling law for first-order transitions:
\begin{equation}
\beta_{C_V^{max}}(L) = \beta^{\infty} - \frac{\log\, q}{L^D \Delta \hat{e}} + \,.\,.\,.
\label{1storderscaling}
\end{equation}
This equation is correct up to higher-order terms in $L$, which become more relevant at small lattice sizes.

The dip between the two peaks in the energy histogram is caused by the suppression of states that live between the
ordered and disordered phases. These can occur as mixed-phase configurations and are characterized by the existence of 
domain walls between the coexisting phases. The presence of such interfaces entails an additional 
free-energy cost proportional to the interfacial tension $\sigma$, which corresponds to the free energy density per surface area 
required for the coexistence of competing ordering behaviors~\cite{Janke03}. 
For a system of size $L^D$, the total domain wall free energy cost scales as $\Delta F \sim \sigma L^{D-1}$. 
Compared to purely ordered and disordered states, this free energy cost causes an exponential suppression of 
mixed-phase configurations by an additional Boltzmann factor $e^{-\beta\sigma L^{D-1}}$. 
This suppression captures the leading finite-size contribution governing the sharpness 
of the double-peak structure in the energy distribution.

\section{Subsystem Symmetries in classical Physics}~\label{sec:subsims}

In the previous section we emphasized the role of symmetries in organizing phases of matter and 
how their breaking occurs in the form of phase transitions. While only transitions by means of spontaneous global symmetry 
breaking were mentioned, there is another class of symmetries denoted as local symmetries.
Most phases of matter related to symmetry breaking mechanisms lie
within the framework of global symmetries, where the symmetry operator has support across the entire D-dimensional system. 
This applies to the majority of lattice models discussed in condensed matter literature, 
which are commonly characterized by the presence of a symmetry-broken phase associated with 
different combinations of $\mathbb{Z}_2$, $U(1)$ and $SU(2)$ global symmetries. 
Well-known examples (both classical and quantum) include the nearest-neighbor Ising models in D-dimensions~\cite{Onsager44,Baxter82,Budrikis24}, 
2D and 3D Heisenberg models~\cite{Kosterlitz73,Manousakis91,Auerbach94,Sachdev11}, and lattice models with mobile degrees of freedom such 
as the Bose- and Fermi-Hubbard models~\cite{Fisher89,Hubbard63,Greiner02,Bloch08}.

On the other hand, it is possible to define theories characterized by local invariances, where the 
symmetry transformations act independently on a bounded subset of the system or at each point in space.
In such theories, the group of allowed local transformations is infinite-dimensional in continuum field
theories and consists of independent group elements associated with each site in lattice models.
Consequently, the size of the symmetry group scales proportionally with the number of degrees of freedom
of the system.

However, these local invariances do not correspond to physically observable changes of the system
under symmetry transformations. Instead, they reflect a redundancy in the description of the theory which
can be described in terms of a locally defined \textit{gauge} field, which ensures invariance under
local transformations. While a global symmetry gives rise to distinct ground states characterized by
different observable properties, applying a gauge transformation allowed by a local symmetry merely changes
the representation of the state without affecting any physical observable. Rather than physical symmetries,
local symmetries should be regarded as mathematical redundancies that, as proven by Elitzur's theorem, 
cannot be spontaneously broken~\cite{Elitzur75}.

Between these two limits lies the class of \emph{subsystem} (or \emph{subdimensional}) symmetries, 
which interpolate between global and local symmetries.
A subsystem symmetry acts nontrivially on an extensively large but lower-dimensional subset of the system. 
In a D-dimensional system composed of a set of degrees of freedom $\Lambda$, the minimal non-empty support of 
an operator associated with a subsystem symmetry transformation spans a $d$-dimensional submanifold $\mathcal C\subset\Gamma$, with $d$<D.
Because such symmetries are neither fully local (gauge redundancies) nor fully global, 
they may in general be spontaneously broken.

Each independent subsystem symmetry introduces a degeneracy in the energy spectrum. Assuming that the number of such symmetries
scales with the linear system size $L$, one can identify $R\propto L^{D-d}$ distinct subspaces $\mathcal C_1 ,...,\mathcal C_R$ 
embedded in $\Lambda$, each invariant under the action of the symmetry group $G_l$.

The full symmetry group can then be factorized as 
\begin{equation}
\mathcal{G}= \bigotimes_{l\in\{1,...,R\}} G_l, 
\end{equation}
and a generic symmetry operation can be written as a composition of subdimensional transformations
$g_{\mathcal C_l} \in G_l$ acting on each subsystem:
\begin{flalign}
g = g_{\mathcal C_1} g_{\mathcal C_2} \cdots g_{\mathcal C_R} \,, \qquad g \in \mathcal G .
\end{flalign}
If each symmetry group $G_l$ gives rise to an $m$-fold degeneracy, the total ground-state degeneracy
scales exponentially but subextensively as $m^{L^{D-d}}$~\cite{Nussinov15}.

\subsubsection*{Dimensional reduction}
In the presence of a subsystem symmetry, the ordering mechanism of the system can be related back to 
the one displayed by a single $d-$dimensional subsystem upon which a single subdimensional 
symmetry transformation can act upon~\cite{Batista05}. 

To see this, consider a local observable $f(\mathbf{S})$, defined on the original D-dimensional discrete
field comprised of the set of degrees of freedom $S_i$ whose behavior is described by the classical Hamiltonian $H$. 
Assume that no individual spin $S_i$ contained in the support $\mathcal C_l$ is invariant under the action of the 
$d-$dimensional gauge symmetry group $G_l$. To see whether the observable $f(\mathbf{S})$ showcases symmetry breaking in the thermodynamic limit, 
it is necessary to compute its equilibrium expectation value:
\begin{flalign}
\left<f(\mathbf{S})\right>  &=  \lim_{\substack {N\rightarrow \infty\\ h\,\rightarrow \,0}}
                                        \left<f(\mathbf{S})\right>_{h,N} =\lim_{\substack {N\rightarrow \infty\\ h\,\rightarrow \,0}}
                                        \frac{\sum_{\{S_1,\dots S_N\}}f(\mathbf{S}) e^{-\beta \left(H(\mathbf{S}) + h \sum_{\mathbf{i}} S_i\right)} } 
                                        {\sum_{\{S_1,\dots S_N\}} e^{-\beta \left(H(\mathbf{S}) + h \sum_{\mathbf{i}} S_i\right)} }\,\,,
\end{flalign}
where a symmetry breaking field $h$ was introduced. The sum runs over all possible configurations $\{S_1,\dots S_N\} = \{S_i\}$.

To isolate the contributions coming from the subset $\mathcal C_j$ corresponding to the support of the subsystem symmetry group $G_j$, 
we relabel the degrees of freedom $S_i$ as $\eta_i$ if $i \in \mathcal C_j$ and $\psi_i$ if $i \notin \mathcal C_j$. 
Since $f(\mathbf{S})$ is taken to be an order parameter of the subsystem symmetry, it is expected to identify the emergence of symmetry-breaking order
within the respective submanifold. Thus, any such order parameter will depend only on the spins in $\mathcal C_j$ and can be written as $f(\mathbf{S})\equiv f(\eta)$. 
Decomposing the set of possible configurations $\{S_i\}$ into $\{\eta_i\} \equiv \eta $ and $\{\psi_i\} \equiv \psi $, 
one can factor out the contribution of $\{\eta_i\}$ from the partition function by defining
\begin{equation}
Z_{\{\psi_i\}} = \sum_{\{\eta_i\}} e^{-\beta \left( H(\psi,\eta) + h \sum_{i \in \mathcal C_j} \eta_i \right) }.
\end{equation}
The equation above can then be rewritten as
\begin{equation}
\langle f(\mathbf{S})\rangle  =  \lim_{\substack {N\rightarrow \infty \\ h\,\rightarrow \,0}} 
                                    \frac{ \sum_{\{\psi_i\}} Z_{\{\psi_i\}} 
                                    e^{-\beta h\sum_{i\notin \mathcal C_j} \psi_i}  
                                    \left(\frac{1}{Z_{\{\psi_i\}}} \sum_{\{\eta_{i}\}} 
                                    f(\eta) e^{ -\beta \left( H(\psi,\eta) +  h \sum_{i \in \mathcal C_j} \eta_{i} \right) }\right) }
                                    { \sum_{\{\psi_i\}} Z_{\{\psi_i\}} e^{-\beta h\sum_{i\notin \mathcal C_j} \psi_i} }\,.
\end{equation}
Identifying $\{\bar{\psi}\}$ as the configuration of spins outside of $\mathcal C_j$ that maximizes the absolute value of the expression in the brackets, 
we obtain the bound
\begin{equation}
\left|\left<f(\mathbf{S})\right>_{h,N}\right|\leq   \left|\frac{1}{Z_{\{\bar{\psi}_i \}}} \sum_{\{\eta_i\}} f(\eta) 
                                                    e^{ -\beta \left( H(\bar{\psi},\eta) + h \sum_{i \in \mathcal C_j} \eta_i  \right)  }\right|.
\label{GenElitzur}
\end{equation}

The absolute expectation value of the observable is bounded by the absolute value of the same observable evaluated with respect to an effective 
$d$-dimensional Hamiltonian $H(\bar{\psi}_{\mathbf i},\eta)\equiv \bar{ H}(\eta)$, obtained by fixing the configuration $\bar{\psi}$
outside of $\mathcal C_j$. The reduced Hamiltonian is \textit{globally} invariant under the symmetry group $G_j$. 
Since the contributions to the Hamiltonian involving only degrees of freedom within $\mathcal C_j$ are left unchanged, the range of 
the interactions between $\eta$ fields remains unchanged in this subdimensional model, with no additional or longer-range interactions emerging.

The inequality underlines a phenomenon referred to as \textit{dimensional reduction}: the symmetry-breaking properties of the subsystem symmetry
in the full D-dimensional system are bounded by those of an effective $d-$dimensional model comprising the spins involved in the subsystem symmetry. 
A fundamental consequence of this conclusion is that, in the absence of larger-dimensional symmetries, the symmetry-breaking properties of the full 
D-dimensional system are constrained by those of the corresponding $d-$dimensional subsystems $\mathcal C_j$. Therefore, symmetry breaking
in the D-dimensional system can only occur if the $d$-dimensional counterpart were to allow it, as interactions between $\mathcal C_j$ and the
rest of the system cannot enable symmetry breaking of the subdimensional symmetries~\cite{Nussinov15}. 
However, this no longer holds if, on top of the subsystem symmetry, the system also displays a global symmetry 
or another subsystem symmetry of larger dimension. In such cases, interactions between the different submanifolds $\mathcal C_j$ will no longer be 
irrelevant compared to those between degrees of freedom contained in the submanifold, as they may contribute to the breaking of the additional symmetry.
In this case, the breaking of the lower-dimensional subsystem symmetries generally requires the simultaneous or prior breaking of the higher-dimensional
symmetries, which couple the different submanifolds $\mathcal C_j$~\cite{Canossa23}. Most importantly, these information does not allow us to infer
the nature of the phase transition in the D-dimensional model by simply studying the transition associated with the breaking of the same symmetry in the 
corresponding $d$-dimensional model.

\subsubsection*{Non-standard first-order scaling}
As mentioned above, the presence of subdimensional symmetries is directly linked to a sub-extensive degeneracy of the ground-state manifold. 
Following the theory of finite-size scaling previously discussed, it is possible to predict how this degeneracy may affect 
the scaling behavior of first-order transitions by plugging the full degeneracy in Eq.~\eqref{1storderscaling}: 
the most dominant contribution to the scaling behavior is then given by:
\begin{flalign}\label{eq:nonstandardscalingderiv}
\beta_{c}(L) = \beta^{\infty} - \frac{\log\, 2^{3L^{D-d}}}{L^D \Delta \hat{e}} + \,.\,.\,. = \beta^{\infty} - \frac{ 3\, \log\,2}{L^{d} \Delta \hat{e}} + \,.\,.\,. 
\end{flalign}
where $D$ is the total dimension of the system and $d$ is the dimension of the submanifold involved in a subsystem symmetry transformation.
This change in the scaling properties of first-order phase transitions assumes that, in presence of an individual transition point, 
all subsystem symmetries are broken simultaneously, leading to a unique ordered phase with extensive degeneracy.

These features have already been explored extensively in the context of compass models~\cite{Canossa23,Nussinov15}. 
However, explicit demonstrations in classical Ising models with subsystem symmetries remain comparatively scarce, with 
only the 3D plaquette Ising model being a well-understood example of such systems~\cite{Johnston17}.
This prompted an investigation of the subsystem ordering, subextensive degeneracy, and modified scaling behavior 
across other subsystem-symmetric 3D Ising models to test whether a unifying framework of $\mathbb{Z}_2$ subsystem symmetry breaking can be identified. 

With this in mind, we will use the Tetrahedral Ising model and the Fractal Ising model as further representative cases for the remainder of the chapter.

\section{$\mathbb{Z}_2$ Subsystem Symmetries in Classical Ising Models}\label{sec:subsym_models}

\subsection{Tetrahedral Ising Model}\label{sec:TIM}
The Tetrahedral Ising model (TIM) is defined on a 3D face-centered cubic (FCC) lattice with one Ising spin $S_i = \pm 1$  
on every lattice site. Considering a system of even linear size $L \in 2\mathbb{Z}$ along each spatial direction, the FCC lattice 
can be decomposed into four interpenetrating cubic sublattices: this can be seen by parametrizing the lattice 
sites as $\mathbf{r} = \mathbf{v} + \mathbf{a}$, where $\mathbf{v} \in (2\mathbb{Z})^3$ and 
$\mathbf{a} \in \{ (0,0,0), (1,1,0), (1,0,1), (0,1,1) \}$ distinguishes the four basis vectors.
The set of all such $\mathbf{r}$ generates the full FCC lattice, with a total of $\frac{1}{2}L^3$ sites.

The Hamiltonian is given by a sum of four-spin interaction terms:
\begin{align}\label{eq:TIM_model}
	H_{\mathrm{TIM}} =  -\sum_{{\mathbf{v}}, {\bf a}^\prime} \left(J_+ \prod_{\bf a} S_{{(\mathbf{v} + {\bf a}^\prime )}+{\bf a}} 
                        + J_-\prod_{\bf a} S_{{(\mathbf{v} + {\bf a}^\prime )}-{\bf a}} \right),
\end{align}
where each spin is shared by four $J_+$ and four $J_-$ tetrahedral interaction terms. 
In this work, we set the interaction strengths to be $J_+ = J_-=1$.

A key feature of this model is the presence of a subextensive set of planar $\mathbb{Z}_2$ subsystem symmetries: 
$H_{\mathrm{TIM}}$ is invariant under flipping all spins of an arbitrary $XY$, $YZ$, or $XZ$-plane.
This is a consequence of the fact that, for each interaction term in Eq.~\eqref{eq:TIM_model}, there are either no or two spins residing within the same plane.

There are in total $3L$ distinct planes in the lattice; however, these are redundant when periodic boundary conditions (PBCs) are imposed.
The $XY$, $YZ$ and $XZ$ planar symmetries are generated, respectively, by the plane-flip operators $g_k^z$, $g_i^x$, and $g_j^y$, each defined as the product of all 
spin variables in the corresponding $k$th $XY$ plane, $i$th $YZ$ plane, and $j$th $XZ$ plane, that is, the sets of spins with fixed coordinates $z=k$, $x=i$, and $y=j$, respectively.
These symmetries, however, are not independent of each other, and are related by the following constraints:
\begin{subequations} \label{eq:TIM_cons}
\begin{gather}
    \prod_{\text{odd }i} g^x_i = \prod_{\text{even }j} g^y_j \cdot \prod_{\text{even }k} g^z_k, \\
    \prod_{\text{odd }j} g^y_j = \prod_{\text{even }k} g^z_k \cdot \prod_{\text{even }i} g^x_i, \\
    \prod_{\text{odd }k} g^z_k = \prod_{\text{even }j} g^x_i \cdot \prod_{\text{even }j} g^y_j.
\end{gather}
\end{subequations}
Hence, only $3L-3$ plane flip operators are independent. This implies that a symmetry-broken phase would have a subextensive number of ground states, with $\log_2{\rm GSD} = 3L-3$.
Another property that follows from the structure of the Hamiltonian is how the connectivity of the domain walls of $H_{\mathrm{TIM}}$ cannot deform arbitrarily, but are instead 
constrained: domain walls can reshape only by changing an even number of excited tetrahedra and must be neutralized within individual planes.

As two degenerate ground states differ by flipping at least $\frac{1}{2}L^2$ spins, i.e., an entire plane of the lattice, no finite-order perturbation can 
connect them in the thermodynamic limit. Therefore, a long-range order accompanying the spontaneous plane-flip symmetry breaking is allowed by dimensional-reduction arguments.
However, such ordering is fundamentally distinct from the one accompanying the breaking of a global symmetry.
As a necessary condition, the magnitude of a physical correlator or an order parameter needs to be preserved under arbitrary plane flips, which excludes the use of a 
conventional local order parameter. To capture this, we consider the following minimal correlator
\begin{equation} \label{eq:TIM_corr}
	G^z_{\mathrm{TIM}}(r) = \frac{8}{L^3} \sum_{\mathbf{v}} \mean{S_{\mathbf{v}} S_{{\mathbf{v}} + \hat{x} + \hat{y}} S_{{\mathbf{v}} + \hat{y} + r\hat{z}} S_{{\mathbf{v}} + \hat{x} + r\hat{z}}},
\end{equation}
where $\hat{x}$, $\hat{y}$, and $\hat{z}$ are unit vectors along their respective axes, and $r \in 2\mathbb{Z}+1$.

$G^z_{\mathrm{TIM}}(r)$ spans an irregular tetrahedron, with two pairs of spins living in two different $XY$-planes separated by a distance $r$.
This is associated with the existence of an order parameter $Q^z_{\mathrm{TIM}}$, constructed so that in the limit $r \rightarrow \infty$, 
$G^z_{\mathrm{TIM}}(r) \sim \left(Q^z_{\mathrm{TIM}}\right)^2$, leading to the following order parameter:
\begin{flalign}\label{eq:TIM_op}
	Q^z_{\mathrm{TIM}} &= \frac{4}{L^3} \sum_{x, y} \mean{q^z_{\mathrm{TIM}}} = \frac{4}{L^3} \sum_{x, y}
	 \mean{\,\left|  \sum_{z} S_{\mathbf{v}} S_{{\mathbf{v}} + \hat{x} + \hat{y}} + S_{{\mathbf{v}} + \hat{y} + \hat{z}} S_{{\mathbf{v}} + \hat{x} + \hat{z}} \right|\,}.
\end{flalign}
The factor $q^z_{\mathrm{TIM}}$ can be viewed as an extended or semi-local ordering moment in an $XY$ plane that represents 
the total moment of the local correlators $S_{\mathbf{v}} S_{{\mathbf{v}} + \hat{x} + \hat{y}}$ and $S_{{\mathbf{v}} + \hat{y} + \hat{z}} S_{{\mathbf{v}} + \hat{x} + \hat{z}}$ 
along an entire $z$-line of FCC unit cells. Despite being defined as a sum over $z$ planes, each of its components is defined on a 1D (linear) support, meaning that $q^z_{\mathrm{TIM}}$
has characteristic dimension $\dim(q^z_{\mathrm{TIM}}) = 1$ and a codimension ${\rm codim}(q^z_{\mathrm{TIM}}) = D-\dim(q^z_{\mathrm{TIM}}) = 2$. 
The fact that the order parameter has nonzero spatial extent and codimension strictly smaller than D reflects the distinction between subsystem symmetries, where pointlike operators 
would transform nontrivially under a plane flip, and global symmetries. This allows us to identify $Q^z_{\mathrm{TIM}}$ as a \textit{sub-dimensional order parameter}, distinguishing it
from conventional global symmetries, which admit the presence of pointlike local order parameters.

One can also use the same logic to construct two other order parameters $Q^x_{\mathrm{TIM}}$ and $Q^y_{\mathrm{TIM}}$: however, upon symmetry breaking, 
the three-fold rotation symmetry $x\rightarrow y \rightarrow z$ in $H_{\mathrm{TIM}}$ remains unbroken due to the constraints in Eq.~\eqref{eq:TIM_cons}, 
making the three order parameters equivalent. This marks a crucial distinction between subsystem-symmetric Ising spin models and constraint-free or other 
sub-dimensional symmetric models, where such order parameters may spontaneously become anisotropic and select a preferred orientation through 
order-by-disorder mechanisms~\cite{Canossa23}.

\begin{figure}
  \centering
  \includegraphics[width=1.\textwidth]{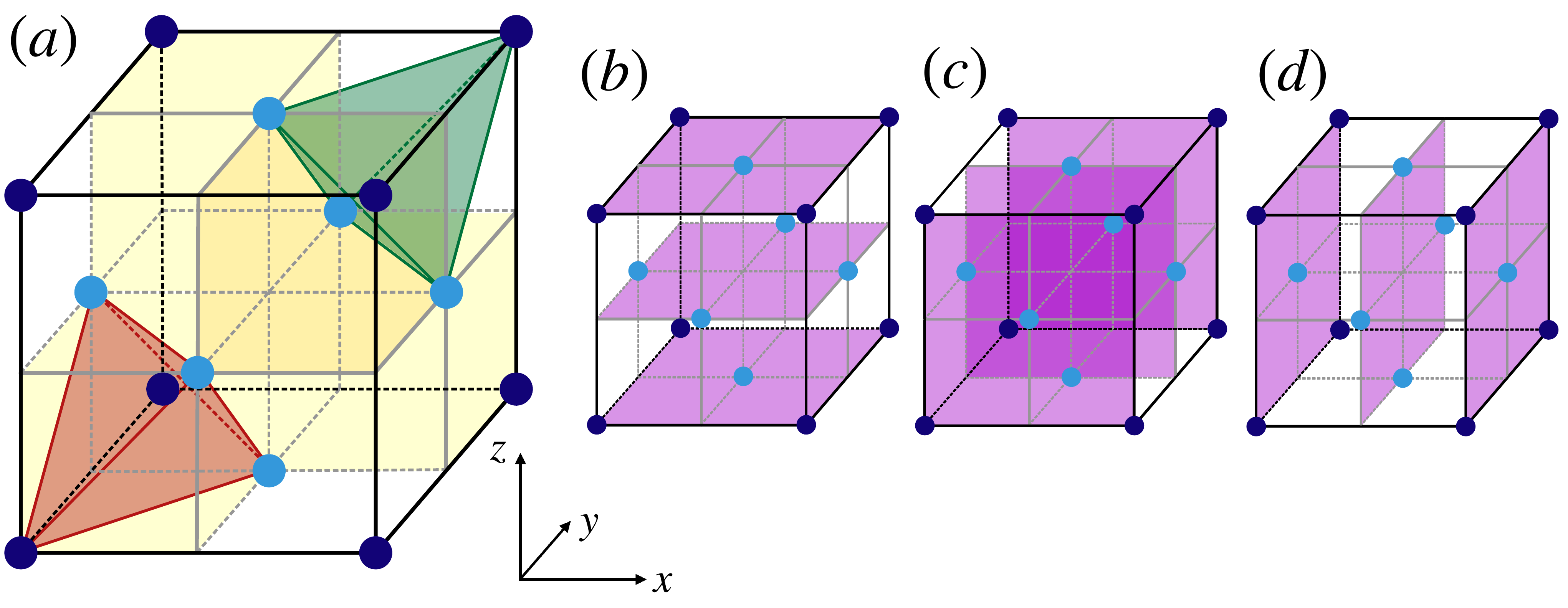}
  \caption[Illustration of the FCC unit cell of the Tetrahedral Ising model.]
        { Illustration of the FCC unit cell of the Tetrahedral Ising model. 
            Blue and cyan circles represent vertices ${\mathbf{v}} = (x, y, z) \in (2\mathbb{Z})^3$ and face centers of an FCC lattice, respectively. 
            The red and green tetrahedra show an example of the $\prod_{\bf a} S_{{\mathbf{v}}+{\bf a}}$ and $\prod_{\bf a} S_{{\mathbf{v}}-{\bf a}}$ 
            interaction terms, respectively, where ${\bf a} \in \left\{ (0,0,0), \, (1,1,0), \, (1,0,1), \, (0,1,1) \right\}$ labels the four FCC sublattices. 
            The entire lattice can be intuitively visualized with small cubes: Each shaded (empty) cube contains a single red (green) tetrahedron interaction.
            (b–d) Representative planar subsystem symmetries acting on $XY$, $YZ$ and $XZ$ planes respectively. 
            In each case, spins within a given plane are flipped collectively while preserving all tetrahedral interaction terms.
          }
  \label{fig:TIM_model}
\end{figure}

\subsection{Fractal Ising Model} \label{sec:FIM}
In the Fractal Ising model (FIM), Ising spins are placed at the vertices of a cubic lattice, with vertex positions represented by ${\mathbf{v}} = \{x, y, z\}\in \mathbb{Z}^3$.
In addition, we define two finite sets of vectors ${\bf a}_1 \in \left\{(0,0,0), \, (1,0,0), \, (0,1,0), \, (0,0,1)\right\}$ and 
${\bf a}_2 \in \left\{(0,0,0), \, (1,1,0), \, (1,0,1), \, (0,1,1)\right\}$ to label an arbitrary spin $S_{\mathbf{v}}$ and three of 
its nearest and next-nearest neighbors, respectively.

The Hamiltonian consists of two four-body interactions on the tetrahedra specified by $\{{\bf a}_1\}$ and $\{{\bf a}_2\}$, as depicted in Fig.~\ref{fig:TIM_model},
\begin{align} \label{eq:FIM_model}
	H_{\rm FIM} = -\sum_{{\mathbf{v}}} \left(J_1 \prod_{{\bf a}_1} S_{{\mathbf{v}}+{\bf a}_1} + J_2\prod_{{\bf a}_2} S_{{\mathbf{v}}+{\bf a}_2} \right),
\end{align}
where each $S_{\mathbf{v}}$ participates in four $J_1$ tetrahedra and four $J_2$ tetrahedra. Like in the previous model, we take $J_1 = J_2 = 1$.
Under PBC, $H_{\rm FIM}$ can exhibit a fractal subsystem symmetry. 
While there is no compact algebraic expression for its fractal symmetry generators and the full ground-state manifold, these can be constructed 
using the polynomial ring formalism introduced in Ref.~\cite{Haah13} to describe translationally invariant systems.

We start by defining the group ring $R=\mathbb{Z}_{2}\left[\Lambda\right]$, where $\Lambda=\{x^{i}y^{j}z^{k}|i,j,k\in\mathbb{Z}_{L}\}$  represents the group
of lattice translations on a lattice of size $L$. The coordinates of each vertex are represented in a multiplicative notation, and PBCs are implemented by
the identifications $x^{L}=y^{L}=z^{L}=1$. The ring $R$ is a set consisting of all polynomials of the form 
\begin{equation}
f =\sum_{(i,j,k) \in\Lambda} a_{ijk}x^{i}y^{j}z^{k}.
\end{equation}
where $a_{ijk}\in\{0,1\}\subset\mathbb{Z}_2$. The polynomial $f \in R$ represents the subset of lattice vertices with coefficients $a_{ijk} = 1$.

The sets $\left\{\mathbf{a}_{1}\right\}$ and $\left\{ \mathbf{a}_{2}\right\}$ corresponding to the four-body interaction terms in Eq.~\eqref{eq:FIM_model}
are represented by the following two polynomials:
\begin{equation}
\varepsilon_{1}=1+x+y+z \quad \text{and} \quad\varepsilon_{2}=1+xy+yz+zx.
\end{equation}

We introduce the notation $X\left(f\right)$ to denote the operation of flipping spins on the vertices specified by $f$. 
Suppose a single spin-flip operator acts at the origin site $f=1$: $X(1)$ will give rise to a set of excitations described by the following polynomials: 
\begin{equation}\label{eq:epsilon}
\overline\varepsilon_{1}=   1+ \overline{x} + \overline{y}+\overline{z}\quad\mathrm{and}\quad\overline\varepsilon_{2} = 
                            1+ \overline{x}\overline{y}+\overline{y}\overline{z}+\overline{z}\overline{x}.
\end{equation}
These represent the spatial inversions of $\varepsilon_{1}$ and $\varepsilon_{2}$, respectively, with $\overline{x}\equiv x^{-1}$, $\overline{y}\equiv y^{-1}$,
and $\overline{z}\equiv z^{-1}$. 

From this, it is possible to identify $\varphi(f)$ as the excitation map associated with the spin-flip operator $X(f)$ defined by an arbitrary polynomial $f$:
\begin{equation}\label{eq:error}
\varphi:f \mapsto\varphi\left(f\right)= \left(\overline{\varepsilon}_{1} f, \, \overline{\varepsilon}_{2} f\right).
\end{equation}
The $R$-linearity in Eq.~\eqref{eq:error} is implied by translation symmetry. The polynomial pair $\left(\overline{\varepsilon}_{1} f, \, \overline{\varepsilon}_{2} f\right)$ 
describes the excitation pattern of the two interaction terms in $H_{\rm FIM}$. In the case $f=1$, $\varphi(1) = \left(\overline{\varepsilon}_{1}, \, \overline{\varepsilon}_{2}\right)$ 
label the locations of the excited $J_1$ and $J_2$ tetrahedra in $H_{\rm FIM}$ due to a single spin flip at $x^0 y^0 z^0$.

With this excitation map in hand, we can now derive the fractal symmetries of the model. Any symmetry is identified by spin-flip configurations $f$ that create no excitations,
meaning $\varphi\left( f \right) =0$. These correspond to the set of polynomials $f$ that constitute the kernel of the map $\varphi$,
\begin{equation}
\ker\varphi\coloneqq\left\{f \in R\;|\;\varphi\left(f\right)=0 \ {\rm mod} \ 
\left(\overline{\varepsilon}_{1}, \ \overline{\varepsilon}_{2} \right) \right\}. 
\end{equation}
This space of polynomials can be identified by finding all solutions $f\in R$ that satisfy the condition $\overline\varepsilon_1 f = \overline\varepsilon_2 f=0$.
To see what these conditions look like, let us express $f=\sum_{k=0}^{L-1} b_k z^k$ as explicit polynomials in $z$ with 
coefficients $b_k\in\mathbb{Z}_2[\bar{x},\bar{y}]/(\bar{x}^L-1,\bar{y}^L-1)$ being polynomials in $\bar{x}$ and $\bar{y}$ (taking into account the PBC constraints).
The first condition can then be written as
\begin{flalign}\label{eq:var_1f=0}
\bar{\varepsilon}_1 f   &= (\bar{z}+t)\sum_{k=1}^{L-1} b_k z^k = \sum_{k=-1}^{L-2}b_{k+1}z^k + \sum_{k=0}^{L-1}b_k t z^k \nonumber \\
                        &= \sum_{k=1}^{L-2}\left(b_{k+1} + b_k t\right) z^k + \left(b_0 + b_{L-1}t\right)z^{L-1},
\end{flalign}
where we used PBC to impose the equivalence $\hat{z}=z^{L-1}$ and we decomposed $\bar{\varepsilon}_1 = 1+\bar{x}+\bar{y}+\bar{z} = \bar{z} + t$, with $t\equiv 1+\bar{x}+\bar{y}$.

For the condition $\bar{\varepsilon}_1 f=0$ to be satisfied, all coefficients in Eq.~\eqref{eq:var_1f=0} must vanish, therefore leading to the conditions
\begin{flalign}
&b_0 = b_{L-1}t, \label{eq:cond1}\\
&b_{k+1}=b_k t \,\,\,\forall k=0,1,\dots,L-2 \label{eq:cond2}.
\end{flalign}
Setting $b_0 \equiv b$, combining these two equations allows us to rewrite the condition in Eq.~\eqref{eq:cond2} as
\begin{equation}
b_k = t^k b.
\end{equation}
Finally, the condition $\bar{\varepsilon}_1f=0$ is satisfied for all polynomials of the form
\begin{equation}\label{eq:rb}
f = b(1+tz+\left(tz\right)^{2}+\cdots+\left(tz\right)^{L-1}),
\end{equation}
with the factor $b\in\mathbb{Z}_2[\bar{x},\bar{y}]/(\bar{x}^L-1,\bar{y}^L-1)$ subject to the constraint
\begin{equation}\label{eq:bt}
b(t^{L}-1)=0.
\end{equation}

We can further refine the form of $b$ by imposing the condition $\overline\varepsilon_2 f = 0$. For convenience, we introduce 
\begin{align}\label{eq:alpha}
\overline{\alpha} & \coloneqq\left(\overline{x}+\overline{y}\right)\overline\varepsilon_1 +\overline\varepsilon_2=
    \overline{y}^{2}+\left(\overline{x}+1\right)\overline{y}+\overline{x}^{2}+\overline{x}+1.
\end{align}
to eliminate the dependence on $\overline{z}$. Provided that $\overline\varepsilon_1 f = 0$, $\overline{\alpha} f = 0$ is 
equivalent to $\overline\varepsilon_2 f = 0$. Following the same trick as before, we can explicitly formulate $b$ 
as a polynomial in $\overline{y}$ with coefficients $c_j\in\mathbb{Z}_{2}\left[\overline{x}\right]/(\overline{x}^{L}-1)$.
Following our latest formulation of $f$ in Eq.~\eqref{eq:rb}, the condition $\overline{\alpha} f =0$ reduces to $\bar{\alpha}b=0$
in the ring where $\bar{y}^L=1$. To facilitate its explicit formulation, let us temporarily remove the PBC condition $y^L=1$: 
the condition above is then equivalent to requiring the divisibility condition $\bar{\alpha}b=q(\bar{y})\left(\bar{y}^L-1\right)$, where $q(\bar{y})$ is some polynomial
in $\bar{y}$. 
Since $\deg_{\overline{y}}(\overline{\alpha}) = 2$, polynomial division implies 
that the remainder must have degree strictly smaller than $2$ in $\overline{y}$, and is therefore linear in $\overline{y}$. This leads to
\begin{equation}
\overline{\alpha}b=\left(c_{1}\overline{y}+c_{0}\right)(\overline{y}^{L}-1),
\end{equation}
where the coefficients $c_0$ and $c_1$ are polynomials in $\overline{x}$. Formally, $b$ can be expressed as
\begin{equation}\label{eq:b}
b=\frac{\left(c_{1}\overline{y}+c_{0}\right)(\overline{y}^{L}-1)}{\overline{\alpha}}.
\end{equation}

\begin{figure}[!t]
	\includegraphics[width=1.\textwidth]{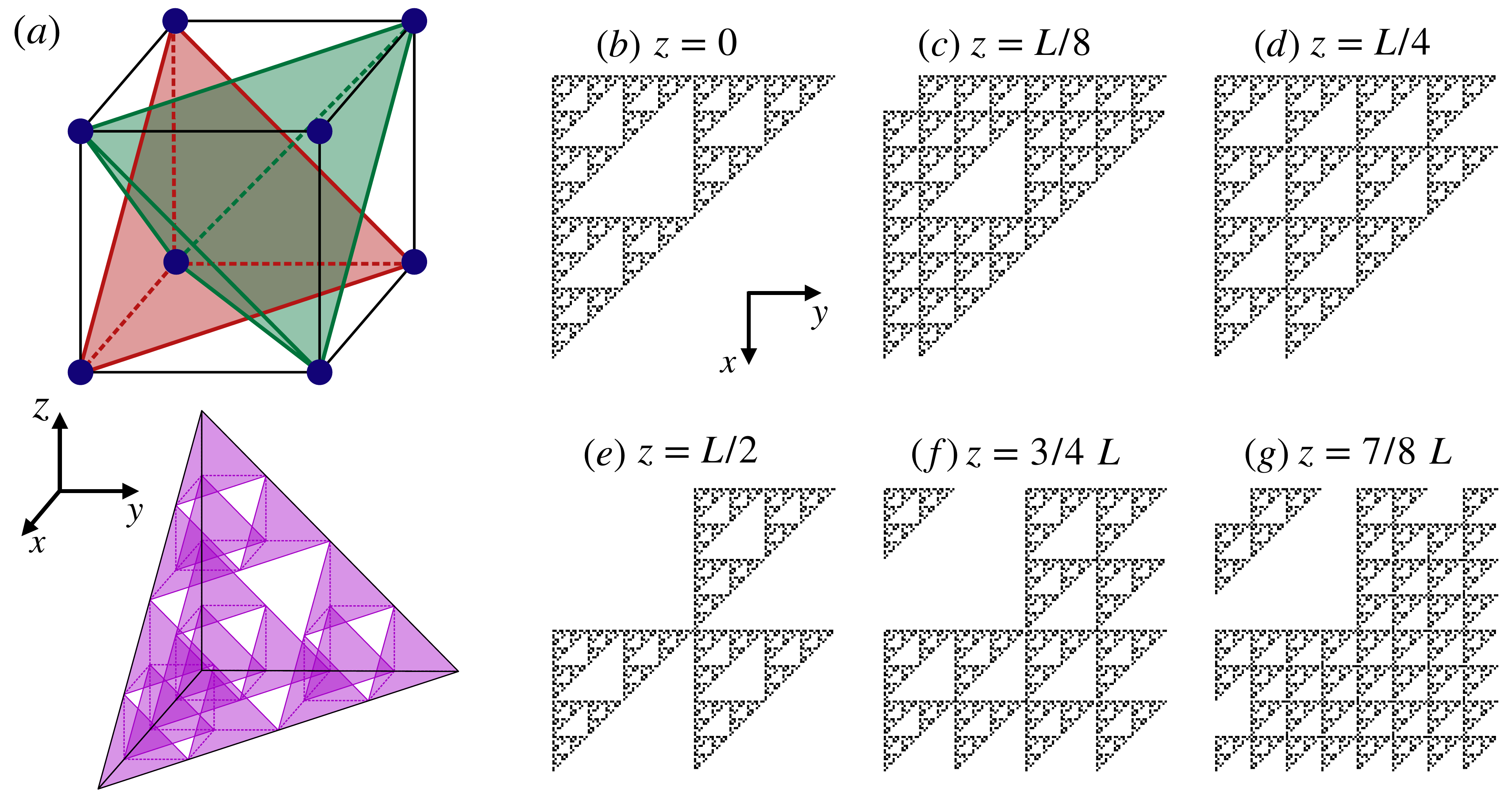}
	\caption[Illustration of the unit cell and symmetries of the Fractal Ising model.]
                {(a) Unit cell of the Fractal Ising model on a simple cubic lattice. Spins (blue spheres) reside on the lattice vertices at positions
                ${\mathbf{v}} = (x, y, z) \in \mathbb{Z}^3$. The red and green tetrahedra represent the nearest-neighboring and next-nearest-neighboring interaction 
            terms $\prod_{{\bf a}_1} s_{{\mathbf{v}}+{\bf a}_1}$ and $\prod_{{\bf a}_2} s_{{\mathbf{v}}+{\bf a}_2}$, respectively, with 
            ${\bf a}_1 \in \left\{(0,0,0), \, (1,0,0), \, (0,1,0), \, (0,0,1)\right\}$ and ${\bf a}_2 \in \left\{(0,0,0), \, (1,1,0), \, (1,0,1), \, (0,1,1)\right\}$.
                The model exhibits a subextensive number of fractal subsystem symmetries. (b-g) Pattern of spin-flips associated with a representative 
                fractal symmetry operator $X\left(f\right)$ corresponding to $\left(c_{0},c_{1}\right)=\left(1,1\right)$ in a periodic system of linear size $L=128$. 
                The flipped spins (black pixels) correspond to the non-zero terms in $f$ and are illustrated through a selection of $XY$-planes along the $z$-direction.}
	\label{fig:FIM_model}
\end{figure}

Having lifted PBC, the operation $X(b)$ flips a set of spins specified by $b$ in some $xy$-plane. Following its expression in Eq.~\eqref{eq:b}, this operation 
generates $\alpha$-type excitations at the two open boundaries $y^0$ and $y^L$, with the excitation patterns at both ends described by $c_{1}\overline{y}+c_{0}$. 
To ensure that excitations at both the $y^0$ end and the $y^L$ end get cancelled when we reinstate the $y$-PBC, the polynomial long division in Eq.~\eqref{eq:b} has 
to produce zero remainder. This allows us to identify further constraints on $c_0(\bar{x})$ and $c_1(\bar{x})$.
One can now pick up two independent polynomials in $\overline{x}$ as the $c_0$ and $c_1$ coefficients and examine if $c_1 \bar{y} + c_0$ is divisible by $\bar{\alpha}$ and 
if Eq.~\eqref{eq:bt} is satisfied. This allows us to generate all the symmetry operators $X(f)$, although some degree of ambiguity is still left.

A most distinctive feature of the Fractal Ising model is that its symmetry group strongly depends on the system size $L$. In particular, only specific $L$ 
support fractal symmetries under PBC. As a consequence, the ground-state degeneracy, which is determined by the number of spin-flip symmetries, i.e., 
${\rm GSD} = \left|\ker\varphi\right|$, is also size-dependent~\cite{Haah11,Aitchison24}. 

The model showcases fractal symmetries under PBC only for system sizes $L = 2^n$, with $n \in \mathbb{Z}^+$.
In such cases, Eq.~\eqref{eq:bt} is automatically fulfilled since $t^L = (1 + \overline{x} + \overline{y})^{2^n} = 1 + \overline{x}^{2^n} + \overline{y}^{2^n} \equiv 1$.  
The choices of $c_{0}$ and $c_{1}$ are then only subject to the requirement that $\left(c_{1}\overline{y}+c_{0}\right)(\overline{y}^{L}-1)$ is divisible by $\overline{\alpha}$. 

To determine the constraints on $c_0$ and $c_1$, let us first define, for convenience, $p=\bar{x}+1$ and $q=\bar{y}+1$: these allow us to rewrite $\bar{\alpha}$ as
\begin{equation}
\bar{\alpha} = q^2 + pq + p^2.
\end{equation}
Since, for $L=2^n$, we have $\bar{y}^{2^n}-1 = (q+1)^{2^n} - 1 =q^{2^n}+1 -1 = q^{2^n}$, $p^{2^n} = \bar{x}^{2^n}-1$ and $q^{2^n} = \bar{y}^{2^n}-1$. 
We can thus rewrite the condition of polynomial divisibility by $\bar{\alpha}$ as
\begin{equation}\label{eq:qLeq}
\left(c_1(q+1)+c_0\right)q^{2^n} \equiv 0 \quad(\text{mod}\, \bar{\alpha}).
\end{equation}
The $\text{mod} \bar{\alpha}$ implies that we are working in the quotient ring $R/(\alpha)$: this allows us to identify multiples of $\alpha$ with zero, thus imposing the relation
\begin{equation}
q^2 + pq + p^2 = 0.
\end{equation}
In Eq.~\eqref{eq:qLeq}, $q$ appears as $q^{2^n}$ and $q^{2^n+1}$: within the quotient ring $R/(\alpha)$, these can be rewritten as
\begin{flalign}
q^{2^n} &= \left(pq + p^2\right)^{2^{n-1}} = (pq)^{2^{n-1}} + (p)^{2^n} = p^{2^{n-1}}q^{2^{n-1}} + 0 = p^{2^{n-1}} \left(pq + p^2\right)^{2^{n-2}} \nonumber \\
        &= p^{2^{n-1}}\left(pq\right)^{2^{n-2}} = p^{2^{n-1}+2^{n-2}}q^{2^{n-2}} = p^{2^{n-1}+2^{n-2}+2^{n-3}}q^{2^{n-3}} \nonumber\\
        &= \dots = p^{\sum_{i=1}^n 2^{n-i} } q = p^{2^n-1}q \quad(\text{mod}\,\bar{\alpha}),         \\
q^{2^n+1} &= p^{2^n-1}q^2 = p^{2^n-1} (pq + p^2) = p^{2^n}(q+p) = 0 \quad(\text{mod}\,\bar{\alpha}).
\end{flalign}
Equation~\ref{eq:qLeq} then reduces to
\begin{equation}\label{eq:secondtolast}
\left(c_1+c_0\right)p^{2^n-1}q=0\quad(\text{mod}\,\bar{\alpha}).
\end{equation}
Since $c_0$ and $c_1$ can be written as polynomials in $\bar{x}$ or, equivalently, $p$, we can expand $c_0+c_1\equiv a_0 + pB(p)$, where $B(p)$ is some polynomial in $p$.
For the condition above to be satisfied, $c_0+c_1$ must be divisible by $p$, meaning $a_0=0$. 

Expanding $p=\bar{x}+1$, we then see that $\left(c_0+c_1\right) = (\bar{x}+1)B(\bar{x}+1)$ is divisible by $\bar{x}+1$, meaning that the evaluation of $c_0+c_1$ at $x=-1$
is $0$. Recalling that the coefficients of all polynomials living in $R$ are defined over $\mathbb{Z}_2$, this means that 
\begin{equation}
c_0(-1)+c_1(-1)=c_0(1)+c_1(1)= \sum_{i=0}^{L-1} a_i (1)^i = \sum_{i=0}^{L-1} a_i = 0 \,\,\text{mod} 2. 
\end{equation}
Thus, the condition in Eq.~\eqref{eq:secondtolast} is equivalent to requiring that $c_0+c_1$ contains an even number of monomials, i.e. $|c_0+c_1|=0 \,\,(\text{mod}\, 2)$.

By taking different combinations of $c_0(x)$ and $c_1(x)$ that satisfy this condition, one can retrieve the corresponding $b$ and the associated $X(f)$ that 
generate a variety of distinct fractal-like symmetry operators, as well as global symmetries.
For instance, $c_0 = c_1 = \sum^{L-1}_{i=0} x^i$ leads to the trivial global symmetry flipping the entire lattice.
Alternatively, we can choose $c_0 = c_1= x^0 = 1$, which leads to a fractal pattern depicted in Fig.~\ref{fig:FIM_model}.

As already stated, the GSD is determined by the number of independent spin-flip symmetries. 
For $L=2^n$, this number is given by the inequivalent choices of $c_0$ and $c_1$ subject to the constraint that $\left|c_{0}+c_{1}\right|$ is even. 
Each polynomial can be expanded as $c_i(\bar{x})=\sum_{j=0}^{L-1} a^{(i)}_j \bar{x}^j$, with $a^{(i)}_j\in\{0,1\}$. While there are $2^L$ possible 
choices for each $c_i$, giving $2^{2L}$ total pairs $(c_0,c_1)$, the parity constraint on $c_0+c_1$ removes one binary degree of freedom. 
This leads to a subextensive GSD for this class of system sizes given by $\left|\ker\varphi\right|=2^{2L-1}$.

When $L \neq 2^n$, the Fractal Ising Hamiltonian realizes different symmetry groups and degeneracies.
While it is not clear whether an analytical expression exists to describe the GSD for all values of $L$, it is possible to
identify several classes of system sizes that have well-defined ground-state degeneracy:
\begin{align}\label{eq:GSD}
\text{GSD}(H_{\mathrel{\text{FIM}}}) = \left|\ker\varphi\right| = 
\begin{cases}
2^{2L-1}, & L=2^{n},\\
2^{2L-5}, & L=4^{n}-1,\\
2^{2^m-1}, & L= 2^{m-1}(2^{n}+1), \\
2^{2^m-1}, & L= 2^{m-1}(2^{2n-1}-1), \\
\end{cases}
\end{align}
with integers $n, m \geq1$.
The first two classes in Eq.~\eqref{eq:GSD} are subextensive degeneracies due to different fractal symmetries, 
both having an exponent linear in $L$. The latter two classes have constant degeneracies for various lattices classified by $m$.
For instance, $L = 3, 5, 7, ...$ ($m=1$) have ${\rm GSD}=2$ and only a trivial global symmetry.
Instead, lattice sizes $L = 6, 10, 18, ...$ ($m=2$) give ${\rm GSD}=8$, indicating the existence of more complicated global symmetries.

To remain within the context of subsystem-symmetric phases of matter, we restrict our attention to the $L=2^n$ subfamily.
Since the set of spins transformed by these symmetry operations depend on choices of fractal generators and the overlap between
symmetry generators is not trivially definable, neither local ordering nor semi-local ordering like the one captured by $Q^z_{\mathrm{TIM}}$ 
can be defined. This puts the Fractal Ising model in stark contrast to systems with global symmetry breaking and models with subdimensional order parameters, 
which admit local, semi-local, or sub-dimensional order parameters that can be understood intuitively through the lens of Landau theory or through dimensional reduction.

The next best thing we can do is to construct a non-local correlator that reflects the fractal symmetry breaking. 
To highlight the distinction discussed above, we will identify it as a {\it fractal order parameter}.
By isotropically scaling the two interaction terms in  $H_{\rm FIM}$, we can define the correlation functions
\begin{subequations}\label{eq:FIM_corr}
\begin{align}
	G_{\rm FIM} (r) & = \frac{1}{L^3} \sum_{\mathbf{v}}\mean{S_{\mathbf{v}} S_{{\mathbf{v}} + r\hat{x}} S_{{\mathbf{v}} + r\hat{y}} S_{{\mathbf{v}} + r\hat{z}} } \\
	G'_{\rm FIM}(r)& = \frac{1}{L^3} \sum_{\mathbf{v}} \mean{S_{\mathbf{v}} S_{{\mathbf{v}} + r\hat{x} + r\hat{y}} S_{{\mathbf{v}} + r\hat{y} + r\hat{z}} S_{{\mathbf{v}} + r\hat{z} + r\hat{x}}},
\end{align}	
\end{subequations}
where $L$ belongs to a non-trivial class in Eq.~\eqref{eq:GSD} and $r=2^k$, with $k=1,2,\dots$.
Both $G_{\rm FIM}$ and $G'_{\rm FIM}$ measure long-range correlations of spins at the four corners of a tetrahedron of exponentially growing size. 
The two correlation functions proposed in Eq.~\eqref{eq:FIM_corr} are equivalent due to the presence of a mirror $\mathbb{Z}_2$ symmetry 
with respect to the plane normal to $\hat{x}-\hat{y}$, along with a $\mathbb{Z}_3$ rotational symmetry about the $\hat{x}+\hat{y}+\hat{z}$ axis.

These correlators allow us to identify long-range order when $G_{\rm FIM}(r),G'_{\rm FIM}(r) \neq 0$ at $r \rightarrow \infty$. 
Noticing that $G_{\rm FIM} = G'_{\rm FIM} = 1$ can only be achieved by ground-state configurations allows us
to use them as order parameters to describe the fractal symmetry breaking of $H_{\rm FIM}$.

\subsection{Through the lens of geometric (self-)duality} \label{sec:self-duality}

Most subsystem-symmetric phases of matter studied so far are typically characterized by an individual 
transition point separating the disordered phase and the subsystem-ordered phase. For Ising spin models, the 
location of this phase transition can be inferred by exploiting the Kramers-Wannier duality
~\cite{Kramers41a, Kramers41b} or, in its general form, the Wegner duality~\cite{Wegner71, Wegner73}.
This relation allows one to rewrite the partition function in terms of dual spin variables living on 
a shifted dual lattice, where the roles of order and disorder degrees of freedom are exchanged.

To see how this duality relation manifests, consider a model with $N_S$ Ising spins described by the following D-dimensional Hamiltonian:
\begin{equation}
H = -J \sum_{C_j} \prod_{i\in\partial C_j} S_i.
\end{equation}
Here, $C_j$ labels the $r$-dimensional coupling object (i.e., a simplex) which represents an interaction 
involving all Ising spins $S_i$ lying on its boundary $\partial C_j$ (e.g., a segment on the 2D 
nearest-neighbor Ising model or a tetrahedron in the 3D Tetrahedral Ising model), with $j=0,...,N_C$ labeling the couplings that characterize the model. 
Each of these couplings contributes to the partition function at $K \equiv \beta J$ through the aligned and anti-aligned Boltzmann factors 
$u_{\pm}(K) = e^{\pm K}$ which appear in the configurations with $\prod_{i\in\partial C_j} S_i = \pm 1$.

From an algebraic point of view, the model can be described in terms of an $N_S \times N_C$ \textit{incidence matrix} with components $\theta_{ij}\in\{0,1\}$,
which indicate whether the spin $S_i$ is involved in the coupling $C_j$, allowing us to rewrite the coupling term $C_j$ as $\prod_{i=0}^{N_S} S_i^{\theta_{ij}}$. 

Let us now introduce a new model of the form
\begin{equation}
H^* = -J^* \sum_{C^*_j} \prod_{\ell\in\partial C^*_j} S^*_\ell,       
\end{equation}
with the same number of couplings, labelled with the same index and referred to by $C^*_j$, 
but a different number of spins, labelled by $S^*_{\ell}$, and a new incidence matrix $\theta^*_{\ell j}$. 
Under Wegner's duality prescription~\cite{Wegner14}, we say that the two models are mutually dual if they satisfy the closure condition,
\begin{equation}\label{eq:closure}
\sum_j \theta_{ij}\theta^*_{\ell j} \equiv 0 (\text{mod}\,2)\quad\forall i,{\ell},
\end{equation}
and the completeness condition
\begin{equation}\label{eq:completeness}
N_{C}-\text{rank}({\theta})-\text{rank}({\theta^*}) = 0.
\end{equation}

The construction of the \textit{dual} lattice can be performed using the method described in Ref.~\cite{Wegner71}. 
With these properties in mind, we can introduce for each $C_j^*$ the dual weights as the 2-component Fourier 
transform of the original weights 
\begin{equation}\label{eq:2FT}
u^*_{\pm}(K) = \left( u_+ (K) \pm u_{-}(K) \right) /\sqrt{2}.
\end{equation}
Under this construction, the partition functions of the original and dual lattice are related by the Normal Factor Graph (NFG) 
Duality theorem~\cite{Forney11}, which, in this context, represents a generalized version of the Kramers-Wannier Duality
~\cite{Kramers41a,Savit80}:
\begin{equation}\label{eq:genKW}
\mathbf{Z}\{u_{\pm}(K)\} = 2^a \,\tilde{\mathbf{Z}}\{u^*_{\pm}(K)\}
\end{equation}
where $a$ is some appropriate factor. It follows that the product $\,\mathbf{Z}\{u_{\pm}(K_1)\}\tilde{\mathbf{Z}}\{u^*_{\pm}(K_2)\}$ 
is invariant under the simultaneous transformations $u_{\pm}(K_1)\rightleftharpoons u_{\pm}^*(K_2)\,,\,\,u_{\pm}(K_2)
\rightleftharpoons u_{\pm}^*(K_1)$: the transition points $K_1,K_2$ of the two models are then given by the fixed 
point condition~\cite{Nishimori10}
\begin{equation}\label{eq:duality}
u_{\pm}(K^C_1)u_{\pm}(K^C_2)=u_{\pm}^*(K^C_1)u^*_{\pm}(K^C_2).
\end{equation}
This fixed-point condition reflects the invariance of the theory under the duality transformation and 
provides a criterion for identifying multicritical points related by duality.

We can now apply this duality to our subsystem-symmetric models. The dual of $H_{\rm TIM}$ is given by placing a dual Ising spin 
$S_{{\bf v}^\star+ {\bf a}}$ at the center of an original $J_+$ tetrahedron $\prod_{\bf a} S_{{\bf v}+{\bf a}}$.
It is convenient to work with the shaded and empty unit cubes in Fig.~\ref{fig:TIM_model}, which are 
exclusively occupied by $J_+$ and $J_-$ tetrahedra, respectively.  
The dual lattice is still an FCC lattice, with a uniform shift of $\frac{1}{2} (\hat{x}, \, \hat{y}, \hat{z})$. 
The dual vertices and face centers are labeled by 
${\bf v}^\star+ {\bf a} = {\bf v}+ {\bf a} + (\frac{1}{2}, \, \frac{1}{2}, \, \frac{1}{2})$. 
An original spin $S_{{\bf v}+ {\bf a}}$ has four neighboring filled cubes, whose centers span a dual 
$J^\star_-$ tetrahedral coupling $\prod_{\bf a} S_{{\bf v}^\star - {\bf a}}$. In the same way, the centers of 
the four empty neighboring cubes of $S_{{\bf v}+ {\bf a}}$ span a dual 
$J^\star_+$ tetrahedral coupling $\prod_{\bf a} S_{{\bf v}^\star + {\bf a}}$.
Thus, the dual Hamiltonian preserves the form of $H_{\rm TIM}$, while the filled and empty cubes 
are swapped on the dual side. We refer to this equivalence between the primal and dual model as self-duality.

The dual of $H_{\rm FIM}$ can be derived in a similar fashion. Placing the dual Ising spin 
$S_{{\bf v}^\star}$ at the center of each cubic unit cell, the dual lattice is a simple cubic lattice 
with vertices given by
$\left\{{\bf v}^\star \vert {\bf v}^\star = {\bf v} + (\frac{1}{2}, \, \frac{1}{2}, \, \frac{1}{2})\right\}$.
The two tetrahedron interactions in Eq.~\eqref{eq:FIM_model} are dual to 
$\prod_{{\bf a}_1} S_{{\bf v}^\star - {\bf a}_1}$ and $\prod_{{\bf a}_2} S_{{\bf v}^\star - {\bf a}_2}$, 
respectively, where the reverse sign of ${\bf a}_1$ and ${\bf a}_2$ merely swaps the incidence of spins in each unit cube.
Hence, the Fractal Ising model is also self-dual. 

Setting $J=1$ so that $K=\beta$, self-duality leads to the following relation between partition functions: 
\begin{align}
	& Z\left(\beta\right) \coloneqq \sum_{\{S\}} e^{-\beta H} = \sum_{\{S\}} \prod_i u(\beta, h_i) \nonumber \\
	& \qquad \Leftrightarrow \sum_{\{S^\star\}} \prod_i u^\star(\beta^\star, h^\star_i) 
                = \sum_{\{S^\star\}} e^{-\beta^\star H^\star} \eqqcolon Z\left(\beta^\star\right), 
\end{align}
where $u(\beta, h_i)=e^{-\beta h_i},u^{(\star)}(\beta^\star, h^\star_i) = e^{-\beta^{(\star)} h^{(\star)}_i}$ 
denote the Boltzmann factors of individual interaction terms $h_i,h^{(\star)}_i$ in the primal and dual model respectively, and 
$i$ is a shorthand label used to indicate said tetrahedra in $H_{\rm TIM}$ or $H_{\rm FIM}$.
Since $h^{(\star)}_i = \pm 1$ is a binary function, the primal Boltzmann factors reduce to 
$u(\beta,\pm 1)\equiv u_{\pm}(\beta)=e^{\pm \beta}$. The corresponding dual weights $u^*(\beta,\pm 1)\equiv u^*_{\pm}(\beta)$ 
are given by the 2-component Fourier transform introduced in Eq.~\eqref{eq:2FT}.

Given the self-duality of the models and provided they only have a single phase transition, $Z\left(\beta\right)$ and $Z\left(\beta^\star\right)$ 
shall experience the same singularity at a common transition point $\beta_c = \beta^\star_c$. Using Eq.~\eqref{eq:2FT}, the fixed-point
condition reduces to
\begin{align} \label{eq:weight_relation}
\frac{u_+(\beta_c)}{u_-(\beta_c)} = \frac{u^\star_+(\beta_c)}{u^\star_-(\beta_c)}.
\end{align}
The solution of Eq.~\eqref{eq:weight_relation} gives the self-dual fixed-point $\beta_c = \frac{1}{2}\ln(\sqrt{2}+1)$,
which coincides with the self-dual fixed-point of the 2D Ising model~\cite{Onsager44}. 

Beyond providing a prediction for the transition point of the models, this notion of self-duality will also be 
relevant in the following chapters of the thesis.

\section{Numerical Methods}\label{sec:NumericalMethods}

Since one is typically interested in the behavior of the system at large lattice sizes, it is nearly impossible to carry out a direct calculation 
of the partition function due to the rapidly growing number of degrees of freedom over which to sum. This issue can be circumvented by simulating 
the model of interest and making it evolve according to classical statistical physics. This section provides an overview of the main algorithms used for 
simulating the models discussed previously. Due to the difficulties associated with the equilibration of frustrated spin systems, 
it is necessary to use additional tools alongside classical Monte Carlo algorithms.

Most simulations performed primarily rely on Markov Chain Monte Carlo (MCMC) methods to sample the canonical distribution dictated by the partition function:
\begin{equation}
P(\mathbf{S}) = \frac{1}{\mathcal{Z}} e^{-\beta H(\mathbf{S})}.
\label{Px}
\end{equation}
Once this stationary distribution is achieved, observables are then estimated as timeseries averages, with statistical uncertainties obtained 
from standard binning/jackknife procedures to account for autocorrelations.

During the simulation, the system is allowed to evolve from a configuration $\mathbf{S}_t$ at time $t$ to a new configuration $\mathbf{S}_{t+1}$ 
at time $t+1$ with a transition probability $T(\mathbf{S}_t \rightarrow \mathbf{S}_{t+1})$, which takes into account 
both the proposal and acceptance probability of the new configuration conditioned on the current configuration of the system.
Since we want the sample configuration to follow a specific stationary distribution at equilibrium, the algorithm must preserve said probability distribution 
and be time-reversible, meaning that when given two adjacent configurations after thermalization, one cannot tell which step came before. 
At any time step $t$ in the simulation, the probability $P'$ of obtaining a configuration $\mathbf{S}$ at time $t+1$ is given by its probability at time $t$, 
together with the net probability influx from all other configurations $\mathbf{S}^*$:
\begin{equation}
P^{'}(\mathbf{S}_{t+1}) = P(\mathbf{S}) + \displaystyle\sum_{\mathbf{S}^*} \left(\,P(\mathbf{S}^*)T(\mathbf{S}_t\rightarrow \mathbf{S}) - P(\mathbf{S})T(\mathbf{S}\rightarrow \mathbf{S}^*) \,\right)
\end{equation}
To obtain a stationary distribution, the second term must cancel out: this leads to the detailed balance condition
\begin{equation}
P(\mathbf{S}^*)T(\mathbf{S}^*\rightarrow\mathbf{S}) = P(\mathbf{S})T(\mathbf{S}\rightarrow \mathbf{S}^*).
\label{detailedbalance}
\end{equation}
Enforcing detailed balance using Eq.~\eqref{Px} for the probabilities, the likelihood of a specific update in comparison to its reverse is given by
\begin{equation}
R_{ij}\vcentcolon = \frac{T(\mathbf{S}_i \rightarrow \mathbf{S}_j)}{T(\mathbf{S}_j \rightarrow \mathbf{S}_i)}=\frac{P(\mathbf{S}_j)}{P(\mathbf{S}_i)} = e^{-\beta(\,H(\mathbf{S}_j) - H (\mathbf{S}_i)\,)}.
\end{equation}
The efficiency of the algorithm strongly depends on how the proposal probability is defined: the acceptance probability then follows from detailed balance.

\subsection{Single-Spin Metropolis update}
Another crucial requirement for the update algorithm is ergodicity: in equilibrium, the Markov chain must be able to reach any configuration 
allowed with nonzero probability in a finite number of Monte Carlo steps. The stochastic element in Markov-chain algorithms appears in the 
process of selecting and accepting a new state. Due to the size of the configuration space, 
it is necessary to find an ergodic method that optimizes the number of accepted updates while keeping the process stochastic. 
To this end, we decided to employ local, single-spin updates, since any spin configuration can be constructed from any other by a sequence of local spin flips.
Other non-local algorithms, such as cluster algorithms, can be more efficient in the case of ferromagnetic second-order phase transitions~\cite{Loison04}, 
helping to alleviate the problem of critical slowing down. However, their applicability is rather limited, and they are generally ineffective for first-order phase transitions, 
where phase coexistence causes supercritical slowing down which cannot be cured through cluster techniques~\cite{Janke03}.
Instead, local single-spin updates are versatile and easily combined with other types of updates, 
which can be additionally employed to help reduce autocorrelation times, especially near phase transitions.
A Monte Carlo sweep consists of \(N\) single-spin flip attempts, with sites chosen at random.

Transition probabilities are implemented using the Metropolis--Hastings algorithm, which provides a general construction of 
Markov chains satisfying detailed balance with respect to the target distribution~\cite{Metropolis49,Metropolis53}.
In the Metropolis--Hastings framework, the transition probability $T(\mathbf{S}_i\rightarrow\mathbf{S}_j)$ 
is written as the product of a proposal probability $\bar T(\mathbf{S}_j|\mathbf{S}_i)$ and 
an acceptance probability $A(\mathbf{S}_i\rightarrow\mathbf{S}_j)$.
For a system with $N$ Ising degrees of freedom and single-spin flip updates,
the update proposal consists of choosing a lattice site uniformly at random, meaning that
$\bar T(\mathbf{S}_j|\mathbf{S}_i)=\bar T(\mathbf{S}_i|\mathbf{S}_j)$,
so detailed balance constrains only the ratio of acceptance probabilities.
In the Metropolis algorithm, the acceptance probability is set to be
\begin{equation}
A(\mathbf{S}_i\rightarrow\mathbf{S}_j) = \min\{1,R_{ij}\} = \min \{1,e^{-\beta(\,H(\mathbf{S}_j) - H (\mathbf{S}_i)\,)}\}.
\end{equation}
This choice satisfies the detailed-balance condition in Eq.~\eqref{detailedbalance}, ensuring that the Markov chain has the
desired Boltzmann distribution as its stationary distribution, and ensures that energetically favorable
updates are always accepted.

\subsection{Extended Ensemble Methods}
Magnetically frustrated systems are characterized by an extensive ground-state degeneracy and a rugged free-energy landscape with several local minima. 
A faithful description of the system requires the simulation to explore this complex landscape according to the probabilities determined by the partition function. 
This task represents a great computational issue when employing Monte Carlo algorithms. This is especially true at low temperatures, where the system is prone
to becoming trapped in local minima where the energy barriers for the acceptance of local updates become higher than the thermal energy scale.

As the system size increases, so does the number of metastable configurations and local minima:
the amount of time required to escape these local minima grows rapidly with the inverse temperature, leading to long autocorrelation times. In addition,
the divergence of the correlation lengths near the phase transition further enhances the autocorrelation times, leading to a critical slowing down.
This is especially true when using local updates, which are further hindered by the large correlation lengths~\cite{Brown87}.

To address these issues, extended ensemble methods have been proposed, such as multicanonical updates and simulated tempering: 
these algorithms facilitate the escape of the system from its local minimum by allowing it access to a larger region of the configuration space. 
This typically involves treating the parameters of the simulation as dynamic variables.

\subsubsection{Parallel Tempering}\label{sec:PT}
Parallel tempering (PT) approaches the problems caused by the rugged energy landscape by heating up the system so that it can avoid being trapped
in local minima~\cite{Hukushima96}.
In PT, $M$ replicas of the system are simulated simultaneously at a fixed set of inverse temperatures $\{\beta^1,\ldots,\beta^M\}$.
In the absence of replica exchange updates, these replicas evolve as independent, parallel processes. 
However, one could propose an exchange between the configurations (or, equivalently, the temperatures) of the replicas at inverse temperatures $\beta^i$ and $\beta^{i+1}$. 
Through repeated exchanges, individual configurations perform a random walk in temperature space, intermittently accessing higher temperatures where 
thermal fluctuations are strong enough to overcome otherwise prohibitive energy barriers, before returning to low temperatures.

In order for this update to be useful, one needs to define the decision function very carefully: automatically swapping configurations, 
or even fixing an acceptance rate to a specific constant value will inevitably break ergodicity. 

To address this issue, it is useful to describe PT as Monte Carlo simulation working in an extended configuration space composed of an 
ensemble of $M$ systems described by their state $\mathbf{S}^i$ and their inverse temperature $\beta^i$. 
The global partition function characterizing this extended ensemble will be given by the product of the 
partition functions of the single replicas: the probability associated with the ensemble configuration 
$\left\{(\mathbf{S}^1,\beta^1),\,...,\,(\mathbf{S}^M,\beta^M)\right\}$ is going to be the product of the distinct probabilities $P(\mathbf{S}^i,\,\beta^i)$.

A replica exchange corresponds to swapping the temperatures associated with two configurations:
\begin{equation*}
\Big\{ \Big( \substack{\mathbf{S}^1\\\beta^1} \Big) , \, ..., \, \Big( \substack{\mathbf{S}^i\\ \beta^i}\Big),\Big(\substack{\mathbf{S}^{i+1}\\ 
\beta^{i+1}}\Big),\,...,\,\Big(\substack{\mathbf{S}^M\\ \beta^M}\Big)\Big\} \longrightarrow \Big\{ \Big( \substack{\mathbf{S}^1\\\beta^1} \Big) , 
\, ..., \, \Big( \substack{\mathbf{S}^i\\ \beta^{i+1}}\Big),\Big(\substack{\mathbf{S}^{i+1}\\ \beta^{i}}\Big),\,...,\,\Big(\substack{\mathbf{S}^M\\ \beta^M}\Big)\Big\}
\end{equation*}
Imposing detailed balance in the extended ensemble yields the ratio of forward and reverse transition probabilities,
\begin{flalign}
&P\left(\,\Big( \substack{\mathbf{S}^i\\ \beta^i}\Big),\Big(\substack{\mathbf{S}^{i+1}\\ \beta^{i+1}}\Big)\,\right)\, 
        T \Big(\, \Big( \substack{\mathbf{S}^i\\ \beta^i}\Big)\substack{\textcolor{white}{\text{}}\\\small{\ce{<=>}}}
        \Big(\substack{\mathbf{S}^{i+1}\\ \beta^{i+1}}\Big) \, \Big) = P\Big(\, \Big( \substack{\mathbf{S}^i\\ \beta^{i+1}}\Big),
        \Big(\substack{\mathbf{S}^{i+1}\\ \beta^{i}}\Big)\,\Big)\, T\Big(\, \Big( \substack{\mathbf{S}^i\\ \beta^{i+1}}\Big)
        \substack{\textcolor{white}{\text{}}\\\small{\ce{<=>}}}\Big(\substack{\mathbf{S}^{i+1}\\ \beta^{i}}\Big)\, \Big) \longrightarrow \nonumber \\
\text{\textcolor{white}{}}\nonumber\\
&\longrightarrow R_{ij}=\frac{T(\beta^i \ce{<=>} \beta^{i+1})}{T(\beta^{i+1} \ce{<=>} \beta^{i})} = e^{-(\beta^{i+1}-\beta^{i})(\,H(\mathbf{S}^{i+1})-H(\mathbf{S}^i)\,)} = e^{-\Delta \beta \Delta H}
\end{flalign}
Following the Metropolis--Hastings prescription, the acceptance probability for a replica exchange is therefore chosen as
\begin{equation}
A = \min\{1,e^{-\Delta\beta\,\Delta H}\}.
\end{equation}
This choice guarantees that the extended Markov chain samples the correct equilibrium distribution and that exchanges 
are only accepted when the energy distributions of neighboring replicas overlap sufficiently.

Since its introduction, PT has been very successful when applied to a wide range of systems,
especially in 3D spin glasses~\cite{Machta09}, but while the simplicity of the algorithm makes it very applicable to
a wide range of canonical simulations, its effectiveness may vary for different models and is strongly 
dependent on the choice of the fixed temperatures at which the replicas are simulated. 
While low temperatures are more interesting, it is necessary to include high temperatures to give each 
replica the possibility to escape from any local minimum. Once the range is set, the spacing 
between adjacent temperatures and the number of replicas must be selected: spacing replicas too far 
apart in temperature space gives low acceptance rates and makes PT less effective. On the other hand, 
using too many temperatures will be both very expensive from a time perspective and disadvantageous 
for the simulation, as the extension of the configuration space offsets the utility of high acceptance 
rates, nullifying the contribution of PT.

A quantitative measure of the performance of a given temperature set is the \emph{round-trip time}, 
defined as the average time required for a replica to travel from $T_{\min}$ to $T_{\max}$ and back. 
Efficient equilibration requires frequent round trips: hence, minimizing the round-trip time is the most
natural optimization criterion.

One way to study how each replica moves in configuration space is by assigning to it an "upness" label: 
if a replica has reached $T_{\min}$ and begins its climb toward $T_{\max}$, it is assigned the upness label $f=1$, 
which it retains until it reaches $T_{\max}$. It then begins its descent toward $T_{\min}$ and is assigned the upness label $f=0$.
At every PT update, the value of the replica residing at temperature $T_i$ is stored; by calculating the average upness $f(T_i)$ one can see how often the replicas visiting that 
temperature are on their way towards $T_{max}$ or $T_{min}$. Ideally, to ensure a uniform number of round trips throughout the replicas, $\langle f(T) \rangle $ 
should be a straight line going from $\langle f(T_{min}) \rangle =1$ to $\langle f(T_{max}) \rangle =0$. Large deviations from a linear behavior indicate the presence of a bottleneck
in temperature space. A systematic optimization of the temperature grid can be achieved via the feedback procedure proposed 
in Ref.~\cite{Katzgraber06} where, starting from an initial set $\{T_i\}$, one measures the upness profile $\langle f(T) \rangle $ and 
redistributes the temperatures iteratively in order to approach a linear behavior.

While this method offers a significant advantage in simulations characterized by continuous or weak first-order transitions,
the effectiveness of PT in the presence of strong first-order phase transitions is 
significantly reduced due to the emergence of a pronounced energy gap near the transition. 
In this regime, a uniform temperature spacing becomes unreliable: either the grid fails to resolve the coexistence region 
adequately, leading to exponentially suppressed inter-phase exchanges, or an excessively dense spacing is required, 
which leads to the acceptance of almost every PT update between replicas whose configuration lie in the same phase 
and therefore severely hinders round-trip time.
For this reason, PT was not employed for the simulations discussed in the present chapter, 
where the strongly first-order character of the transitions leads to poor replica exchange efficiency. 
Nevertheless, we include this method as it will play a central role in the numerical analysis presented in Chapter~\ref{chap:fractons_in_QEC}.

\subsubsection{Multicanonical Monte Carlo}\label{sec:MUCA}

As discussed above, first-order phase transitions are characterized by a large free-energy barrier separating coexisting phases. 
This barrier leads to an exponential suppression of mixed-phase configurations with increasing system size, making tunneling 
events between phases exceedingly rare in canonical Monte Carlo simulations. As a result, simulations at fixed temperature may 
become trapped for very long times in one of the metastable phases, preventing reliable sampling of the equilibrium distribution.
Multicanonical Monte Carlo (MUCA) methods provide a controlled way to overcome this limitation by deliberately modifying the 
sampling weights in order to enhance the probability of visiting intermediate-energy configurations~\cite{Janke08}. 
The central idea is to replace the canonical Boltzmann distribution by an effective ensemble, where the energy histogram 
is approximately flat over a chosen interval, thereby enabling frequent tunneling between the ordered and disordered phases.

We employ MUCA algorithms with the aim of achieving a flat energy distribution in a sufficiently large interval of possible energy
values covering the mean energy in the ordered ($E_o$) and disordered phase ($E_d$).
To do so, we define the following generalized MUCA partition function
\begin{equation}
\label{eq:Z_MC_0}
Z_{\rm MUCA}(\beta) = \sum_{E} \rho(E) e^{-\beta E - g(\beta, E)},
\end{equation}
where $\rho(E)$ denotes the density of states at energy $E$, and $g(\beta,E)$ is an energy-dependent reweighting function 
that has yet to be determined. The role of $g(\beta,E)$ is to compensate for the strong energy dependence 
of $\rho(E)e^{-\beta E}$ near a first-order transition. In particular, a flat energy distribution over an interval 
$[E_o,E_d]$ is obtained if
\begin{equation}
e^{g(\beta,E)} \propto \rho(E)e^{-\beta E} \qquad \text{for } E \in [E_o,E_d],
\end{equation}
so that all energies in this range are sampled with comparable probability.

Since the density of states is generally unknown a priori, we first need a method to determine the MUCA weights. 
In practice, this can be done by initializing $g(\beta,E)=0$ for all energies and refining the weights through the following iterative
procedure. First, one performs a Monte Carlo run using the currently available set of weights. From the simulation, one can identify the 
energy distribution of configurations explored by the system through the averaged energy histogram
\begin{equation}
H(E) = \frac{1}{N_R}\sum_{j=1}^{N_R} h_j(E).
\end{equation}
From this, one can modify the MUCA weights with the following update formula:
\begin{equation}
g(\beta,E) \rightarrow g(\beta,E) + \ln H(E) - \big\langle \ln H(E') \big\rangle_{E'},
\end{equation}
where $\langle \cdots \rangle_{E'}$ denotes an average over the target energy interval to which the weight $g(\beta,E)$ is applied.
This update suppresses energies that were oversampled in the previous iteration and enhances those that were undersampled.

These two steps are then repeated until the resulting energy histogram becomes approximately flat over the interval $[E_o,E_d]$. 
To ensure that all intermediate states are sampled appropriately in the final MUCA distribution,
we consider $H(E)$ to have achieved an adequate flatness if
\begin{equation} 
\left|H(E)- \langle H(E') \rangle_{E^\prime} \right|/\langle H(E') \rangle_{E^\prime} \lesssim 0.1
\end{equation}
for three consecutive iterations.
At this stage, the learned reweighting function $g(\beta,E)$ produces a nearly uniform MUCA energy distribution $H_{\rm MUCA}(E)$.

Once the MUCA weights have been determined at a reference inverse temperature $\beta$ that lies near the transition,
we can use the learned weight to adjust the Boltzmann factor at any nearby temperature $\beta^\prime$ 
by setting 
\begin{equation}
g(\beta^\prime, E) = g(\beta, E) + (\beta - \beta^\prime) E.
\end{equation}
This allows for an efficient simulation within a finite temperature region around $\beta$.
Due to statistical uncertainty in the estimation of the initial weights $g(\beta, E)$,
the use of the reweighted $g(\beta^\prime, E)$ does not lead to a perfectly flat
energy distribution for $\beta^\prime \neq \beta$, but it still allows for an efficient sampling of
both ordered, disordered, and mixed-phase configurations that lie within the relevant energy interval $[E_o, E_d]$.

Since physical observables are defined with respect to the canonical ensemble, expectation values measured 
in the MUCA ensemble must be reweighted accordingly. 
For an observable $\mathcal{O}$, the canonical expectation value can be recovered through
\begin{flalign}
\big\langle \mathcal{O} e^{g(\beta,E)} \big\rangle_{\rm MUCA}
&= \frac{1}{Z_{\rm MUCA}} \sum_{E'} \mathcal{O}(E')\, \rho(E')\, e^{-\beta E'} \nonumber\\
&= \langle \mathcal{O} \rangle \frac{Z}{Z_{\rm MUCA}},
\end{flalign}
where the normalization factor is given by
\begin{equation}
\frac{Z}{Z_{\rm MUCA}} = \sum_{E'} H_{\rm MUCA}(E')\, e^{g(\beta,E')}.
\end{equation}

\begin{figure}[!t]
\centering
\includegraphics[width =.7\textwidth]{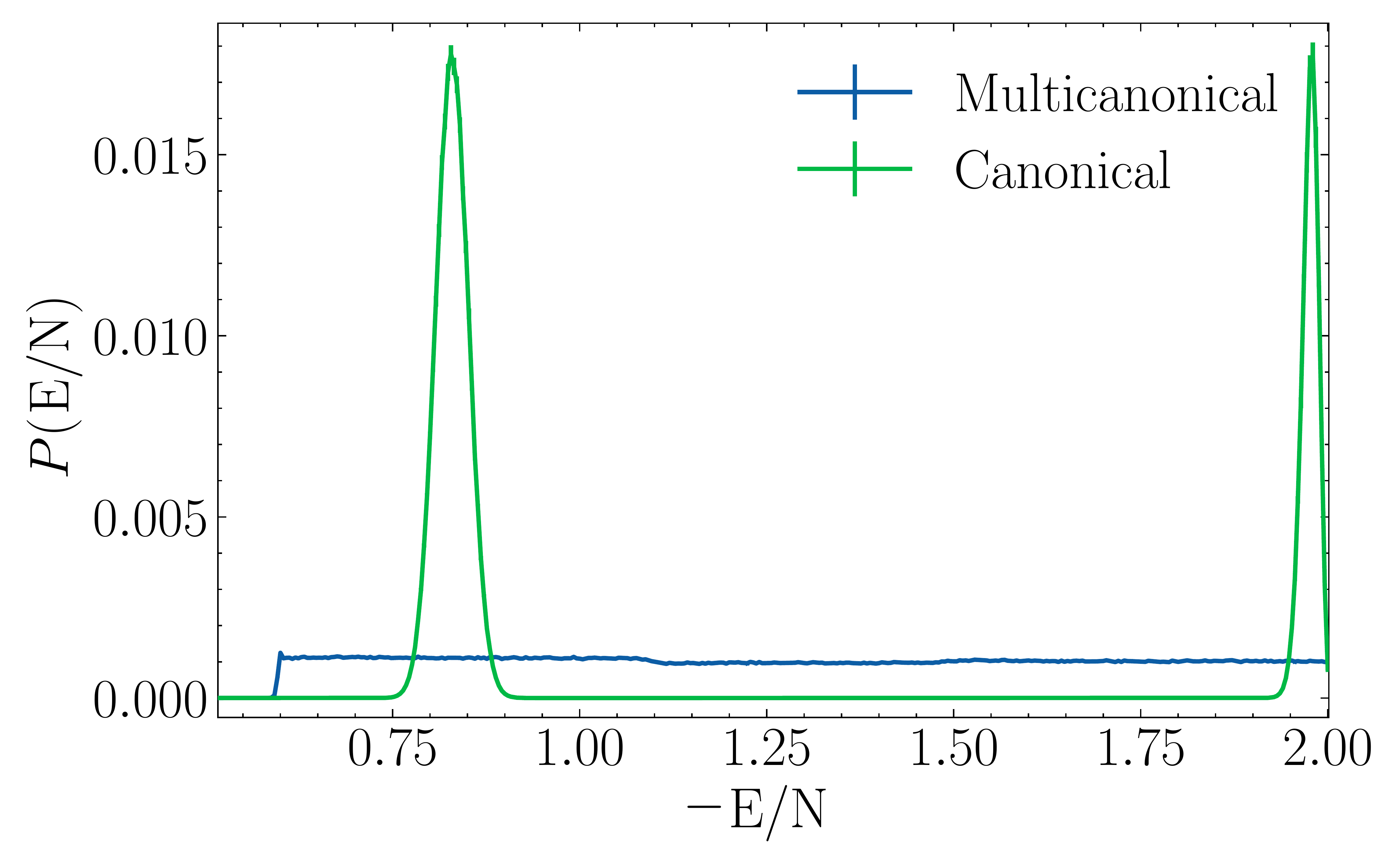}
\caption[Exemple of the target flat energy histograms at the transition point obtained through MUCA simulations.]
        {Example of the flat energy histograms at the estimated transition point obtained through the multicanonical simulation of the Tetrahedral Ising model. 
The flat distribution denotes the system's ability to visit mixed-phase configurations, thus facilitating the
exploration of both phases and overcoming supercritical slowing down. From these results, the canonical reweighted distribution can then be derived.}
\label{fig:QLDPC_Codes_map.pdf}
\end{figure}
Finally, we note that MUCA simulations explore a substantially enlarged region of configuration space compared to canonical simulations. 
As a consequence, long Monte Carlo runs are required to obtain accurate estimates of thermodynamic observables, particularly for large system sizes.
This can, however, be circumvented by running multiple replicas in parallel and then performing ensemble averages over them, allowing us to 
trivially extend the sample size. 

By overcoming the exponential suppression induced by large free-energy barriers between coexisting phases, the multicanonical method 
enables a reliable numerical study of the first-order phase transitions discussed in this chapter.

\section{Data Collection and Data Analysis}\label{sec:DCDA}

In the following section, we define the main observables that were investigated in this work and discuss the methods 
used to sample and analyze these quantities in the Monte Carlo simulations. The contents of this section
are relevant for the derivation of the results presented both in this chapter and the last chapter of the thesis.

\subsection{Statistical analysis of observables}\label{SANO}

The primary quantities obtained from Monte Carlo simulations are the expectation values $\langle \mathcal{O} \rangle$ 
and their associated variances for the energy and relevant order parameters. 
These quantities are of particular importance for estimating the heat capacity and the susceptibilities of order parameters, 
as defined in Sec~\ref{sec:1.1}. 

Due to the finite length of Monte Carlo simulations, exact expectation values cannot be obtained. 
Instead, measurements collected during the simulation are used to estimate the expectation value of an observable $O$ as
\begin{equation}
\langle O \rangle 
= \sum_{\mathbf{S}_i \in \mathcal S} P(\mathbf{S}_i)\,O(\mathbf{S}_i)
\approx 
\langle O \rangle_{\rm MC}
= \frac{1}{N_{\rm MC}} \sum_{j=1}^{N_{\rm MC}} O(\mathbf{S}_j),
\end{equation}
where $O(\mathbf{S}_j)$ denotes the value of the observable measured at the $j$-th Monte Carlo step after an initial equilibration period, 
and $N_{\rm MC}$ is the total number of recorded measurements.

The variance of the observable can be estimated analogously as
\begin{equation}
\sigma_{O}^2 
= \langle O^2 \rangle - \langle O \rangle^2
\approx 
\langle O^2 \rangle_{\rm MC} - \langle  O \rangle_{\rm MC}^2.
\label{variance}
\end{equation}
The statistical uncertainty on the estimated mean value is quantified by the standard error,
\begin{equation}
\Delta O
= \sqrt{
\frac{1}{N_{\rm MC}(N_{\rm MC}-1)}
\sum_{j=1}^{N_{\rm MC}}
\left( O_j - \langle O \rangle_{\rm MC}\right)^2 }.
\end{equation}

Since the autocorrelation time is not zero and becomes rather large near a phase transition, the errors on these quantities 
are generally underestimated due to correlations between the samples.
To account for this effect, statistical uncertainties were estimated using a binning analysis. 
The data were successively grouped into bins of increasing size, and the variance between bins was monitored. 
Once the bin size exceeds the autocorrelation time, correlations between bins become negligible, and the estimated 
error converges to the true statistical uncertainty~\cite{Bernd04}. 
This procedure also allows for a reliable estimate of the uncertainty associated with $\sigma_{ O}$ itself.

While binning provides an efficient method to estimate means, variances, and autocorrelation times without requiring storage of the full timeseries, 
it becomes less suitable for error propagation in composite observables, such as higher-order cumulants. 
In such cases, resampling techniques like the jackknife method offer a more reliable alternative.

\subsection{Computing $T_c(L)$ through interpolation}\label{Interpolation}

To perform a finite-size scaling analysis of the models, estimates of the pseudo-critical temperature $T_c(L)$ 
and the corresponding peak value of the susceptibility $\chi_{ O}^{\rm max}(L)$ are required.
These quantities were extracted from numerical data by fitting the susceptibility and heat-capacity 
curves in the vicinity of their maxima using the functional form
\begin{equation}
f(T) = a^* \exp\!\left[b\,|T-T_c^*|^{d}\right],
\end{equation}
where $a^*$, $b$, $T_c^*$, and $d$ are fitting parameters. 
The parameters $a^*$ and $T_c^*$ are interpreted as estimates of the peak height and pseudo-critical temperature, respectively.

While all these fits were compatible according to the $\rchi$ squared test, the fluctuations in the estimate of $T_c(L)$ 
obtained when adding or removing points from the fit were larger than the uncertainties derived from the fit itself
(typically 2-3 times larger). For this reason, only the estimates of $T_c(L)$ given by the best fit were used, while the
uncertainties were determined using the aforementioned fluctuations.

\subsection{Histograms and reweighting}\label{Histograms}

For each observable $O$, it is possible to construct a corresponding histogram with a finite number of bins.
Whenever a measurement of $O$ is recorded, the counter corresponding to the bin whose range contains the measured value is incremented by one. 
After normalization, the resulting histogram provides an estimate of the probability distribution of $O$ associated with the equilibrium ensemble.
The statistical uncertainty of a normalized bin is given by $\sqrt{n}/N$, where $n$ is the number of measurements in the bin and $N$ is the total number of measurements.
The shape of these histograms can be used to determine the nature of the phase transition, 
to identify which order parameter is able to describe it, and to infer whether the samples faithfully capture the canonical distribution.

Particular attention was given to energy histograms. The probability to measure a specific energy $E$ in a replica running at inverse 
temperature $\beta_0 = 1/ k_B T_0$ can be written as
\begin{equation}
P_{\beta_0}(E) \propto \rho (E) e^{-\beta_0 E},
\label{prob_density_function}
\end{equation}
where $\rho(E)$ denotes the density of states at energy $E$. 
The normalization constant is irrelevant when evaluating expectation values, since for any function $f(E)$ one has
\begin{equation}
\langle f(E) \rangle_{\beta_0}
= \frac{\sum_E f(E)\, P_{\beta_0}(E)}{\sum_E P_{\beta_0}(E)}.
\label{mean_function}
\end{equation}

From the energy histograms it is also possible to derive the density of states $\rho(E)$ associated with each energy interval.
In the case of a first-order phase transition, the presence of an energy gap manifests itself as a region of strongly 
suppressed probability between two peaks corresponding to the ordered and disordered phases. 
Analogous features can also be observed in the histogram of an appropriate order parameter.
This density of states can be used to evaluate the energy distribution at any other inverse temperature $\beta$ by simply replacing 
$\beta_0$ with $\beta$ in Eq.~\eqref{prob_density_function}.
Equivalently, the new probability distribution can be obtained through the following reweighting:
\begin{equation}
P_{\beta} \propto \rho(E) e^{-\beta E} = \rho (E) e^{-\beta_0 E} e^{-(\beta-\beta_0)E} \propto P_{\beta_0}(E)e^{-(\beta-\beta_0)E}.
\end{equation}
An energy histogram away from phase transition can be described by a Gaussian distribution centered around its expectation value. 
In a numerical simulation with a limited number of measurements, the resulting histogram will have a limited extension, with the bins 
far from the Gaussian center being nearly or completely empty, giving large statistical relative errors in the estimates of $P_{\beta_0}(E)$ 
at the tails of the histograms. 
Because of this, the reweighting procedure can accurately estimate the probability distribution for a limited range of $\beta$ close to 
$\beta_0$. As a rule of thumb, the reweighting is reliable as long as the location of the peak in the reweighted distribution is within 
$\approx 2\sigma$ from the center of the original distribution. The energy histogram reweighting technique is especially useful 
near the transition point, where the diverging susceptibilities are accompanied by a broadening of the energy distribution, giving access to a wider reweighting range.

For first-order transitions, the reweighting range near the transition temperature is further extended: since the system 
explores both phases, the histogram measured at any temperature close to $T_c$ is given by two gaussians centered at the energy 
of the ordered phase $E_o$ and disordered phase $E_d$ respectively, with the height of the two peaks depending on the distance 
from $T_c$. As long as both peaks are clearly visible, reweighting to temperatures in the region of the transition is possible.

To test whether a simulation has given reliable results, the energy histograms of different temperatures 
$T_1$ and $T_2$ can be cross-referenced by reweighting the distribution at $\beta_1$ to obtain the histogram at $\beta_2$ and vice versa. 
As long as the two temperatures lie within reweighting range and enough measurements were taken, each bin of the reweighted histograms 
is expected to lie approximately within the statistical uncertainty of the original one.

This method is especially useful to verify whether the expectation values measured near the transition point are reliable: inconsistencies 
between the reweighted histograms and the originals mean that the simulation at one or more of the compared temperature points is either not thermalized or based on too few measurements. 
By reweighting energy histograms in proximity to the heat capacity peak $T_{{C_V}^{\text{max}}}$, one can estimate the temperature at which:
\begin{itemize}
\item the histogram is at its maximum spread for second-order transitions;
\item the two peaks of the histogram in a first-order transition have the same weight.
\end{itemize}
While in second-order transitions this method does not seem to be more useful than a simple analysis of the susceptibility 
peaks, for first-order transitions it gives a reliable estimate of the transition temperature: since at the phase transition the ordered 
and disordered phases are explored with the same frequency, $\beta_c(L)$ will be the one at which the two peaks have the same weight. 
This is estimated by reweighting the histograms to the temperature $\beta^{\text{eqw}}$ at which, after splitting the distribution at 
the energy minimum lying between the peaks, the integrals of the left- and right-hand sides of the distribution have the same value.
It is also possible to use the temperature $\beta^{\text{eqh}}$ at which both peaks have the same height: given the relation between 
the behavior of the energy histogram and the heat capacity, the leading contribution to the scaling behavior 
of $\beta^{\text{eqh}}(L)$ and $\beta^{\text{eqw}}(L)$ is the same as that of $\beta_{C_V^{\text{max}}}(L)$ up to multiplicative constants~\cite{Janke08}.

To obtain a significant estimate of the transition temperatures, the aforementioned methods are employed using an average 
of the reweighted distributions taken from various temperature points in proximity of $T_c^{\text{eqh}}$ and $T_c^{\text{eqw}}$. 
The deviations are inferred by comparing these estimates to the ones that would be obtained from each separate histogram.

\subsection{Binder cumulant}
The Binder cumulant is a diagnostic tool widely used for identifying the presence and nature of a phase transition and, 
for first-order transitions, obtain an additional estimate of the transition temperaturel~\cite{Binder98}.
Given an observable $O$ that serves as an order parameter for the considered model, one can construct higher-order moments of its distribution, 
in particular $\left< O^2\right>$ and $\left< O^4\right>$. These quantities define the fourth-order Binder parameter,
\begin{equation}\label{BC}
B_{O} = 1- \frac{\left< O^4\right>}{3\left< O^2\right>^2},
\end{equation}
which is directly related to the kurtosis of the order parameter's distribution.
Effectively, $B_{O}$ analyzes the deviation from the normal distribution, where $\left< O^4\right>/\left< O^2\right>^2=3$ and $B=0$. 

In the case of a first-order phase transition, the probability distribution of $O$ becomes bimodal near the transition, 
corresponding to the coexistence of ordered and disordered phases. 
In the thermodynamic limit, this leads to a pronounced minimum of the Binder cumulant located at the transition temperature. 
For finite system sizes, the minimum is smoothed out; its depth depends on the separation and relative weights of the two peaks of the distribution, 
which become increasingly sharp and well separated as the lattice size grows. 
As a consequence, the ratio $\langle O^4 \rangle / \langle O^2 \rangle^2$ increases, leading to a deeper minimum of $B_{O}$.
The pseudo-critical temperature associated with the minimum of the Binder cumulant is known to exhibit the same leading 
finite-size scaling behavior as the susceptibility peak~\cite{Janke93}.
The pseudocritical temperatures are then estimated through polynomial fitting of the data points near the minima of the Binder parameters; 
their uncertainties are estimated from the size of the standard errors of adjacent data points and the deviation 
of $T_c(L)$ arising when adding/removing points to the fit.

For second-order transitions, the behavior of the Binder cumulant is quite different, since the distribution of any observable describing 
the transition is always composed of one peak and there is no minimum of $B_{O}$ near the transition temperature. 
However, since the kurtosis can easily be rewritten in terms of the variance as $\left< O^4\right> / \sigma_{O}^4\,$, $\,B_{O}(\beta)$ 
will not be an analytic function in the thermodynamic limit due to diverging susceptibilities~\cite{Janke03}.

Being a dimensionless quantity, the value of the Binder cumulant at the true critical point is expected to be independent of the system
size up to subleading finite-size corrections. The presence of a unique crossing in the Binder ratio can therefore be used to confirm whether 
the order parameter $O$ shows signs of a continuous phase transition and identify the critical temperature, 
while its absence could indicate the presence of a crossover rather than a continuous phase transition~\cite{Parisi98}.

\subsection{Correlation functions}\label{corrfuncdef}
The analysis of the correlation function can give useful insights into the ordering pattern displayed by the system and the behavior 
of the correlation length near the transition. Before deriving its definition for the models discussed in this thesis, we point out that
this quantity becomes relevant primarily around second-order phase transitions, where diverging correlation lengths allow us to infer 
the nature of the phase transition. This makes the correlation length less relevant for this chapter, as all models discussed were shown to 
have only strong first-order transitions and no critical scaling is to be expected. We include its definition nevertheless, as 
its analysis will become an important tool for the numerical analyses performed in Chapter~\ref{chap:fractons_in_QEC}.

Recall the definition of the disconnected correlation function:
\begin{equation}
G(\mathbf{r}) = \displaystyle\sum_{\substack{\mathbf{r}_i,\mathbf{r}_j\\\mathbf{r}_i-\mathbf{r}_j=\mathbf{r}}} 
                \left( \left<\,O (\mathbf{r}_i) O (\mathbf{r}_j)\,\right> \right).
\end{equation}
On a lattice with cubic unit cell in D dimensions, this can be decomposed into Fourier modes:
\begin{equation}
G(\mathbf{r}) = \frac{1}{L^3} \displaystyle \sum_{\mathbf{k}} \widetilde{G}(\mathbf{k}) e^{i\mathbf{k}\cdot\mathbf{r}},
\end{equation}
where $k_j=2\pi n_j / L\,$, $\,n_j\in \mathbb N$. In continuous phase transitions and very soft first-order transitions, 
the amplitudes for long-wavelength modes of the spin-spin correlation function in the disordered phase can be effectively written as~\cite{Janke08}
\begin{equation}
\widetilde{G}(\mathbf{k}) = a \left( \displaystyle \sum_{i=1}^{D} 2\left( 1-\text{cos}\,k_i\right)+ m^2\right)^{-1} 
                \substack{|\mathbf k| \rightarrow 0\\ \approx\\\text{\textcolor{white}{-}}} \frac{a}{\mathbf{k}^2 + m^2},
\label{Gk}
\end{equation}
where $a$ and $m$ are temperature-dependent parameters.
For $\left|\mathbf{r}\right|\gg 1$, the spin-spin correlation function takes the form
\begin{equation}
G(\mathbf{r}) \propto \left|\mathbf{r}\right|^{-(D-1)/2} e^{-m\left|\mathbf{r}\right|}.
\label{Grscaling}
\end{equation}
From this equation, it is possible to identify the correlation length $\xi \equiv 1/m$.
For very small $\left|\mathbf{k}\right|$, it is possible to rewrite (\ref{Gk}) as
\begin{equation}
\widetilde{G}(\mathbf{k})^{-1} = \frac{1}{a}\left(\displaystyle \sum_{i=1}^{D} 2\left(1-\text{cos}\,k_i\right) + 
                                    m^2\right) \simeq \frac{1}{a}\left(\sum_{i=1}^{D} k_i^2 + m^2\right) = c_1 \left|\mathbf{k}\right|^2 + c_0,
\label{Gkscaling}
\end{equation}
where $\,c_1 = 1/a\,,\,\, c_0=m^2/a$. Using this relation, the correlation length can be derived by calculating 
$\widetilde{G}(\mathbf{0})$ and $  \widetilde{G}(\mathbf{k}_{\text{min}})$, where $\left|\mathbf{k}_{\text{min}}\right| = 2\pi/L$.

Solving for $\xi = 1/m = \sqrt{c_1/c_0}$ gives the following solution:
\begin{equation}
\xi = \frac{L}{2\pi} \left(\frac{\widetilde{G}(\mathbf{0})}{\widetilde{G}(\mathbf{k}_{\text{min}})}-1\right)^{1/2}.
\label{corrlength}
\end{equation}
This gives us an estimate of the spin-spin second-moment correlation length.

For a second-order phase transition, $\xi/L$ is expected to converge towards a constant value when approaching the 
true critical point in the same fashion as other adimensional quantities such as the Binder's ratio. For this reason, the presence of a 
crossing point in $\xi/L$ between curves at different lattice sizes can be used to estimate $T_c$. This becomes particularly useful
in Chapter~\ref{chap:fractons_in_QEC} to infer the existence of a second-order transition where a unique crossing point is expected, a crossover where no crossing 
can occur, and a first-order transition, where crossings are neither forbidden nor expected~\cite{Binder86,Binder87,Janke03}.

\section{Numerical Results}\label{sec:nodis_results}

We now present numerical evidence for the thermal phase structure of the Tetrahedral Ising model (TIM) and the Fractal Ising model (FIM).
The objective of this analysis is to determine whether each model exhibits a single thermally driven transition separating 
a high-$T$ disordered regime from a low-$T$ subsystem-ordered regime. In addition, we want to verify whether the low-$T$ phases of 
TIM and FIM realize the planar and fractal ordering patterns predicted from their respective subsystem symmetries. 
Finally, we test whether the finite-size shift of the transition temperature follows the non-standard first-order scaling implied by the 
subextensive ground-state degeneracy, consistent with the simultaneous breaking of all subsystem symmetries.

\begin{figure}[t]
\centering
\includegraphics[width =1.\textwidth]{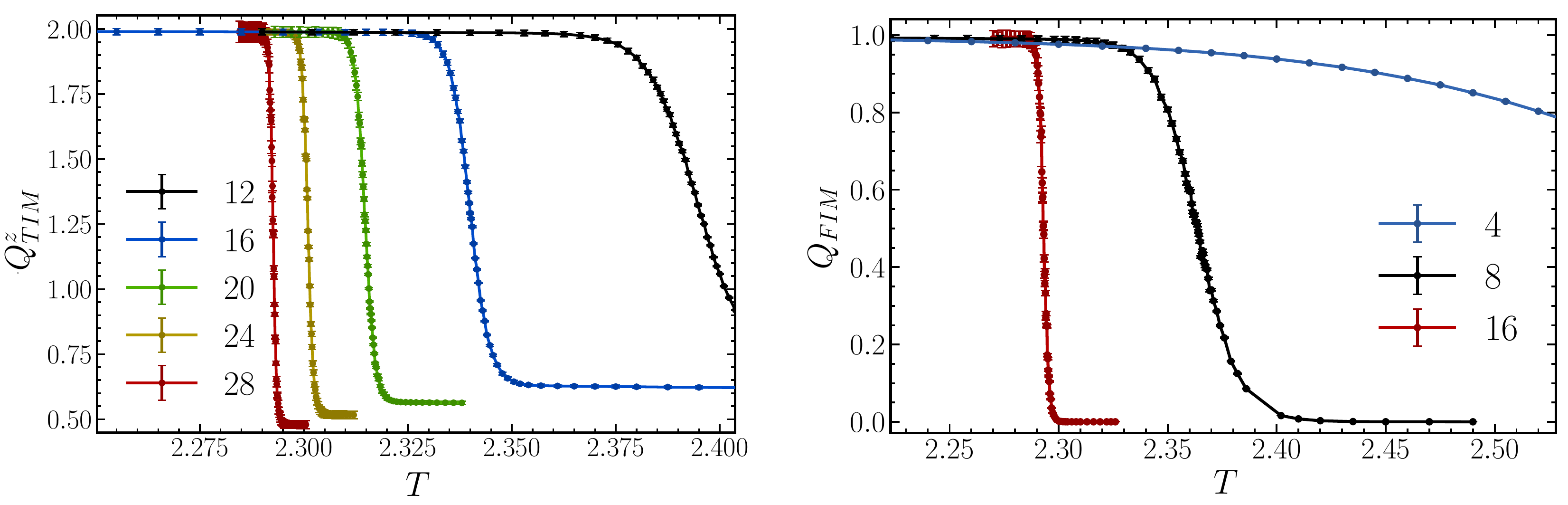}
\caption[Behavior of the order parameters of the TIM and FIM as a function of temperature for different system sizes.]
        { Behavior of the order parameters of the TIM (left) and FIM (right) as a function of temperature for different system sizes $L$. 
          In both models, the order parameter exhibits an abrupt jump to its low-temperature saturation value just below the respective transition temperature, 
          reflecting the strongly first-order character of the transition. For the FIM, saturation of the fractal order parameter is observed only for system sizes 
          belonging to the $L=2^n$ family, which support non-trivial fractal subsystem symmetries. Other lattice sizes either display only trivial global ordering 
          or remain effectively disordered within the accessible system sizes.}
 \label{fig:nodis_orderparams}
\end{figure}

Having established all the relevant features and observable quantities of our subsystem-symmetric models, we can perform our analysis using 
the numerical tools discussed in Sec.~\ref{sec:NumericalMethods}.
The behavior of each model for a system of size $N$ at a given temperature $T$ is determined by studying the energy density $e=E/N$, 
its probability distribution $P(e)$, specific heat $C_V$, susceptibility $\rchi$ of the model's order parameters as defined in Sec.~\ref{sec:1.1}, 
as well as the Binder cumulant $B$ associated with the order parameter as defined in the previous section. The order parameters used to verify the
subsystem ordering for the TIM and FIM are, respectively, $ Q^z_{\mathrm{TIM}}$ and $Q_{\rm FIM}\equiv G_{\rm FIM} (r=1)$ 
as defined in Eq.~\eqref{eq:TIM_op} and Eq.~\eqref{eq:FIM_corr}. 

The first-order character of the transition is diagnosed via the double-peak structure of $P(e)$ and is cross-checked by the Binder cumulant of the appropriate order parameter.
Finally, we extract the finite-size pseudocritical temperatures $\beta_c(L)$ using histogram reweighting and peak locations, and compare their scaling with the prediction
from Eq.~\eqref{eq:nonstandardscalingderiv}.
Through preliminary Parallel Tempering Monte Carlo simulations, it was observed that both $H_{\rm TIM}$ and $H_{\rm FIM}$ experience 
a very strong first-order phase transition already at the smallest available system sizes $L=4$. This behavior was accompanied by a 
strong suppression of metastable states, which prevented any sort of phase mixing in proximity of the presumed transition temperature,
and large hysteresis, which rendered the localization of the transition point impractical within realistic time scales. These
problems made the use of Parallel Tempering inefficient due to poor replica exchange acceptance rates and slow equilibration across 
metastable sectors. To circumvent the problem, we resorted to the multicanonical Monte Carlo algorithms described in Sec.~\ref{sec:MUCA}
by learning weights that produce a nearly flat energy distribution. Further details on the numerical simulations are reported in Appendix~\ref{app:no_disorder}.

\begin{figure}[t]
\centering
\includegraphics[width =1.\textwidth]{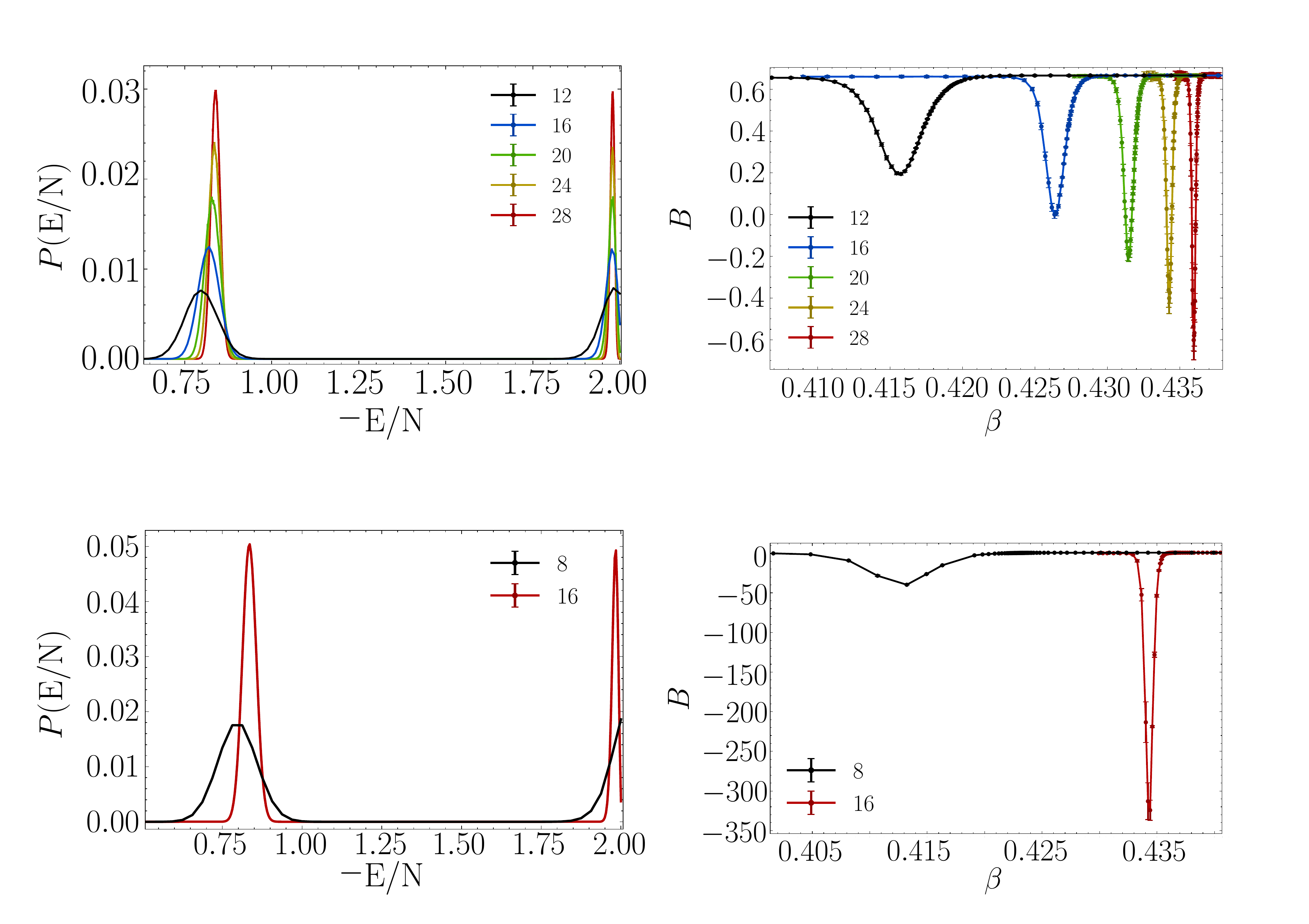}
\caption[Numerical results of the energy histograms and Binder cumulants for TIM and FIM.]
        {       Energy histograms (left column) and Binder cumulants (right column) for TIM (first row) and FIM (second row), 
                computed with multicanonical simulations in the proximity of the phase transitions. The behavior of the dips in 
                the Binder cumulants and the double peaks in the energy histograms for growing lattice sizes confirm the presence 
                of a strong first-order phase transition in both models.}
 \label{fig:TIM-FIM_plots}
\end{figure}
The use of multicanonical methods allowed us to verify the saturation of the order parameters of TIM and FIM. In both models, the order parameter reaches its maximum 
value in the ordered phase, confirming that the ordered phase is accompanied by the emergence of the corresponding planar or fractal subsystem ordering.

Throughout the simulations, we were also able to verify the existence of a unique first-order phase transition in both models, as shown in Fig.\ref{fig:TIM-FIM_plots}. 
While the TIM consistently exhibits a sharp first-order transition across all system sizes, the FIM displays a much sharper transition
already at small sizes such as $L=8$. This property, coupled with 
the presence of strong finite-size effects at lower system sizes and the restriction to the $L=2^n$ family, severely limits the number of 
system sizes that can be explored.

\begin{figure}[t]
  \centering
  \includegraphics[width=0.5\textwidth]{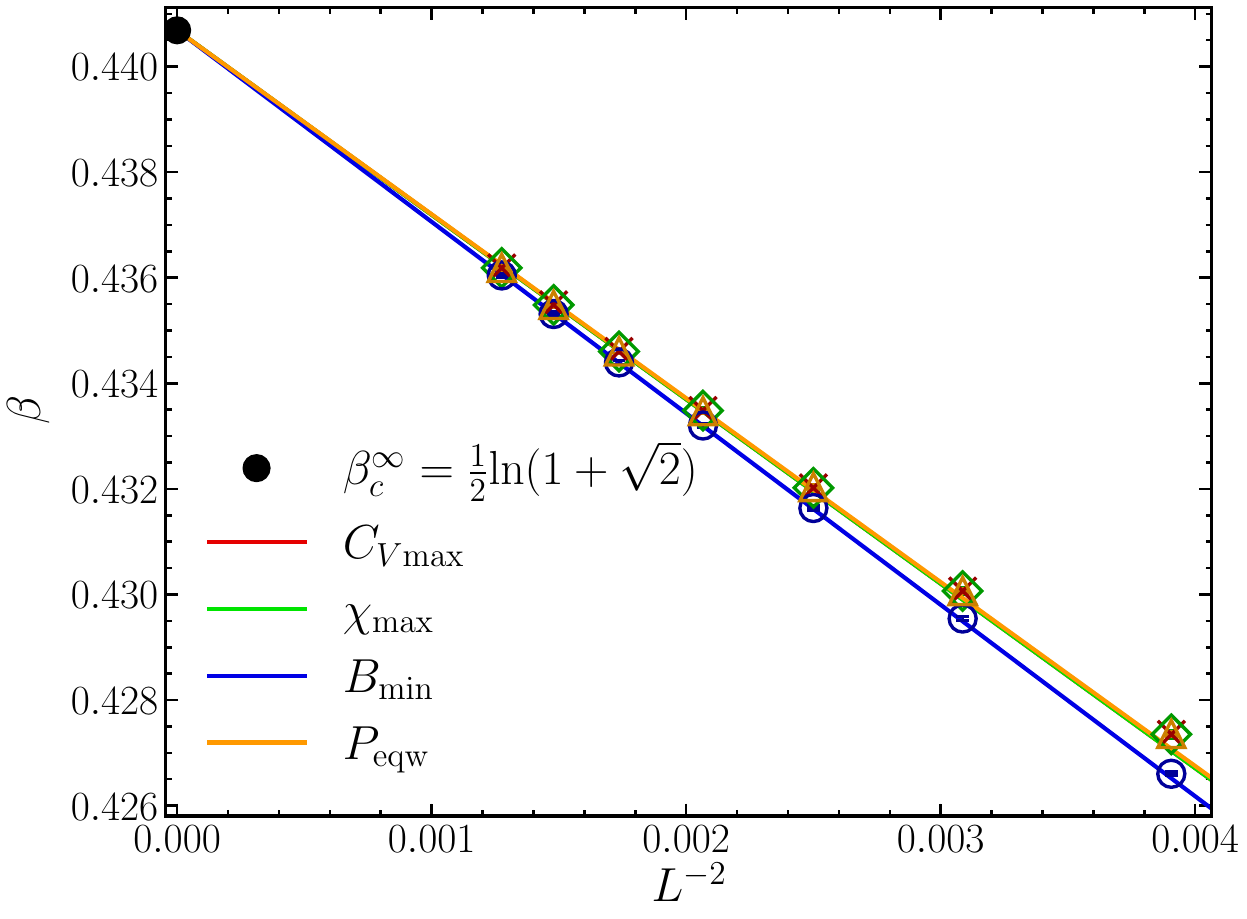}
  \caption[Anomalous scaling analysis for the Tetrahedral Ising model.]
        {Anomalous scaling for the Tetrahedral Ising model. Simulations are performed on lattices with 
  $L = 16, 18, ..., 28$ ($N = \frac{1}{2}L^3$). Finite-size transition temperatures $\beta_c(L)$ are determined 
  from the minimum/maximum peaks of specific heat $C_V$, susceptibility $\chi$, and Binder cumulant $B$ and the equal-weight 
  peaks of energy histogram $P(E)$. The thermodynamic $\beta_c^{\infty}$ is fixed by self-duality. All fits 
  collapse onto the expected $\frac{1}{L^2}$ scaling.}
  \label{fig:TIM_scaling}

\end{figure}

To test whether this finite-temperature ordered phase is breaking the full set of subsystem symmetries, we look for the
subextensive degeneracy of the ordered phase that accompanies subsystem symmetry breaking. This is reflected by the
first-order scaling behavior of the transition temperature $\beta_c(L)$. From the $2^{3L-3}$ and
$2^{2L-1}$ ground-state degeneracies obtained for the TIM and FIM, we obtain the following anomalous scaling behavior:
\begin{equation} 
\beta_{c}(L) \sim \beta_c^{\infty} + \frac{ b}{L^{D-d}} + \mathcal{O}\left(\frac{1}{L^{D-d-1}}\right),
\end{equation}
where $b$ is a constant prefactor and $d$ reflects the power of $\log_2 GSD\sim L^d$. Thus, we expect both TIM and FIM 
to display an $\frac{1}{L^2}$ scaling. Since both models are characterized by a unique transition point, we can use the self-dual 
fixed-point condition derived in Sec.~\ref{sec:self-duality} to fix $\beta_c^{\infty}=\frac{1}{2}\text{ln}\left(\sqrt{2}+1\right)$. 
This scaling behavior was indeed confirmed for the TIM as shown in Fig.~\ref{fig:TIM_scaling}.

\begin{figure}[t]
  \centering
  \includegraphics[width=1\textwidth]{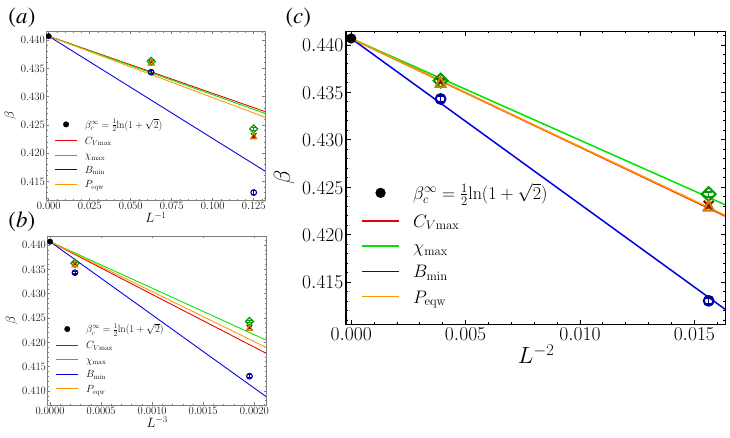}
  \caption[Non-standard finite-size scaling analysis for the Fractal Ising model.]
        {Non-standard scaling for the Fractal Ising model, obtained using system sizes $L = 8, \, 16, \, \infty$. The three points are
        compatible only with a fit following the $\frac{1}{L^2}$ scaling law (c) and exclude the conventional $\frac{1}{L}$ 
        fit (a) and a $\frac{1}{L^3}$ fit (b).}
  \label{fig:FIM_scaling}
\end{figure}

The same feature is less straightforward to demonstrate for the FIM due to the limited number of system sizes available. Simulations of intermediate
system sizes that do not belong to the $L=2^n$ family were shown to have both different ordering and finite-size scaling behaviors,
which is consistent with the structural difference of their respective symmetries and, consequently, the different symmetry breaking patterns. 
Thus, using the same $\beta_c^{\infty}=\frac{1}{2}\ln(\sqrt{2}+1)$ given by self-duality arguments, the scaling analysis could only be performed using
$L\in\{8,16,\infty\}$, with others being either too large for practical simulations or too small for a meaningful fit. 
Nevertheless, this minimal set of data points allows us to identify the expected non-standard scaling by comparing the behavior of the 
transition temperature to different $\frac{1}{L^{D-d}}$ scalings: 
as shown in Fig.~\ref{fig:FIM_scaling}, the behavior of $\beta_c(L)$ is in agreement only with the theoretical fit corresponding to $d=1$,
thus confirming the fractal subsystem symmetry breaking and the subextensive ground-state degeneracy in the ordered phase of the FIM. Given this agreement, one can
expect that this behavior will also hold for the sequence of system sizes $L=4^n-1$ that also has subextensive GSD and showcases a strong first-order
transition at the only accessible system size $L=15$. For the sequences $L=2^{m-1}(2^n+1)$ and $L=2^{m-1}(2^{2n-1}-1)$, where $m$ is a fixed integer, 
the FIM can be shown to have distinct global-symmetry generators and degeneracies that depend only on $m$.
As in the other families, strong first-order transitions were detected, but scaling analysis is, once again, not feasible due 
to a lack of available system sizes. Nevertheless, the persistence of a strong first-order transition across these families supports the expectation 
that they converge to the same thermodynamic limit, albeit with distinct finite-size corrections reflecting their different symmetry structures.

\section{Discussion and Concluding remarks}

The results presented in this chapter, together with previous work on 3D subsystem-symmetric spin models, allow us to establish a coherent 
thermodynamic picture associated with subsystem symmetry breaking. As showcased in the previous chapter, both the Tetrahedral Ising model and the 
Fractal Ising model exhibit a single temperature-driven transition separating a high-temperature disordered regime from a low-temperature subsystem-ordered phase. 
In both cases, the transition is strongly first-order, as evidenced by phase coexistence in the energy histograms and pronounced Binder cumulant dips.
The ordered phases are characterized by the emergence of planar and fractal ordering patterns, respectively, consistent with the subsystem symmetry structure of each model. 
Most interestingly, the finite-size shift of the pseudocritical temperature follows the anomalous scaling $\beta_c(L) - \beta_c^{\infty} \propto L^{-(D-d)}$ 
predicted by the subextensive ground-state degeneracy associated with the simultaneous breaking of the subsystem symmetries. 

A natural question emerging from these observations is whether the strongly first-order character of the transition is accidental or represents a fundamental feature 
of subsystem symmetry breaking. This question is closely related to the ongoing search for critical behavior associated with the emergence of subsystem ordering,
which has so far escaped categorization in the form of universality classes. The evidence accumulated so far suggests the latter. In addition to the TIM and FIM studied 
here, the 3D plaquette Ising model~\cite{Johnston14,Johnston17}, which possesses planar subsystem symmetries, also exhibits a single first-order transition associated 
with subsystem ordering. Similar behavior persists in 3D compass models~\cite{Gerlach15,Canossa23}, where, 
unlike the aforementioned Ising spin models, subsystem symmetry breaking coexists with magnetic frustration and is driven by order-by-disorder phenomena.
Regardless of the type of ordering and spin degree of freedom, all these models are characterized by a unique, discontinuous thermal phase transition.

At first, these observations suggest that $\mathbb{Z}_2$ subsystem symmetry breaking in short-range 3D models typically favors first-order behavior.
However, no general theorem establishes this conclusion for arbitrary 3D subsystem-symmetric Ising Hamiltonians. 
Dimensional reduction bounds provide a useful organizing principle for what kinds of subsystem symmetries can be broken at finite temperature~\cite{Batista05,Nussinov15},
most notably precluding the spontaneous breaking of $d=1$ (line-like) subsystem symmetries for short-range interactions, while allowing $d\ge2$ subsystem symmetries to break,
but do not constrain the order of the resulting transition.

One structural ingredient that distinguishes these models from conventional global symmetry breaking is the subextensive but exponentially large number of ground-state sectors related by the subsystem transformations. 
In the ordered regime, the number of symmetry-related sectors scales exponentially with $L^{D-d}$: approaching the transition thus requires the redistribution of statistical weight among macroscopically many distinct 
ordered configurations. However, extensive or subextensive degeneracy alone does not forbid continuous phase transitions in 3D~\cite{Racz83}, and the mere presence of $\mathbb{Z}_2$ subsystem symmetries is not 
sufficient to enforce discontinuity, a property that becomes evident in anisotropic limits of these models, where the effective dimensional reduction (e.g., decoupled or weakly coupled layers) restores conventional 
critical behavior~\cite{Mueller17}. The isotropic subsystem-symmetric models considered here therefore occupy a nontrivial regime in which degeneracy, subsystem geometry, and interlayer coupling collectively shape 
the thermodynamic transition into a first-order regime, a phenomenology that was shown to persist across models with fundamentally different subsystem symmetry geometry.

In the absence of a general analytic classification, the robust emergence of first-order signatures in these models suggests that discontinuity may be a structurally favored outcome of subsystem symmetry breaking 
in 3D. Whether subsystem symmetry breaking in a fully interacting, short-range 3D Ising model can occur through a continuous phase transition, and long-range subsystem ordering can be associated with 
genuine critical behavior, remains an open question.

What has been presented in this chapter lays the foundation for the remainder of this thesis. Beyond their intrinsic uniqueness in the context of classical spin physics, the high-temperature expansion of 
these subsystem-symmetric Ising models provides a natural route to a class of systems exhibiting unconventional topological order, commonly referred to as fracton models. 
Starting with a specific characterization of fracton phases of matter, in Chapter~\ref{chap:fracton_models} we present the relation between $\mathbb{Z}_2$ fracton models and subsystem symmetries in classical Ising models. 
The thermodynamic behavior of the classical models will also reappear in Chapter~\ref{chap:fractons_in_QEC}, where we demonstrate how the same statistical-mechanical structure governs the response of quantum fracton 
models to random noise, establishing a direct connection between classical Ising physics and quantum error correction.

%% file: mainmatter/theory/fracton_models.tex
\chapter{Fracton Phases of Matter}
\label{chap:fracton_models}

In Chapter~\ref{chap:subsystem_symmetries}, we focused on classical Ising models endowed with subsystem symmetries, studying how the codimension and geometry of the symmetry 
support define ordering phenomena, degeneracies, and finite-size behavior. A central message of that analysis is that subsystem symmetries impose 
structured constraints on allowed configurations and on the way defects can be created and moved. The purpose of the present chapter is to show how 
the same structures reappear in the quantum setting in the form of lattice gauge theories and how they naturally connect to a family of exotic 
phases whose elementary excitations have restricted mobility, collectively referred to as \emph{fracton phases of matter}.

The term "fracton" has been used in several distinct contexts over the years. Historically, it first appeared in the study of vibrational modes on 
fractal networks~\cite{Toulouse83}, and it was later reintroduced to describe gapped three-dimensional spin models exhibiting topological order 
with immobile or subdimensionally mobile excitations~\cite{Vijai15}.
In the modern condensed-matter literature, the word is frequently employed as an umbrella term for systems in which the motion of excitations 
is constrained by conservation laws that go beyond a simple global charge conservation, including effective dipole, higher-multipole, 
or fractal constraints~\cite{Vijai16,Nandkishore19}. This broad usage can sometimes obscure which properties are universal and which are model-dependent.
For the sake of clarity, we adopt the following working definition: a fracton phase is a phase of matter in which elementary charge excitations exhibit 
restricted mobility, in the sense that no strictly local operator can translate an isolated charge without producing additional excitations.

This restricted mobility can arise in several ways, but a unifying principle is that the low-energy theory effectively enforces additional conservation laws. 
Motion of an isolated charge would necessarily change the corresponding conserved quantity, and is therefore forbidden unless accompanied by the creation, 
annihilation, or coordinated motion of other excitations.

A useful high-level way to organize fracton phases is in terms of the gauge structure that captures their low-energy constraints and the geometrical structure
of their associated conservation laws.
While non-abelian generalizations of fracton phases have also been explored~\cite{Song19,Prem19,Prem19_3,Shirley19,Tu21}, much of the modern literature 
has focused on two large and complementary settings where abelian fractonic behavior has been studied in detail: higher-rank $U(1)$ symmetric-tensor gauge 
theories~\cite{Pretko17,Pretko18,Gromov19} and gapped, exactly solvable $\mathbb{Z}_2$ lattice gauge theories~\cite{Vijai16}.

For the sake of disambiguation, in Sec.~\ref{sec:U1gauge} we first give a brief overview of the $U(1)$ tensor-gauge framework to establish how restricted mobility
directly follows from higher-moment conservation laws. In Sec.~\ref{sec:Z2gauge} we turn our attention to the gapped $\mathbb{Z}_2$ setting to show how fracton 
topologically ordered phases can be derived from classical subsystem-symmetric Ising models, such as the ones discussed in Chapter~\ref{chap:subsystem_symmetries}, through the mechanism
of subsystem symmetry gauging, with particular emphasis on the impact that lattice and symmetry geometry have on the distinct kinds of restricted mobility that fracton phases can exhibit.
These themes will be carried forward in Chapter~\ref{chap:fractons_in_QEC}, where the same models are analyzed as candidate quantum memories.

\section{Fractons in $U(1)$ Gauge Theories}~\label{sec:U1gauge}

In high-energy and condensed-matter physics alike, matter and interactions can be described by locally defined fields.
In certain scenarios, these fields have a degree of redundancy in their definition which does not affect the overall equations of motion:
this redundancy is referred to as a gauge invariance and can be written as a transformation of the local field as a function of an 
external gauge field. Being a redundancy in the definition of the theory, gauge-dependent properties of the gauge field are unobservable, 
while all physical observables are gauge invariant. Gauge invariance is an organizing principle enforcing 
local conservation laws, constraining dynamics, and naturally encoding long-range entanglement~\cite{Kogut79}.
A paradigmatic example is the classical theory of electrodynamics, where the dynamics of the two relevant fields, 
electric and magnetic, is described by the Lagrangian density
\begin{equation}
    \mathcal{L} = \frac{1}{2} \left(\mathbf{E}^2 - \mathbf{B}^2\right),
\end{equation}
with the electric and magnetic field vectors $\mathbf{E}$, $\mathbf{B}$ satisfying the two homogeneous Maxwell equations
\begin{equation}
    \nabla\cdot \mathbf{B} = 0,\qquad\qquad \nabla\times\mathbf{E} + \partial_t\mathbf{B}=0.
\end{equation}
This allows us to write them in terms of a $4$-vector field 
$A_{\nu} = \left(-\phi(\bf{r},t), A_i(\bf{r},t)\right)$
composed of the scalar potential $\phi$ and the vector potential $A_i$,
with $\nu=(0,1,2,3)$ and $i=(1,2,3)$ labelling vector components.
This can be used to describe the electric and magnetic field as functions of $A_\nu$:
\begin{equation}\label{EB_fields}
   \mathbf{B} = \nabla \times \mathbf{A},\qquad\qquad \mathbf{E}= - \nabla\phi -\partial_t \mathbf{A}.
\end{equation}
For simplicity, we introduce the $4$-derivative $\partial_{\mu}=(-\partial_t,\nabla)$ 
and adopt the convention that indices are raised using the Minkowski metric with signature $(-,+,+,+)$, 
such that $\partial^{\mu}=(+\partial_t,\nabla)$ and $A^{\mu} = \left(\phi(\mathbf{r},t), A_i(\mathbf{r},t)\right)$~\cite{Schwartz14}.
This allows us to write the Lagrangian density in terms of the vector field as
\begin{flalign}
    \mathcal{L}&=\frac{1}{2}\left[\left(\partial_t\mathbf{A}+\nabla\phi\right)^2 - \left(\nabla\times\mathbf{A}\right)^2\right]\nonumber\\
               &=-\frac{1}{4}\left(\partial_{\nu} A_{\mu} -\partial_{\mu} A_{\nu}\right) \left(\partial^{\mu} A^{\nu} -\partial^{\nu} A^{\mu}\right)\nonumber\\
               &=-\frac{1}{4}F_{\mu\nu}F^{\mu\nu}
\end{flalign} 
where $F_{\mu\nu} = \partial_{\mu}A_{\nu} - \partial_{\nu}A_{\mu}$.
In the presence of a (complex) charge matter field $\Phi$ with charge density $\rho=\Phi^{\dagger}\Phi$ and a non-dynamical current density $\mathbf{J}$, this becomes:
\begin{equation}
    \mathcal{L} = -\frac{1}{4}F_{\mu\nu}F^{\mu\nu} - J_{\mu}A^{\mu},
\end{equation}
where $J_{\mu}=(\rho, \mathbf{J})$.
The gauge field $A_{\mu}$ is not uniquely defined and can be changed by the total derivative 
of some real, smooth, single-valued scalar function $\alpha(\bf{r},t)$: the physical properties of the theory 
are invariant under gauge transformations of the form:
\begin{equation}
    A_{\mu}(\bf{r},t) \longrightarrow A_{\mu}(\bf{r},t) + \partial_{\mu} \alpha(\bf{r},t)
\end{equation}
The first term of the Lagrangian density is evidently invariant under such transformations; while the second term is not explicitly invariant, 
one can prove that it does not change the total action due to the conservation laws which directly follow from the Euler-Lagrange equations of motion:
\begin{equation}
\partial_{\nu}\frac{\partial \mathcal L}{\partial(\partial_{\nu}A_{\mu})} - \frac{\partial\mathcal L}{\partial A_{\mu}}=0 \longrightarrow \partial_{\nu}F^{\nu\mu} = J^{\mu}
\end{equation}
From this, one can apply a total derivative to obtain
\begin{equation}
\partial_{\nu}\partial_{\mu}F^{\nu\mu}=\partial_{\mu}J^{\mu}\longrightarrow \partial_{\mu}J^{\mu}=0,
\end{equation}
which ensures invariance of the action upon gauge transformations with an arbitrary choice of $\alpha$. 
This is a typical example of a rank-1 (i.e. vector) $U(1)$ gauge field theory: the gauge invariance is directly
tied to the presence of a local conservation law in the theory. In this case, this boils down to the Gauss law
which, collecting the definitions given so far, can be written as:
\begin{equation}\label{Gauss1}
\nabla\cdot \mathbf{E}(\mathbf{x},t)=\partial_i E^i=\rho(\mathbf{x},t)
\end{equation}
From this, we can infer that the presence of a gauge invariance is always related to a
local conservation law, which implicitly defines a conserved charge (or charge density).

The operator $\partial_iE^i$ takes on a much more important role than simply constraining the charge density: 
in the framework of gauge theories, it represents a Gauss law constraint operator $G(\mathbf{x},t)$. One can 
prove that this acts as a generator for all local $U(1)$ gauge transformations in the form of a unitary operator~\cite{Wen04}
\begin{equation}
    U\left[\alpha\right]= \text{exp}\,\left(i\int d^3x\,\alpha(\mathbf{x})G(\mathbf{x})\right), \qquad G(\mathbf{x}) = \partial_iE^i(\mathbf{x}) - \rho(\mathbf{x})
\end{equation}
The low-energy sector of the theory is given by field configurations which obey the 
$\partial_i E^i=0$ condition: in the vacuum, the model boils down to
\begin{flalign}
    \mathcal{L} = -\frac{1}{4}F_{\mu\nu}F^{\mu\nu}.
\end{flalign}
In the Lorentz gauge $\partial_{\mu} A^{\mu}=0$, the equations of motion for such system become
\begin{equation}
    \left(-\partial_t^2 + \nabla^2\right) A_{\nu}=0,
\end{equation}
meaning that the solutions to the equations of motion will be massless, gapless plane waves with 
dispersion relation $\omega^2 = |\mathbf{k}|^2$. 
Any configuration with non-zero charge density of the emergent electric field 
$\rho\neq0$ is characterized as having higher-energy excitation.

The Gauss operator $G$ enforces that no local operator can create a net charge in the system and 
determines the energy sector in which the effective description of the theory lies.
However, one may go a step further and ask: what happens when the gauge field itself 
has a tensorial rank greater than 1? How would the conservation laws of higher-rank gauge fields look like? 
Could these give rise to qualitatively different dynamical constraints arising from higher multipole moments of the gauge charge excitations?
Without jumping to rank-$N$ theories with arbitrarily large $N$, we explore a minimal extension to these higher-rank gauge theories
by considering real rank-2 tensor gauge fields $A_{ij}$. 
The generalization of the Gauss law in Eq.~\eqref{Gauss1} from a rank-1 gauge theory to a rank-2 tensor gauge field $A_{ij}$, 
obtained by enforcing the conservation of total charge, is expected to take the following form:
\begin{equation}
\partial_i\partial_j E^{ij}=\rho,
\end{equation}
where $E^{ij}$ is some appropriate electric field tensor~\cite{Pretko18}. 

Let us start by defining a generalized gauge transformation 
\begin{equation}
    A_{ij} \rightarrow A_{ij} + \partial_i\partial_j \alpha(\mathbf{x}), 
\end{equation}
with $A_{ij}$ being a real tensor field and $\alpha(\mathbf{x})$, once again, a real scalar function.
We can immediately see that, since $\partial_i\partial_j \alpha$ is symmetric under the exchange of indices $i\leftrightarrow j$,
it follows that only the symmetric part of $A_{ij}$ transforms nontrivially: decomposing $A_{ij}$ into its symmetric and antisymmetric components,
\begin{equation}
    A_{ij} = A_{(ij)} + A_{[ij]},    
\end{equation}
one finds that only $A_{(ij)} \rightarrow A_{(ij)} + \partial_i\partial_j \alpha$, while the antisymmetric part $A_{[ij]}$ will show no gauge redundancy.
In a scalar gauge theory, any antisymmetric field component will simply decouple from the gauge field and can be gapped out. 
This is generally true as long as we don't go into the territory of two-form field theories and remain in theories with point-like sources
(i.e. with 1-form gauge fields~\cite{Kalb74,Wen04,Pretko17}). We will therefore restrict to gauge theories with symmetric rank-2 gauge fields.

For the construction, we use the same logic that relates the gauge invariance to the source-free Gauss law and require that~\cite{Pretko17}
\begin{equation}
    \partial_i\partial_j E^{ij}=0.
\end{equation}
This low-energy version of the Gauss law will be trivially generalized once one analyzes the action of the gauge transformation
on the total action or an arbitrary state of a generic model with such a gauge field~\cite{Pretko18}.  
More generally, the electric tensor can be defined as any symmetric tensor derived as a canonical conjugate to $A_{ij}$
in the Hamiltonian formalism, meaning it satisfies the condition
\begin{equation}
    \left[A_{ij}(\mathbf{x}),E_{kl}(\mathbf{y})\right] = \frac{i}{2}\left(\delta_{ik}\delta_{jl} + \delta_{il}\delta_{jk}\right)\delta(\mathbf{x}-\mathbf{y})
\end{equation}
This constraint for the low-energy description of the theory can be enforced energetically in an emergent gauge theory by introducing an energy penalty of the 
form $U (\partial_i\partial_j E^{ij})^2$, which projects low-energy states onto the gauge-invariant subspace in the $U\rightarrow \infty$ limit~\cite{Pretko17}.
One possible definition of the electric and magnetic field tensors that satisfies our requirements and can be seen as a direct generalization of Eq.~\eqref{EB_fields} are~\cite{Pretko18} is: 
\begin{equation}
B_{ij} = \epsilon_{ikl}\partial^k A^{l}_{j}, \qquad \qquad E_{ij} = \partial_t A_{ij} - \partial_i\partial_j \phi,
\end{equation}
where $\phi$ is a scalar potential of the theory and $\epsilon_{ikl}$ is the Levi-Civita tensor. From this, we can write a generalized Maxwell-like Lagrangian
by introducing the matter field $\Phi$, which, once again, transforms as $\Phi\rightarrow e^{i\alpha(\mathbf{x})}\Phi$ and its associated static current tensor $J^{ij}$:
\begin{equation}
    \mathcal{L}=\frac{1}{2}\left(E_{ij}E^{ij}-B_{ij}B^{ij}\right) + A_{ij}J^{ij} - \phi \rho.
\end{equation}
As one can see, all elements of the Lagrangian are gauge invariant; imposing gauge invariance leads to our predicted generalized Gauss law for high-energy configurations:
\begin{equation}
    \partial_i\partial_j E^{ij}=\rho.
\end{equation}
One can integrate the equation above over a volume $V$ to get
\begin{equation}
Q = \int_{V} d^3x \rho = \int_{V} d^3 x \partial_i\partial_j E^{ij} = \oint_{\partial V} d\hat{n}_i \partial_j E^{ij}
\end{equation}
\begin{equation}
P = \int_{V} d^3x \,\mathbf{x} \rho = \int_{V} d^3 x \,x^{i} \partial_j\partial_k E^{jk} = \oint_{\partial V} d\hat{n}_j \left(x^i\partial_k E^{jk} - E^{ij} \right)
\end{equation}
Both the charge and the dipole moment associated with the matter field can be written as fluxes of corresponding charge and dipole current operators through the surface $\partial V$.
Assuming that we are working in closed systems, with a field decaying sufficiently fast at infinity, this implies the existence of a \textbf{local dipole moment conservation law} on top of the already present local charge conservation.
This additional constraint has significant implications for the dynamics of the excitations: since the motion of any isolated charge would change the local dipole moment, 
this implies that no charge can move at all on its own without creating or destroying an additional dipole. 
This gives us a formal field-theoretic characterization of \textit{fractons}: symmetric tensor gauge theories naturally encode local 
conservation laws whose resulting constraints on charge and dipole motion give rise to fracton phases; 
the charges of these gauge theories serve as canonical examples of fracton excitations, which can only move as bound states that preserve the total dipole moment of the system~\cite{Pretko17,Pretko18,Gromov19}.

More generally, a rank-$n$ U(1) symmetric tensor gauge field enforces the conservation of all multipole moments up to order $n-1$, 
extending the notion of local conservation laws beyond charge to encompass dipole, quadrupole, and higher moments.
These higher-rank gauge theories thus represent a new frontier in our understanding of constrained quantum dynamics, 
emergent locality, and exotic entangled phases of matter. While a rigorous derivation can be found in Ref.~\cite{Gromov19}, our
interest will remain within the confines of rank-2 tensor gauge theories.

Most of the subtleties regarding the possible dynamical features of fractons lie in the kinetic terms encoded by the current tensor $J^{ij}$ 
that couples to the gauge field, whose form can be quite generic and strongly dependent on the specific scenario considered.
A fully covariant formulation of a generalized rank-2 symmetric gauge field theory was accurately 
derived in Ref.~\cite{Pretko18} by appropriately gauging the field theory describing a gapped matter field $\Phi$ with a global charge and dipole moment conservation.
While the entire derivation is beyond the scope of this thesis, a very interesting takeaway can still be drawn from this generalized approach:
starting with the matter field alone, the assumption that our theory has global charge conservation (i.e. is invariant under global rotations by a phase $\alpha$)
and global dipole moment conservation (i.e. $\int d^3 x \,\,|\Phi|^2\mathbf{x}=\text{const}$, under which $\Phi$ transforms as $\Phi \longrightarrow e^{i\lambda \cdot \mathbf{x}}\Phi$),
the lowest-order covariant derivative operators would take the form
\begin{equation}
\Phi \partial_i\partial_j\Phi - \partial_i\Phi\partial_j\Phi.
\end{equation}
Under symmetry transformation, this factor picks up a phase $e^{i2\alpha}$. To preserve the symmetry, the lowest-order possible Lagrangian will take the form
\begin{equation}
\mathcal{L} = |\partial_t\Phi|^2 - m^2|\Phi|^2 - \delta |\Phi \partial_i\partial_j\Phi - \partial_i\Phi\partial_j\Phi|^2.
\end{equation}
The last term in the theory captures a core feature of fracton models:
as one would expect from a dipole-moment conserving theory, the smallest creation operators allowed are quadrupolar, 
regardless of the system dimensionality, geometry or whether the theory is in a continuum or discrete spatial support. 
Expanding this term to higher orders, one can obtain various mechanisms enabling fracton motion.

At this stage, the intuition gleaned from the pure gauge theory formulation remains somewhat abstract.
While the field theory description given above captures the essential idea of kinematic constraints of isolated charges, 
a complete microscopic understanding of the underlying mechanism associated with the emergence of fracton phases across different
realizations remains an active area of research. This is further complicated by the model-specific ways in which constrained fracton dynamics can manifest.
Over the past two decades, a number of models have displayed this behavior
~\cite{Jimenez02,Elmatad09,Castelnovo10,Keys11,Biroli13,Chleboun14,Pretko17_2,Slagle17,Albert17,Prem18,Bulmash18,Prem19,Pai19,Gromov20,Seiberg20,Seiberg21,You22,Hirono24}, 
with some having been observed long before the term \textit{fracton} was first coined~\cite{Xu06,Pankov07,Xu10}, ranging from 
topologically protected spin systems to higher-rank lattice gauge theories. Each provides a different perspective on how 
similar sets of conservation laws emerge from different local microscopic mechanisms and lattice geometries, 
ultimately leading to different mobility restrictions and spectral properties.
In the remainder of this chapter, we will shift our attention to another class of fracton models that is directly related to 
the subsystem symmetries discussed in the previous chapters. For more details on the aforementioned classes of fracton models, we refer the reader to the references listed above.

\section{Fractons in Exactly Solvable Models}\label{sec:Z2gauge}

So far we always discussed symmetric rank-2 $U(1)$ gauge theories, where the continuity of the gauge group ensures that, in the absence
of external couplings inducing a Higgsing of the gauge degrees of freedom, the low-energy excitations of the system remain gapless, 
as any mass term would explicitly break gauge redundancy. 
While it gives one point of access to the field of fractons, it overlooks an important detail: 
the vast majority of fracton models studied in the literature emerge as gapped, exactly solvable, topologically ordered spin liquid phases 
~\cite{Chamon05,Haah11,Yoshida13,Haah13,Vijai15,Vijai16,Slagle17_2,Slagle18,Shi18,Shirley18,Ma18_2,Schmitz18,Nandkishore19}. As a matter of fact, the term was first coined 
to describe this group of models which, unlike the family of $U(1)$ fracton 
models outlined in the previous section, originally emerged in the context of discrete $\mathbb{Z}_2$ LGTs.

These two "families" of fracton models, which can very comfortably be labelled by their respective gauge group, have vastly different aspects of interest. 
The key features of $U(1)$ fracton theories lie in the type of correlations, excitations and dynamics
that can emerge as a consequence of the interplay between gauge redundancy and higher-moment symmetry constraints~\cite{Pretko17_2,Prem19,Pai19,Gromov20,Seiberg20,You22,Hirono24}.
In contrast, gapped fracton models represent a peculiar type of topologically ordered phase of matter, where the interplay between geometrical
and topological features of the system and its symmetries leads to very distinctive dynamical constraints and topologically ordered states with 
extensively large ground-state manifolds, giving them a non-trivial relevance in the context of quantum error correction.

Like with $U(1)$ gauge theories, $\mathbb{Z}_2$ fracton models can be derived with an analogous gauge principle as the one outlined in Ref.~\cite{Pretko18}:
the core differences between the two families of fracton models lie in the different framework underlying $U(1)$ field theories and $\mathbb{Z}_2$ LGTs,
the global symmetries that characterize the models prior to gauging, and how the gauging procedure gives rise to different types of generalized Gauss laws
which enforce local dynamical constraints and fractonic behavior. Another important distinction lies in the energy spectrum of the gauge field excitations: 
whereas in $U(1)$ theories we find a tensor gauge field with a continuum of modes that can stimulate
the already present matter-field excitations (i.e. our fractons), in the $\mathbb{Z}_2$ theories the gauge field takes discrete values ($\pm1$) 
and we have topological flux excitations that are either at $0$ energy or have a finite energy cost and accompany the creation of additional fractons.
Nevertheless, in both $U(1)$ and $\mathbb{Z}_2$ fracton models, fractons emerge as gapped gauge charge excitations of the theory; 
the qualitative distinction between the two families lies instead in the spectrum of the gauge field modes.

So far, there have been three primary approaches used for deriving $\mathbb{Z}_2$ fracton models:
\begin{itemize}
    \item \textbf{Gauging subsystem symmetries.} Starting from an Ising spin model with a series of subsystem $\mathbb{Z}_2$ 
          symmetries with specific geometrical structures and gauging the subsystem symmetries by introducing 
          a $\mathbb{Z}_2$ gauge field~\cite{Vijai16}.
    \item \textbf{Foliation / coupled-layer construction.} Start from stacks of decoupled two-dimensional topological layers 
          (e.g., toric-code or double-semion sheets) arranged along multiple orientations, then condense extended composites 
          to "glue" the layers together. The resulting three-dimensional phase is defined up to adding/removing such 2D layers 
          and finite-depth local unitaries. 
    \item \textbf{Higgsing $U(1)\!\to\!\mathbb{Z}_2$.} Coupling the gauge field of a $U(1)$ symmetric rank-2 gauge theory 
          to a charge-2 matter field can, in some specific cases, effectively reduce the gauge group to $\mathbb{Z}_2$~\cite{Bulmash18,Ma18}.

\end{itemize}

The coupled-layer/foliation framework has been remarkably successful at relating a variety of $\mathbb{Z}_2$ fracton 
models and explaining their shared structure in the form of embedded 2D layers and generalized renormalization moves
~\cite{Ma17,Shirley18,Shirley19,Shirley19_2,Prem19_2,Wang23}. 
This construction was shown to give a remarkably intuitive description of most 3D fracton models with 
an underlying planar symmetry structure; however, this turns out to limit its scope and rules out all models 
whose restricted mobility is given by more complicated conservation laws.

It must also be mentioned that the possibility of deriving certain fracton models by Higgsing a $U(1)$ symmetric gauge theory into a $\mathbb{Z}_2$ theory 
by coupling it with an external charge-2 matter field was explored~\cite{Bulmash18, Ma18_2,Hermele21}; this method was successful in 
deriving only one of the various types of models known so far, as the possibility of achieving the desired 
$U(1)\rightarrow \mathbb{Z}_2$ Higgsing strongly depends on the way the conservation laws of the original $U(1)$
theory manifest.
We will instead focus on the gauging route. This method was first introduced by Franz Wegner in 1971 as a way to reformulate Ising models in terms 
of theories with an exact local invariance, leading to the formulation of new models capable of hosting phases of matter whose order cannot be 
detected by local observables and which therefore lie beyond the Ginzburg–Landau symmetry-breaking paradigm~\cite{Wegner71}. These phases were 
later understood within the framework of lattice gauge theories~\cite{Kogut79} and characterized as exemplary spin-liquid models exhibiting topological order~\cite{Wen90}. 
In the following sections, we review the basic structure of $\mathbb{Z}_2$ lattice gauge theories, introduce the gauging framework 
and show how it can be used to construct and analyze exactly solvable $\mathbb{Z}_2$ fracton phases, with special attention to how
nontrivial mobility constraints on the excitations can emerge~\cite{Wen17}.

\subsection{$\mathbb{Z}_2$ Lattice Gauge Theories: an introduction}\label{sec:LGT}

In the context of condensed matter physics, the concepts discussed in the previous sections acquire 
a discretized formulation when working on a lattice: the spatial continuum is replaced by a set of 
sites connected by bonds, and fields are assigned to these discrete elements. Matter degrees of 
freedom — be it spinless particles, fermions, or bosons — reside on the lattice sites, while the 
gauge fields are typically defined on the links connecting them. The local redundancy of the theory 
is then encoded in gauge constraints acting at each site, which play the role of a discrete Gauss law. 
These constraints ensure that physical observables are gauge invariant and that the low-energy 
Hilbert space consists only of states satisfying the local conservation laws. This way, the 
lattice gauge theory  reproduces the essential features of its continuum counterpart (i.e. 
discretized spatial degrees of freedom, local conservation laws, gauge redundancy, 
and long-range correlations) while remaining amenable to microscopic modeling.

This framework has proven remarkably fruitful in condensed matter physics, where
gauge theories naturally emerge in systems whose microscopic Hilbert space is restricted 
by local constraints arising from strong interactions or frustration. 
Quantum spin liquids are a paradigmatic example: in these phases, the absence of classical 
long-range order allows the low-energy degrees of freedom to fractionalize into emergent quasiparticles
whose collective dynamics are governed by an emergent lattice gauge theory (LGT).
Owing to their rich spectra of long-range entanglement and local correlations, 
spin liquids provide a natural setting for exploring novel phases of matter~\cite{Savary16}.
Most quantum spin-liquid phases are effectively described by matter fields coupled to emergent 
gauge fields—direct analogues of high-energy gauge theories, but arising purely from the internal 
dynamics of the spin system~\cite{Savary16,Capponi25,Pretko17}. 
This correspondence has made LGT not only a unifying language but also a powerful tool 
for identifying and modeling strongly correlated quantum materials.

Before starting, a fair question to ask, especially from a reader with a high-energy physics background, who might 
initially think of $\mathbb{Z}_2$ LGTs as simple toy models compared to continuum gauge theories which are
typically relevant for lattice QED or QCD, would be: why focus on $\mathbb{Z}_2$ LGTs? 
Within the condensed matter context, these are among the simplest gauge theories on a lattice that can exhibit
instances of deconfinement and topological order while remaining analytically tractable and
physically relevant, from their use as effective low-energy description of strongly correlated 
systems~\cite{Sachdev19}, to the construction and study of paradigmatic models with topological phase transitions~\cite{Wegner71}. 
These settings, among many others, prompted significant theoretical and experimental progress in the realization of LGT dynamics 
on quantum simulation platforms in the last several years~\cite{Halime20}.

As will be seen throughout this work, interest in the construction of $\mathbb{Z}_2$ LGT also arises from the quantum computation
perspective: certain subfamilies of such theories, such as the toric code or the fracton codes discussed in the following sections, 
provide paradigmatic frameworks for encoding and protecting quantum information in topologically ordered states~\cite{Homeier21}. 
On top of this, the ability to prepare and probe these gauge models on digital and analog quantum devices has made them a 
major target of contemporary quantum simulation efforts~\cite{Lumia22}.

Following the standard methods used in high-energy physics, our approach to the construction of LGTs will be to gauge the global symmetry of a given system.
As previously stated, we will restrict ourselves to discussing $\mathbb{Z}_2$ LGTs with fixed particle number, 
without dwelling on more general cases. For a broader introduction to the topic, see Refs.~\cite{Kogut75,Kogut79,Burrello15}.

\subsection{$\mathbb{Z}_2$ LGT from gauging global symmetries}\label{sec:Wegnergauging}

In the following section, we present an explicit construction of LGTs via the mechanism of 
"\textit{symmetry gauging}". This approach of "promoting" a global symmetry to a local one is borrowed from high-energy physics
and applied in condensed matter as a way of designing novel quantum phases of matter. Since we are also interested in the development 
of $\mathbb{Z}_2$ gauge theories, we will start by introducing the gauging procedure using its simplest example, 
the 2D Ising model, and describe a generalized notation that can be used for any subsequent model.

The overall process of deriving a LGT with topological order through gauging can be performed either by gauging a system with global $\mathbb{Z}_2$ symmetry
and introducing gauge-invariant terms derived from higher-order contributions in perturbation theory, or by performing the gauging procedure on
two independent systems with global $\mathbb{Z}_2$ symmetries coupled by the introduction of a gauge field. While the first approach relies on
perturbation theory, one may equivalently define the derived models with arbitrary choices of parameters. However, this distinction is important, 
as it clarifies why models such as the toric code are often described as having two independent
sets of commuting $\mathbb{Z}_2$ constraints (electric and magnetic), rather than as a single gauged Ising model. 
This structure, which is characteristic of CSS codes, will reappear in the fracton models discussed later.

Start with a classical Ising spin model with an Ising matter field $S$ such that, for any site $i$, $S_i=\pm 1$ (one could start 
from a quantum Hamiltonian, but this allows us to generalize the following description to the previous chapter).
For each spin, introduce a set of multispin coupling objects $O_i^{(a)}[S] = \prod_{l\in \partial O}S_l$, where $a=1,\dots,e$ labels different interaction terms
and $\partial O$ represents the set of Ising spins belonging to the multispin interaction. 

We first quantize the classical Ising model by introducing quantum fluctuations in the form of a transverse field, relabeling $S_i \longrightarrow \tau_i^Z$. 
The quantum Hamiltonian of this model will take the form
\begin{equation}\label{eq:H0}
    H_{0}=-J\sum_{i,a}O_i^{(a)}[\tau^Z] - h\sum_i \tau^X.
\end{equation}
For the 2D Ising case, this reduces to $H_0 = -J\sum_{\langle i,j\rangle} \tau^Z_i \tau^Z_j - h\sum_i \tau^X$.
One can see that the system is invariant under a global $\mathbb{Z}_2$ spin flip operator $\prod_i \tau^X_i$.
To maintain generality, let us say that the system is invariant under a global $\mathbb{Z}_2$ symmetry generated by a symmetry operator of the form
$\prod_{i\in\Sigma} \tau^X_i$ where $\Sigma$ is some manifold comprising an extensive subset of spins.

To promote the symmetry to a local $\mathbb{Z}_2$ gauge invariance, we need to introduce local degrees of freedom 
that are able to account for the energy changes that individual spin flips would cause. 
This can be done by adding a quantum \textbf{nexus field} $\sigma_{i,(a)}$ (or gauge field) at each multispin coupling $O_i^{(a)}$.
Since we are gauging a $\mathbb{Z}_2$ symmetry, the nexus field takes the form of local Ising degrees of freedom, remaining 
consistent with the $\pm 1$ eigenvalues of the multispin operators $O_i^{(a)}$.
We introduce a minimal coupling between the original matter field and the gauge field, as well as an additional transverse field for the nexus field.
With these additional gauge-field degrees of freedom, the minimally coupled gauged Hamiltonian of the 2D Ising model takes the form
\begin{flalign}\label{Hamiltonian_gauge-matter}
    H &= -J \sum_{i,a}  \sigma^Z_{i,(a)} O_i^{(a)}[\tau^Z] - h\sum_i \tau^X_i - t\sum_{i,a} \sigma^X_{i,(a)} \\
    &= -J \sum_{\langle i,j \rangle} \sigma^Z_{ij,(a)} \tau^Z_i \tau^Z_j - h\sum_i \tau^X_i - t\sum_{i,a} \sigma^X_{i,(a)} 
\end{flalign}
For concreteness, we place the $\mathbb{Z}_2$ gauge fields $\sigma_{i,(a)}$ on the interaction terms $O_i^{(a)}$ (which for the 2D Ising model
correspond to nearest-neighbor links).
One can immediately see that the global $\mathbb{Z}_2$ symmetry has been promoted to a set of $\mathbb{Z}_2$ local symmetries,
generated by symmetry operators of the form 
\begin{equation}
   G_i = \tau^X_{i} A_i ,\qquad \text{where }
   A_i = \prod_{\{j,a\}: i\in \partial O_j^{(a)} } \sigma_{j,(a)}^{X},\qquad \left[G_i,H\right]=0.
\end{equation}
In the definition of $A_i$, $\partial O_j^{(a)}$ runs over all the nexus-field spins $\sigma_{j,(a)}\,$ 
which correspond to the multispin interactions $O_j^{(a)}$ for which $\left[O_j^{(a)},\tau^X_i\right]\neq 0$: 
this allows us to identify the nexus field introduced in our gauging procedure with the gauge field of our newly formulated 
LGT, with $A_i$ representing the $\mathbb{Z}_2$ \textbf{nexus charge operator}.
In the case of the 2D transverse field Ising model, this will take the form of the 4-body operator 
$A_i= \prod_{j: \,\langle i, j\rangle} \sigma_{ij}^X $.
 
Similarly to how we require the charge and current tensors to respect a Gauss law in the continuum $U(1)$ 
field theory, we require the gauge invariance in the LGT to hold not only for the Hamiltonian $H$ but also 
for the physical states of the system $\ket{\psi}$: this step is essentially where the global symmetry 
$\mathcal{O}$, which up to this point has simply been broken down into a set of local symmetries 
generated by $G_i$, is promoted to a local redundancy. In the Hamiltonian formalism, this invariance 
is implemented by the following \textit{Gauss law constraint}:
\begin{equation}
G_i \ket{\psi} = \ket{\psi}\,,\qquad A_i\ket{\psi}=\tau_i^X \ket{\psi}.
\end{equation}
The \textbf{Gauss operators} $G_i$ generate local gauge transformations and define the $\mathbb{Z}_2$ 
gauge charge through their eigenvalues. This acts as a projector of the enlarged Hilbert space 
onto its gauge-invariant subspace through the projector operator $P=\prod_i (1+G_i)/2$: this projection 
enforces the Gauss law constraint at every site, whereas physical violations of this constraint 
correspond to localized gauge charge excitations.
Within this subspace, the action of $\tau_i^X$ and $A_i$ is the same, meaning $P\tau_i^X P = P A_i P$. 
More interestingly, one can easily see that $P\tau_i^Z P=0$: this is expected, as any residual action 
of $\tau_i^Z$ would otherwise flip the Gauss law eigenvalue, i.e. change the $\mathbb{Z}_2$ 
gauge charge of our LGT.

Imposing the generalized Gauss law on Eq.~\eqref{Hamiltonian_gauge-matter} yields the \textbf{field-nexus Hamiltonian}
\begin{equation}\label{Hamiltonian_Gauss_law_fix}
H = -J\sum_{i,a} \sigma^Z_{i,(a)} O_i^{(a)}[\tau^Z] - h\sum_i A_i - t\sum_{i,a} \sigma^X_{i,(a)}.
\end{equation}
This represents a $\mathbb{Z}_2$ LGT coupled to an Ising matter field in the same form as
the Fradkin-Shenker model~\cite{Fradkin79}. This implies that, in spatial dimensions D$\geq 2$, this model admits a 
deconfined phase which is separated by a phase boundary from the confined/Higgs regime.
Since $\tau_i^Z$ commutes with our new Hamiltonian and its action within the constrained Hilbert space 
is effectively null (meaning, $H$ is block-diagonal in $\tau_i^Z$), 
we are free to explicitly "gauge out" the matter-field contributions by restricting ourselves to a submanifold of the 
Hilbert space where each local $\tau_i^{Z}$ operator has a fixed eigenvalue $\pm1$, 
which in turn fixes the value of the multispin interaction terms $O_i^{(a)}[\tau_i^Z]$.
Restricting ourselves to the Hilbert space superselection sector comprising the set of states $\ket{\psi}$ 
that satisfy the condition $\tau_i^Z\ket{\psi}=+\ket{\psi}\,\forall i$, the Hamiltonian becomes
\begin{equation}\label{Hamiltonian_Gauge_fix}
H = -J\sum_{i,a} \sigma^Z_{i,(a)}  - h\sum_i A_i - t\sum_{i,a} \sigma^X_{i,(a)}.
\end{equation}
To obtain an exactly solvable gauge theory, we now introduce gauge-invariant $\mathbb{Z}_2$ \textbf{flux operators}.
Assuming $J/h \ll 1$ and going to higher orders in perturbation theory, one can generate a variety of
flux terms for the gauge field in the form of products of $\sigma_i^Z$, which we will represent as $B^{(k)}_i$. 
In order for our LGT to be exactly solvable, we require the charge operators $A_i$ to be the generators 
of local gauge transformations in the resulting $\mathbb{Z}_2$ LGT (meaning $[A_i,H]=0$). 
These can be built from the product of $\sigma^Z$ corresponding to local constraints of the original matter field, i.e. the
independent minimal sets of coupling terms whose product is either identity (or a product of on-site symmetry generators), meaning
they satisfy the condition $\prod_{(i,(a))}O_i^{(a)}[\tau^Z]=1$. 
Crucially, these minimal sets of coupling terms are constructed to be invariant under the original global $\mathbb{Z}_2$ symmetry: 
this ensures that, in the resulting theory, the flux contributions satisfy the condition $[B^{k}_i,A_j]=0 \,\,\forall i,j$. 
Enforcing the condition $B^{k}_i\ket{\psi}=\ket{\psi}$ means restricting the Hilbert space of $\{\sigma\}$ to that of all 
domain wall configurations in the ordered phase of the original Hamiltonian~\cite{Vijai16}. 

In the case of the 2D transverse field Ising (2DTFI) model, any 1,2 or 3-body term would anticommute with some charge operator and would therefore
not be gauge invariant. The minimal set of couplings that correspond to identity is given by the bonds between the four vertices of a plaquette:
\begin{equation*}
\prod_{i\in P} O_i [\tau^Z] = \tau^Z_1 \tau^Z_2\,\, \tau^Z_2 \tau^Z_3 \,\,\tau^Z_3 \tau^Z_4 \,\,\tau^Z_4 \tau^Z_1 = 1.
\end{equation*}
Upon minimal coupling, the matter-field contributions cancel inside these symmetric products, leaving behind the well-known 4-body plaquette term 
for toric code $B_i = \sigma^Z_{i,i+\hat{x}}\sigma^Z_{i,i+\hat{y}} \sigma^Z_{i+\hat{x},i+\hat{x}+\hat{y}}\sigma^Z_{i+\hat{y},i+\hat{x}+\hat{y}}$.
The final Hamiltonian describing our effective LGT takes the form
\begin{equation}\label{eq:H_LGT}
H_{\text{LGT}} = - K\sum_{i,k}B_i^{(k)} - h\sum_i A_i - t \sum_{i,a}\sigma_{i,(a)}^X.
\end{equation}
At this point, the model has been reduced to a pure $\mathbb{Z}_2$ LGT, with $A_i$ enforcing Gauss law 
constraints and $B_i^{(k)}$ encoding magnetic flux terms.

As reviewed in Ref.~\cite{Fradkin79}, gauging acts as a duality map between phases: if the ungauged model admits a symmetry-broken phase
and a trivial symmetric phase, these will be respectively mapped to a trivial paramagnetic phase and a deconfined symmetry-broken phase in
the gauged model. This can be seen from the model described in Eq.~\eqref{eq:H_LGT}: if we were to set $K=h\gg t$, the resulting low-energy effective 
theory would enter the deconfined phase of a $\mathbb{Z}_2$ LGT. In the case of the 2DTFI model, this leads to the well-known toric-code Hamiltonian~\cite{Kogut79}.
If, instead, right after introducing Eq.~\eqref{Hamiltonian_Gauss_law_fix} we were to set $J/h \gg 1$, the emergence of any additional $\sigma_{i,(a)}^X$
excitation would require a non-trivial energy cost, leading to a confined phase.
In the absence of a transverse field, one can exactly diagonalize every term in the Hamiltonian and identify the excitations of 
the resulting model as gapped $\mathbb{Z}_2$ electric charges and magnetic flux excitations (sometimes referred to as $e$ and $m$ monopoles), 
obtained by flipping the eigenvalues of individual $A_i$ and $B_i^{(k)}$.

The way these excitations emerge depends on locality and the structure of the Gauss operators.
Suppose that we were to act on our newly defined LGT with an arbitrary operator $U$ with finite support acting 
on the gauge fields $\sigma_{i(a)}$: the action of $U$ changes the $\mathbb{Z}_2$ charge of a given $A_i$ 
(or $B_i^{(k)}$) operator if $[A_i,U]\neq 0$ ($[B_i^{(k)},U]\neq 0$). The set of sites at which such eigenvalue 
flips occur is therefore determined entirely by the local structure of the Gauss law generators and the support of $U$.
This property has fundamental implications depending on the geometry and topology of the model.
The toric code model is a primary example of this: like all homological codes~\cite{Bombin06}, the
Gauss operators of the toric code are identified with the boundary and coboundary
of the gauge field lattice: on a 2D manifold, this ensures that, for a string-like excitation $U$, 
electric and magnetic excitations are always created as pairs located at the boundary sites $\partial U$ of 
said string (i.e. its endpoints). This feature follows directly from the geometric incidence structure 
of the Gauss law constraints and their local support on the lattice.

While the pairwise creation of excitations follows from the geometrical properties of the Gauss law generators,
the global charge parity conservation law that accompanies it originates independently from the global $\mathbb{Z}_2$ 
symmetry of the original model. Before gauging, the symmetry of the model could be represented by means 
of a global operator written as $\prod_{j\in Q}\tau_j^X$: the symmetry is manifested by the fact that 
each interaction term $O_i^{(a)}$ is invariant under said symmetry, meaning
\begin{equation}
\left[O_i^{(a)}[\tau^Z],\prod_{j\in Q}\tau_j^X\right]=0\,\,\forall i,a.
\end{equation}
In the newly defined $LGT$, $\tau_i^X$ is mapped to the nexus charge operators $A_i$ and $O_i^{(a)}$ is mapped to
a single-spin operator $\sigma_{i,(a)}^Z$. Since the gauging procedure preserves the algebraic 
relations between operators~\cite{Fradkin79}, the operators in the newly defined LGT satisfy the commutation
relation
\begin{equation}\label{eq:commrelGauss}
\left[\sigma_{i,(a)}^Z,\prod_{j\in Q}A_j\right]=0.
\end{equation}
For this to be generally valid for all $i,a$, it is necessary that $\prod_{j\in\Sigma}A_j=1.$ In the toric code,
this implies that $\prod_i A_i=1$, meaning that (1) not all charge operators are independent of each other, (a property which,
as we will see in Sec.~\ref{sec:Topocodes}, is fundamental for topological codes in quantum computing), and (2) the global
electric charge parity must be conserved, i.e., $\prod_i A_i \ket{\psi}=\ket{\psi}$.
This constraint on the nexus-field operators can be viewed as a form of \textit{redundancy} in the definition of the charge operators 
which directly results from the $\mathbb{Z}_2$ symmetry of the original model under periodic boundary conditions. 
Most importantly, because this is a redundancy rather than a symmetry of the gauged model, any degeneracy it introduces in the spectrum 
is \textit{not} associated with spontaneous symmetry breaking and hence cannot be probed by a local order parameter. Instead, depending on the construction, 
such degeneracy may reflect the topological features of the underlying manifold on which the model is defined. 
This procedure can be applied to arbitrary lattice geometries and a wide class of symmetries, but only
in specific cases does it yield a topologically ordered phase.
In the following section, we will show how the interplay of the resulting Gauss law constraints 
inherited from the underlying global symmetries will play a central role in the construction of fracton LGTs.

\subsection{$\mathbb{Z}_2$ fracton order from gauging subsystem symmetries}\label{sec:gauging}

As illustrated by the toric code, the form of the Gauss operators determines the allowed patterns of 
excitation creation and motion; in that case, the Gauss law structure enforces pairwise creation of 
electric charges and permits their free propagation.
More generally, different choices of Gauss law constraints lead to qualitatively distinct mobility 
restrictions on gauge charges. Since fracton phases are primarily characterized by constrained dynamics, 
simply gauging a global $\mathbb{Z}_2$ symmetry of a generic spin model is not sufficient to obtain a fracton theory.
Additional conditions regarding the ungauged model and the Gauss law constraints that follow are required in 
order to restrict the mobility of the resulting gauge-charge excitations. In the literature, people typically distinguish the fracton phases of matter in two distinct families: \\
\,-\,\textbf{Type-I fracton phases}, where elementary excitations either (1) remain mobile along lower-dimensional
submanifolds of the system or (2) are individually immobile, but certain bound states of such excitations can be transported 
across the system by means of finite-support operators. 
\\
\,-\,\textbf{Type-II fracton phases}, in which all nontrivial excitations are strictly immobile and no local or finite-support operators 
are capable of transporting either individual excitations or their bound states without the creation of additional excitations.
\\
The elementary excitations in these models can be separated in \textbf{sub-dimensional particles}, which are mobile only within a particular submanifold and 
\textbf{fractons}, particles that cannot be individually moved but whose bound states might have some degree of mobility, and are 
therefore said to have \textit{fractionalized mobility}. In the following, we focus primarily on models exhibiting the latter type of excitation, 
drawing on the results reported in Ref.~\cite{Vijai16}.

We begin by formulating the necessary conditions that the ungauged spin model must satisfy in order for the gauged LGT to 
admit a topologically ordered phase with fractons. First, it is necessary to clarify what we mean by "topological order": while this defining property 
of models with topological order is given by a ground-state degeneracy that depends solely on the topology of the manifold the system lives on~\cite{Kitaev03},
the fundamental properties that accompany the term and give rise to such behavior are local indistinguishability of the topologically ordered phase and the 
existence of deconfined excitations that accompany it.

Consider the translationally invariant Hamiltonian $H_0$ in Eq.~\eqref{eq:H0} defined in D-dimensional space, with $l$ lattice sites 
and $m$ interaction terms $O_i^{(a)}\left[\tau^Z\right]$ per site ($i\in\{1,\dots,l\},a\in\{1,\dots,m\}$).
Imposing local indistinguishability means that any local operator that commutes 
with $H_{\text{LGT}}$ (in the absence of transverse fields) can only emerge as a product of $A_i$ and $B_i^{(k)}$ operators.

The validity of this statement for $\sigma^Z$-type operators immediately follows from the fact that the $B_i^{(k)}$ 
are explicitly defined as the set of all local $\sigma^Z$-type operators commuting with the nexus charge operators: local indistinguishability
in the $Z$-sector is therefore proven by construction.
This is not the case when considering $\sigma^X$-type operators. The exact characterization of local indistinguishability in the $X$-sector
relies on the algebraic formulation of the gauged LGT as an exact chain complex relating the local matter-field degrees of freedom, 
the interaction terms $\{O_i^{(a)}\}$ and the constraints imposed by the flux operators.
While a more detailed analysis can be found in Ref.~\cite{Vijai16} and a rigorous mathematical derivation is available in Ref.~\cite{Haah13}, 
we will focus on the key geometrical requirements for local indistinguishability and their implications for the construction of 
topologically ordered phases via gauging. To this end, we require an algebraic criterion ensuring that all local operators commuting with 
the Hamiltonian are generated by charges and fluxes.

The ungauged theory admits a set of algebraically independent, finite-support interaction terms $O_i^{(a)}$: these can be represented in the form 
of Laurent polynomials over the finite $\mathbb{F}_2$ field. Identifying a site $i$ as the origin of the unit cell, any lattice displacement described
as a vector in the form $(a,b,c)$ can be described using a polynomial of the form $x^a\,y^b\,z^c$. Each interaction term $O_i^{(a)}$ can be expressed
as a Laurent polynomial by adding up the terms associated to each of its Pauli operators. In this notation, a two-body interaction of the form 
$\tau_i^Z \tau_{i+\hat{x}}^Z$ will correspond to the polynomial $f=1+x$. This gives us a formulation of the interactions within the unit cell which
incorporates the lattice translational invariance of the model. A set of interactions $O_i^{(a)}$ will span a set of polynomials $f_{(a)}$.
The Buchsbaum-Eisenbud theorem~\cite{Buchsbaum73} implies that the mapping between local operators in the ungauged Ising model and a product of
charge/flux operators in the gauged LGT is exact if and only if the system of polynomial equations $f_{(a)}=0$ identifies an algebraic variety (its zero-locus) 
that lives in less than D-1 dimensions~\cite{Haah13,Vijai16}. This is also referred to as the \textit{Codimension condition}, as it implies that the
zero-locus must have codimension $\geq(2)$.

A trivial example is given by the 2D Ising model which, for each site $i$, has two algebraically independent interactions of the 
form $O_i^{(1)}=\tau_i^Z \tau_{i+\hat{x}}^Z$, $O_i^{(2)}=\tau_i^Z \tau_{i+\hat{y}}^Z$: this leads to two polynomials of the form $1+x$, $1+y$ which impose
two constraints on the D-dimensions and the corresponding zero-locus variety is identified by a (D-2)-dimensional manifold.
This condition is also satisfied by all Ising models with nearest-neighbour interactions in D$\geq2$ dimensions.
Physically, this condition is typically satisfied if the ungauged model $H_0$ contains two independent interaction terms per lattice site. This has two relevant implications:
\,(1)\, any local spin-flip operator of the form $\prod \tau^X$ corresponds to a specific product of nexus charge operators $\prod A_i$;
\,(2)\, the ungauged model admits local identity relations in the form of products of $O_i^{(a)}$; these translate into a set of flux operators $B_i^{(k)}$.

However, this is not enough to ensure that any local operator $M=\prod \sigma^X$ which commutes with all $B_i^{(k)}$ is necessarily a product of $A_i$ operators.
Such an operator must correspond to some flip of interaction terms in the ungauged model which satisfies all locality constraints and gives rise to a valid domain wall. 
This implies that it can be represented as a spin-flip operator of the form $\widetilde{M}=\prod \tau^X$: this product will have a corresponding support 
$\text{supp}(\widetilde{M})$, scaling with system size through some effective dimension $d_s<$D, and domain wall 
$\partial\widetilde{M}_X=\{O_i^{(a)}[\tau^Z]\,|\,\{O_i^{(a)},\widetilde{M}_X\}=0\}$. Since the gauging procedure maps any finite-support $\tau^X$ operator to a 
finite product of nexus charge operators, the resulting $\widetilde{M}=\prod \tau^X$ cannot have finite support: this means that $\widetilde{M}$ is strictly 
non-local and, consequently, must have $d_s>0$.

Using the algebraic definition of topological order, it is shown in Ref.~\cite{Haah13_2} that the existence of local $\sigma^X$ 
operators commuting with the Hamiltonian is not excluded if the ungauged model 
admits subsystem symmetries whose supporting manifolds have dimension strictly less than D-1. 
Therefore, a further necessary condition for local indistinguishability and the realization of topological order after gauging is that the ungauged 
D-dimensional model must not possess subsystem symmetries acting on manifolds of dimension smaller than D-1~\cite{Vijai16}. We will refer to this as 
the \textit{Subsym dimensionality condition}.

Let us now focus on the requirement that the ungauged model must satisfy in order to enforce
mobility constraints on the gauge charges. To determine this, we notice that the creation and propagation of excitations in the 
$X$ sector of the resulting LGT is solely determined by the structure of the local interactions in the ungauged model. Preserving the notation of the previous section, 
one can represent any possible operator that can create and move excitations around the lattice as
\begin{equation}
W \equiv \prod_{(i,a)\in \Omega} \sigma_{i,(a)}^Z.
\end{equation}
$W$ creates nexus charges (or annihilates previously existing ones) on all sites $i$ where $\{W,A_i\}=0$. The excitation pattern corresponds to the set of 
spin-flips created by its ungauged counterpart:
\begin{equation}
\widetilde{W} = \prod_{(i,a)\in \Omega} O_i^{(a)}\left[\tau^Z\right]
\end{equation}
It follows that, if the ungauged system were to have two-body interaction terms in each of its unit cells, it would always be possible to move a single charge in 
the corresponding LGT with zero energy cost. In order for fractons to emerge in the gauged LGT, individual nexus charge excitations must be \textit{immobile}. 
This requirement can be rephrased as the \textit{fracton Condition}~\cite{Vijai16}: the topological excitations in the 
gauged model are fractons if and only if the classical spin Hamiltonian contains no binomial terms, i.e. no product of any interaction terms 
$O_i^{(a)}\left[\tau^Z\right]$ can be written as a two-body $\tau^Z$ term.

The \textit{fracton} condition, together with the \textit{Subsym dimensionality} and the \textit{Codimension} condition required for local indistinguishability, 
provide a foundation for the geometrical properties a classical Ising spin model must satisfy for the corresponding LGT to host fractons.
The authors in Ref.~\cite{Vijai16} identified a straightforward method to formulate fracton models by applying the gauging procedure on models with subsystem symmetries. 
In line with the definition provided in Chapter~\ref{chap:subsystem_symmetries}, subsystem symmetries consist of a subextensively large set of independent symmetry operations, each acting on a
set of degrees of freedom whose support $\Gamma$ lives in an extensive $d$-dimensional submanifold of the D-dimensional model that hosts it, with $0<d<$D.

The procedure of gauging subsystem symmetries is identical to the one formulated in the previous section, with a few fundamental differences. First of all,
from each of the  $\mathbb{Z}_2$ subsystem symmetries of the original model, a $\mathbb{Z}_2$ charge conservation law 
involving the nexus-field operators $A_i$ located within the corresponding submanifold $\Sigma$ will follow. This means that, starting from a 
classical model with a subextensive number of $\mathbb{Z}_2$ symmetries, the respective LGT will possess the same number of charge conservation laws.
Secondly, the ground-state degeneracy of the topologically ordered gauged model will no longer be exclusively determined by the genus of the supporting manifold. 
Each subsystem symmetry of the ungauged model can be represented with a corresponding symmetry operator $S_{\Gamma} = \prod_{i\in\Gamma}\tau_i^X$ which
commutes with all interaction terms $O_i^{(a)}$. Since gauging preserves commutation relations, the operators in the dual theory satisfy the commutation relation:
\begin{equation}
\left[\sigma^Z_{i,(a)},\prod_{j\in\Sigma}A_j\right]=0\,\,\,\forall i,a.
\end{equation}
Much like in Eq.~\eqref{eq:commrelGauss}, this condition enforces the redundancy $\prod_{j\in\Sigma} A_j=1$ on the nexus charge operators
within the submanifold $\Sigma$, reducing the number of independent charge operators by the amount of independent subsystem symmetries $k_A$.
If one also considers the redundancy relations between the flux operators, which we will label as $k_B$, the model's topological ground-state degeneracy is given by
$GSD=2^{N - (M-k_A - k_B)}$, where $N$ is the total number of nexus spins and $M$ is the total number of charge and flux operators. In cases where $N=M$, this
leads to a degeneracy that scales with the number of symmetries of the ungauged model as $GSD=2^k$. Therefore, a subextensive ground-state degeneracy of the classical
ungauged model will lead to a subextensive topological degeneracy of the gauged fracton model. It is precisely the combination of local indistinguishability
and the inherent dependence on the geometrical structure of the subsystem symmetries that places fracton models beyond the conventional paradigm of 
topologically ordered phases of matter.

The formulation given above leaves a lot of freedom for exploring different ways to formulate fracton phases of matter through the gauging of subsystem-symmetric models.
$0$-dimensional subsystem symmetries (or, more precisely, gauge "symmetries") cannot be further gauged and are therefore automatically excluded as mentioned in Sec.~\ref{sec:subsims}~\cite{You18}. 
Furthermore, it is worth noting that no topological phases of matter can emerge in one spatial dimension~\cite{Chen11}.
The 2D and 3D models with $d=1$ subsystem symmetries (i.e. linear symmetries) primarily discussed in the literature that satisfy the aforementioned
\textit{fracton} condition do not allow for the existence of a topologicaly ordered phase. 
The simplest realization of such a symmetry on a square or cubic lattice can be achieved using a minimal coupling term which enforces a line symmetry along each lattice direction, 
such as with a 4-body plaquette term in 2D, which leads to the Xu-Moore model~\cite{Xu04}:
\begin{equation}
H_{\text{plaq}} = -J\sum_i \tau_i^Z \tau_{i+\hat{x}}^Z \tau_{i+\hat{y}}^Z \tau_{i+\hat{x}+\hat{y}}^Z - h \sum_i \tau_i^X
\end{equation}
or an eight-body cubic interaction in its 3D analogue.
This model admits independent $\mathbb{Z}_2$ symmetries generated by products of $\tau^X$ along straight horizontal and vertical lines. However, gauging these 
line symmetries does not produce a deconfined phase: at $T=0$, the models at $J/h \gg 1$ and $J/h \ll 1$ turn out to give the same line-symmetry broken phase and 
trivial paramagnetic phase of the ungauged model. In other words, gauging maps the Xu-Moore model into itself, thus yielding no topologically ordered regime. 
In particular, no local magnetic flux operators can be generated from higher-order perturbation theory, reflecting the fact that in models with 1D subsystem 
symmetries, flux operators can only be defined non-locally across entire closed 1D loops.

More generally, translation invariant 2D models composed of commuting Pauli operators cannot give rise to fracton phases in 2D~\cite{Aasen20,Haah21}. 
The remainder of the models studied in the literature either follow a similar description~\cite{Xu04,You18} or admit a phase with subsystem symmetry protected 
topological (SSPT) order that are mapped into an SSPT fracton phase upon gauging~\cite{Yoshida16,Devakul18,You18,Shirley18,Shirley19}. Unlike conventional 
topologically ordered phases, where we expect the symmetries of the model to become local gauge degrees of freedom, SSPT phases preserve their subsystem symmetries 
and are characterized by non-local order and short-range entanglement, with no deconfined excitations.
Conventional topological fracton phases begin to emerge in 3D models characterized by $d=2$ subsystem symmetries. 
While several constructions of such models exist, most published works center around a representative set of models introduced in Ref.\cite{Vijai16}, which correspond to the
gauging of the models discussed in Chapter~\ref{chap:subsystem_symmetries}. To be consistent with the "type-I" and "type-II" categorization of $\mathbb{Z}_2$ fracton models, we will separate them according to the nature 
of their subsystem symmetry, which is reflected in the mobility of the resulting fracton excitations.

\subsection{Type-I Checkerboard fracton model}\label{sec:checkerboard}
There is a variety of 3D fracton models characterized by charge excitations lying at the corners of a two-dimensional surface ~\cite{Chamon05,Vijai15,Vijai16,Shirley18,You20}, 
generated by a \textit{membrane} operator of the form $W= \prod_{i\in A}\sigma^X_i$, where $A$ denotes the set of qubits lying on said surface. In some cases, these can be derived 
by gauging planar subsystem symmetries in classical Ising models. The range of such fracton models is quite broad, as it encompasses different methods for constructing such phases 
and, consequently, different ways in which topological order is presented and different ways in which excitations can propagate through the lattice~\cite{Dua19,Aasen20}.

Among these, two models serve as canonical examples of, respectively, self-dual and non-self-dual type-I
fracton models: the Checkerboard model and the X-cube model on a cubic lattice. Both were originally derived in Ref.~\cite{Vijai16} by applying 
the gauging procedure to Ising models with planar subsystem symmetries satisfying the conditions outlined in the previous section. 
While the X-cube model has been extensively studied in recent years~\cite{Slagle17_2,Brown20,Song22,Zhou22}, following the discussions in Chapter~\ref{chap:subsystem_symmetries}, 
in this section we primarily focus on the derivation and characterization of the Checkerboard model.

Let us take the TIM on the FCC lattice discussed in Sec.~\ref{sec:TIM} as our starting point. Relabeling 
the classical Ising degrees of freedom $S$ with quantum spins $\tau$ and adding a transverse field, the ungauged Hamiltonian takes the following form:
\begin{flalign}\label{eq:TIM_model_rewrite}
H_{\rm TIM} &=  -\sum_{{\bf v}, {\bf a}^\prime} \left(\eta^+ \prod_{\bf a} \tau^Z_{{(\bf v + {\bf a}^\prime )}+{\bf a}} + 
                                                    \eta^- \prod_{\bf a} \tau^Z_{{(\bf v + {\bf a}^\prime )}-{\bf a}} \right) 
                - h \sum_{{\bf v}, {\bf a}^\prime} \tau^X_{{(\bf v + {\bf a}^\prime )}} \\
            &=  - \sum_i \left(\eta^+ \tau_i^Z\tau^Z_{i+\hat{x}+\hat{y}}\tau^Z_{i+\hat{x}+\hat{z}}\tau^Z_{i+\hat{y}+\hat{z}} +
                               \eta^- \tau_i^Z\tau^Z_{i-\hat{x}-\hat{y}}\tau^Z_{i-\hat{x}-\hat{z}}\tau^Z_{i-\hat{y}-\hat{z}} \right)  
                - h \sum_{i} \tau^X_{i}, \nonumber
\end{flalign}
where $i$ is an index running over every site of the FCC lattice. Imposing periodic boundary conditions and setting $\eta^+=\eta^-=\eta$, 
a system of size $L$ is characterized by a set of $3L-3$ independent planar subsystem symmetries.

\begin{figure}[!tb]
\centering
\includegraphics[width =1.\textwidth]{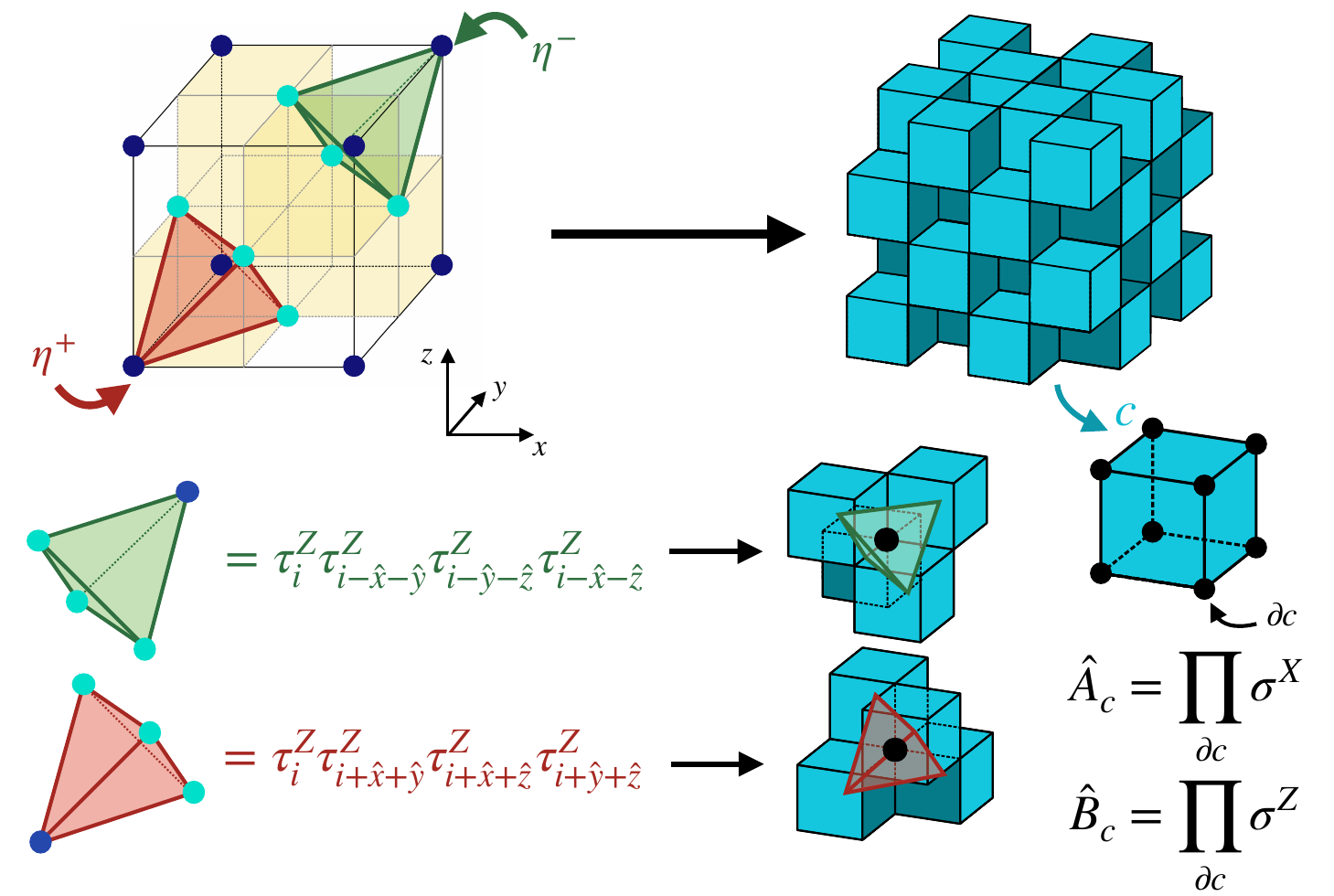}
\caption[Schematic representation of the ungauged tetrahedral Ising model on FCC lattice and the corresponding gauged 3D Checkerboard model on a cubic lattice.]
    {Schematic representation of the ungauged tetrahedral Ising model defined on an FCC lattice and the corresponding gauged 3D Checkerboard model defined on a cubic lattice.
Red and green tetrahedra, respectively residing in the yellow-shaded and unshaded regions of the FCC lattice, are mapped onto nexus-field qubits located at the vertices of the cubic lattice.
The matter-field spins, originally residing at the vertices and face centers of the FCC lattice, are mapped onto cubes arranged in a checkerboard pattern.
Cubes corresponding to matter degrees of freedom participating in a four-body coupling meet at the vertex occupied by the nexus-field qubit associated with that coupling.
Due to the self-duality of the model, each colored cube in the checkerboard lattice supports both an $A_c$ and a $B_c$ operator.
With the FCC unit cell defined as shown and the checkerboard unit cell taken to be a single cube, an FCC lattice of linear system size $\frac{L}{2}$ maps onto a cubic lattice of linear size $L$.}
\label{fig:Checkerboardmapping.pdf}
\end{figure}

Upon gauging, nexus-field degrees of freedom $\sigma_j$ are introduced at the center of each tetrahedral interaction, which we identify with $O_j[\tau^Z]$: since each unit cube in the original
lattice contains only one tetrahedral interaction, the resulting set of $\sigma_j$ lies at the vertices of a cubic lattice with a unit cell that is half the size of the FCC lattice.
Introducing a minimal coupling between the $\eta_j$ interaction terms and the corresponding $\sigma_j$, the resulting Hamiltonian takes the form
\begin{equation}
H= -\eta\sum_j \sigma^Z_j O_j[\tau^Z] - h\sum_i \tau^X_i.
\end{equation}
Each $\tau^Z_i$ participates in eight interactions, corresponding to four $\eta^+_j$ and four $\eta^-_j$-type tetrahedral couplings. Denoting by $\{\eta_j\}:i\in\partial\eta_j$
the set of interaction terms containing site $i$, 
we next define at each site $i$ a nexus charge operator $A_i$ of the form $A_i = \prod_{\{\eta_j\}:i\in\partial\eta_j} \sigma^X_j$.
These correspond to cubic eight-body operators which recur every other unit cell of the cubic lattice in a 3D checkerboard-like fashion as shown in Fig.~\ref{fig:Checkerboardmapping.pdf}.
Introducing the Gauss operator $G_i=\tau_i^X A_i$, we see that $[\tau_i^XA_i,H]=0$: we can then substitute $\tau^X_i$ with the corresponding nexus charge operators $A_i$.
One can see that, for a given site $i$ of the FCC lattice, the product of the interaction terms corresponding to the cubic cells which share $i$ as a vertex corresponds to identity: 
this allows us to identify the flux term of the gauge field as $B_c= \prod_{\{\eta_j\}:i\in\partial\eta_j} \sigma^Z_j$. 
The resulting LGT is the Checkerboard model, which for $\eta=h=1$ is given by
\begin{flalign}\label{eq:Checkerboard_Hamiltonian}
H_{\text{CB}} &= - \sum_c A_c - \sum_c B_c, \\
&\text{where}\,\,\,A_c=\prod_{j\in\partial c} \sigma^X_j,\,B_c=\prod_{j\in\partial c} \sigma^Z_j \nonumber
\end{flalign}
Here, we introduced the index $c$ to label each cubic cell of the newly defined model whose (0,0,0) vertex (the bottom-left-front corner) is located at a lattice site $(x,y,z)$ satisfying the
condition $x+y+z = 0\, (\text{mod} 2)$, with the notation $\partial c$ used to indicate the qubits residing at the vertices of said cubic cell.

For a system of linear size $2L$, stabilizers occupy every other cube, meaning that there are $L$ stabilizers per linear dimension. To satisfy both periodic boundary conditions (PBC) and the checkerboard structure, 
the model is defined only for even linear system sizes. Neighboring stabilizers always share two qubits, guaranteeing that all stabilizers commute with 
each other and that the energy spectrum and ground-state degeneracy can be determined analytically, with the ground states satisfying $\langle A_c\rangle=\langle B_c\rangle=1 \,\,  \forall c$.

It is important to point out that the model is invariant under the exchange $\sigma^X \leftrightarrow \sigma^Z$, which implies that the $A_c$ and $B_c$ 
stabilizers share the same excitation spectrum and dynamics. This property is directly related to the notion of \textbf{self-duality} characterizing the classical Tetrahedral Ising model 
and follows directly from the Wegner prescription that accompanies both the gauging procedure used to derive the quantum model and the construction used to derive the dual Ising model discussed 
in Sec.~\ref{sec:self-duality}. This property will become particularly relevant in the next chapter.

Reflecting the planar symmetries of the ungauged model, the Checkerboard model is characterized by a $\mathbb{Z}_2$ charge conservation law for each $IJ$-plane of cubes 
(where $IJ = xy$, $yz$ or $xz$), accompanied by a redundancy that takes the form
\begin{equation} 
\prod_{c\in IJ-\text{plane}} A_c =\prod_{c\in IJ-\text{plane}} B_c =\mathcal{I}.
\end{equation} 
Any excitation emerging above the ground state must satisfy these $\mathbb{Z}_2$ planar conservation laws. Due to the geometry of the checkerboard distribution, 
the application of an individual $\sigma^X$ or $\sigma^Z$ operator on a given qubit creates four gapped elementary excitations, i.e. fractons, 
associated with the $B_c$ or $A_c$ operators acting on the cubes sharing that qubit.

Since each cubic unit cell lies at the intersection of an $xy$, a $yz$ and an $xz$ plane, the mobility of each fracton excitation is constrained by three independent 
$\mathbb{Z}_2$ planar charge conservation laws. Since moving said fracton by itself in any direction would break these conservation laws, 
\textit{individual fractons are strictly immobile}.

\begin{figure}[!tb]
\centering
\includegraphics[width =1.\textwidth]{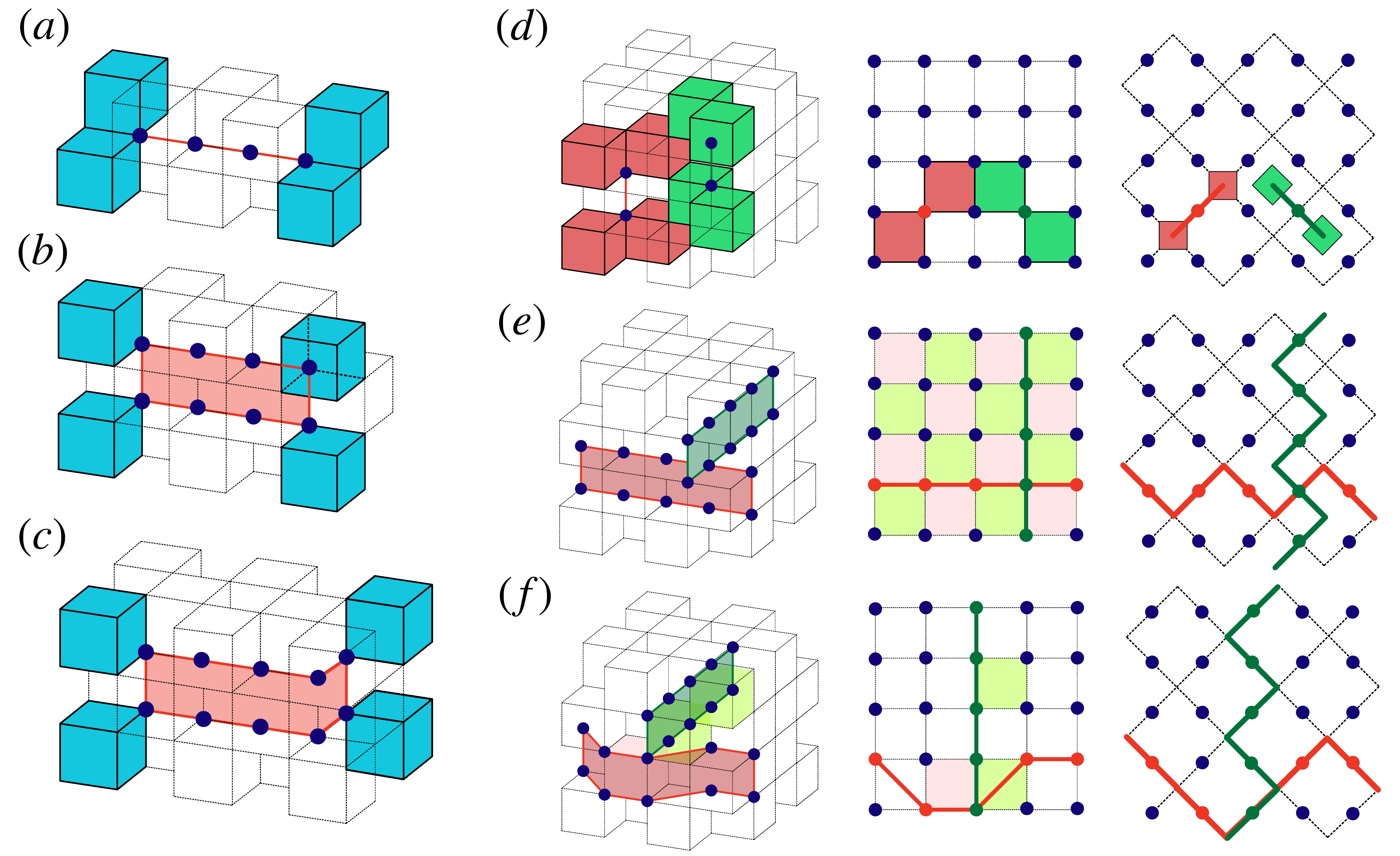}
\caption[Representation of the excitation mobility in the Checkerboard model.]
    {Excitation mobility in the Checkerboard model.
(a) Fracton excitations can be created in groups of four by a single-qubit operator and subsequently separated using a rigid one-dimensional string operator.
(b) Fractons can alternatively be created at the boundaries of a rigid membrane operator lying in an $IJ$ plane.
(c) A fracton dipole with even separation along a given direction $\hat{k}$ and dipole moment aligned along said direction can move freely along $\hat{i},\hat{j}\perp \hat{k}$. 
    This motion results in fracton excitations appearing at the boundaries of a flexible membrane, which may be viewed as a stack of flexible string operators.
(d) For a system of linear size $L=4$, the correspondence between fracton excitations and anyons in the 2D toric code is illustrated by introducing $X$-type fractons 
    on the first and third cube planes and $Z$-type fractons on the second and fourth cube planes (left column). Projecting these excitations onto the $IJ$ plane at $k=2$, 
    where both $\sigma^X$- and $\sigma^Z$-type string operators act (middle column), shows that the resulting configuration is equivalent to the rotated square lattice 
    representation of the 2D toric code, with $X$-type and $Z$-type excitations residing on vertices and plaquettes, respectively (right column).
(e) As in the 2D toric code, these excitations can be braided across the lattice to generate non-local, non-commuting membrane operators that wrap around the system boundaries.
(f) These membrane operators can be continuously deformed by applying $A_c$ operators on the second plane of cubes (light pink) and $B_c$ operators on the third plane of cubes (light green), 
    corresponding to planes in which the $\sigma^X$- and $\sigma^Z$-type membranes reside, respectively. Their projection onto the $k=2$ $IJ$ plane yields a checkerboard lattice 
    (panel~(e), middle column), which maps directly onto the action of vertex and plaquette operators in the rotated 2D toric code.
}    

\label{fig:Checkerboardexcitations.pdf}
\end{figure}

However, bound pairs of fractons exhibit some degree of constrained mobility. This can be seen by treating a pair of fractons 
as a "fracton dipole", characterized by a dipole moment $\textbf{d}$ given by the vector connecting the center of the two corresponding cubes. Such a dipole can move along any 
$\hat{x}$-, $\hat{y}$-, or $\hat{z}$-direction orthogonal to $\textbf{d}$. 
For example, a dipole of adjacent, edge-sharing fractons living in the same $IJ$-plane can be moved with no energy cost along a rigid straight line $\ell\perp IJ$ 
by applying the string operator $\prod_{n\in \ell}\sigma^{X/Z}_n$. Similarly, two fractons in an $IJ$-plane previously 
separated by a string along the $\hat{j}$-direction can be moved in pairs along the $\hat{k}$-direction by applying adjacent parallel $j$-strings, forming a rigid membrane along 
the $JK$-plane generated by a membrane operator $\mathcal{M}$ as shown in Fig.~\ref{fig:Checkerboardexcitations.pdf}.
Moreover, if the fractons were separated this way by stacking an even number of strings, one can obtain fracton dipoles whose dipole moments reside entirely along the $\hat{k}$-axis.
These dipoles can move freely along the $\hat{i}$- and $\hat{j}$-directions by extending the membrane by applying $\sigma^{X/Z}$ to adjacent rigid strings of qubits oriented along $\hat{k}$. 
Alternatively, this motion can be seen as being generated by even stacks of adjacent, flexible $\ell$-strings, forming a flexible membrane that permits motion within the $IJ$-plane, 
as shown in Fig.~\ref{fig:Checkerboardexcitations.pdf}(b, c).

Alternatively, one can leverage the flexibility of the membranes generated by two stacks of $\ell$-strings in any given $IJ$ plane and the planar mobility of the dipole excitations that accompany
it to gain a more intuitive understanding of this degeneracy.
For concreteness, consider a system of linear size $L=4$ along the $\hat{z}$ direction: one may define a non-local operator corresponding to a distance-2 membrane $\mathcal{M}^X$ of $\sigma^X$ operators wrapping around
the lattice across the $\hat{x}$ (or $\hat{y}$) boundaries and a distance-2 membrane $\mathcal{M}^Z$ of $\sigma^Z$ operators wrapping across the $\hat{y}$ (or $\hat{x}$) boundaries. Staggering these two membranes
by one lattice spacing along the $\hat{z}$ direction, they will intersect exactly at one qubit and anticommute. One can deform these membrane operators by applying $A_c$ and $B_c$ 
operators corresponding to cubes lying within the plane in which the membrane is embedded, as well as extend these membranes beyond their original plane.

While these membrane operators act on their corresponding fracton dipole as a whole, their action on the plane in which both $\mathcal{M}^X$ and $\mathcal{M}^Z$ act nontrivially 
effectively corresponds to a braiding of individual $X$-type and $Z$-type fracton excitations. This can be seen by noticing that, while the qubits lie on the vertices of the lattice, the
$X$-type and $Z$-type fracton excitations lie on adjacent planes separated by one lattice spacing in the $\hat{z}$ direction. Projecting these fractons on the $xy$-plane where $\mathcal{M}^X$ 
and $\mathcal{M}^Z$ can meet, their positions relative to the qubits on the plane exactly correspond to those of the rotated 2D toric code~\cite{Bombin07,Stephens14}. 
This shows that the mobility and braiding properties of the fracton dipoles generated by distance-2 membrane operators $\mathcal{M}^X$, $\mathcal{M}^Z$ staggered by one lattice spacing, as shown 
in Fig.~\ref{fig:Checkerboardexcitations.pdf}(e,f), are the same as those in the rotated 2D toric code, as these are primarily determined by the topological features of the planar subsystem manifold. 
This behavior, which can be loosely thought of as a sort of "topological dimensional reduction", is a defining feature of several fracton models with planar subsystem symmetries. 
This relation becomes more intuitive once one starts exploring the landscape of fracton models that can be constructed by coupling layers of 2D toric code through foliation procedures~\cite{Ma17,Pretko20}.

This relation between the Checkerboard model and the 2D toric code has crucial implications for the degeneracy and stability of the fracton phase. 
Having inherited the planar conservation laws and the corresponding redundancies, the Checkerboard model on a lattice of $L^3$ qubits exhibits a ground-state degeneracy of $2^{6L-6}$. 
The most straightforward way one can go from one ground state to another is by applying non-local rigid string operators that wrap across the lattice in any
of the three spatial directions. Any such closed string operator $S^{X/Z}$ would commute with all $A_c$ and $B_c$, as it would share two qubits with every cube it crosses, thus representing
a non-local operator which satisfies the condition $[H,S]=0$. 

The existence of these string-like operators plays a central role in determining the finite-temperature behavior of the model.
At any $T>0$, topological order in the 2D toric code is lost due to the proliferation of local perturbations which break the non-local order. More generally, topological order is disrupted whenever the coupling to a thermal 
bath can lead to thermal processes (occurring as a sequence of local Pauli operators) which result in the implementation of non-local operators that mix the different topological ground-state sectors. 
Such processes impose a minimum energy barrier $\Delta E$ that must be surpassed in order for local perturbations to give rise to non-local operators.

More formally, let $\ket{\psi_0}$ be a ground state of the model 
and $\ket{\psi_1}=S\ket{\psi_0}$ another, topologically distinct ground state, where $S$ is any of the non-local operators connecting different topological sectors. 
We can define the set of all possible dynamical processes that generate, move
and eventually annihilate excitations such that the system evolves from the state $\ket{\psi_0}$ to $\ket{\psi_1}$ as $\mathcal{D}_{\bar{S}}$. We then define the energy barrier as 
\begin{equation}
\Delta E = \min_{S,\mathcal{D}_{\bar{S}}}\,\left(\max_k (E_{(k)}-E_0)\right),
\end{equation}
where $E_0$ is the ground-state energy and $E_{(k)}$ is the energy of the $k^{\text{th}}$ intermediate state of a given dynamical process in $\mathcal{D}_{\bar{S}}$.
For a system coupled to a bath at a sufficiently low temperature $T$, collective non-local perturbations remain exponentially suppressed and can occur only through sequences of local perturbations.
At fixed energy barrier $\Delta E$, the rate at which these events occur is enhanced by the number of distinct dynamical processes which connect the two ground states while satisfying the condition $\max_k(E_{(k)}-E_0)\leq \Delta E$. 
Taking this into account, the timescale $\tau$ for processes with energy barrier $\Delta E$ to take place is determined by the Arrhenius law~\cite{Roberts20,Aitchison24}
\begin{equation}
\tau \sim e^{\beta (\Delta E -T\Delta S)}.
\end{equation}

A fundamental requirement for finite-$T$ stability of topological order is that the lifetime of such non-local processes diverges with the system size. 
This requires that the free-energy cost associated with thermally activated dynamical processes implementing non-local operators grows without bound as the system size 
increases, with the suppression of such events caused by the diverging $\Delta E$ dominating over the entropic proliferation of distinct processes~\cite{Haah13,Bombin13}.

An immediate consequence of this result is that any topologically ordered phase whose ground states can be connected by non-local string-like operators has a finite energy barrier independent of system size
and, consequently, cannot admit topological order at any finite temperature.
For such operators, the energy cost comes solely from the initial creation of the excitations at the endpoints of the string, while their separation can always be performed through a succession of local
single-qubit operators that can be performed with no additional energy cost. This mechanism applies to all translationally invariant topologically ordered models with one particle per site in 2D and 3D, such as 
the toric code and type-I fracton models such as the X-cube and Checkerboard model.

\subsection{Type-II Haah's code fracton model}\label{sec:Haah}

In contrast to Type-I fracton phases, whose restricted mobility of excitations is compatible with the presence of rigid string- or membrane-like non-local operators, 
Type-II fracton models are characterized by a more drastic departure from conventional topological order: they admit no string-like non-local operators of any kind. 
In these systems, point-like excitations are strictly immobile, and any operator capable of separating or transporting them necessarily has support on a fractal 
subset of the lattice. The paradigmatic example of such behavior is Haah’s cubic code~\cite{Haah12}, a three-dimensional stabilizer code whose logical operators 
exhibit fractal geometry rather than string or membrane structure.

From a constructive viewpoint, Type-II fracton phases can be understood as arising from the gauging of classical models endowed with fractal subsystem symmetries.
Unlike planar or linear subsystem symmetries, fractal symmetries act nontrivially on self-similar sets of sites that lack a well-defined lower-dimensional manifold 
description. Gauging these fractal symmetries yields quantum phases in which gauge charges inherit fractal conservation laws from the underlying fractal symmetries of the ungauged model. 
A consequence of such conservation laws is that no operator with bounded cross section can separate a pair of point excitations without producing additional defects. 
This obstructs string- or membrane-like logical operators and forces the minimal nontrivial logicals to have fractal support.

The presence of $X$-type and/or $Z$-type string-like non-local operators in topologically ordered 3D codes is not limited to the Checkerboard model: most known 3D 
models characterized by topological order,
such as the 3D toric code~\cite{Nussinov08}, the 3D color code model~\cite{Bombin07_2}, the Chamon model~\cite{Chamon05} and, most importantly, in all type-I 
fracton models derived by planar subsystem symmetry gauging
such as the aforementioned Checkerboard model and X-cube model~\cite{Song22}. Focusing on the latter fracton models, the presence of topological string-like operators is a direct 
consequence of the geometric locality of charge and flux operators required to achieve local indistinguishability inherited from 2D models~\cite{Bravyi09}. 

This constraint can be avoided if one were to depart from the more conventional planar subsystem symmetries and instead focus on other types of symmetries 
whose support still satisfies the \textit{codimension} condition: fractal symmetries. To see what sort of models one may obtain in the presence of fractal-like 
conservation laws, we can apply the gauging procedure on the Fractal Ising model introduced in Sec.~\ref{sec:FIM}. Substituting
the classical spins $S$ for quantum qubits $\tau$ with an added transverse field, its interaction Hamiltonian on the cubic lattice becomes
\begin{flalign}\label{eq:FIM_model_rewrite}
	H_{\rm FIM} = -\sum_i\left(\eta^+ \tau^Z_{i}\tau^Z_{i+\hat{x}}\tau^Z_{i+\hat{y}}\tau^Z_{i+\hat{z}} + \eta^- 
            \tau^Z_{i}\tau^Z_{i+\hat{x}+\hat{y}}\tau^Z_{i+\hat{y}+\hat{z}}\tau^Z_{i+\hat{x}+\hat{z}}\right) - h\sum_i \tau^X_{i}.
\end{flalign}
Unlike the Checkerboard model, the model is characterized by two distinct interaction terms in each unit cell: upon gauging, these are mapped into two distinct nexus 
field degrees of freedom $\sigma_j$, $\mu_j$ residing at the vertices of the dual cubic lattice of the original model. For simplicity, let us identify the nexus 
field label $j$ running over the cubes of the dual lattice with the matter-field label $i$ (this means labelling $\sigma_i$ as the interaction term 
residing in the cube that has $\tau_i$ at its $(0,0,0)$ corner and, in the dual lattice, identifying $\tau_i$ as the center of the
cube that has the spin $\sigma_i$ at its $(111)$ corner).

After introducing a coupling between the newly added nexus field and the corresponding interactions, we get
\begin{equation}
H = -\sum_i\left( \eta^+ \sigma_i^Z \tau^Z_{i}\tau^Z_{i+\hat{x}}\tau^Z_{i+\hat{y}}\tau^Z_{i+\hat{z}} + \eta^- \mu_i^Z \tau^Z_{i}\tau^Z_{i+\hat{x}+\hat{y}}
        \tau^Z_{i+\hat{y}+\hat{z}}\tau^Z_{i+\hat{x}+\hat{z}}\right) - h\sum_i \tau^X_{i}.
\end{equation}
Similarly to the Checkerboard model, each $\tau_i^Z$ participates in eight interactions: identifying the $\eta^+$-type and $\eta^-$-type interaction terms at site $i$ 
as $O_i^1$ and $O_i^2$ respectively, $\tau_i^Z$ is shared in $4$ $O^{(1)}$ interactions, which correspond to the $\sigma$-type nexus-field degrees of freedom 
$\sigma_i$, $\sigma_{i-\hat{x}}$, $\sigma_{i-\hat{y}}$, $\sigma_{i-\hat{z}}$, and four $O^{(2)}$ interactions corresponding to the $\mu$-type nexus-field spins 
$\mu_i$, $\mu_{i-\hat{x}-\hat{y}}$, $\mu_{i-\hat{y}-\hat{z}}$, $\mu_{i-\hat{x}-\hat{z}}$. From this, it is possible to define the nexus charge operators
\begin{equation}
A_i=\sigma_i^X\,\sigma_{i-\hat{x}}^X\,\sigma_{i-\hat{y}}^X\,\sigma_{i-\hat{z}}^X\,\mu_i^X\,\mu_{i-\hat{x}-\hat{y}}^X\,\mu_{i-\hat{y}-\hat{z}}^X\,\mu_{i-\hat{x}-\hat{z}}^X.
\end{equation}
Imposing the local Gauss law constraints, we can substitute $\tau_i^X$ with the newly defined $A_i$, which inherit the fractal structure of the subsystem symmetries from 
the original model in the form of $\mathbb{Z}_2$ fractal charge conservation laws.

The flux operators are identified at the eighth-order expansion in the interaction terms in the form of
\begin{equation}
B_i=\sigma_{i-\hat{x}-\hat{y}-\hat{z}}^Z\, \sigma_{i-\hat{x}}^Z\, \sigma_{i-\hat{y}}^Z \,\sigma_{i-\hat{z}}^Z \,\mu_{i-\hat{x}-\hat{y}-\hat{z}}^Z 
        \,\mu_{i-\hat{x}-\hat{y}}^Z \,\mu_{i-\hat{y}-\hat{z}}^Z \,\mu_{i-\hat{x}-\hat{z}}^Z.
\end{equation}
Associating each operator with the cube $c$ supporting it, one obtains Haah’s cubic code~\cite{Haah11}:
\begin{equation}\label{eq:Haah's_Hamiltonian}
H_{\text{Haah}} = -\sum_c A_c - \sum_c B_c,
\end{equation}
where the index $c$ runs through every cubic cell of the new lattice.

The first notable feature of this newly defined LGT is the presence of a $\mathbb{Z}_3$ rotational symmetry about the $\hat{x}+\hat{y}+\hat{z}$ axis, 
together with a mirror $\mathbb{Z}_2$ symmetry with respect to the plane normal to $\hat{x}-\hat{y}$.
\begin{figure}[!b]
\centering
\includegraphics[width =1.\textwidth]{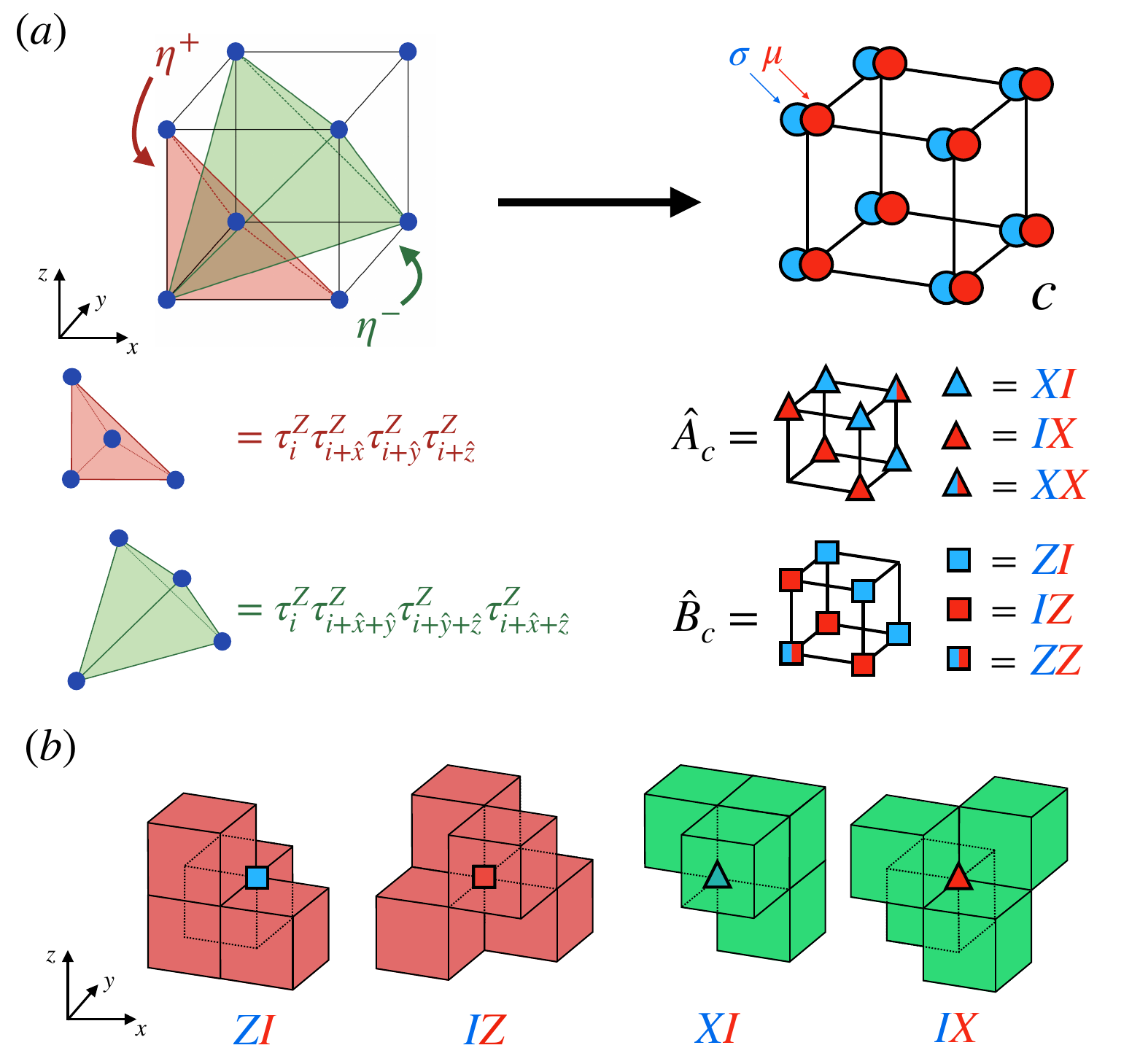}
\caption[Schematic representation of the derivation of Haah's cubic code and its excitation patterns.]
{(a) Schematic representation of the ungauged Fractal Ising model on a cubic lattice and the corresponding gauged Haah's cubic code. The two interaction terms associated with each cubic unit 
cell are mapped onto a pair of qubits, $\sigma$ (red), $\mu$ (blue), located at the vertices of the dual cubic lattice. Gauging gives rise to the eight-body charge and flux operators $A_c$ and $B_c$.\\
(b) Acting with a single-qubit Pauli operator (from left to right: $\sigma^{Z},\mu^Z,\sigma^X,\mu^X$) on a ground state creates four fractons arranged in the same tetrahedral geometry of the interactions in the original ungauged model. 
The positions of the charge excitations relative to the $X$-type Pauli operators that create them is identical to that of the flux excitations created by $Z$-type operators up to spatial inversion.}
\label{fig:HaahsCodemapping.pdf}
\end{figure}
In addition, the model is invariant under the simultaneous application of spatial inversion and the exchanges $\sigma^X \leftrightarrow \mu^Z$ and 
$\sigma^Z \leftrightarrow \mu^X$. Since the stabilizer operators are agnostic with respect to whether a $\sigma$ or $\mu$ Pauli operator is applied, 
the resulting $X$-type and $Z$-type excitation maps are identical up to spatial inversion. We can therefore say that Haah's code is self-dual under a 
generalized $X \leftrightarrow Z$ transformation in the same sense as the Checkerboard model, a feature which will be revisited in Chapter~\ref{chap:fractons_in_QEC}.

As discussed in Sec.~\ref{sec:FIM}, for linear system sizes $L=2^n, n\in\mathbb{N}$ with PBC, the ungauged model has a set of $2L-1$ 
independent fractal subsystem symmetries generated by an operator of the form $S=\prod_{i\in\mathcal{F}} \tau_i^X$ and defined on the set of Ising spins 
distributed in a 3D fractal-like support $\mathcal{F}$ constructed with the Sierpinski rule. From each of these symmetries, the gauged model inherits 
a $\mathbb{Z}_2$ charge conservation law on the subsets of cubic charge operators $A_c$ corresponding to the matter-field qubits residing in $\mathcal{F}$.
Due to the generalized duality, a similar subset of $B_c$ operators residing on the corresponding inverted manifold $\tilde{\mathcal{F}}$, so that
\begin{equation}
\prod_{c\in\mathcal{F}}A_c = \mathcal{I},\,\,    \prod_{c\in\tilde{\mathcal{F}}}B_c = \mathcal{I},
\end{equation}
which, for a cubic lattice of size $L^3$, leads to $4L-2$ total redundancies and a $2^{4L-2}$-fold topological ground-state degeneracy.

A single-qubit $X$ and $Z$ Pauli operator creates four excitations on each lattice site as shown in Fig.~\ref{fig:HaahsCodemapping.pdf}. 
It is immediate to see that these excitations are immobile fractons, with a much more restricted mobility compared to those generated through planar 
subsystem symmetry gauging, as neither individual nor bound pairs of fractons can be moved with the action of a product of Pauli operators without 
creating additional excitations. While their dynamics was originally characterized in full detail in Ref.~\cite{Haah11}, one can develop an understanding 
of the fracton mobility in Haah's code by leveraging the fact that excitations map generated by single-qubit Pauli operators is described by the same Laurent 
polynomials used to represent the original couplings of the ungauged model and the charge and flux operators in the gauged model. To see this, we first 
define $R = \mathbb{Z}_2[x,y,z][\Gamma]$ as the group ring of polynomials over $\mathbb{Z}_2$ with periodic boundary conditions in each direction, with 
$\Gamma=\{x^iy^jz^k|i,j,k\in\mathbb{Z}_L\}$. Next, we use the Laurent polynomial $f(x,y,z)\in R$ to represent any local operator $O$ in a translation 
invariant fashion, constructed by summing the set of monomials $x^iy^jz^k$ corresponding to the real space displacement $(i,j,k)$ of qubits comprising 
the support of $O$ with respect to a unit cell origin site $(0,0,0)$: $f(x,y,z) = \sum_{(i,j,k)\in\text{supp}(O)} x^i y^j z^k$. Since the $B_c$ operators 
consist of both $\sigma^Z$ and $\mu^Z$ Pauli operators as shown in Fig.~\ref{fig:HaahsCodemapping.pdf}, we define the Laurent polynomial
\begin{equation}
f(x,y,z)=1+x+y+z,\,\,\,g(x,y,z)=1+xy+yz+xz
\end{equation}
to describe the set of $\sigma^Z$ and $\mu^Z$ that appear in the $B_c$ stabilizer at a given cubic cell. The same polynomials can be used to describe 
the composition of charge operators $A_c$ in terms of $\sigma^X$ and $\mu^X$ Pauli operators since they follow the same structure up to lattice inversion. 
Comparing these expressions with Eq.~\eqref{eq:epsilon} and with the excitation patterns shown in Fig.~\ref{fig:HaahsCodemapping.pdf}, 
one finds a direct correspondence between the structure of the operators, the coupling geometry of the ungauged Fractal Ising model and, most importantly 
the excitation geometry, that is, the relative positions of the fractons that are created by $\sigma^{X},\mu^{X}$ (or $\sigma^Z,\mu^Z$) operators, and how the 
action of these operators on different lattice sites propagates the excitations throughout the lattice.
This equivalence allows us to apply the same method used to derive the fractal symmetries of the ungauged model in order to identify 
the non-local operators that connect topologically distinct ground states of Haah's code. This implies that, while individual excitations are strictly immobile, 
a collective motion is possible by moving them along a fractal manifold by applying, at each step, a set of single-qubit Pauli operators determined by the 
Sierpinski rule: this gives an iterative procedure to separate the excitations to arbitrary distances. 
Performing this $L$ times allows the excitation pattern to wind non-trivially around the periodic boundary conditions and return to their point of origin, where 
the excitations annihilate. The resulting operator acts non-trivially within the ground-state manifold while creating no excitations. As a consequence, 
the non-local logical operators of Haah’s code possess the same fractal support as the subsystem symmetries of the classical ungauged model.

This model represents a paradigmatic example of type-II fracton models, as it was originally conceived to construct a topologically ordered model 
with no string-like logical operators~\cite{Haah11,Aitchison24}. Instead, the presence of fractal-like operators implies that, for a given separation $d$ between 
fractons, their separation cannot be further increased by applying a bounded set of Pauli operators, in contrast to models with string-like non-local operators such 
as the Checkerboard model. Moreover, the Sierpinski rule ensures that there exists no sequence of single-qubit operators capable of moving an individual 
fracton or a fracton dipole without incurring an energy cost. Starting from a minimal separation $d=1$ upon creation, the application of Pauli 
operators determined according to the Sierpinski rule allows fractons to be separated by distances $d=2^j$, where $j\in\mathbb{N}_0$. Owing to the fractal geometry, 
the total number of Pauli operators $w$ needed to apply a non-local operator connecting different ground states on a cubic code of size $L=2^n$ scales superlinearly 
as $w=O(L^\alpha)$ with $\alpha>1$. 
Numerical studies have shown that the energy barrier associated with such processes scales as $\Delta E=O(\log L)$
~\cite{Bravyi11,Aitchison24}: starting from a ground state, the number of excitations one must create to construct a set of fractons with separation $d$ by 
sequentially applying single-qubit Pauli operators scales sublinearly with $d$. This also implies that any sequence of local Pauli operations that maps one ground 
state to a topologically distinct ground state must pass through intermediate configurations containing a number of excitations that diverges logarithmically with $L$.

The thermodynamic implications of this logarithmic energy barrier were rigorously analyzed in Ref.~\cite{Bravyi11,Haah13}, where it was demonstrated that the 
competition between the logarithmic energy barrier and the combinatorial proliferation of thermally accessible excitation configurations leads to a memory lifetime 
that grows polynomially with the system size, $\tau(L,T)\sim L^{c\beta}$, although only for sufficiently low temperatures and only up to a temperature-dependent 
cutoff length $L^*(\beta)\lesssim e^{\beta/3}$, beyond which entropic effects dominate and the lifetime no longer increases with system size.
In the thermodynamic limit, the sublinear growth of $\Delta E$ is therefore insufficient to prevent the eventual mixing of topologically distinct ground states at 
finite temperature, implying that the fracton topological order of Haah's code does not persist at finite $T$~\cite{Bravyi11,Haah13}. Nevertheless, this represents 
the first example of a topologically ordered 3D model without string-like non-local operators, which results in the existence of a finite-$T$ regime in which, 
below a certain temperature-dependent size $L^*(\beta)$, the characteristic lifetime of the fracton phase increases with system size, a qualitative difference in the robustness 
against thermal fluctuations when compared to conventional 3D topologically ordered models.


\section{Concluding Remarks}

This chapter had two complementary goals. The first was to clarify the broad notion of fractonic behavior by emphasizing how restricted fracton dynamics
follows from the emergence of conservation laws that extend beyond ordinary charge conservation. Despite the wide variety of models and phenomena associated 
with the term, we summarized their distinctive properties through a categorization into $U(1)$ and $\mathbb{Z}_2$ subfamilies, which are naturally described in the form of
symmetric-tensor and $\mathbb{Z}_2$ gauge theories, respectively.  While the proposed categorization does not account for other, more general fracton LGTs such as aperiodic, 
non-abelian, and SSPT fracton phases~\cite{Song19,Prem19,Prem19_3,Shirley19,Tu21}, it highlights the fundamental contrast between the broad landscape of fracton models 
discussed in the literature and the subclass of phases that are conventionally understood to possess fracton topological order.

The second goal, which is most directly tied to the broader narrative of this thesis, was to connect the subsystem symmetries to gapped fracton phases with topological order. 
Following our results from Chapter~\ref{chap:subsystem_symmetries}, this chapter draws out the connection between the 
subsystem-symmetric phases of matter and the construction of fracton topological order through the generalized gauging procedure explored in Ref.~\cite{Vijai16}. 
Focusing on the Checkerboard model and Haah's code, we highlight how different classes of subsystem symmetries in the underlying classical spin 
models give rise to distinct flavors of fracton topological order, whose properties are strongly shaped by the symmetry structure, most notably by the presence 
or absence of string-like logical operators. The ground-state degeneracy and excitation mobility of the resulting fracton phases were shown to depend nontrivially on
both the topology of the subsystem supporting the original symmetries and the geometrical features of the model 
and its interactions. We also discussed excitation dynamics and showed how it directly follows from the geometrical properties of the subsystem symmetries 
in the respective ungauged models.

Beyond their status as unconventional topologically ordered phases of matter, fracton models have attracted significant attention in the context of quantum error correction,
primarily as candidate quantum memories. This interest emerged for a variety of reasons: first of all, their topological ground-state degeneracy allows for the encoding of a 
subextensive number of logical qubits, a tantalizing property to take into consideration when trying to develop scalable quantum memories. Another point of interest is that
the restricted mobility of excitations raises the possibility that local errors may be unable to propagate freely across the system, a feature that could lead
to a more efficient error correction. This is further highlighted by the absence of string-like logical operators and the presence of non-trivial 
energy barriers in Haah's code, which have been proposed as mechanisms that could further enhance its robustness against local noise.
Following these motivations, the next chapter investigates the Checkerboard model and Haah's code in the presence of noise
and to assesses their viability as stabilizer quantum memories.


%% file: mainmatter/theory/fractons_in_QEC.tex
\chapter{Fracton codes in Quantum Error Correction}
\label{chap:fractons_in_QEC}

The previous chapters explored classical subsystem-symmetric Ising models, their gauged quantum counterparts, and examined how symmetry geometry
shapes the excitation structure and mobility constraints of fracton phases. In this chapter, we will look at the gapped fracton models introduced earlier within 
the framework of quantum error correction (QEC).

The connection between topologically ordered phases of matter and quantum error correction has been a cornerstone of quantum computation since the earliest stages of its development~\cite{Preskill98,Nielsen00,Delgado02}.
Such models are characterized by Hamiltonians composed of local commuting constraints, whose ground-state subspace can be used to encode quantum information in the form 
of non-local degrees of freedom, while physical qubit errors manifest as localized excitations. Local constraints, excitations, and topologically 
non-trivial operators acquire a direct interpretation in terms of stabilizers, syndrome measurements and logical operators.

The investigation of fracton stabilizer models within this context began with the error analysis of the X-cube model~\cite{Song22}, which was shown to achieve a code-capacity threshold higher 
than those reported for previously studied three-dimensional topological stabilizer codes under independent Pauli noise, including the 3D toric code and the 3D color code. 
In this chapter, we complement this result by studying the behavior of the self-dual Checkerboard model and Haah's code presented in Chapter~\ref{chap:fracton_models},
investigating whether these codes exhibit comparable or better error resilience in the presence of noise.

A central theme of this chapter is the structural relation between CSS stabilizer codes and duality in classical Ising models. 
Rather than addressing the full dynamical problem of memory lifetimes under circuit-level noise, we focus on optimal code-capacity thresholds under independent Pauli noise, 
namely the maximum physical error rate below which decoding succeeds with probability approaching one in the thermodynamic limit. 
In this regime, decoding admits a direct formulation in terms of the statistical mechanics of classical Ising models with quenched disorder~\cite{Dennis02,Nishimori07}. 
In particular, we argue that, for classes of CSS codes whose $X$- and $Z$-decoding problems are related by a generalized Kramers--Wannier duality, the corresponding error 
thresholds are expected to nearly saturate $H(p_X^{\mathrm{th}}) + H(p_Z^{\mathrm{th}}) \le 1$. This duality-based mechanism provides a unifying explanation for the 
near-saturation commonly observed in conventional topological codes.

With this perspective in place, we turn to fracton stabilizer codes in Sec.~\ref{sec:Fractons_in_QEC}. 
The self-dual Checkerboard model and Haah's code do not fit naturally into the conventional topological and homological code paradigm shared by the models studied so far and exhibit strongly 
constrained excitation dynamics and unconventional logical-operator geometry. Their constrained excitation dynamics and unconventional logical-operator geometry make them a stringent test of 
whether the duality-based threshold-saturation picture extends beyond conventional topological codes. We numerically estimate the optimal threshold for the Checkerboard code and assess its compatibility with 
the generalized duality prediction. Finally, we discuss the expected behavior of Haah's code in light of its finite-size constraints and fractal logical-operator structure.

The chapter is organized as follows. In Secs.~\ref{sec:FTQC} and~\ref{sec:CSS}, we review the stabilizer formalism and the CSS construction, with emphasis on geometric 
and homological codes. In Sec.~\ref{sec:SMmapping}, we derive the quantum-to-classical statistical-mechanical mapping that relates decoding to a disordered Ising model. 
In Sec.~\ref{sec:threshold-KW}, we formulate the duality-based fixed-point condition underlying the threshold relation. Finally, in Sec.~\ref{sec:Fractons_in_QEC}, 
we apply this framework to the Checkerboard and Haah fracton codes, presenting a numerical estimate the optimal threshold of the Checkerboard code, and assess its 
compatibility with the duality-based threshold relation. We then use these results to discuss the expected behavior of Haah's code.

\section{Fault-tolerant Quantum Computation}\label{sec:FTQC}

\subsection{Primer on QEC and the stabilizer formalism}
A central challenge in quantum computation is the development of efficient and robust
quantum error-correction (QEC) methods capable of suppressing error accumulation~\cite{Shor96,Knill97}.
The central goal of any QEC protocol is to detect and correct quantum errors without disturbing the encoded information.
In the case of individual qubits, whose state is 
most commonly described by a vector in the Bloch sphere representation and of the form 
$\ket{\psi}=\alpha\ket{0}+\beta\ket{1}= \cos\theta\,\ket{0}+e^{i\phi} \sin\theta\,\ket{1}$,
the action of single-qubit physical errors can be described through a quantum channel which can be expanded in the Pauli basis as
\begin{equation}
E=e_0 I + e_1 X + e_2 Y + e_3 Z.
\end{equation}
More generally, an arbitrary noise channel can be represented by Kraus operators, each of which can be expanded in the Pauli basis.

A single qubit is susceptible to a continuum of possible errors, and one cannot directly measure 
its state in order to diagnose them, since such a measurement would collapse the state and destroy the encoded quantum information. 
Together with the no-cloning theorem, this rules out any straightforward error-diagnosis procedure. 
One then needs to find a non-destructive method to detect and correct errors~\cite{Nielsen00}.
The key insight underlying modern QEC is that, with a suitable
encoding, we can significantly simplify the type of errors one has to correct. 
Instead of tracking a continuum of possible error channels on individual qubits, 
we can borrow a page from classical error correction and encode 
the logical quantum information onto a larger set of qubits, thereby reducing error correction to the protection of a distinguished subspace $\mathcal Q$ and its codewords.
The basic idea, introduced by Peter Shor in 1995~\cite{Shor95} and formalized 
by Knill and Laflamme in 1997~\cite{Knill97}, is to encode information into a distinguished subspace $\mathcal Q$ of the many-body Hilbert space. 
The essential idea is to define $\mathcal Q$ as the joint eigenspace of a set of commuting observables and to diagnose errors 
by measuring them non-destructively. After an error occurs, these measurements project the state into a syndrome subspace 
labelled by a discrete set of eigenvalues, thereby reducing the continuum of possible physical errors to a discrete set.

To make this more concrete, suppose we want to encode $k$ qubits of information in $n$ physical
qubits (i.e. any sort of physically realizable two-level system), so that 
$\text{dim}\left(\mathcal{H}\right) = 2^{n},\,\text{dim}\left(\mathcal{Q}\right) = 2^{k}$.
The code space $\mathcal{Q}\subset \mathcal{H}$ is defined as the simultaneous $+1$ eigenspace of
a family of commuting projectors $\{P_i\}$, meaning that any encoded logical state $\ket{\psi}$ satisfies
$P_i \ket{\psi} = \ket{\psi} \forall i$. In stabilizer codes, one starts from a commuting set of $n$-qubit Pauli 
operators $\mathcal{S}=\{g_i\}$ called stabilizer generators, with each $g_i$ defining a projector $P_i=\frac{1}{2}(I+g_i)$. 
The code space $\mathcal Q$ is then defined as the joint $+1$ eigenspace of all $g_i$ and serves as the 
effective Hilbert space of the encoded logical degrees of freedom.

The set of stabilizers $g_i$ generates the stabilizer group 
$\mathcal{S} = \langle g_1,\dots,g_m\rangle \subset \mathcal{P}_n$, where $\mathcal P_n$ denotes the $n$-qubit Pauli group.
From this, one can define the normalizer group
$N(\mathcal{S}) = \{ P \in \mathcal{P}_n \,\mid\, [P,S]=0 \ \text{for all } S\in\mathcal{S}\}$ as the set of all Pauli 
operators that commute with every operator contained in the stabilizer group.
While elements of $N(\mathcal{S})$ preserve the code space $\mathcal{Q}$, those in 
$N(\mathcal{S})\setminus \mathcal{S}$ act non-trivially within $\mathcal{Q}$ and define the
\emph{logical operators} of the code. 

When a physical error $E$ acts on a code state $\ket{\psi}$, if $[E,g_i]\neq0$ for 
at least one stabilizer generator, the corrupted state $E\ket{\psi}$ will lie in a different stabilizer eigenspace 
(subspace of $\mathcal{H}$ with eigenvalues given by the different  syndrome sector labelled by 
the eigenvalues $\{s_i\}=\{\pm 1\}$). Measuring the stabilizer syndrome $\{s_i\}$ is 
equivalent to applying the set of projectors $P_{i,s_i} = \frac{1}{2}(I + s_i g_i)$, 
projecting the state onto $(\prod_i P_{i,s_i})E\ket{\psi}$. 
This procedure reduces error diagnosis to a discrete syndrome $\vec s$, from which one chooses a recovery 
operator that returns the state to the code space, up to stabilizers and possible logical ambiguity~\cite{Gottesman98}.

Within this framework, our ability to detect errors and preserve the encoded quantum information depends 
primarily on the choice of stabilizer generators. To be able to distinguish 
two error operators $E_a,E_b$, the syndrome measurement must be able to tell apart the action of $E_a$ on 
a given code basis state $\ket{\psi_i}$ from the action of $E_b$ on any other basis state $\ket{\psi_j}$. 
A quantum code $\mathcal Q$ is able to distinguish any pair of errors $E_a$, $E_b$ belonging to the
set of distinguishable error configurations $\mathcal{E}=\{E\}$ if, for any $\ket{\psi_i},\ket{\psi_j}\in\mathcal Q$, 
it satisfies the Knill-Laflamme condition:
\begin{equation}
\bra{\psi_i}E_a^\dagger E_b\ket{\psi_j}
= C_{ab}\,\braket{\psi_i}{\psi_j},
\label{eq:Knill-Laflamme}
\end{equation}
where $C_{ab}$ is some Hermitian matrix~\cite{Knill97}.
Equivalently, $P E^{\dagger}_a E_b P = C_{ab}P$, where $P$ is the projector onto $\mathcal Q$.
This condition ensures that errors either map the code space into mutually orthogonal 
syndrome subspaces or act identically on all logical states within the code space. 
As a result, correctable errors fall into equivalence classes: errors that have the same action on the code space can be reversed by the same recovery operation.

Since logical operators commute with all stabilizers, they do not change the measured 
syndrome and are therefore undetectable by construction. Errors that contain a nontrivial 
logical operator correspond to \emph{logical errors}, as they change the encoded information in a 
way that cannot be distinguished, using syndrome information alone, from the action of an intended 
logical gate. Consequently, logical errors cannot be diagnosed and reversed by the error-correction procedure. 
The objective of QEC is therefore to design codes in which the probability of logical 
errors is strongly suppressed, while any other error configuration generated by a product of Pauli errors 
is detectable after projection onto a measured stabilizer sector, and the corresponding recovery operation 
can be implemented without introducing an additional undetectable logical error in the process.

Within the stabilizer formalism, there are a few observations which can guide us towards the construction of better QEC codes:
\begin{itemize}
\item In most architectures, physical noise is local, typically affecting only one or a few nearby qubits at a time. Within the stabilizer framework, such errors can be analyzed through their Pauli expansion;
\item Because any single-qubit operator can be expanded in the Pauli basis $\{I,X,Y,Z\}$, a code that corrects all single-qubit Pauli errors also corrects arbitrary single-qubit errors~\cite{Gottesman98};
\item Since $Y=iXZ$, we can say that, if our code is able to independently correct single-qubit $X$ and $Z$ Pauli 
   errors, it will implicitly be able to correct single-qubit $Y$ errors~\cite{Nielsen00}. 
\end{itemize}

Although physical noise may be described by arbitrary single-qubit operators, syndrome extraction 
effectively reduces it to a product of Pauli errors, with the 
post-measurement state lying in a syndrome subspace where the action of the physical error is 
equivalent to that of a definite Pauli operator $X,Y,Z$ (or product thereof).

\subsection{Introduction to CSS codes}\label{sec:CSS}

While the stabilizer framework already imposes strong algebraic constraints, 
the space of possible stabilizer codes on $n$ qubits is still enormous, 
and most such codes have little to no practical use. To characterize how efficient and useful 
an error-correcting code can be, it is standard to label it using a set of primary parameters $[[n,k,d]]$, 
respectively denoting the number of physical qubits $n$, the number of encoded logical qubits 
$k$, and the code distance $d$, defined as the minimal weight (number of qubits 
acted on) of a Pauli operator that maps one valid codeword to a different one. Equivalently, 
the code distance represents the least number of qubits one needs to act on in order to apply
a logical operator on the system. Following our observations made thus far, one can conclude that
a code of distance $d$ can detect all error configurations of weight $w<d$. 

The typical bottleneck in designing stabilizer 
codes is to maximize the encoding rate $k/n$ while keeping $d$ as large as possible. 
In addition, we would like the code family to be scalable, in the sense that $d$ grows with $n$ and, 
ideally, the encoding rate $k/n$ remains finite as $n\to\infty$. This requirement is a crucial condition
to achieve fault-tolerant quantum computation: the \textbf{quantum threshold theorem}~\cite{Aharonov99}
states that, for a local noise model where physical errors occur at a given fixed rate $p$ below 
some threshold rate $p_{\mathrm{th}}$, as $n,d\rightarrow \infty$ the logical error rate can be 
made arbitrarily small by increasing the code size, allowing arbitrarily long quantum computations 
with only polylogarithmic overhead in $n$. To make such schemes implementable, one needs codes 
whose stabilizer structure and logical-gate implementation preserve locality and prevent uncontrolled error propagation.

Within the kingdom of stabilizer codes, we define quantum LDPC ($q$LDPC) codes as those
for which the parity-check matrices have bounded row and column weight, so that each stabilizer acts on $O(1)$ qubits and each qubit participates in $O(1)$ stabilizers as $n$ grows.
This ensures that syndrome extraction introduces a bounded amount of physical errors, meaning that a round of stabilizer measurement does not propagate physical
errors from one qubit to the rest of the system. On top of that, it has been shown that, 
depending on the $q$LDPC design, one can achieve finite asymptotic encoding rate 
($k/n>0$ in the limit $n\rightarrow \infty$) and macroscopic code distance 
$d=\Omega(n^\alpha)$ with $\alpha\in(0,1]$~\cite{Breuckmann21,Panteleev21,Panteleev22}.

A particularly important and widely discussed subclass of $q$LDPC codes is provided by the 
Calderbank--Shor--Steane (CSS) construction~\cite{Calderbank96}. This relies on the fact that
bit-flip and phase-flip errors can be diagnosed independently using separate sets of $Z$- and $X$-type stabilizers.
This separation allows us to construct a quantum code using two classical error-correction codes: one responsible for detecting $X$ errors and the other for $Z$ errors.
From this intuition, we introduce $C_X$ and $C_Z$ as two binary linear codes of length $n$ used for detecting $Z-$ and $X-$ type errors respectively. 
These identify their code space through a set of parity constraints (which, in this context, correspond to stabilizer generators): 
$m_X$ stabilizers of $X-$type and $m_Z$ stabilizers of $Z-$type. Note that, in the presence of errors, this construction implies 
that $C_X$ detects $Z-$type errors that do not commute with $X-$type stabilizers and $C_Z$ detects $X-$type errors that 
do not commute with $Z-$type stabilizers. The two codes can be represented with two binary parity-check matrices
\begin{equation}
H_X\in \mathbb{F}_2^{m_X\times n},\quad H_Z\in \mathbb{F}_2^{m_Z\times n}
\end{equation}
where each row of $H_X$ specifies the support of an $X$-type stabilizer
and each row of $H_Z$ specifies the support of a $Z$-type stabilizer.
The CSS construction associates to this pair a quantum code provided 
the classical codes satisfy the orthogonality condition
\begin{equation}\label{eq:orthogonality}
C_Z^\perp \subseteq C_X \quad\Longleftrightarrow\quad H_X H_Z^{\mathsf T} = 0 \pmod 2,
\end{equation}
which ensures that the $X$ and $Z$ stabilizer generators of the two codes commute. 
Each row $h$ of $H_X$ defines an $X$-type stabilizer
\[
g_h^{(X)} = \bigotimes_{i=1}^n X_i^{\,h_i},
\]
and each row $h$ of $H_Z$ defines a $Z$-type stabilizer
\[
g_h^{(Z)} = \bigotimes_{i=1}^n Z_i^{\,h_i}.
\]
The stabilizer group is generated by all $g_h^{(Z)}$ and $g_h^{(X)}$, and the corresponding code space is their 
joint $+1$ eigenspace. The number of logical qubits of the newly constructed $\mathrm{CSS}(C_X,C_Z)$ code is
\begin{equation}\label{eq:k_CSS}
k = n - \mathrm{rank}(H_X) - \mathrm{rank}(H_Z),
\end{equation}
and the distance $d$ is given by $d = \min\{d_X,d_Z\}$, where $d_X$ (and $d_Z$) is the minimum weight of a Pauli $X$-type (and $Z$-type) logical operator that 
commutes with all stabilizers. 

In this construction, the measurement of $Z$- and $X$-type stabilizer operators allows us to detect $X$- and $Z$- errors respectively.
This separation of $X$- and $Z$-type error detection, 
together with the ability to leverage classical coding theory for $C_X$ and $C_Z$, has made
CSS codes the backbone of most practical quantum error-correcting architectures~\cite{Gottesman98,Moses23,Bluvstein24,Google25}: all one needs to do
is now figure out how to efficiently combine two classical linear codes.

\subsection{From CSS to topological stabilizer codes}\label{sec:Topocodes}

Among CSS codes, the family of topological stabilizer codes stands as 
the one most largely favored by the quantum physics community due to its
formulation being derived from well-understood topological LGTs. 
These were shown to be particularly promising candidates to achieve 
fault-tolerant quantum computation. Prominent examples include the 2D toric 
(and surface) code~\cite{Dennis02,Kitaev03,Wang03,Satzinger21,Krinner22,Zhao22} 
and color codes~\cite{Bombin06,Katzgraber09,Bombin12}, which have both
comparatively high error thresholds and the ability to transversally implement
Clifford gate operations via constant-depth circuits, independently of system
size. 

The family of topological CSS codes first originated with the introduction of 
Kitaev's toric code~\cite{Kitaev03}: for this reason, we will use it as an example to
outline the fundamental properties of such codes.

The model is defined on a 2D square lattice with a qubit on each edge and periodic boundary conditions (PBC),
and is described by an exactly solvable gapped stabilizer Hamiltonian of the form
\begin{equation}
H = - J_X \sum_{v} A_v - J_Z \sum_{p} B_p,\qquad A_v = \prod_{i\in\partial v}\sigma^X_i,\quad B_p = \prod_{i\in\partial p}\sigma^Z_i.
\end{equation}
Here, $A_v$ and $B_p$ are referred to as star/vertex and face/plaquette operators, with $v$ and $p$ respectively 
labelling the vertices and the plaquettes/faces of the square lattice and $\partial v$,$\partial p$ indicating the set of qubits 
living on the boundary of each vertex and plaquette.

With this construction, one can easily see that all components in the Hamiltonian commute with each other, making it exactly solvable
in the absence of transverse fields. For a system of linear size $L$ with $2L^2$ edges (and qubits),
$L^2$ vertices and $L^2$ plaquettes, the ground-state subspace of the system is defined as 
\begin{equation*}
\mathcal{Q}=\{\ket{\psi}|A_v \ket{\psi}=+1\,\ket{\psi},\,\, 
B_p \ket{\psi}=+1\,\ket{\psi} \forall v,p\}.
\end{equation*}
The simplest state within $\mathcal{Q}$ that can be constructed by taking the state $\ket{0}^{\otimes 2L^2}$ and applying the stabilizer projector product
\begin{equation}
\ket{\psi} \propto \prod_v\left(1+A_v\right)\prod_p\left(1+B_p\right) \ket{0}^{\otimes 2L^2} = \prod_v\left(1+A_v\right) \ket{0}^{\otimes 2L^2}.
\end{equation}
This allows us to effectively describe the ground state as the superposition of all possible products of vertex operators $A_v$: representing the action
of each $A_v$ as a square "loop" in the dual lattice centered at $v$, the ground state described above is typically described in the literature as a string-net condensate.
The same logic can be applied for $B_p$ operators if we were to start with the $[\frac{1}{\sqrt{2}}(\ket{0}+\ket{1})]^{\otimes 2L^2}=\ket{+}^{\otimes 2L^2}$ state.

Although each stabilizer involves a different set of qubits and the number of stabilizer constraints defining our subspace 
is equal to the number of qubits in the system, one can show that they are not all linearly independent, but are instead
related by the condition
\begin{equation}
\prod_v A_v = I,\qquad \prod_p B_p = I.
\end{equation}
These relations coincide with the generalized lattice gauge theory formulation discussed 
in Sec.~\ref{sec:LGT}, and can be interpreted as a global $\mathbb{Z}_2$ Gauss-law constraint.
In fact, if we were to restrict our attention solely on eigenstates of the Hamiltonian, 
the effective description of our theory would be given by a $\mathbb{Z}_2$ LGT.
This has three major implications:
\begin{itemize}
\item In the presence of excitations, the Gauss law enforces a $\mathbb{Z}_2$ charge conservation for both the $X$ and $Z$ sectors on all eigenstates of the Hamiltonian $\ket{\phi}$: 
      \begin{equation*}
        \prod_v A_v \ket{\phi}= \ket{\phi},\qquad \prod_p B_p\ket{\phi} = \ket{\phi}
      \end{equation*}
      This implies that excitations for the LGT describing the model always emerge in pairs. This becomes immediate to see once we point out that every qubit 
      lies at the boundary of two $A_v$, $B_p$ stabilizers, meaning that any single-qubit $\sigma^X,\sigma^Z$ operator anticommutes with a pair of
      stabilizers. We will refer to excitations of $A_v$, $B_p$ as $e$ and $m$ quasiparticles. 
\item The global relations above reduce the number of independent stabilizer constraints, which in turn determines the ground-state degeneracy 
      (and, consequently, the number of logical qubits). Through a simple counting argument, we find that the dimension of any Hilbert-space eigenspace 
      where the $A_v$ and $B_p$ operators have a given set of eigenvalues $\{\vec{s}_A,\,\vec{s}_B\}$ is
      \begin{equation}
      \log_2 \,  \text{dim}(H_{\{\vec{s}\}}) = 2L^2 - 2(L^2 -1) = 2,
      \end{equation}
      corresponding to two logical qubits.
\item The operators comprising the normalizer of the group generated by $\{A_v,B_p\}$ can be identified in two ways, both rooted in the topological properties of the model's support.
      Upon noticing that the ground-state manifold is invariant upon the application of any product of stabilizer operators and that, leveraging PBC, the product
      of stabilizer operators along the horizontal or vertical direction is effectively equivalent to applying two closed strings of 
      $\sigma^X$ (or $\sigma^Z$) operators. On the other hand, since adjacent plaquettes and adjacent vertices share two qubits, one can see
      that a single string of such operators is enough to avoid giving rise to any excitation: this allows us to formulate the 
      two logical $X_L^{1,2}$ and $Z_L^{1,2}$ as any closed, homologically non-trivial (i.e. non-contractible to a $0$-measure object) loop operators 
      that can be defined on the torus and that are linearly independent up to stabilizer operators.
      An alternative way to identify these is by noticing that operators in the normalizer group can be constructed by
      creating a pair of $e$ or $m$ excitations and, by moving them around the lattice, finding a non-trivial way to bring them back together 
      and annihilate them, which again brings us to the homologically non-trivial loops around the PBC.
\end{itemize}

This description neatly carries over to our previously introduced CSS stabilizer formalism: 
identifying the $A_v$ and $B_p$ operators as $X-$ and $Z-$type stabilizer operators, the ground-state subspace
of the model becomes our code space $\mathcal{Q}$ in which one can encode $k=2$ logical qubits in the Toric Code's ground-state manifold with
our logical operators being $X_L^{1,2}$, $Z_L^{1,2}$. Given the structure of the logical operators, the code distance $d=L$ can 
be made arbitrarily large by sending the system to the thermodynamic limit. From this, one can also open the boundary conditions 
to obtain the surface code, which allows for was shown to be a much more approachable method to manipulate the encode quantum informations.

The toric code, and in particular its interpretation as an effective $\mathbb{Z}_2$ lattice gauge theory, provides a useful illustration 
of the broader class of topological CSS codes, which we can characterized as $q$LDPC codes derived from 
exactly solvable Hamiltonian models with geometrically local stabilizers of bounded support (i.e., each stabilizer acts on a finite 
number of qubits independent of system size) and gapped, topologically ordered ground-state manifolds. More importantly, 
their topological ground-state degeneracy is exactly what allows one to encode a fixed number $k$ of logical qubits non-locally,
giving a topologically protected code space.

Given this broad characterization of topological CSS codes outlined above, it is useful to explicitly relate the main formalisms 
through which these codes are constructed and analyzed. This will clarify how the different perspectives employed in the 
literature connect to one another, and will provide a common framework for situating the fracton codes studied later in this work.

\subsubsection*{Formulations of topological CSS codes}\label{sec:Codefamilies}

Topological CSS codes admit several equivalent formulations, each highlighting different structural relations between the 
$X-$ and $Z-$type stabilizers, which allow us to predict the code's properties. The ones most prominently used are:

\begin{enumerate}
  \item the \emph{Pauli stabilizer Hamiltonian} description, which can be used to gain an intuition of the models from a condensed-matter physics perspective~\cite{Gottesman98,Kitaev03};
  \item the \emph{CSS parity–check matrix} description outlined above, which allows for an algebraically formal characterization of the code's properties~\cite{Calderbank96};
  \item the \emph{hypergraph–product} construction, which provides a general method to build CSS $q$LDPC codes through the tensor-product of pairs of classical codes, including non-local and non-geometric codes~\cite{Tillich14};
  \item the \emph{chain–complex homological} description, where qubits are associated with cells of a lattice and stabilizers are defined at boundaries and coboundaries of the underlying cell complex~\cite{Dennis02}.
\end{enumerate}

The Stabilizer Hamiltonian formalism was used implicitly throughout the construction of the fracton codes in Chapter~\ref{chap:fracton_models} and was made explicit 
in Sec.~\ref{sec:Topocodes}: the code is identified with a Hamiltonian model identified with the set of commuting stabilizer operators 
$\{A_v,B_p\}$ whose ground-state manifold is identified with the code space.

Collecting these stabilizer operators into binary parity--check matrices $H_X$ (and $H_Z$), with each row of $H_X(H_Z)$ corresponding to the support of an $X$-type( 
$Z$-type) stabilizer $A_v$($B_p$), we obtain the CSS parity-check description outlined in Sec.~\ref{sec:CSS},
with the commutativity of all stabilizers translating into the CSS orthogonality condition $H_X H_Z^{\mathsf T} = 0 \pmod 2$.

Remaining in the CSS parity-check matrix description, the Hypergraph-product formulation gives a method to construct pairs of 
$(H_X,H_Z)$ matrices that satisfy the CSS orthogonality constraint. Fundamentally, this construction is achieved by taking
two classical linear codes with parity-check matrices $H_1 \in \mathbb{F}_2^{m_1 \times n_1}$, $H_2 \in \mathbb{F}_2^{m_2 \times n_2}$,
and taking their hypergraph product to define the CSS code through the block matrices
\begin{align}
H_X &= 
\begin{pmatrix}
H_1 \otimes I_{n_2} & I_{m_1} \otimes H_2^{\mathsf T}
\end{pmatrix}, \qquad
H_Z &= 
\begin{pmatrix}
I_{n_1} \otimes H_2 & H_1^{\mathsf T} \otimes I_{m_2}
\end{pmatrix},
\end{align}
which automatically satisfy the $H_X H_Z^{\mathsf T}=0 \pmod 2$ condition.
This method can also be used to construct topological codes. For example,
the 2D toric code can be derived by taking $H_1 = H_2$ to be the $(L-1)\times L$ 
parity–check matrix of the length-$L$ cyclic repetition code: the corresponding hypergraph
product code is a $[[2L^2,2,L]]$ CSS code which is locally equivalent to
the toric code on a closed $L\times L$ torus~\cite{Tillich14}. This method was also shown to give rise 
to novel types of aperiodic fracton codes~\cite{Tan25} as well as non-geometric quantum LDPC codes 
whose lack of geometric locality in the choice of $(H_1,H_2)$ enables the construction of code families 
with finite asymptotic encoding rate and scalable code distance~\cite{Panteleev21,Breuckmann21}.

While the parity-check and hypergraph-product viewpoints make the commutation constraints explicit, they do not directly encode any information
regarding the geometric relation between $X$- and $Z$-type stabilizers that underlies many topological and fracton codes. This structure is naturally
captured through the homological point of view by a chain-complex formulation and, as we will see, makes explicit a relation between 
the structure of certain families of CSS codes and the duality relation of classical Ising spin models.

To make this more explicit, we start from the two-dimensional setting of~\cite{Bombin07_3}.
A cellular decomposition of a closed surface $\Sigma$ is obtained by embedding a graph $\Gamma$:
this consists of a set of vertices and edges that define $0$- and $1$-cells, while the connected components of $\Sigma\setminus\Gamma$
define a set of $2$-cells (faces). We denote the corresponding $\mathbb{Z}_2$ vector spaces by
$\mathcal{C}_0(\Sigma)$, $\mathcal{C}_1(\Sigma)$, and $\mathcal{C}_2(\Sigma)$.

The incidence relations of the cellulation are captured by the boundary maps
\begin{equation}
\partial_2 : \mathcal{C}_2 \to \mathcal{C}_1,\qquad \partial_1 : \mathcal{C}_1 \to \mathcal{C}_0,
\end{equation}
which send each face to the $\mathbb{Z}_2$ sum of its boundary edges and each edge to 
the sum of its two endpoints. The composition of these two linear maps satisfies the identity
\begin{equation}
\partial_1 \partial_2 = 0,
\end{equation}
reflecting the closedness of the 2D faces of the cellulation. 
The cellular structure defines the chain-complex
\begin{equation}
\mathcal{C}_2(\Sigma) \xrightarrow{\partial_2} \mathcal{C}_1(\Sigma) \xrightarrow{\partial_1} \mathcal{C}_0(\Sigma).
\end{equation}



Following this construction, one can define two classical homological codes
by placing physical bits on the edges ($1$–cells). A binary configuration 
$z\in \mathcal{C}_1(\Sigma)$ can then be used to represent a subset of edges affected by an error.
By introducing at every vertex ($0$-cell) a binary parity-check $A_v$, the boundary map $\partial_1$ returns the set of 
vertices incident to an odd number of selected edges, thus acting as a 
local parity–check operator enforcing even incidence at each vertex. Conversely
one can define another classical code by introducing $B_p$ binary parity-checks on
the faces ($2$-cells). Representing a binary configuration of edges $x\in \mathcal{C}_1(\Sigma)$,
the (transpose) coboundary map $\partial_2^{T}$ defines parity checks on the faces, detecting whether 
the edge configuration $x$ has odd incidence to any face. This way, the cellulation
determines a pair of classical linear codes built from vertex and face incidence relations captured 
by the boundary and coboundary maps~\cite{Bombin06}. Since $\partial_1 \partial_2 = 0$ holds identically, 
the parity-check matrices corresponding to these two codes automatically satisfy the CSS orthogonality constraint 
introduced in Sec.~\ref{sec:CSS}. Promoting these classical parity checks to commuting Pauli stabilizers 
yields a homological CSS code.



This construction can be extended naturally to D dimensions. By employing a cellulation of the system where qubits reside on $d$-cells, 
one type of stabilizers (say, $X-$type $A$ stabilizers) is defined on the $(d-1)-$cells, and the other ($Z-$type $B$ stabilizers) defined 
on $(d+1)$-cells, to these we can associate the vector spaces $\mathcal{C}_d^{\mathcal Q}$,$\mathcal{C}_{d-1}^{A}$,$\mathcal{C}_{d+1}^{B}$. 
It is then possible to define a homological CSS stabilizer code through the chain-complex
\begin{equation}\label{eq:chain}
\mathcal{C}_{d+1}^{B} \xrightarrow{\partial_{d+1}^B} \mathcal{C}_d^{\mathcal Q} \xrightarrow{\partial_{d}^A} \mathcal{C}_{d-1}^{A},
\end{equation}
with $\partial_{d}^{A}\partial_{d+1}^{B}=0$ ensuring commutativity of stabilizers.

A key structural consequence of this construction is that the geometric incidence relations in Eq.~\eqref{eq:chain}
match the generalized Wegner duality introduced in Sec.~\ref{sec:self-duality}.
This becomes evident by identifying $(d-1)$-cells with primal Ising spins and $(d+1)$-cells with the dual spins, with
the $d$-cells then acting as coupling objects for both models. In this language, the boundary map $\partial_d^A$ and 
coboundary map $(\partial_{d+1}^B)^{\mathsf T}$ correspond to the two incidence matrices $\theta_{ij}$, $\theta_{\ell j}$ 
capturing the connectivity of the Ising models in Sec.~\ref{sec:self-duality}. 
The closure condition from Eq.~\eqref{eq:closure} is equivalent to the chain-complex property $\partial_{d}^A\partial_{d+1}^B=0$
and is therefore satisfied by construction, while the completeness condition in Eq.~\eqref{eq:completeness} reduces to 
\begin{equation}\label{eq:complet_1}
\text{dim}\,\mathcal{C}_d^{\mathcal{Q}}-\text{rank}(\partial_d^A)-\text{rank}(\partial_{d+1}^B) = 0,
\end{equation}
where we identified the rank of the incidence matrices with the rank of their respective boundary maps
\begin{flalign*}
&\mathrm{rank}(\partial_d^A)=\dim\,\mathrm{Im}(\partial_d^A),\qquad \text{where }\mathrm{Im}(\partial_d^A) = \left\{\, \partial_d^A x \;\middle|\; x \in \mathcal C_d^{\mathcal Q} \right\} \subseteq \mathcal C_{d-1}^{A}, \\
&\mathrm{rank}(\partial_{d+1}^B)=\dim\,\mathrm{Im}(\partial_{d+1}^B),\,\,\,\,\text{where }\mathrm{Im}(\partial_{d+1}^B)= \left\{\, \partial_{d+1}^B z \;\middle|\; z \in \mathcal C_{d+1}^{B} \right\} \subseteq \mathcal C_{d}^{\mathcal Q}.
\end{flalign*}
Since $\dim\,\mathcal C_d^{\mathcal Q} = \mathrm{rank}(\partial_d^A) +\dim\,\mathrm{ker}(\partial_d^A)$, Eq.~\eqref{eq:complet_1} can be written as
\begin{equation}
\dim\,\mathrm{Im}(\partial_{d+1}^B) = \dim\,\mathrm{ker}(\partial_d^A).
\end{equation}
However, due to Eq.~\eqref{eq:chain}, $\mathrm{Im}(\partial_{d+1}^B) \subseteq \mathrm{ker}(\partial_d^A)$: this, together with the condition above, changes the completeness condition into requiring that
\begin{equation}\label{eq:completeness_homo}
\mathrm{Im}(\partial_{d+1}^B) = \mathrm{ker}(\partial_d^A).
\end{equation}

This parallelism allows us to make one fundamental conclusion: any two classical homological codes whose parity check matrices are related by a chain-complex of the form shown in 
Eq.~\eqref{eq:chain} and satisfy the condition in Eq.~\eqref{eq:completeness_homo} are mutually dual under the algebraic Wegner prescription.
While the stabilizer Hamiltonian formalism establishes a fundamental link between CSS codes and exactly solvable $\mathbb{Z}_2$ quantum spin liquids, the geometric structure encoded in 
their boundary maps provides the first step toward understanding how code structure influences the stability of the encoded information in the presence of noise, particularly when this geometric duality can be exploited. 
We will develop this connection further in Sec.~\ref{sec:SMmapping}, where the interplay between QEC and the notion of duality will become more explicit.


\begin{figure}[!t]
\centering
\includegraphics[width =1.\textwidth]{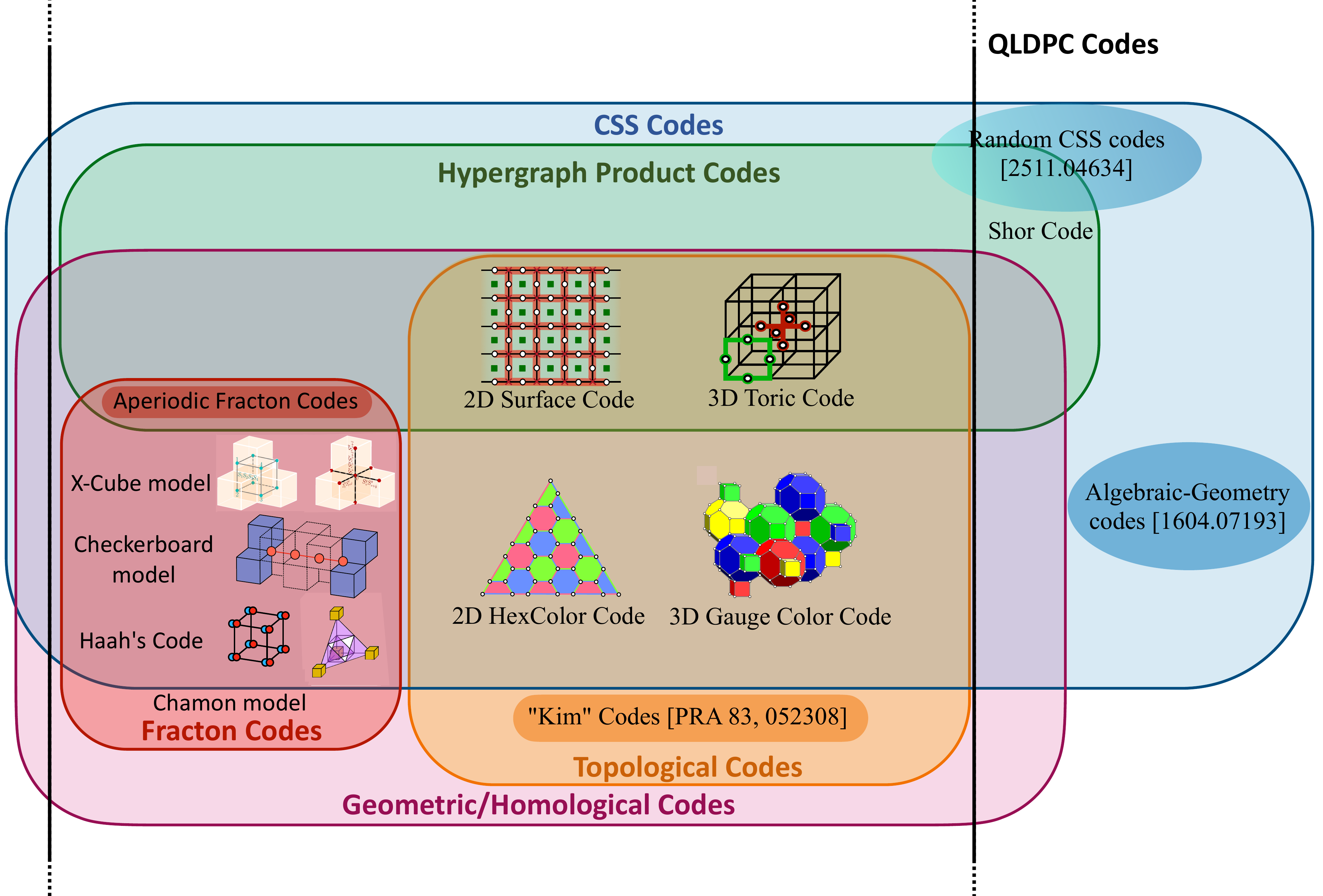}
\caption[Map of the different QEC codes mentioned in this work and their classification among the main code families discussed.]
{Map of the different QEC codes mentioned in this work and their classification among the main code families discussed.}
\label{fig:QLDPC_Codes_map.pdf}
\end{figure}

\newpage
\subsection{Fault-tolerant decoding and the error threshold relation}\label{sec:Decoding}

\subsubsection{The Threshold Theorem}
The notion of code distance, combined with the Knill-Laflamme condition stated in Eq.~\ref{eq:Knill-Laflamme}, allows 
us to find different crucial aspects of code scalability. Suppose we are given two error configurations 
$E_a$,$E_b$ of weight $w_a,w_b \le t$, with $t=\lfloor (d-1)/2 \rfloor$: in case these have different 
syndromes (i.e. they map our code space into different stabilizer subspaces), we are automatically able 
to distinguish them and find a distinct recovery operation for each of them. Problems arise when the 
syndrome extraction is not able to do so: this can only happen if $E_a$ and $E_b$ only differ by a product 
of stabilizer operators (which keeps them in the same equivalence class) or by the action of a logical operator
(which affects our encoded state). Now suppose that, in the act of correcting an error configuration $E_b$, we 
performed the correction of $E_a$ instead: the overall action on our encoded state is to apply $E_a^{\dagger} E_b$.
Since $w(E_a),w(E_b)\le \lfloor (d-1)/2 \rfloor$, it follows that $w(E_a^{\dagger}E_b)< w(E_a)+w(E_b) \le (d-1)$: 
the correction operation cannot introduce a logical operation; this also implies that any two indistinguishable
error configurations of weight $\le \lfloor (d-1)/2 \rfloor$ must belong to the same error equivalence class.
The objective, then, is to devise a quantum code for which the errors accumulated between successive correction steps 
remain small compared to the code distance, so that they can still be corrected reliably.
However, we also have to take into account that both physical gate operations, including our parity check 
measurements themselves, are subject to noise, thus representing another potential source of error. This 
means that, regardless of the encoding method, CSS stabilizer codes that rely on active decoding methods 
will inevitably suffer from inaccuracies. For this reason, fault tolerance is not expected to outright
prevent the emergence of errors, but instead requires that error correction introduce, on average, fewer 
errors than it removes~\cite{Shtetl}. 
The central question then becomes whether this can be achieved asymptotically for a given code
and, if so, for what physical error rates. This essentially depends on three ingredients: 
the noise model at hand, the code and the \emph{decoder} used to infer the error configuration from the measured syndrome.

To understand this, it is useful to consider not a single code, but a family of scalable stabilizer codes 
$\{\mathcal{Q}_n\}_{n\in\mathbb{N}}\equiv \{ \llbracket n,k_n,d_n \rrbracket \}$, all with similar construction but increasingly 
large blocklength $n$ and code distance $d_n$ (Examples include topological codes such as the surface code and color code). 
We place the family $\mathcal{Q}_n$ inside a simple stochastic noise model: during a single error-correction cycle, each 
physical qubit independently suffers a nontrivial Pauli error with probability $p$, and no error with probability $1-p$.
Let $E$ denote the resulting error operator and $w(E)$ its Hamming weight. On top of that, let us assume that our code is 
non-degenerate so that, through the considerations made above, we can construct a decoder that corrects \emph{all} error patterns of 
weight at most $t_n = \lfloor (d_n-1)/2 \rfloor$. Then, we define the \emph{logical failure probability} $P_{\text{fail}}(n,p)$
as the probability that said decoder outputs a recovery operation that induces a logical error on the encoded state.
Logical failure in a single correction cycle can then only arise from error configurations with weight $w(E) > t_n$.

Intuitively, increasing the code distance enlarges the minimum number of physical faults $t_n$ required to induce a logical 
error during error correction. If $t_n$ grows sufficiently fast compared to the typical amount of errors introduced by 
physical noise, logical failures become exponentially unlikely~\cite{Knill98}. For local stochastic noise with physical error rate 
$p$, this condition is satisfied below a threshold value $p_{\mathrm{th}}$: in this regime, the probability 
that an error configuration $E$ exceeds the correctable weight $t_n$ vanishes in the large-code limit, 
i.e. $\Pr[w(E) > t_n] \xrightarrow[n\to\infty]{} 0$. In this regime, typical error configurations are correctable with 
high probability, even though the average error weight scales extensively with system size~\cite{Aharonov99}.

This property is known as the \textit{threshold theorem}. Rigorous proofs of a non-zero
threshold are typically obtained on a case-by-case basis for specific code families and
noise models~\cite{Dennis02,Li25}. More generally, fault-tolerant constructions based on
code concatenation show that scalable quantum error correction is possible in settings
where logical errors require an increasing number of suitably correlated physical faults
and error propagation can be controlled~\cite{Aharonov97,Knill98,Aharonov99}. Despite its
simplicity, this heuristic formulation captures the central message of fault-tolerant
quantum computation: \textit{quantum information can be protected arbitrarily well provided the
physical noise strength lies below a fixed constant}. The exact value of this threshold depends
on several factors, primarily on the choice of error-correcting code, the type of gate operations 
being implemented, the noise model taken into account, and the type of decoder that is used to perform
error correction.

A \emph{fault-tolerant error threshold} is therefore defined as a constant $p_{\mathrm{th}} > 0$ such that,
$\forall p < p_{\mathrm{th}},\,\, \lim_{n \to \infty} P_{\mathrm{fail}}(n,p) = 0$.
Equivalently, for noise rates below the threshold, one can make the logical error rate
arbitrarily small by choosing $n$ (and therefore $d_n$) sufficiently large.
Thus, the threshold arises from the asymptotic suppression of high-weight error events 
relative to the growing error-correcting capability of the scalable code family.
If a scalable code family admits (1) a growing distance $d_n \to \infty$, (2) a decoder 
whose logical failure probability decays exponentially in $d_n$ for $p < p_{\mathrm{th}}$, 
and (3) a universal set of fault-tolerant logical operations,
then fault-tolerant quantum computation becomes possible with arbitrarily low logical error.
One simply chooses $n$ large enough to suppress logical faults below any desired target.
This framework encompasses any scalable QEC architecture.

Of course, this behavior is not guaranteed for arbitrary codes or noise models. One must assume that the code family is scalable
and that the noise is composed of local contributions. In this context, locality means that each physical fault acts on only a bounded number 
of qubits, and that error propagation through the circuit is itself bounded, as a constant number of faulty gates can only affect a constant-size neighborhood determined by the gate set.
Under such assumptions, scalable stabilizer codes admit the possibility of a nonzero threshold. These conditions, while intuitive, become a fundamental 
constraint when trying to develop a fault-tolerant implementation of a universal gate set necessary for quantum computation.

The arguments above make two idealizations: syndrome extraction and logical gates are treated as black boxes 
with an effective error rate $p$, and error correction cycles are assumed to be independent. In a complete fault tolerant 
construction, one must explicitly implement stabilizer measurements and encoded gates by local circuits, and analyze
how physical faults propagate through these processes. For geometrically local CSS codes such as surface and color 
codes, this can be recast as a percolation problem in space–time, leading to rigorous and numerical thresholds 
for local stochastic noise~\cite{Dennis02,Raussendorf07}.

It is also necessary to keep in mind that this is a generic definition and not very realistic from an experimental point of view, 
as different sources of error will contribute with different error probabilities: the notion of error threshold is therefore
a fundamental theoretical parameter used to assess the performance of a QEC code in the case of homogeneous probability of all error events.

\subsubsection*{Noise Models}\label{sec:Noise_Model}
While most of the theory discussed in this thesis is generally applicable to arbitrary noise models, the error-correction
process relies on the choice of a specific quantum error channel. Our findings are obtained assuming an independent, identically
distributed noise model (i.i.d.), where $X$ and $Z$ errors occur independently on each qubit with probability $p_X$ and $p_Z$. 
This is described by the following error channel:
\begin{flalign}
\mathcal{E}(\rho)
= &(1 - p_X)(1 - p_Z)\,\rho
+ p_X(1 - p_Z)\,X\rho X \\
& + (1 - p_X)p_Z\,Z\rho Z
+ p_X p_Z\,Y\rho Y. \nonumber
\end{flalign}
This noise model is not intended to represent the full complexity of a physical device. Even in the absence of gate operations, 
platforms such as cold-atom arrays, trapped ions, or superconducting circuits experience a range of idle-qubit errors, each described by a different 
quantum channel (including energy relaxation, dephasing, leakage) that generally do not factorize into independent $X$ and $Z$ flips. 
Capturing these effects realistically would require a hardware-specific noise model and a decoder adapted to the correlated error structure that emerges 
at the circuit level.

What is then the purpose of using an i.i.d. noise model? Despite its simplicity, the i.i.d. model allows us to isolate the intrinsic properties of the code and decoder, 
without conflating them with hardware-specific noise mechanisms. The ability to statistically separate $X$ and $Z$ error events allows us to split the error 
correction procedure of CSS codes into two independent decoding procedures: this structural decomposition means that the syndrome information for 
bit-flip and phase-flip errors resides in separate syndrome spaces and depends only on the parity-check matrices of the respective classical codes. 
While, in reality, thresholds obtained under i.i.d. noise offer an upper bound on the achievable performance of more realistic circuit-level noise, 
they are nevertheless an optimal starting ground to benchmark a candidate CSS code while leveraging the combinatorial structure of the classical codes, 
allowing us to catch a glimpse of how the intrinsic geometrical properties of the code are reflected in its performance. 
This preliminary analysis is fundamental for the construction of more efficient quantum codes.

\subsubsection*{Quantum Hamming bound and the error threshold relation}

While the threshold theorem ensures fault-tolerant quantum computation below a certain finite physical error rate, 
it leaves open a fundamental question: how do encoding rate and error resilience trade off against each other?\\
The CSS code construction allows us to formulate a hard bound on the relationship between encoding rate and error threshold that follows
from the Hamming bound defined for classical, non-degenerate binary codes~\cite{Hamming50}.

Consider a CSS code $\mathrm{CSS}(C_X,C_Z)$ constructed from two binary linear codes $C_X$ and $C_Z$ of length $n$, with distances $d_X$ and $d_Z$, 
and satisfying the CSS commutation condition $C_Z^\perp \subseteq C_X$. The code encodes $k = k_X + k_Z - n$ logical qubits, where $k_X \equiv \dim(C_X)$ 
and $k_Z \equiv \dim(C_Z)$. From standard classical coding theory~\cite{Hamming50} and assuming non-degeneracy of the classical codes, we can use the 
same arguments outlined in the previous section and say that each of these codes is guaranteed to correct up to $t_X$ and $t_Z$ errors respectively.
Let
\begin{equation}
t_X = \left\lfloor\frac{d_X-1}{2}\right\rfloor,
\qquad
t_Z = \left\lfloor\frac{d_Z-1}{2}\right\rfloor,
\end{equation}
so that the code is guaranteed (in the worst-case sense) to correct any $X$ error configuration of weight $w(E_X)\le t_X$ and any $Z$ error of weight $w(E_Z)\le t_Z$.
Let us define the sets of all $X$-type and $Z$-type error configurations of weight up to $t_X$ and $t_Z$ respectively as $\mathcal E_X$ and $\mathcal E_Z$. Their cardinalities are
\begin{equation}
|\mathcal E_X| = \sum_{j=0}^{t_X} \binom{n}{j},
\qquad
|\mathcal E_Z| = \sum_{j=0}^{t_Z} \binom{n}{j}.
\label{eq:EXEZ}
\end{equation}
For large enough $t_{X/Z}$, this is exactly the volume of a Hamming ball of radius $t_{X/Z}$. For $t \le n/2$, this volume respects the following upper bound:
\begin{equation}
\sum_{j=0}^{t} \binom{n}{j} \le 2^{\,n H(t/n)},\quad\,\,\text{where}\,\,H(x) = -x\log_2 x -(1-x)\log_2(1-x).
\label{eq:hamming_entropy_bound}
\end{equation}
In a CSS code, the $Z$-type stabilizer checks provide $n-k_Z$ independent syndrome bits for diagnosing $X$ errors, hence there are at most $2^{\,n-k_Z}$ distinct $X$-error syndromes.
If the code is to correct \emph{all} $X$ errors of weight $w(E_X)\le t_X$, the number of distinct correctable $X$ error patterns cannot exceed the number of available syndromes. Similarly,
the $X$-type stabilizer checks provide $n-k_X$ syndrome bits for diagnosing $Z$ errors. These are represented by the following constraint:
\begin{equation}
\sum_{j=0}^{t_X}\binom{n}{j} \le 2^{\,n-k_Z},\qquad\sum_{j=0}^{t_Z}\binom{n}{j} \le 2^{\,n-k_X}.
\label{eq:css_hamming_X}
\end{equation}
Multiplying the two and using the fact that $k=k_X+k_Z-n$ yields the CSS quantum Hamming bound
\begin{equation}
\sum_{j=0}^{t_X}\binom{n}{j}\sum_{j=0}^{t_Z}\binom{n}{j} \leq 2^{\,n-k}.
\label{eq:css_qhamming}
\end{equation}
Introducing the encoding rate $R\equiv k/n$, one obtains
\begin{equation}
R \leq 1 - \frac{1}{n}\log_2\!\left(\sum_{j=0}^{t_X}\binom{n}{j}\right)
      - \frac{1}{n}\log_2\!\left(\sum_{j=0}^{t_Z}\binom{n}{j}\right).
\label{eq:R_finite_n}
\end{equation}
Using the entropy bound \eqref{eq:hamming_entropy_bound} gives the following form:
\begin{equation}
R \leq 1 - H(t_X/n) - H(t_Z/n),
\label{eq:Hammingbound1}
\end{equation}
This inequality expresses a fundamental trade-off between the encoding rate and the error resilience of CSS codes: no CSS code can correct
these many $X$ and $Z$ independent errors while simultaneously encoding an arbitrarily large amount of logical qubits. On the other hand, it also
leaves open the possibility of the existence of good QEC codes (i.e. codes with non-zero encoding rate)~\cite{Panteleev21}.

Suppose now that each of the $n$ physical qubits has a probability $p_X$ and $p_Z$ of having an $X$ or $Z$ error. In the limit of large code size $n$, the typical 
error weight will be sharply concentrated around its mean: $\langle w(E_{X/Z})\rangle \approx n p_{X/Z}$, with relative fluctuations of the
order $\mathcal{O}(1/\sqrt{n})$. As mentioned when formulating the threshold theorem, the error correction procedure starts to break down when
the weight of the average error configuration surpasses $t_{X/Z}$. Heuristically, this happens when $np_{X/Z}$ exceeds $t_{X/Z}$, which is 
equivalent to saying that, as $n \rightarrow \infty$, at the error threshold we have $t_{X/Z} \approx n p^{\mathrm{th}}_{X/Z}$. 
Plugging this in the inequality above, we get the following \textbf{error threshold relation}:
\begin{equation}
R\leq 1 - H(p^{\mathrm{th}}_{X}) - H(p^{\mathrm{th}}_{Z}),
\end{equation}
which, in the limit of zero encoding rate $R\rightarrow 0$, reduces to 
\begin{equation}\label{eq:Thresholdbound}
H(p^{\mathrm{th}}_{X}) + H(p^{\mathrm{th}}_{Z}) \leq 1.
\end{equation}
This result represents a fundamental constraint on the achievable i.i.d. noise thresholds for CSS codes and will be the centerpiece of 
the discussions in the rest of this work.

\section{Statistical nature of decoding}\label{sec:statmech}
To decode an error configuration in a CSS code, one is given an error syndrome $s$ determined by the outcomes of the stabilizer measurements.
From this syndrome, one has to infer a compatible error configuration $E$ and apply a corresponding recovery operation to restore the encoded information.
By construction, the syndrome cannot distinguish an error configuration $E$ from any configuration obtained from it by multiplication by a stabilizer operator.
Nevertheless, this does not represent a problem, since correcting $E$ or $gE$, where $g=\prod_i A_i \prod_j B_j$ is a product of stabilizers, has the same effect on the code space.
However, the syndrome $s$ also cannot distinguish $E$ from $E\lambda$, where $\lambda \in \mathcal L$ and $\mathcal L$ is generated by the logical Pauli operators $\{X_L^1,Z_L^1,\dots,X_L^k,Z_L^k\}$.
Decoding is therefore intrinsically probabilistic: the most appropriate strategy is therefore to separate all error configurations
compatible with the syndrome, $\{ E \mid P(E \mid s) > 0 \}$, into equivalence classes distinguished by the action of these logical operators.
For a representative error $E$, these equivalence classes are labeled by logical operators $\lambda$ and described as
\begin{equation} 
\overline{E \lambda} = \left\{\lambda \left(\prod_i S_i^{n_i}\right) E \;\middle|\; n_i \in \{0,1\} \right\}, \qquad \lambda \in \mathcal{L}.
\end{equation}

The most effective decoding method then consists in (1) finding a suitable error configuration $E$ given the error syndrome $s$ and (2)
identifying the most probable logical error class
\begin{equation}
\lambda^\star = \text{argmax}_{\lambda \in \mathcal{L}} P(\overline{E\lambda} \mid s).
\end{equation}
Once the most likely error class is found, one simply has to apply the correction that solves any error configuration within said class.

This approach to decoding, also known as \textbf{maximum-likelihood} decoding, represents the most reliable method as well as
the most expensive one, as an adequate computation of $P(\overline{E\lambda} \mid s)$ is exponentially expensive with the number of stabilizers and, consequently,
with the system size. This can be done by sampling the space of error configurations.
Since the possible error configurations that match a given syndrome only differ by the action of logical operators and stabilizer operator, 
by associating an Ising degree of freedom $\sigma_i$ to each applicable stabilizer one can approach this sampling problem as the simulation of a Random-Bond Ising (RBI) model.
This observation lays the foundations for the Statistical Mechanical (SM) mapping. In this section, we derive this mapping for the case of i.i.d. noise models and discuss 
the main insights that the statistical-mechanical perspective provides about QEC codes.

\subsection{Quantum-to-Classical SM mapping}\label{sec:SMmapping}

Consider a CSS stabilizer code defined on a set of physical qubits $Q$. 
The $X$-type and $Z$-type stabilizers are specified by the parity-check matrices $H_X$ and $H_Z$. 
Each row of $H_X$ and $H_Z$ defines a commuting stabilizer operator, which we denote by $A_{\mathcal S_X}$ and $B_{\mathcal S_Z}$, respectively.
Here, $\mathcal{S}_{X/Z}$ 
serves both as a label for the individual stabilizer and as an identifier for its anchor cell in a lattice description. 
The supports of these operators on the lattice are denoted by $\partial \mathcal{S}_X, \partial \mathcal{S}_Z \subset Q$, i.e. the set of
qubits onto which the stabilizer corresponding to $\mathcal{S}_{X/Z}$ acts. 

The logical code space is the common $+1$ eigenspace of all stabilizers and can equivalently be viewed as the ground-state subspace of the stabilizer Hamiltonian
\begin{align}\label{eq:Check_Ham}
&H= - \sum_{\mathcal{S}_X} A_{\mathcal{S}_X}
   \;-\; \sum_{\mathcal{S}_Z} B_{\mathcal{S}_Z},\\
&\text{where }\,\,A_{\mathcal{S}_X} = \prod_{i \in \partial \mathcal{S}_X} \sigma_i^{x}, 
\qquad B_{\mathcal{S}_Z} = \prod_{i \in \partial \mathcal{S}_Z} \sigma_i^{z}. \nonumber
\end{align}

For i.i.d. error models on CSS codes, $X$ and $Z$ errors can be treated separately: 
we therefore focus on error configuration $E$ corresponding to the set of qubits $\eta\subseteq Q$ affected by the corresponding type of Pauli error ($X$ or $Z$). 
When the system is initialized in a ground state of $H$, an error generally drives it into an excited state in which a subset 
of the stabilizers acquires eigenvalue $-1$. The resulting pattern of violated stabilizer eigenvalues defines the \emph{error syndrome} $s$.

Because the exact configuration $\eta$ is typically unknown, the syndrome $s$ is used to estimate the most likely
logical error class. As mentioned before, two error configurations $\eta$ and $\eta'$ are considered equivalent if 
they differ only by an element of the stabilizer group. 
This equivalence breaks down if the configurations differ by a logical operator $\lambda$. Consequently, for a given 
syndrome $s$ and a compatible error configuration $\eta$, errors are grouped into equivalence classes $[\eta + \lambda]$. 
The probability of a syndrome is given by the sum over these classes: $\mathrm{Pr}(s) = \sum_{\lambda} \mathrm{Pr}([\eta + \lambda])$, 
and the most likely logical error class is the one with the highest total probability.

Error correction is asymptotically reliable if there exists a unique most-probable class $[\eta^*]$ such that, 
in the thermodynamic limit, $\mathrm{Pr}([\eta^*]) \to \mathrm{Pr}(s)$~\cite{Dennis02}. This condition holds only 
when physical error rates remain below a threshold probability $p^{\text{th}}$, which is expected by the threshold 
theorem.

Consider an error configuration $\eta\subseteq Q$ that is compatible with a given observed syndrome $s$. 
Knowing the error probability $p$ of having a single-qubit $X$ (or $Z$) error, one can compute the probability 
of the error configuration as $P(\eta)=\prod_{\ell\in\eta}p \prod_{\ell\notin\eta}(1-p)$, where $\ell$ labels individual qubits, and the products run 
over all the qubits in the specified subsets of $Q$.

For any other error configuration $\eta'=\eta + V$ obtained by further applying Pauli error operators to the set of qubits 
$V\subseteq Q$, we get
\begin{flalign}\label{eq:stat1}
P(\eta + V)& =P(\eta)\,\frac{P(\eta+V)}{P(\eta)}= P(\eta)\prod_{\ell\in V\cap\eta}\frac{1-p}{p}\;
\prod_{\ell\in V\setminus\eta}\frac{p}{1-p}.
\end{flalign} 
Two configurations belong to the same equivalence class iff they differ by a product of stabilizers $A_{\mathcal{S}_X}$ 
(or $B_{\mathcal{S}_Z}$): in that case, the possible flip set $V$  corresponds to the symmetric difference of the supports 
$\partial \mathcal{S}_X$ (or $\partial \mathcal{S}_Z$). This allows us to express the probability distribution of all error 
configurations belonging to the same equivalence class through a statistical mechanical partition function of a classical spin model. 

To make this correspondence explicit, it is convenient to parametrize each equivalence class by the choice of stabilizers that are 
applied to a fixed reference configuration $\eta$. Since multiplying by a stabilizer flips precisely the qubits in its support, 
any configuration in the same equivalence class can be generated by selecting a subset of stabilizers and taking the symmetric 
difference of their supports. To parameterize which stabilizer is applied to the initial configuration, we start by placing 
an Ising degree of freedom $\sigma_{\mathcal{S}_{X/Z}}=\pm 1$ at the center of each stabilizer $\mathcal{S}_{X/Z}$. 
Then, for each possible physical qubit error $X_i/Z_i$, define the set of stabilizers that apply said operator as 
$\mathcal N_{X/Z}^i\equiv \{\mathcal{S}_{X/Z} \,|\,i\in \partial{\mathcal{S}_{X/Z}}\}$. If a collection of stabilizers is applied, 
their net action on qubit $i$ is nontrivial whenever an odd number of stabilizers in $\mathcal N^i_{X/Z}$ are chosen. 
This parity relation is captured by $\prod_{j \in \mathcal N^i_{X/Z}} \sigma_j$: this can be represented as a coupling term between 
the $\sigma_j$ which correspond to stabilizers that share said operator on qubit $i$.

For a given initial error configuration $\eta$, the total probabilities of all configurations in the same equivalence class are 
captured by the following partition function:
\begin{equation}\label{eq:StatMechPartition}
\mathcal{Z}_{X/Z}(\eta,T)\,=\, \sum_{\{\sigma\}} \text{exp}\left[ \beta \sum_{i \in Q} \eta_i \prod_{j\in \mathcal{N}^i_{X/Z}}\sigma_j \right],
\end{equation}
where we introduced $j$ as a more comfortable label for the stabilizers and their respective Ising degrees of freedom.
Additionally, the coupling strengths $\eta_i=\pm 1$ depend on whether or not the qubit $i$ was part of the original error configuration 
(i.e, if $i\in\eta,\,\eta_i=-1$, else $\eta_i=+1$).

The proposed classical model accurately captures the statistical properties of the error model given by Eq.~\eqref{eq:stat1} once we relate 
the effective inverse temperature $\beta$ to the single-qubit error probability $p$ via the Nishimori condition:
\begin{equation} \label{eq:NL}
e^{-2\beta_N} = \frac{p}{1-p}. 
\end{equation} 
The weight associated with the state for which $\sigma_j=+1\,\forall j$ can be identified with an (unnormalized) probability for the error 
configuration $\eta$: one can then see that the relative probability of any configuration $\eta+V$ is given by the Ising spin state with 
$\sigma_j=-1$ at the sites corresponding to the stabilizers that are applied to change the error configuration.
Furthermore, it can be shown that the partition function $\mathcal{Z}(\eta,T)$ depends only on the error equivalence class of $\eta$, 
consistent with its correspondence to $\mathrm{Pr}([\eta])$.

\subsection{Decodability and phase transitions} \label{sec:OptimalErrorThreshold} 

One can infer the most likely logical error by comparing the relative probabilities of different error classes. 
For an independent $X$ (or $Z$) error model with a homogeneous error rate $p$, the probability ratio of two equivalence 
classes distinguished by a logical operator $\lambda$ is reflected by the ratio of the partition functions of the model 
with error configuration $\eta$ and $\eta+\lambda$, which is equivalent to the free energy cost of applying $\lambda$ onto the system:
\begin{equation}
\frac{\mathrm{Pr}\left(\left[\eta+\lambda\right]\right)_{\left( X/Z,p \right)}}{\mathrm{Pr}\left(\left[\eta\right]\right)_{\left( X/Z,p\right)}} = 
\frac{\mathcal{Z}_{X/Z}(\eta,\beta_N)}{\mathcal{Z}_{X/Z}(\eta+\lambda,\beta_N)} \equiv e^{-\beta_N \delta F_{\eta,\lambda}(\beta_N)}.
\end{equation}
To determine the appropriate correction, it is necessary to identify the most probable equivalence class $[\eta_{s}^{*}]_{X(Z)}$ 
such that, in the thermodynamic limit, $\Pr\left([\eta_{s}^{*}]_{X(Z)}\right) \to \Pr(s)$. For this to occur, the free energy 
difference $\delta F_{\eta,\lambda}$ associated with the addition of a logical operator must diverge in the thermodynamic limit. 

Since applying a logical operator corresponds to flipping a set of couplings across the entire lattice of the associated Ising model, 
the condition $\delta F_{\eta,\lambda} \to \infty$ implies that introducing a domain wall incurs a free energy cost proportional to its size. 
This condition is met only if the associated Hamiltonian supports an ordered phase at finite temperature~\cite{Dennis02}. 

The existence of an ordered phase at the temperature $\beta_N$ corresponding to the error probability $p$ ensures that, 
in the thermodynamic limit, the error model can always be decoded for error rates below the threshold. 
Thus, \textit{the optimal error threshold can be inferred by studying the order-disorder phase transition of a classical RBI model with quenched disorder}.

\begin{figure}[!t]
\centering
\includegraphics[width =.55\textwidth]{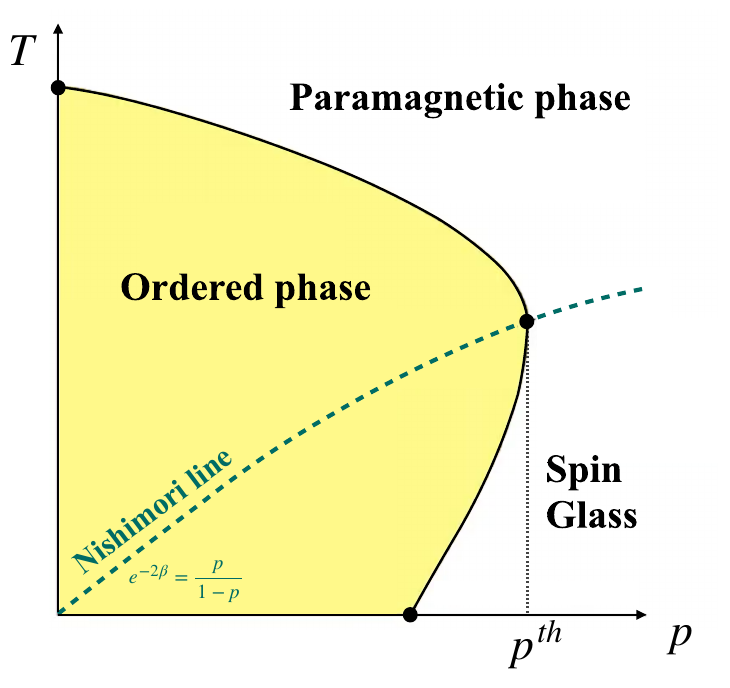}
\caption[Exemplary phase diagram of commonly studied RBI models with quenched disorder.]
      {Exemplary phase diagram of commonly studied RBI models with quenched disorder. The model is expected to have a tricritical point 
at the intersection between the phase boundary and the Nishimori line, which corresponds to the error threshold of the respective i.i.d. noise model.}
\label{fig:Disorder_diagram}
\end{figure}

To summarize, the SM mapping translates the decoding problem in the language of disordered spin systems by converting:
\begin{center}
\begin{tabularx}{\textwidth}{
    >{\raggedleft\arraybackslash\hspace{0pt}}X
    @{\hspace{1em}}
    c
    @{\hspace{1em}}
    >{\raggedright\arraybackslash\hspace{0pt}}X
}
Stabilizer operators
& $\longrightarrow$ &
Ising degrees of freedom $\sigma_j$
\\[0.6em]

Qubit degrees of freedom
& $\longrightarrow$ &
Couplings between Ising spins
\\[0.6em]

Pauli error configuration $\eta$
& $\longrightarrow$ &
Coupling strengths $\eta_i$
\\[0.6em]

Decodability below threshold
& $\longrightarrow$ &
Ordered phase below a threshold disorder rate at a given temperature
\end{tabularx}
\end{center}
\vspace{1.5cm}

\subsection{Decodability in terms of disordered spin systems}

This mapping has a variety of physical implications which allow us to transpose our understanding of disordered spin 
systems to the context of QEC decoding.

First of all, the existence of a decodable regime - ensured by the threshold theorem for scalable QEC codes~\cite{Aharonov99} - 
implies that the RBI models derived through the SM mapping of such codes are characterized by a region in parameter space associated to an ordered phase. 
This correspondence also works in the opposite direction~\cite{Dennis02,Chubb18}: to see this, consider the classical spin Hamiltonian obtained from the SM mapping of 
a CSS($C_X,C_Z$) stabilizer code relevant for the decoding of, say, $X$-errors. In the absence of disorder ($p=0$), 
the Hamiltonian reduces to a homogeneous Ising-type model whose interaction structure is fixed by the binary parity–check matrix $H_{Z}$ 
of the code $C_Z$. 

Now suppose that this model supports a finite-$T$ phase transition separating a high-$T$ disordered phase from a 
low-$T$ ordered one with a positive free-energy cost for the extended domain-wall excitations (i.e. extensive free-energy penalty 
associated to the application of a non-trivial logical operator). In the limit of large system size, the introduction of an infinitesimal 
amount of quenched disorder will partially destabilize the ordered phase but will not destroy the long-range order~\cite{Harris74,Imry75,Vojta06}, 
leading to a continuous change of transition temperature $T_C(p)$ as the disorder is increased.
Consequently, while the critical behavior of the system may still be subject to change in presence of weak disorder, 
there must exist a finite region in $p-T$ space where the ordered phase persists beyond the $p=0$ axis.

As the disorder strength is increased, frustration eventually destroys the long-range ordered phase. Beyond this point, the low-temperature regime is no longer decodable and 
may become glassy or strongly frustrated, depending on the structure of the associated disordered Ising model~\cite{Imry79}.
This high-disorder region is identified as the regime in which maximum-likelihood decoding can no longer be performed reliably.
While the shape of the curve defining the boundaries of the region of $p-T$ space with a well-defined ordered phase depends on the nature 
of the RBI model that is being studied, said curve must intersect the Nishimori line (NL) given by Eq.~\eqref{eq:NL} at least once at
some finite value of $p$ as shown in Fig.~\ref{fig:Disorder_diagram} due to its continuity.

While, in the context of QEC, the NL represents the only region of $p-T$ space in which the disordered RBI model
accurately reproduces the error statistics relevant to decoding, in disordered spin physics it represents a special locus along which the 
quenched disorder distribution is exactly matched to the Boltzmann weight of the classical model~\cite{Nishimori07}. 
This property highlights a deeper significance of the NL as being a region of parameter space characterized by an \textit{enhanced symmetry}
between thermal and disorder fluctuations which constrains the critical behavior of any transition point lying at the intersection between
the NL and the phase boundary. This can be seen by looking at the joint ensemble probability distribution
\begin{equation}
P(\sigma,\eta) = P(\eta)P(\sigma | \eta) \propto \text{exp}\left[ \beta \sum_{i \in Q} \eta_i 
                \prod_{j\in \mathcal{N}^i}\sigma_j + \frac{1}{2}\log\left(\frac{1-p}{p}\right) \sum_{i\in Q} \eta_i \right].
\end{equation}
It is possible to see that, along the NL, this takes the form
\begin{equation}
P(\sigma,\eta) \propto \text{exp}\left[ \beta \sum_{i \in Q} \eta_i \left(1+ \prod_{j\in \mathcal{N}^i}\sigma_j \right) \right].
\end{equation}
Factorizing the contributions of the first term and the second term in the exponent, we can see that the change in probability caused 
by flipping an Ising spin, dictated by its Boltzmann weight $e^{-\beta E}$, for a fixed bond configuration is identical to the 
probability change of flipping the bonds which share said spin, dictated by the error probability $p$ in the second term: the joint distribution 
is therefore invariant under such transformation along the NL. This manifests as a $\mathbb{Z}_2$ gauge symmetry with local transformations of the form
\begin{equation}
\sigma_j \longrightarrow \tau_j \sigma_j \qquad \eta_i \longrightarrow \eta_i \prod_{j\in\mathcal{N}^{i}} \tau_j \,,\qquad \tau_j=\pm 1,
\end{equation}
where we introduced a local symmetry generator $\tau_j$ for each Ising spin and corresponding couplings. 
This enhanced symmetry renders the NL an invariant 
manifold under renormalization-group transformations~\cite{Doussal88,Iba99}, as any effective coupling 
that could be generated through coarse-graining is bound to preserve the symmetry~\cite{Doussal89}.
As a consequence, if the ordered-disordered phase boundary intersects the NL at a continuous transition, 
the critical behavior at the intersection is governed by a fixed-point constrained to lie on the NL. 
Due to the relevance of perturbations corresponding to both deviations along and transverse to the NL,
the fixed-point is said to be multicritical~\cite{Doussal89}. 

In this sense, the existence of an error threshold corresponds to the notion that, for those CSS code families admitting 
a decodable phase, the corresponding SM mapping supports a multicritical fixed-point located at the
intersection between the ordered-phase boundary and the NL.  Conversely, the existence of such a crossing 
point on the NL implies the existence of a finite decodable region in parameter space.
Such a fixed point is expected for all CSS codes whose SM mapping has a finite-$T$ phase transition 
in the disorder-free limit.

\subsection{Threshold relation and the Kramers-Wannier duality}\label{sec:threshold-KW}

Much of the QEC literature concentrates on topological and geometric/homological CSS codes, with higher-dimensional 
generalizations of the toric code and surface code having gained a lot of attention in recent years in search of 
efficient ways to perform non-Clifford logical operations in a locality-preserving manner. 
The interest in these codes is also supported by the fact that their parity checks are both LDPC and geometrically 
local, making syndrome extraction achievable by means of constant-depth local circuits.
An additional advantage of geometric/homological CSS codes lies in the duality structure relating their $C_X$ and 
$C_Z$ classical codes. As reviewed in Sec.~\ref{sec:Codefamilies}, these codes follow a cellular chain-complex 
construction, in which the $X-$ and $Z-$stabilizers live on mutually dual cellulations 
(the primal lattice and its dual). As a consequence, the two Ising models derived through the SM mapping 
of $X$- and $Z$-error decoding satisfy, by construction, a generalized Kramers-Wannier (KW) duality.

Following the intuition from Ref.~\cite{Takeda05}, we show how this KW duality can be leveraged to derive an approximate 
fixed-point condition relating the multicritical points of these dual Ising models in presence of disorder,
and explain how this leads to the near-saturation of the bound from Eq.~\eqref{eq:Thresholdbound} for broad families 
of homological CSS codes.
Recall how the dual lattice structure for a given $d$-dimensional Ising model was defined in Sec.~\ref{sec:self-duality},
as well as the resulting fixed-point condition that was derived by leveraging the invariance of the theory under duality transformation.
One can make a similar conclusion in presence of quenched disorder using the replica trick. Consider $n$ replicas 
of the disordered model characterized by the same Hamiltonian
\begin{equation}
H = -J \sum_{C_j} \eta_j \prod_{i\in\partial C_j} \sigma_i,
\end{equation}
where $\eta_j=-1$ with probability $p$. One can define an averaged Boltzmann weight for each of the cases where the 
spins associated with an individual coupling cell $\partial C$ are antialigned for $0\leq k \leq n$ of the $n$ replicas and are aligned in 
the remaining $n-k$ replicas. Averaging over the possible values of $\eta_C$, one gets:
\begin{flalign}
x_k(p,K) & =\langle \,\,\text{exp}\left[ K \eta_C  \sum_{\alpha=1}^n \prod_{i\in\partial C}\sigma_i^{\alpha}\right]\,\,\rangle \nonumber\\
& = p e^{(n-2k)K} +(1-p) e^{-(n-2k)K} \nonumber \\
& = p u_+^{n-k}u_-^{k} + (1-p) w_+^{n-k}w_-^{k}
\end{flalign}
where we introduced the index $\alpha$ to distinguish different replicas, reused the shorthand notation $u_{\pm}(K)$ 
for the Boltzmann factors in the replicas with $\eta_C=+1$ and introduced $w_{\pm}(K)\equiv e^{\mp K}$ for the replicas 
with $\eta_C=-1$. To take into account all possible spin configuration of the $n$-replicated systems, the 
disorder-averaged partition function will be a function of all these Boltzmann factors:
\begin{equation}
\langle Z^n\rangle_{\text{dis}} \equiv Z_n\{x_0 (p,K),x_1 (p,K), \,\dots\,x_n (p,K) \}.
\end{equation}
Keeping in mind that $u_{\pm}$ and $w_{\pm}$ correspond to different bond configurations and therefore undergo 
different Fourier transformations, we can define the dual Boltzmann factors on the dual lattice as the $n$-fold 
2-component Fourier transform,
\begin{flalign}\label{eq:dualBoltzmann}
x^*_{k}(p,K) &= \mathcal{F}\left( p u_+^{n-k}u_-^{k} + (1-p) w_+^{n-k}w_-^{k} \right) \\
&= p (u_+^*)^{n-k} (u_-^*)^{k} + (1-p) (w_+^*)^{n-k}(w_-^*)^{k}   \nonumber \\
&= 2^{-n/2}\left(p + (-1)^k (1-p)\right)\left(e^K+e^{-K}\right)^{n-k}\left(e^K-e^{-K}\right)^{k}. \nonumber 
\end{flalign}
The disorder generalization of Eq.~\eqref{eq:genKW} directly follows:
\begin{equation}
Z_n\{x_0,x_1, \,\dots\,x_n \} = 2^{\bar{a}} Z_n\{x_0^*,x_1^*, \,\dots\,x_n^* \}
\end{equation}
Since the multicritical point is expected to lie along the NL given by Eq.~\eqref{eq:NL}, where the enhanced 
symmetry reduces the fixed-point conditions to a one-parameter problem, the replicated local weights $\{x_k\}$ 
cannot, in general, satisfy the full set of self-duality equations $x_k=x_k^{*}$ simultaneously in the presence 
of disorder, as the resulting system of equations is overconstrained.

Restricting our analysis to the Nishimori line, which in our generalized formulation is given by the condition 
$e^{-2K_N}= p/(1-p)$, Eq.~\eqref{eq:dualBoltzmann} yields an exact hierarchy among the dual Boltzmann factors, 
distinguished by the parity of $k$:
\begin{equation}\label{eq:hierarchy}
\frac{x_{k}^*}{x_0^*}=(1-2p)^{k}\,\,\text{(for even }k),\qquad
\frac{x_{k}^*}{x_0^*}=-(1-2p)^{k+1}\,\,\text{(for odd }k).
\end{equation}
For $0<p<1$, the $k>0$ sectors are suppressed in magnitude relative to the principal factor, with $|x_{k}^*/x_0^*|\leq(1-2p)^2<1$. 
Following the standard replica approach, we set $K=K_N(p)$ and impose the \emph{principal-factor}
condition only on the $k=0$ sector. This gives~\cite{Takeda05,Nishimori07,Song22}
\begin{equation}
x_0(p_1^C,K_1^C)x_0(p_2^C,K_2^C) \approx x_0^*(p_1^C,K_1^C)x_0^*(p_2^C,K_2^C).
\end{equation}
The inclusion of subleading sectors ($k>0$) would slightly shift the estimate while leaving the leading condition 
as the dominant constraint. Nevertheless, it was shown in several cases to yield the closest approximation of the 
true multicritical point~\cite{Hinczewski05,Hasenbusch07,Masayuki09,Kubica18,Gattringer18,Song22}. 
Explicitly, this results in
\begin{equation}\label{eq:duality}
H(p_1^C) + H(p_2^C) \approx 1,
\end{equation}
where $H(p) := -p \,\log_2 p - (1-p)\,\log_2 (1-p)$.
This coincides with the upper limit of the i.i.d. error threshold bound
$H(p_X^{\text{th}}) + H(p_Z^{\text{th}}) \leq 1$.
Although the derivation of Eq.~\eqref{eq:duality} relies on the principal-factor
approximation, it works surprisingly well for several known codes as summarized in Table~\ref{tab:threshold}.
This duality-based perspective thus provides a unifying explanation for the systematic saturation of the optimal error
threshold bound in CSS codes whose i.i.d. noise models can be mapped to mutually dual Ising spin models. 

We can take this observation one step further by combining this result with the homological construction of CSS codes. 
As discussed in Sec.~\ref{sec:Codefamilies}, homological CSS codes are constructed from two classical homological codes
related through the chain-complex structure described in Eq.~\ref{eq:chain}. If, in addition, these two classical codes 
satisfy the Wegner completeness condition of Eq.~\eqref{eq:completeness_homo}, their SM mapping yields 
a pair of dual Ising models, which can be expected to approximately satisfy Eq.~\eqref{eq:duality}.
This analysis suggests the existence of a structural principle, namely that homological CSS codes and, more generally, 
any CSS code whose independent $X$- and $Z$-noise models admit a SM mapping to mutually dual Ising models, 
are expected to nearly saturate the optimal error-threshold bound. This expectation is, thus far, consistent with the results shown 
in Table~\ref{tab:threshold} and explains why all listed topological CSS codes, each admitting a chain-complex homological 
description, saturate said bound. 

The natural question that follows is whether this duality-based mechanism extends beyond conventional topological codes admitting
a homological construction. To address this, we turn our attention to $\mathbb{Z}_2$ fracton models.

This same threshold bound saturation was recently shown to be satisfied by the X-cube model, a primary example of a
non-self-dual 3D fracton model which admits a chain-complex description~\cite{Song22} and, much like the Checkerboard model and Haah's code, 
can be seen as the gauged counterpart of a classical model, i.e.\ the Plaquette Ising model~\cite{Vijai16}.
Here, we instead focus on self-dual fracton models, with particular emphasis on the Checkerboard model, which will serve as a stringent test 
of whether this duality-based conjecture is indeed generally valid in 3D, and whether the unconventional excitation structure and 
subdimensional constraints characteristic of fracton phases introduce qualitative deviations from this behavior.

\begin{table}[!t]
\centering
\setlength{\tabcolsep}{9pt}
\renewcommand{\arraystretch}{1.3}
\begin{tabular}{@{}lccc@{}}
\toprule
\hline
Code & $p_X^{\text{th}}$ & $p_Z^{\text{th}}$ & $H(p_X^{\text{th}}) + H(p_Z^{\text{th}})$\\
\hline
2D Surface code & $0.1094(2)$~\cite{Dennis02} &  $0.1094(2)$~\cite{Dennis02} & $0.9962(9)$  \\
2D Color code   & $0.109(2)$~\cite{Katzgraber09} & $0.109(2)$~\cite{Katzgraber09} & $0.994(9)$ \\
3D Toric code   & $0.2327(3)$~\cite{Ozeki98} & $0.033(4)$~\cite{Ohno04} & $0.99(2)$\\
\vspace{2mm}
3D  Color code   & $0.276$~\cite{Kubica18} & $0.019$~\cite{Kubica18} & $0.986$ \\
     X-Cube code  & $0.152(4)$~\cite{Song22} & $0.075(2)$~\cite{Song22} & $1.00(1)$\\
{\bf Checkerboard code} & $??$   & $??$ & $??$ \\
{\bf Haah's code} & $??$   & $??$ & $??$ \\
\hline
\bottomrule
\end{tabular}
\caption{Optimal code capacity thresholds of representative 2D and 3D known CSS codes whose noise models admit a duality relation. 
            All of these are shown to saturate the code capacity threshold inequality, in agreement with the fixed-point condition 
            given by the generalized KW duality.}
\label{tab:threshold}
\end{table}

\section{Optimal Error Threshold for Self-Dual Fracton Models}\label{sec:Fractons_in_QEC}

\subsection{Checkerboard and Haah's code as CSS codes}
Having laid out all the relevant notation, we can finally frame the fracton models derived in Chapter~\ref{chap:fracton_models}
within the context of QEC.

Both the Checkerboard model and Haah's code can be viewed as CSS stabilizer codes in the sense described in Sec.~\ref{sec:Topocodes}, 
with their stabilizer Hamiltonians given by Eq.~\ref{eq:Checkerboard_Hamiltonian} 
and Eq.~\ref{eq:Haah's_Hamiltonian} respectively.

The subextensive degeneracy of their ground-state manifold makes these models capable of encoding a subextensive number
of logical qubits. From the Checkerboard model one can construct a code family with $L\in 2\mathbb{Z}$, encoding 
$6L-6$ logical qubits with code distance $L$, each associated with a pair of anticommuting logical operators $X_L$ and $Z_L$
built from string-like operators. 
Haah's code is constrained by the discontinuous dependence of the ground-state degeneracy on system size described 
in Eq.~\ref{eq:GSD}: we therefore focus on the $L=2^n,\,n\in\mathbb{N}$ family, 
with $4L-2$ encoded logical qubits and respective fractal logical operators whose structure can be derived with the 
method described in Sec.~\ref{sec:FIM}.

In both fracton models, the occurrence of a single-qubit $Z-$type ($X-$type) error on an encoded state flips 
the eigenvalue of four cube stabilizer operators $A_c$ ($B_c$), which correspond to our fracton 
excitations. The resulting syndrome excitations obey the same planar and fractal charge-conservation constraints 
described in Secs.~\ref{sec:checkerboard} and~\ref{sec:Haah}.

Since both the Checkerboard model and Haah's code are self-dual, their error thresholds under i.i.d. noise 
satisfy the relation $p_X^{\text{th}}=p_Z^{\text{th}}\equiv p^{\text{th}}$. The rest of our analysis will be 
centered on the identification of this threshold value for the Checkerboard model and the implications of our 
results in relation to the generalized KW duality hypothesis discussed in the previous section.

\subsection{Error threshold of fracton codes through the SM mapping}

The behavior of these fracton models is analyzed using the SM mapping introduced in Sec.~\ref{sec:SMmapping}. 
Comparing the quantum-to-classical SM mapping with the Wegner gauging procedure outlined in Sec.~\ref{sec:Wegnergauging}, 
one finds that, for stabilizer codes obtained via gauging, applying the SM mapping to the i.i.d. $Z$-noise model effectively 
inverts the gauging construction, up to addition of random-bond disorder in the resulting classical model. 
Consequently, in the presence of noise, the SM mapping of the Checkerboard 
and Haah's codes reduces to the classical Tetrahedral and Fractal Ising models with quenched disorder, respectively:
\begin{flalign}
H_{\rm TIM}^{SM} &=  - \sum_i \left(\eta_i^+ \sigma_i \sigma_{i+\hat{x}+\hat{y}} \sigma_{i+\hat{x}+\hat{z}}\sigma_{i+\hat{y}+\hat{z}} +
                               \eta_i^- \sigma_i \sigma_{i-\hat{x}-\hat{y}} \sigma_{i-\hat{x}-\hat{z}}\sigma_{i-\hat{y}-\hat{z}} \right).  \\
H_{\rm FIM}^{SM} &=  - \sum_i \left(\eta_i^+ \sigma_i \sigma_{i+\hat{x}} \sigma_{i+\hat{y}}\sigma_{i+\hat{z}} +
                               \eta_i^- \sigma_i \sigma_{i+\hat{x}+\hat{y}} \sigma_{i+\hat{x}+\hat{z}}\sigma_{i+\hat{y}+\hat{z}} \right).
\end{flalign}
In the notation of the SM mapping, $\sigma_i=\pm 1$ indicate the Ising spins of the model and $\eta_i^{\pm}$ represent the coupling strengths, whose value
depends on the corresponding error configuration of the original model. Since both the classical models and their gauged counterparts are self-dual, these 
Hamiltonians also capture the statistics of the i.i.d.\ $X$-noise model.
Using the i.i.d.\ noise model introduced in Sec.~\ref{sec:Decoding}, each qubit experiences a physical $Z$ (or $X$) error independently with probability $p$. 
For a fixed error rate $p$, a random error configuration $E$ is generated by assigning each coupling constant the value $\eta_i^{\pm} = -1$ with probability 
$p$ and $\eta_i^{\pm} = +1$ with probability $1-p$.

As discussed in Sec.~\ref{sec:OptimalErrorThreshold}, if the error rate $p$ lies below the optimal threshold $p^{\text{th}}$, the random-bond Ising (RBI) model 
obtained via the SM mapping of the corresponding stabilizer code exhibits an ordered phase that contains the Nishimori inverse temperature
$\beta_N(p)=\frac{1}{2}\log{\frac{1-p}{p}}$ in the thermodynamic limit. Keeping the disorder rate of the bond configuration $p$ fixed, the presence of this 
ordered phase is accompanied by a phase transition at a temperature $T_c(p)$ satisfying $T_c(p) > T_N(p)=1/\beta_N(p)$. 
Diagnosing the existence of such a phase transition requires a finite-size scaling analysis. 
This becomes non-trivial in the presence of quenched disorder, whose effect on first-order transitions can be understood qualitatively through the Imry-Ma argument~\cite{Imry79}. 
For systems exhibiting a strong first-order transition in the disorder-free limit, quenched disorder lowers the free-energy cost of forming interfaces in the thermodynamic limit. 
This is captured by a deviation of the domain wall free energy cost from its disorder-free scaling behavior $\Delta F\propto L^{D-1}$ and in an overall softening of the first-order transition.
While in D$\leq 2$ spatial dimensions this is known to eliminate first-order behavior altogether~\cite{Aizenman89}, its effect on 3D models is not generally predictable analytically
Nevertheless, in many 3D Ising models with quenched disorder, a sufficiently strong softening of the first-order transition is observed, so that the critical behavior near 
the transition becomes effectively continuous~\cite{Cardy99,Chatelain01,Song22}. When the disorder becomes excessively strong, the transition disappears altogether, giving way to a crossover regime.
\vspace{3cm}
\newpage

\subsection{Numerical analysis of the optimal error threshold for the Checkerboard Model}

\subsubsection{Disorder averaging}

A disorder realization at error rate $p$ is generated by starting from the fully ferromagnetic configuration, $\eta_i^\pm=+1$ for all $i$, 
and independently flipping each coupling with probability $p$.

To perform a finite-size scaling analysis at a fixed error rate $p$, we first need to estimate the thermodynamic expectation values of different observables in the 
presence of quenched disorder. Since the accessible system sizes are finite, a single disorder realization samples only a limited subset of the full distribution of error configurations. 
Physical observables for a system of size $L$ at temperature $T$ are therefore obtained by averaging over multiple disorder realizations generated according to the error rate $p$. 
To do this, we first compute the thermal expectation value $\langle O \rangle_T$ on each disorder realization. The disorder-averaged observable is then obtained by averaging these 
quantities over independent realizations. This second averaging step is essential at finite $L$, as different disorder realizations may yield significantly different values of 
$\langle O \rangle_T$, particularly near criticality where self-averaging may fail. Reliable scaling analysis therefore requires sufficiently large disorder samples~\cite{Aharony96}.
The numerical simulations are organized into independent datasets labeled by the pair $(L,p)$, where $L$ denotes the system size and $p$ the error rate.

Due to the large number of numerical simulations required to obtain a disorder-average for a single $(L,p)$ data point, the multicanonical MC methods used in Chapter~\ref{chap:subsystem_symmetries} are not a viable option, 
as the weight learning procedure would have to be performed for each individual simulation. Leveraging the expected softening of the first-order transition, we instead decided to simulate each individual random-bond
configuration using Parallel Tempering MC simulations discussed in Sec.~\ref{sec:PT}. To optimize the efficacy of the update, for each set of disorder simulations at a given disorder rate $p$ and system size $L$ we choose
a fixed set of temperature values that ensure that, in most simulations, each replica is able to explore the whole temperature range and yields an approximately linear average upness curve. 
This is a necessary condition to obtain meaningful disorder averages while still ensuring that the majority of replicas reach equilibrium. 

\begin{figure}[t]
\centering
\includegraphics[width =1.\textwidth]{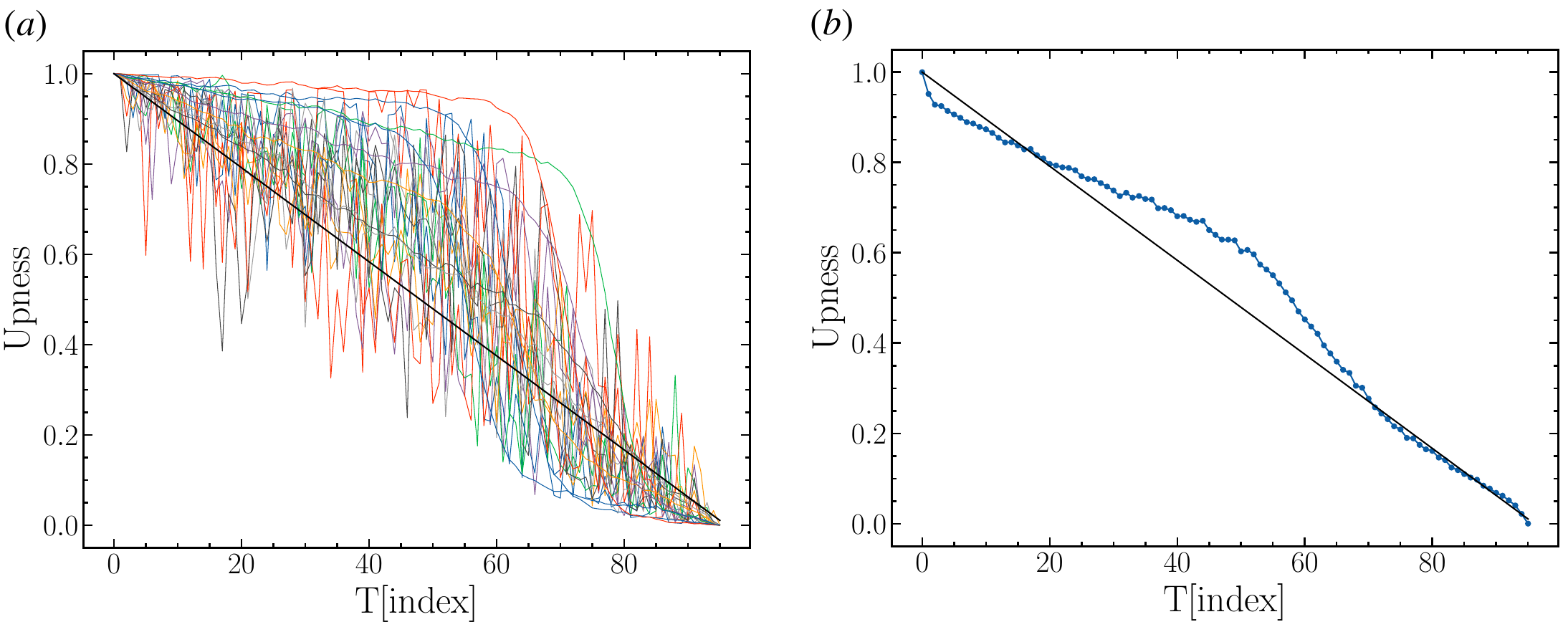}
\caption[Exemplary upness curves for Tetrahedral Ising model in presence of quenched disorder.]
      {Exemplary upness curves for TIM, taken from the disorder simulations at ($L=5,p=0.107$). (a) Since each disorder configuration contains a different set of interaction constants, their energy landscapes will differ non-trivially,
thus leading to significantly different optimal temperature allocations. (b) The distribution of temperature points is chosen such that the average of the upness curves across the different simulations 
is approximately linear.}
\label{upness_disorder}
\end{figure}

Each disorder realization was simulated using single-spin Metropolis-Hastings updates in conjunction with PT, with a temperature distribution tailored to each $(L,p)$ dataset 
to reduce autocorrelation times. Despite the presence of quenched disorder near the error threshold, the model still exhibits clear signatures of a first-order transition 
(see Fig.~\ref{fig:090}), which led to long simulation runtimes. 

Equilibration of each disorder configuration was assessed using several complementary diagnostic methods. As a first test, we analyzed the energy timeseries of each simulation.
The timeseries is first partitioned into logarithmic bins defined by intervals $[2^{\tau},2^{\tau+1}-1]$, with $\tau \in \mathbb{N}$. After averaging the measurements within each bin, 
the resulting binned timeseries was used to infer equilibration. More precisely, the system is considered to have reached equilibrium when the values of the last three bins are within error 
bars of each other~\cite{Bombin12,Song22}. 

While this criterion reliably detects equilibration at low temperatures, it is excessively strict when studying temperatures near the phase transition, where the exploration of both ordered and disordered
phases may lead to a replica being stuck in one of the two phases through the time interval associated to a specific bin, thus leading to the rejection of an otherwise equilibrated
disorder realization due to an undersampling of individual bins. 

In most cases, a qualitative analysis of $O$, $\rchi_{O}$ and $C_V$ of an individual replica was sufficient to infer whether a disorder simulation had failed to reach equilibrium, 
or if the equilibrium distribution was undersampled.
When these two methods turned out to be inconclusive, we performed histogram reweighting tests of the energy distributions at nearby temperatures~\cite{Janke08}. The simulation is considered 
equilibrated if, for most pairs of inverse temperatures $\beta$ and $\beta '$ near the transition, their energy histograms agree within error bars after applying to the distribution of the histogram
at temperature $\beta$ the weights $g(E) = e^{-(\beta ' - \beta)E}$. Agreement between reweighted histograms indicates that the replicas collectively sample a consistent equilibrium 
distribution across temperature space, even if individual realizations might occasionally remain trapped in a single phase.

Taking all these considerations into account, the system still required a remarkable amount of Monte Carlo sweeps to reach equilibration. 
The details of the numerical simulations are reported in Appendix~\ref{app:disorder}.

\subsubsection{Numerical results}

Since both the Checkerboard and Haah's code are self-dual, the generalized Kramers-Wannier duality derived in Sec.~\ref{sec:threshold-KW} places the optimal 
error threshold near $p^{\text{th}}\approx 0.11$. This value serves as a starting reference point for our numerical analysis. In this numerical analysis, 
we focus our attention on the Checkerboard model and, consequently, on the disorder analysis of the TIM with quenched disorder. 

After a preliminary round of simulations used to approximately define the location of the error threshold, the relevant disorder interval was 
restricted to $p\in[0.105,0.110]$. To achieve a precision on the threshold value comparable to that of previous 
studies~\cite{Katzgraber09,Kubica18,Song22}, it was sufficient to simulate three data points: the lower bound, the upper bound and an intermediate conservative 
estimate of the possible location of the threshold, which was fixed at $p^{\text{mid}}=0.107$. Since the transition behavior was shown to change qualitatively between a 
first-order regime and a crossover regime, simulating additional disorder values would not meaningfully improve the statistical robustness of the threshold estimate. 
The reported simulation set thus represents the minimal dataset that captures all qualitative changes relevant for locating the tricritical point while keeping 
the computational effort within feasible limits.

\begin{figure}[t]
\centering
\includegraphics[width =1.\textwidth]{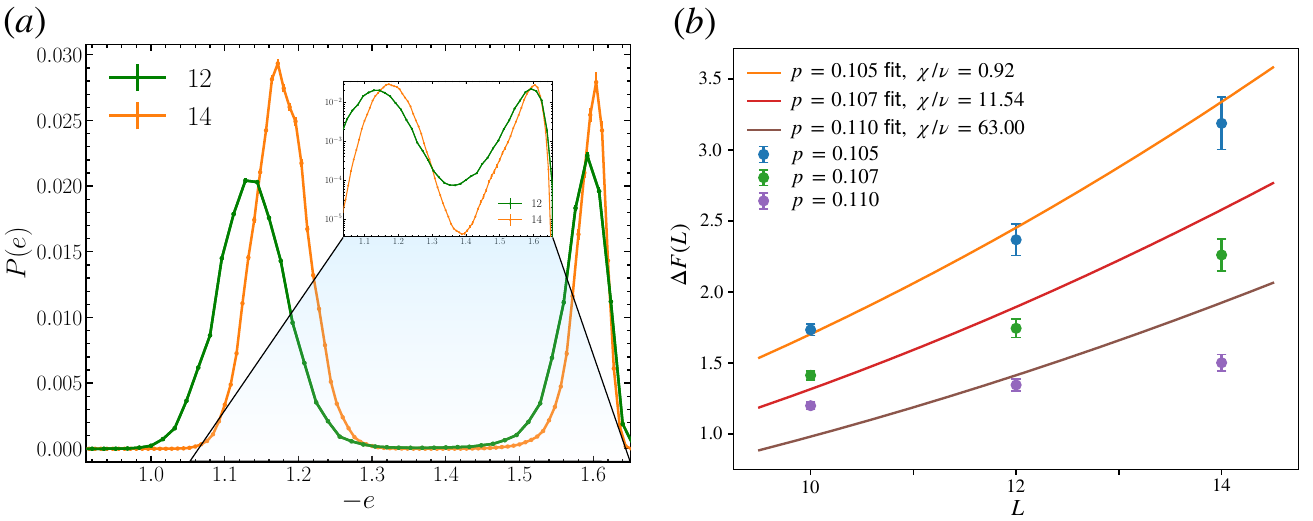}
\caption[Energy histograms and domain-wall free energy cost of the Tetrahedral Ising model.]
      {(a) Energy histograms at the transition temperature for a representative disorder value below threshold ($p = 0.090$). 
            The sharpening of the double-peak structure with increasing system size is consistent with a first-order transition.
            However, a double-peak structure alone is not conclusive, as finite-size effects may also produce two peaks whose separation decreases with growing $L$.
            The defining signature of a first-order transition is the progressive suppression of configurations in the energy range between the peaks, which becomes more pronounced with increasing system size (see inset).
            This behavior serves as a reference for the expected scaling below the error threshold.         
            (b) Domain-wall free energy cost as a function of system size, obtained from disorder-averaged data at the transition of different $(L,p)$ points.
            For each disorder strength, the data are fitted to the $L^2$ scaling expected for a first-order transition using a $\rchi^2$ analysis.
            Deviations from this scaling quantify the disorder-induced softening of the transition, allowing us to estimate the disorder value beyond 
            which the transition can no longer be classified as first-order.}
\label{fig:090}
\end{figure}

The relevant physical quantities such as order parameters, correlation lengths and susceptibilities are the same as those defined in Sec.~\ref{sec:NumericalMethods}
for the disorder-free limit. While these still allow us to identify the subsystem ordering, the presence of quenched disorder softens the first-order features in all of these quantities. 
Furthermore, most of them do not allow for a clear distinction in the changes between first-order, second-order and crossover features caused by small increments of the disorder rate. 

For this reason, our qualitative analysis of the transition primarily focuses on the study of the disorder-averaged energy density histogram 
\begin{equation}
      P(e)=\langle\delta(e-e')\rangle,
\end{equation}      
and the second-moment correlation length
\begin{equation}\label{eq:corrlength_2}
\xi_L = \frac{1}{2 \sin\left(|\textbf{k}_{\text{min}}/2|\right)} \left( \frac{\tilde{G}(\mathbf{0})}{\tilde{G}(\textbf{k}_{\text{min}})} - 1 \right)^{1/2},
\end{equation}
where $\tilde{G}(\textbf{k})$ represents the Fourier transform introduced in Sec.~\ref{corrfuncdef} of the spatial correlator $G(r)$ of the TIM defined in Eq.~\ref{eq:TIM_corr}.
We were also able to quantify the softening of the first-order nature of the transition by observing the free energy barrier, which can be derived from the energy histograms as
\begin{equation}
\Delta F (L) = -\log{\frac{P_{\text{min}}}{P_{\text{max}}}},
\end{equation}
and is expected to scale with system size proportionally to $L^{D-1}$=$L^{2}$ in the limit of a strong first-order transition.
Other relevant quantities, such as heat capacity, order parameter susceptibility and binder cumulants are reported in Appendix~\ref{app:disorder}.

The first-order nature of the phase transition is identified by the characteristic sharpening of the double-peak behavior in the energy histograms. 
From Fig.~\ref{fig:090}(a) it is possible to see that, despite the presence of quenched disorder, the first-order features of the transition remain significantly sharp
even in proximity of the error threshold. This allows us to identify three different regimes in the threshold analysis: a first-order transition regime below threshold, a crossover regime above threshold and 
a softened transition regime in proximity of the error threshold.

Let's start with the lower bound of our chosen interval, at $p=0.105$. In this regime, the double-peak behavior of the disorder averaged energy histograms at the transition point of their 
respective system sizes tend to sharpen as $L$ increases. At the same time, the gap between them becomes deeper, indicating a suppression of intermediate configurations. This behavior 
is consistent with that of a first-order transition and is further confirmed by the free energy barrier which is compatible with an $L^2$ scaling as shown in Fig.~\ref{fig:090}(b). 
The correlation length shows a crossing between the two curves at larger system sizes, with the smallest of the three never crossing due to strong finite-size effects, a feature that 
persists even at much lower error rates. 
This crossing is not necessarily expected in a first-order transition, but is nevertheless consistent with the relaxation of the finite-size effects in a first-order transition and, 
most importantly, rules out the eventuality of a crossover regime, where a crossing should not occur at all. 
As we will see shortly, these results indicate that $p=0.105$ constitutes the upper bound of the first-order transition regime. For larger error rates, the first-order signatures are progressively softened, 
most clearly reflected in the scaling behavior of the domain-wall free energy.

\begin{figure*}[!tb]
\centering
  \includegraphics[width=1\textwidth]{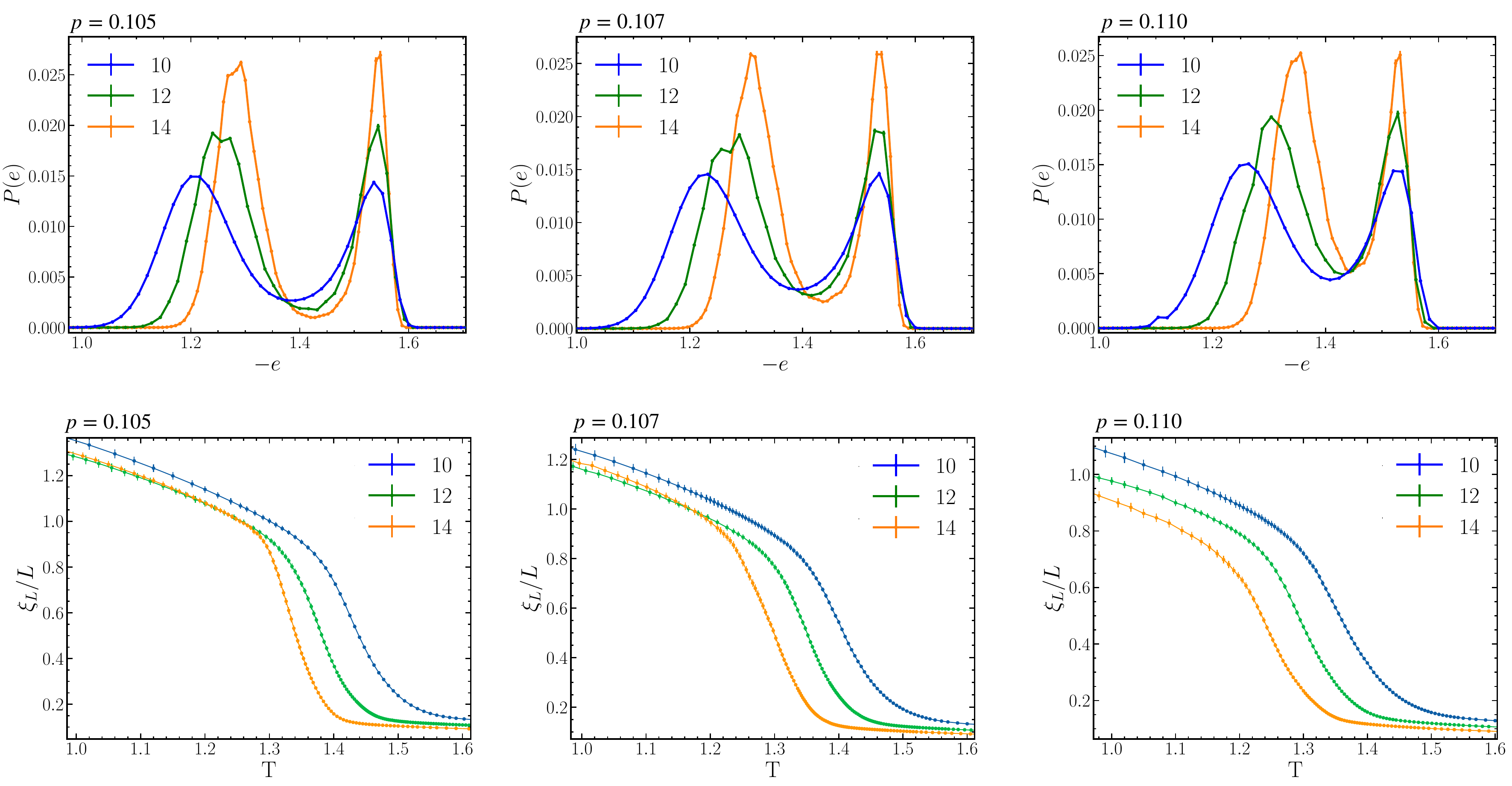}
  \caption[Disorder-averaged energy histograms and correlation lengths of the Tetrahedral Ising model, taken at different disorder values and 
            for various system sizes.]
      {Disorder-averaged energy histograms (top row) and correlation lengths (bottom row) of TIM, taken at different disorder values and 
            for various system sizes. While the energy histograms present a typical first-order double-peak behavior at all 
            disorder values due to strong finite-size effects, the presence of a first-order transition at $p=0.105$ is made 
            obvious by the sharpening of the two peaks and a deepening of the minimum between them as the system size grows, 
            a feature that can still be observed, though less prominently, at $p=0.107$. This lowering of the minimum 
            corresponds to a growing domain wall free energy cost as the system size increases which, in the thermodynamic limit, 
            determines the presence of a first-order transition. On the other hand, at $p=0.110$ the minimum goes up with system 
            size, indicating that the first-order features of the system are slowly disappearing. This is further highlighted by 
            the sharp absence of any crossing in the correlation lengths regardless of system size, another indicator of a 
            crossover regime. The sharp contrast between these behaviors allows us to identify the ordered and disordered phases 
            of the model, and estimate the optimal error threshold to sit in between $0.107$ and $0.110$.}
  \label{fig:disorderplots}
\end{figure*}

Conversely, for $p=0.110$ and above, the double-peak behavior in the energy histogram is still visible at finite system sizes. However, their sharpness visibly softens as the system size increases, and
the suppression of intermediate states weakens with increasing $L$. On top of this, Fig.~\ref{fig:090}(b) shows how the behavior of the free energy barrier sharply deviates from the $L^2$ scaling that
characterizes first-order transitions. Together, these observations rule out the existence of a first-order transition in the thermodynamic limit: instead, one can expect that, in the thermodynamic limit, 
the two peaks in the energy histogram will slowly merge into a single peak. This leaves two possible scenarios: either the system has entered a continuous, second-order transition regime, 
or the model has reached a crossover regime. This is where the correlation length becomes a crucial quantity: as shown in Fig.~\ref{fig:disorderplots}, the clear lack of crossing in the correlation length 
rules out the possible presence of a continuous phase transition, thereby confirming that the system is indeed in a crossover regime. We can then conclude that the remaining double-peak behavior 
observed at finite $L$ is due to the strong finite-size effects affecting the model. This serves both as a characteristic example of the expected behavior in the disordered regime 
as well as a reference for comparison with lower error rates.

Finally, we turn to the intermediate disorder value $p=0.107$. Much like at $p=0.105$, the characteristic double-peak first-order feature of the energy histogram is still visible. Unlike what we observed at 
$p=0.105$ and $p=0.110$, the suppression of intermediate states was neither sharpened nor softened appreciably for different system sizes: instead, the minimal between the peaks lowered only 
slightly as $L$ increases, indicating a substantial reduction of the interfacial free-energy cost compared to that at $p=0.105$. 
The domain wall free energy clearly deviates from the $L^2$ scaling, which is consistent with the softening of the first-order features. However, this does not exclude the persistence of a first-order
phase transition in the thermodynamic limit, nor does it rule out the possibility that the model might actually relax into a second-order phase transition. 
Nevertheless, Considering that the correlation length between the curves at the two larger system sizes still showcase a crossing point, we can use the same arguments made for $p=0.105$ to rule out the
possibility of a crossover regime and argue that the system is still in the decodable phase, albeit being in very close proximity of the true error threshold. We thus identify this point as lying
in the softened transition regime.

Combining these results, we are able to perform an estimate of the optimal error threshold of the Checkerboard model. Since $p^{\text{th}}$ can be safely placed between $0.105$ and $0.110$,
we assume a symmetric uncertainty over the plausible interval $[0.105,0.110)$. Consistently with our numerical results, we take $p=0.107$ as a conservative estimate of the true error threshold and set
the uncertainty to be half the interval width, $\delta p = 0.0025$, which rounds up to $\delta p=0.003$. From this, we estimate the error threshold for the Checkerboard model to be $p^{\text{th}}=0.107(3)$.
Plugging this value in Eq.~\ref{eq:duality} gives $H(p^{\text{th}}_X)+H(p^{\text{th}}_Z) = 0.98(2)$, a result which nearly saturates the optimal error threshold bound as predicted by our generalized KW duality.

\vspace{1.5cm}

\subsection{Considerations about Haah's code}

From the observations obtained thus far, every code studied thus far whose $X-$ and $Z-$noise models are characterized by the mutual duality relation defined in Sec.~\ref{sec:self-duality} were shown to 
saturate the optimal error threshold bound. While the validity of this threshold saturation relies on the \emph{principal-factor} condition introduced in Sec.~\ref{sec:threshold-KW}, 
this property is shown to hold not only for the well-known 2D and 3D Ising models related to toric and color code, but also for all subsystem-symmetric 3D models studied thus far. 
Evidence of the former is provided by our disorder analysis of the Tetrahedral Ising model, while the latter was explored more in detail in Ref.~\cite{Song22}, where the disorder analysis of the 
mutually dual Plaquette Ising model and the Ashkin-Teller model (both derived from the noise models of X-cube fracton model) were shown to also satisfy this threshold saturation hypothesis. 
This strongly suggests that the threshold-saturation hypothesis extends naturally to fracton models.

One immediate question to ask is then whether these results can also be extended to the Fractal Ising model and, consequently, Haah's code. The most straightforward way to address this would be to 
perform an analogous finite-size scaling analysis on the FIM with quenched disorder. This approach, however, is severely limited by the fact that the FIM is characterized by different symmetries and scaling
behavior for specific families of system sizes classified in Eq.~\ref{eq:GSD}. In addition, the numerical analysis is further restricted by the presence of strong finite-size effects, 
which severely affect any system size $L<5$, while computational limitations render system sizes $L>15$ unattainable. These constraints prevent a controlled finite-size scaling disorder analysis, 
particularly for the $L=2^n$ family, for which relevant, adequately accurate disorder-averaged results could be obtained only for $L=8$.

Instead, we propose a heuristic argument explaining how the conclusions obtained for the Tetrahedral Ising model naturally extend to the Fractal Ising model. Recall that, in the disorder-free limit, both models
exhibit a single order-disorder phase transition that persists in the thermodynamic limit. Despite the anomalous finite-size scaling that accompanies subsystem symmetry breaking, both transitions
are strongly first-order, with the domain-wall free energy cost and energy gap between ordered and disordered phases of the FIM being comparable, and in fact larger than those observed in the TIM.
The existence of this transition implies the presence of a stable ordered phase that persists under sufficiently weak quenched disorder, thereby tracing a phase boundary in the $p$–$T$ plane as shown in Fig.~\ref{fig:Disorder_diagram}. 

In the presence of quenched disorder, the principal-factor condition follows from the dominance of the exact hierarchy among the dual weights derived in Eq.~\ref{eq:hierarchy} along the Nishimori line. 
This approximation is reasonable provided that the multicritical point at the intersection between the phase boundary and the Nishimori line is governed by a single thermodynamic transition rather than by 
the emergence of a spin-glass phase with competing pure states~\cite{Takeda05,Queiroz06}. While this can be the case in systems where a spin-glass phase is present and therefore replica symmetry breaking may take place~\cite{Binder86},
in the present cases both the TIM and FIM exhibit a single order–disorder transition in the disorder-free limit, without evidence of magnetic frustration, competing glassy phases or any extensive degeneracy beyond the 
subsystem-symmetric structure discussed previously. Consequently, in the presence of weak quenched disorder, the first-order transition of the Fractal Ising model 
is expected to soften in a similar fashion to that of the Tetrahedral case, leading to a single phase boundary and a unique crossing point with the Nishimori line, where the multicritical point is expected to reside.
Having shown that the principal-factor condition is still satisfied in all non-glassy 3D Ising models with subsystem symmetry breaking studied thus far (both self-dual and non-self-dual), we find no structural reason 
to expect that the FIM could be characterized by properties that would invalidate the dominance of the principal replica sector. We therefore expect that, like the Checkerboard model, the 
Fractal Ising model also satisfies the principal-factor condition and exhibits near-saturation of the optimal error threshold bound.

\section{Discussion}

The numerical results derived in the previous section, combined with the results already available in the literature, provide strong evidence for the validity of the error threshold bound saturation hypothesis formulated 
in Sec.~\ref{sec:self-duality}. In the context of fracton models, threshold saturation has been shown numerically for the self-dual Checkerboard model investigated in this work, and for the non-self-dual X-cube model 
already studied in the literature~\cite{Song22}, and is expected to hold for the self-dual Haah's code as well. 
More broadly, self-dual fracton codes currently represent the sole known examples of 3D self-dual CSS codes arising from models with intrinsic topological order. 
This property, coupled to the fact that non-self-dual codes necessarily have unbalanced error thresholds, with either $p_{\mathrm{th}}^X$ or $p_{\mathrm{th}}^Z$ being below the optimal $11\%$ value, makes the Checkerboard model (and, 
potentially, Haah's code) the model with the highest optimal error thresholds in 3D, surpassing those of conventional topological codes like 3D toric code ($3.3\%$), 3D color code ($1.9\%$) and the lower threshold of the 
non-self-dual X-cube model ($7.5\%$).

From these results, three fundamental lines of inquiry immediately arise. The first concerns the regime of validity of the fixed-point condition derived from Kramers-Wannier duality. While this question is primarily of a 
classical nature, its direct correspondence with CSS codes highlights the deep, still not fully explored interplay between classical spin models and QEC. A more in-depth understanding of this duality structure and exploration
of more exotic, classical Ising models may provide guiding principles for the design of more resilient and potentially computationally relevant quantum codes.

The remaining questions pertain to the possible applications of fracton models in QEC. In particular, it is natural to ask whether their subextensively degenerate ground-state manifold and the restricted 
mobility of their excitations can translate into structural advantages for scalable quantum memories. For a linear system size $L$, the scaling of logical and physical qubits is summarized in Table~\ref{tab:encoding_rates}: 
from the point of view of scalability, all the fracton models discussed so far possess an encoding rate that goes to zero in the limit of large system sizes. Nevertheless, a quantitative comparison between their encoding 
rates highlights a peculiar utility of the Checkerboard model. In order to encode the same number of logical qubits $k$, the Checkerboard model turns out to have a better encoding rate compared to other fracton models, thus 
representing the most efficient quantum memory among the three. Its potential is further highlighted once we see that its encoding rate is six times larger than that of 2D toric code, a property which, combined with its 
saturated optimal error threshold, makes the Checkerboard model a particularly promising candidate for a topological quantum memory. A remaining challenge is then the identification of a canonical and geometrically 
natural choice of orthogonal logical $X$ and $Z$ operators that enables the independent readout of the individual logical qubits stored in the Checkerboard model.

\begin{table}[!t]
\centering
\setlength{\tabcolsep}{9pt}
\renewcommand{\arraystretch}{1.3}
\begin{tabular}{ccc}
\toprule
\hline
Code              & $k$ & $n$ \\
2D Toric code   &  $2$   & $2\,L^2$ \\
3D X-cube model &   $6L-3$    & $3\,L^3\sim k^3\frac{1}{6^3}$\\
3D Checkerboard model &  $6L-6$     & $L^3\sim k^3\frac{3}{6^3}$  \\
3D Haah's code ($L=2^n$ family) & $4L-2$ & $2\,L^3\sim k^3\frac{2}{4^3}$ \\
\hline
\bottomrule
\end{tabular}
\caption{Logical number of qubits $k$ and physical number of qubits $n$ as functions of linear system size $L$ for different CSS code families}
\label{tab:encoding_rates}
\end{table}

These considerations naturally extend to the potential use of fracton codes as platforms for logical quantum computation. From the perspective of fault-tolerant gate implementation, their relevance is severely constrained by two properties. 
First of all, it is possible to prove that the Checkerboard model is capable of hosting a transversal implementation of the entire Clifford group, a set of logical operators generated by the minimal set 
$C=\{\mathrm{CNOT}, H, S\}$, where
\begin{equation}
H = \frac{1}{\sqrt{2}}
\begin{pmatrix}
1 & 1 \\
1 & -1
\end{pmatrix}, \qquad
S =
\begin{pmatrix}
1 & 0 \\
0 & i
\end{pmatrix}, \qquad
\mathrm{CNOT} =
\begin{pmatrix}
1 & 0 & 0 & 0 \\
0 & 1 & 0 & 0 \\
0 & 0 & 0 & 1 \\
0 & 0 & 1 & 0
\end{pmatrix}.
\end{equation}
Both the Checkerboard model and Haah's cubic code are CSS codes, which always allow for a transversal implementation of CNOT gates between code blocks~\cite{Gottesman98}, and are self-dual, 
implying the possibility of transversally implementing the $H$ gate, which maps the $X$ and $Z$ stabilizers into each other (up to a lattice inversion in the case of Haah's code). 
On top of that, the $X$- and $Z$-stabilizer structures of the Checkerboard model are identical: for every cube $c$, the $X$-type stabilizer $A_c$ and the $Z$-type stabilizer $B_c$ act on the same set of eight qubits. 
Since each $A_c$ has weight $8$, applying the transversal phase gate $S^{\otimes n}$ yields
\begin{equation}
S^{\otimes n} A_c S^{\otimes n \dagger}
= i^{8} A_c B_c
= A_c B_c,
\end{equation}
where we used the conjugation relations $S X S^\dagger = i X Z$ and $S Z S^\dagger = Z$, together with the fact that $8 \equiv 0 \pmod{4}$. 
Moreover,
\begin{equation}
S^{\otimes n} B_c S^{\otimes n \dagger} = B_c.
\end{equation}
Therefore, the transversal operator $S^{\otimes n}$ preserves the stabilizer group and implements a logical phase gate on the encoded qubits.

However, by the Eastin-Knill theorem~\cite{Knill09}, no quantum error-correcting code can realize a universal set of logical gates using only transversal operations: consequently, the Checkerboard model 
cannot support a universal set of transversal gates, and in particular cannot realize any non-Clifford gate needed for universality.

More generally, even in cases where a fracton code might admit transversal implementations of certain non-Clifford gates, an additional structural limitation arises from the global nature of locality-preserving 
logical operations in topologically degenerate stabilizer codes. Transversal and constant-depth logical gates act collectively on the full encoded subspace and typically do not allow independent addressing 
of individual logical qubits~\cite{Bravyi13}. Overcoming this limitation would require either abandoning purely transversal implementations in favor of alternative fault-tolerant schemes or modifying the boundary 
conditions in a controlled manner to isolate logical sectors. Such modifications, however, may alter the encoding rate or threshold properties of the models and therefore affect their practical utility as scalable 
quantum memories.

Finally, an important open question concerns the role that fractonic excitation constraints may play in realistic, circuit-level noise models. The threshold analysis performed throughout this work assumes static, 
uncorrelated noise acting independently on physical qubits. In this regime, we showed how the decoding problem reduces to a combinatorial inference task determined by the boundary relations between the qubit support
and the stabilizer group: in this sense, the restricted mobility of excitations plays no direct role, since errors are treated as independent configurations rather than dynamically generated defects. 
However, the problem changes significantly when moving to the case of circuit-level noise, where both qubit measurement and gate operations are subject to errors. In this setting, faults are detected through
the creation and temporal propagation of stabilizer excitations across successive rounds of syndrome extraction~\cite{Fowler12}. Errors are therefore generated by dynamical processes and acquire both spatial and temporal correlations.
Within the viewpoint of SM mapping discussed in this chapter, the decoding problem of 3D fracton models under circuit-level noise is mapped to the simulation of an effective 4D Ising model with quenched disorder,
obtained by stacking copies of the 3D model, each associated with a syndrome-extraction round, and introducing inter-layer couplings that encode the possible gate and measurement error events~\cite{Dennis02,Chubb18}.

A systematic investigation of fracton codes under circuit-level noise may unveil whether such codes could be characterized by protective mechanisms that could further highlight their suitability as robust quantum memories. 
However, this direction remains largely unexplored, primarily due to the substantial computational complexity associated with decoding 3D models under circuit-level noise, since the simulation of an effectively 4D 
disordered spin system presents a formidable analytical and numerical challenge. Understanding whether the immobility of fractonic excitations translates into enhanced resilience against temporally correlated noise 
therefore remains an open and compelling problem, one that may ultimately determine the practical advantage of fracton codes beyond their favorable threshold properties under idealized noise assumptions.

%% file: mainmatter/conclusion.tex
\chapter{Conclusion and Outlook}
\label{chap:conclusion}

In this thesis, we studied the properties of two different subsystem-symmetric classical spin models, the Tetrahedral and Fractal Ising models, reviewed their relation to the 
topologically non-trivial class of quantum fracton models, the Checkerboard model and Haah's cubic model, and investigated their performance as candidate QEC codes. 
From these results, we established a unified picture connecting the observation of unconventional, finite-temperature ordering in the aforementioned classical spin models, 
the emergence of fracton topological order through a generalized gauging procedure and the viability of the resulting phases as QEC codes.

In the classical context, the Tetrahedral Ising model and the Fractal Ising model provide representative self-dual examples of the wider class of 3D subsystem-symmetric 
models~\cite{Johnston14,Johnston17,Gerlach15}. Despite the different geometries of their couplings and symmetry operators, planar in one case and fractal in the other, both models were shown to exhibit a 
single temperature-driven phase transition separating a high-temperature disordered phase from a low-temperature phase with long-range subsystem order. 
The ordered phases of the two models are respectively characterized by the emergence of planar and fractal ordering patterns, reflecting the underlying subsystem symmetry structure of each model. 
In both cases, the transition is strongly first-order, and its finite-size behavior follows the anomalous scaling of the pseudocritical temperature, $\beta_c(L) - \beta_c^{\infty} \propto L^{-(D-d)}$, 
expected in the presence of a subextensively large number of symmetry-related ground-state sectors. These results place the two models within a broader and increasingly coherent phenomenology of subsystem symmetry breaking in three dimensions.

This phenomenology is particularly striking due to its persistence across various types of models with subsystem symmetry breaking, despite the fact that the underlying subsystem symmetry breaking mechanism may differ substantially. 
In compass models, subsystem symmetries are typically accompanied by geometric frustration, and their breaking is generally driven by order-by-disorder effects~\cite{Gerlach15,Canossa23}.
By contrast, in the Ising models studied here, as well as in other 3D Ising models such as the Plaquette Ising model~\cite{Johnston17}, frustration plays no role. Nevertheless, the symmetry-breaking pattern
and the associated phase diagram remain the same: given the lowest-dimensional subsystem symmetry of the model that can be broken, the system appears to favor
a unique, discontinuous thermal phase transition associated with the simultaneous breaking of the entire set of subsystem symmetries. This suggests that the first-order character is not coincidental, 
but may instead be a robust consequence of the interplay between subsystem geometry, subextensive degeneracy, and isotropic couplings in 3D systems.

Although the accumulated evidence from the Tetrahedral and Fractal Ising models, together with previous work on plaquette Ising model and subsystem symmetric compass models, strongly points to first-order behavior as 
a structurally favored outcome in short-range 3D subsystem-symmetric models, no general theorem has yet been found to enforce this conclusion. Dimensional reduction arguments provide powerful 
constraints on which subsystem symmetries may be spontaneously broken at finite temperature, but they do not determine the order of the transition~\cite{Batista05,Nussinov15}. 
Likewise, macroscopic ground-state degeneracy has been known to be compatible with the existence of continuous phase transitions~\cite{Racz83}, thus suggesting that subextensive or 
even extensive degeneracy alone is not sufficient to preclude criticality. In addition, the presence of $\mathbb{Z}_2$ subsystem symmetries is not sufficient to enforce discontinuity, 
a property that becomes evident in anisotropic limits of these models, where the effective dimensional reduction (e.g., decoupled or weakly coupled layers) restores conventional 
critical behavior~\cite{Mueller17}. The isotropic subsystem-symmetric models considered here therefore occupy a nontrivial regime in which degeneracy, subsystem geometry, and interlayer coupling 
collectively shape the thermodynamic transition into a first-order regime, a phenomenology that persists across models with fundamentally different subsystem symmetries. The question of 
whether a fully interacting, isotropic, short-range 3D Ising model with subsystem symmetry can sustain a genuine continuous transition into a subsystem-ordered phase therefore remains open.

Beyond their classical context, $\mathbb{Z}_2$ subsystem-symmetric Ising models provide a practical route to the construction of lattice gauge theories with fracton phases. Through a generalized gauging 
procedure, it can be shown that classical subsystem symmetries satisfying appropriate geometrical conditions lead to $\mathbb{Z}_2$ fracton LGTs whose excitation structure and mobility, 
topologically non-trivial operators, and ground-state degeneracy directly depend on the geometry of the couplings and subsystem symmetries of the ungauged models. This was shown explicitly through the derivation of 
two paradigmatic fracton codes, the Checkerboard model and Haah’s cubic code. 

These $\mathbb{Z}_2$ LGTs derived from classical spin models are particularly relevant in the context of QEC. Fracton models in particular have attracted considerable interest as candidate quantum memories 
due to their subextensive degeneracy, a property that follows directly from the topological and geometrical features of their support and allows for the encoding of a subextensive number of logical qubits. 
At the same time, the restricted mobility of their excitations raises the possibility of enhanced robustness against local noise. 

The analysis developed in this thesis sharpens this connection by quantitatively characterizing the error resilience of fracton codes through the investigation of their code-capacity threshold under i.i.d. $X$ and $Z$ noise. 
This was achieved by exploiting the statistical-mechanical mapping, which allows us to relate the behavior of quantum CSS codes in the presence of qubit errors to the thermodynamic properties of classical Ising models with quenched disorder.  
Within the broader framework of homological and fracton CSS codes, this mapping can be viewed as an ungauging procedure, relating the noise models
of the quantum LGTs to the statistical properties of the Ising models from which they are derived. 

By performing large-scale numerical simulations of the random-bond Tetrahedral Ising model, we identified a code-capacity threshold of $p_{\text{th}}=0.107(3)$ for the Checkerboard model. This constitutes the first instance 
of a 3D CSS code nearly saturating the theoretical limit of $p_{\text{th}}\approx 0.11$ allowed by the optimal threshold bound. This result elevates the Checkerboard model to one of the most promising candidates for 
fault-tolerant quantum memory among previously studied 3D CSS codes, a distinction that becomes even clearer when comparing its encoding rate with that of other 2D and 3D topological or fracton CSS codes.

More broadly, the Checkerboard model represents the latest addition to the growing family of CSS codes whose $X$- and $Z$-noise models approximately satisfy the fixed-point relation conjectured for 
mutually dual Ising models. The remarkable accuracy of this duality-based prediction is plausibly related to the self-averaging nature of both the ordered and disordered phase, 
as well as to the absence of a spin-glass phase along the Nishimori line, which ensures that the self-consistent replica analysis yields reasonable descriptions. The present work therefore provides 
further support for the hypothesis that any CSS stabilizer code whose noise models map to Ising models that are mutually dual under the Wegner prescription tends to approximately saturate the threshold relation 
predicted by the fixed-point condition discussed in Sec.~\ref{sec:threshold-KW}.

Beyond its conceptual relevance, this result also carries clear practical significance. Exact threshold calculations are notoriously resource-intensive and, for most codes defined on three or more spatial dimensions, often prohibitively expensive. 
The consistency with which the duality fixed-point condition approximately reproduces the thresholds of the CSS codes studied so far therefore makes it a powerful predictive tool. This is especially significant in our analysis of
fracton codes, where the threshold bound saturation has now been numerically verified in both the Checkerboard and X-cube models, thus providing representative examples of self-dual and non-self-dual fracton codes, respectively. 
In light of these results, it is therefore natural to conjecture that the Fractal Ising model also satisfies the principal-factor condition and that the code-capacity threshold of self-dual Haah’s code likewise lies in close proximity to 
the optimal value $p_{\text{th}}\approx 0.11$.

While the threshold analysis presented here provides fundamental insight into the construction of quantum codes with near-optimal error thresholds, and clarifies how fracton codes fit within the broader framework of CSS codes,
the question of whether the constrained dynamics of fracton excitations can play a decisive role under more realistic noise models remains open. While the error models studied in this work correspond to static, uncorrelated error distributions,
the introduction of the temporal axis through repeated rounds of syndrome extractions leads to error distributions with both spatial and temporal correlations, in which error propagation can be treated as a dynamical process. 
In such a setting, the constrained mobility of fracton excitations may in principle give rise to an enhanced robustness against temporally correlated noise, potentially reflected in an elevated error threshold relative to 
other 3D CSS codes. However, because the decoding problem for 3D fracton codes under circuit-level noise maps onto the simulation of an effective 4D disordered Ising model, a direct numerical investigation of this regime 
remains far beyond the reach of current large-scale simulations.

In conclusion, this thesis has established a unified connection between classical Ising spin models, generalized $\mathbb{Z}_2$ lattice gauge theories, and the corresponding CSS stabilizer codes. 
By moving beyond the conventional topological codes and focusing instead on subsystem-symmetric Ising models and their respective fracton codes, we have shown how classical Ising models and 
stabilizer CSS codes derived via gauging share the same underlying statistical-mechanical description. In particular, we have highlighted how the noise models associated with these classes of CSS codes are naturally described by mutually 
dual disordered spin systems and therefore tend to approximately saturate the threshold relation predicted by duality. More generally, this work shows how ideas from classical statistical mechanics, lattice gauge theory, and quantum error 
correction are linked by a common structural framework in which symmetry, geometry, disorder, and topology jointly determine both the properties of their phases of matter and the error resilience of the encoded quantum information. 
This perspective may thus serve both as a guiding principle for the construction of new CSS codes and as a foundation for predicting and enhancing error resilience by leveraging the physics of disordered spin systems.

%% file: backmatter/appendices/appendices.tex
\chapter*{Appendix}
\markboth{\textsc{Appendix}}{\textsc{Appendix}}
\addcontentsline{toc}{chapter}{Appendix}
\clearpage
\clearpage

\section{Numerical details for the disorder-free simulations}\label{app:no_disorder}

In this appendix we summarize the implementation details of the multicanonical Monte Carlo simulations used for the results of Chapter~1. 
A methodological description of the MUCA scheme is given in Sec.~\ref{sec:MUCA}.

For both the Tetrahedral and Fractal Ising models, canonical simulations near the transition temperature do not exhibit reliable tunneling 
between the ordered and disordered phases, already for the smallest system sizes considered. 
To restore phase mixing and ensure that the system is able to sample all canonically relevant regions of the configuration space, 
we employed multicanonical sampling in an energy window covering the energy interval $[E_o,E_d]$ between the ordered and disordered phases.

\begin{table}[!b]
\renewcommand{\arraystretch}{1.5}
\centering
\begin{tabular}{p{1.3cm} p{1.3cm} p{1.3cm} p{1.3cm} p{1.8cm}}
\toprule
\hline 
$L$ & $N_R$ & $N_{WL}$ & $N_{T}$ & $N_{S}$ \\
\midrule
\multicolumn{5}{l}{\bf Tetrahedral Ising model}\\
12 & 64 & $10^5$ & 64 & $5.5\times 10^7$  \\
14 & 64 & $10^5$ & 64 & $8.8\times 10^7$  \\
16 & 64 & $10^5$ & 64 & $10^8$  \\
18 & 64 & $5\times 10^5$ & 64 & $2.5\times 10^8$  \\
20 & 64 & $10^6$ & 64 & $2.9 \times 10^8$  \\
22 & 64 & $10^6$ & 64 & $5.5 \times 10^8$  \\
24 & 64 & $2\times 10^6$ & 64 & $7.7\times 10^8$  \\
26 & 64 & $2\times 10^6$ & 64 & $10^9$  \\
28 & 64 & $3\times 10^6$ & 64 & $10^9$  \\
\midrule
\multicolumn{5}{l}{\bf Fractal Ising model}\\
4 & 64 & $10^5$  & 64 & $4\times 10^6$  \\
8 & 64 & $10^5$  & 64 & $10^7$  \\
16 & 64 & $3\times 10^5$ & 64 & $1.2 \times 10^9$  \\
\hline
\bottomrule
\end{tabular}
\caption{Simulation parameters for the weight-learning procedure and the subsequent multicanonical simulations. $N_R$ denotes the number of independent simulations 
         from which histogram averages at the inverse temperature $\beta$ are taken in the weight learning process. $N_{WL}$ is the number of sweeps used 
         for thermalization and sampling of each replica in the last three iterations, meaning that the final flat energy histograms at $\beta$ are obtained from a total 
         of $N_{WL}\times N_R$ samples. Multicanonical simulations are then run at $N_T$ different temperatures for $N_S$ Monte Carlo sweeps at each temperature.}
\label{tab:MUCA_sim_params}
\end{table}

The MUCA weights $g(\beta,E)$ were determined iteratively at a reference inverse temperature $\beta$ close to the transition. 
Starting from $g(\beta,E)=0$, we performed $N_R$ independent simulations of each model at a reference inverse temperature $\beta$, with 
update probabilities determined by the multicanonical weights defined through
\begin{equation}
Z_{\rm MUCA}(\beta) = \sum_{E} \rho(E) e^{-\beta E - g(\beta, E)}.
\end{equation}
Using an adequate number of single-spin heat-bath updates, we then measured the energy histogram $h_j(E)$ of each replica. 
From the averaged histogram
\begin{equation}
H(E)=\frac{1}{N_R}\sum_{j=1}^{N_R} h_j(E),
\end{equation}
the weights were updated according to the prescription of Sec.~\ref{sec:MUCA},
\begin{equation}
g(\beta,E)\rightarrow g(\beta,E)+\ln H(E)-\bigl\langle \ln H(E')\bigr\rangle_{E'}.
\end{equation}
This procedure was repeated until the histogram was sufficiently flat in the target interval, according to the criterion
\begin{equation}
\frac{\left|H(E)-\langle H(E')\rangle_{E'}\right|}{\langle H(E')\rangle_{E'}} \lesssim 0.1
\qquad \forall\, E\in[E_o,E_d],
\end{equation}
for three consecutive iterations.

After convergence of the weight-learning stage, the resulting $g(\beta,E)$ was used for production runs. 
To simulate nearby temperatures $\beta'$, we used the shifted weights
\begin{equation}
g(\beta',E)=g(\beta,E)+(\beta-\beta')E,
\end{equation}
which remain sufficiently accurate to sample the full coexistence region efficiently, although the corresponding histograms are no longer exactly flat for $\beta'\neq\beta$.

Expectation values in the canonical ensemble were obtained from the MUCA simulations by the standard reweighting procedure described in Sec.~\ref{sec:MUCA}. 
The simulation parameters are collected in Table~\ref{tab:MUCA_sim_params}. Since multicanonical sampling explores a much broader region of configuration space 
than canonical sampling, large numbers of Monte Carlo sweeps are required to obtain accurate estimates of the relevant observables.

\section{Numerical details for the simulations with quenched disorder}\label{app:disorder}
In the following, we report the details regarding the numerical simulations of the TIM with quenched disorder. 
For each dataset labeled by $(L,p)$, we generated $N_d$ independent disorder realizations by starting from the 
ferromagnetic model and flipping each coupling independently with probability $p$. 
Each realization was then simulated independently using parallel tempering Monte Carlo with single-spin Metropolis updates. 
For a fixed $(L,p)$, all realizations were simulated using the same temperature grid, optimized such that, on average, each configuration 
was able to explore the full temperature range and yielded an approximately linear disorder-averaged upness.
Thermal expectation values were first computed separately for each disorder realization and subsequently averaged over the $N_d$ samples. 
Equilibration times were estimated using the diagnostics described in the main text; the corresponding simulation parameters are summarized in Table~\ref{tab:simparams}.

The simulations were run on the LRZ KCS cluster~\cite{LRZ_Linux_Cluster} and used over $7\times 10^6$ CPU hours in total, with an average of $\approx 10^6$ CPU hours per $(L,p)$ dataset. 
Due to the steep computation times required, computation was restricted to the minimal
set of $(L,p)$ datapoints required to determine the error threshold. The system sizes simulated lie in the range $L\in[10,14]$: lower system sizes
were affected by excessively strong finite-size effects, which made them irrelevant for scaling purposes. On the other hand, larger system sizes were 
computationally unapproachable due to the exponential growth of the simulation times required. 

\begin{table}[!b]
\renewcommand{\arraystretch}{1.}
\centering
\begin{tabular}{c@{\hspace{1cm}}c@{\hspace{1cm}}c@{\hspace{1cm}}c@{\hspace{1cm}}c@{\hspace{1cm}}}
\toprule
$p$ & $L$ & $N_{d}$ & $N_{T}$ & $\tau$ \\
\midrule
$0.090$ & $12$ & $200$ & $96$ & $21$\\
\vspace{0.2cm}
$0.090$ & $14$ & $200$ & $96$ & $21$\\
$0.105$ & $10$ & $500$ & $64$ & $20$\\
$0.105$ & $12$ & $500$ & $128$& $22$\\
\vspace{0.2cm}
$0.105$ & $14$ & $500$ & $96$ & $22$\\
$0.107$ & $10$ & $500$ & $96$ & $21$\\
$0.107$ & $12$ & $500$ & $128$& $22$\\
\vspace{0.2cm}
$0.107$ & $14$ & $500$ & $128$& $23$\\
$0.110$ & $10$ & $500$ & $96$ & $22$\\
$0.110$ & $12$ & $500$ & $96$ & $23$\\
$0.110$ & $14$ & $500$ & $96$ & $23$\\
\bottomrule
\end{tabular}
\caption{Simulation parameters for the data shown in Figs.~\ref{fig:090} and \ref{fig:disorderplots} for the random-bond TIM. 
        Here, $N_d$ denotes the number of independent disorder realizations used for disorder averaging at fixed system size $L$ and disorder value $p$. 
        For each realization, $N_T$ temperatures were simulated in parallel, with PT swap updates between adjacent replicas proposed, on average, every 10 Metropolis sweeps. 
        The parameter $\tau$ specifies the equilibration scale in sweeps, such that $2^\tau$ is the disorder-averaged equilibration time. 
        Each disorder realization was run for approximately $2^{\tau+1}$ sweeps, with the first $2^\tau$ sweeps used for thermalization 
        and the remaining $2^\tau$ sweeps for measurements.}
\label{tab:simparams}
\end{table}

%% file: frontmatter/thesis.bib
@article{Canossa23,
  title = {Hybrid symmetry breaking in classical spin models with subsystem symmetries},
  author = {Canossa, Giovanni and Pollet, Lode and Liu, Ke},
  journal = {Phys. Rev. B},
  volume = {107},
  issue = {5},
  pages = {054431},
  numpages = {8},
  year = {2023},
  month = {Feb},
  publisher = {American Physical Society},
  doi = {10.1103/PhysRevB.107.054431},
  url = {https://link.aps.org/doi/10.1103/PhysRevB.107.054431}
}

@article{Canossa24,
  title = {Exotic symmetry breaking properties of self-dual fracton spin models},
  author = {Canossa, Giovanni and Pollet, Lode and Martin-Delgado, Miguel A. and Song, Hao and Liu, Ke},
  journal = {Phys. Rev. Res.},
  volume = {6},
  issue = {1},
  pages = {013304},
  numpages = {12},
  year = {2024},
  month = {Mar},
  publisher = {American Physical Society},
  doi = {10.1103/PhysRevResearch.6.013304},
  url = {https://link.aps.org/doi/10.1103/PhysRevResearch.6.013304}
}

@article{Canossa25,
  title = {Error resilience of fracton codes and near saturation of code-capacity threshold in three dimensions},
  author = {Canossa, Giovanni and Pollet, Lode and Martin-Delgado, Miguel A. and Song, Hao and Liu, Ke},
  journal = {Phys. Rev. B},
  volume = {113},
  issue = {10},
  pages = {104204},
  numpages = {12},
  year = {2026},
  month = {Mar},
  publisher = {American Physical Society},
  doi = {10.1103/2y2z-hgfn},
  url = {https://link.aps.org/doi/10.1103/2y2z-hgfn}
}

@article{Ising25,
  author  = {Ising, Ernst},
  title   = {Beitrag zur Theorie des Ferromagnetismus},
  journal = {Zeitschrift für Physik},
  volume  = {31},
  number  = {1},
  pages   = {253--258},
  year    = {1925},
  doi     = {10.1007/BF02980577}
}

@article{Onsager44,
  title = {Crystal Statistics. I. A Two-Dimensional Model with an Order-Disorder Transition},
  author = {Onsager, Lars},
  journal = {Phys. Rev.},
  volume = {65},
  issue = {3-4},
  pages = {117--149},
  numpages = {0},
  year = {1944},
  month = {Feb},
  publisher = {American Physical Society},
  doi = {10.1103/PhysRev.65.117},
  url = {https://link.aps.org/doi/10.1103/PhysRev.65.117}
}

@article{Kadanoff66,
  title = {Scaling laws for ising models near ${T}_{c}$},
  author = {Kadanoff, Leo P.},
  journal = {Physics Physique Fizika},
  volume = {2},
  issue = {6},
  pages = {263--272},
  numpages = {10},
  year = {1966},
  month = {Jun},
  publisher = {American Physical Society},
  doi = {10.1103/PhysicsPhysiqueFizika.2.263},
  url = {https://link.aps.org/doi/10.1103/PhysicsPhysiqueFizika.2.263}
}

@article{Wilson71,
  title = {Renormalization Group and Critical Phenomena. I. Renormalization Group and the Kadanoff Scaling Picture},
  author = {Wilson, Kenneth G.},
  journal = {Phys. Rev. B},
  volume = {4},
  issue = {9},
  pages = {3174--3183},
  numpages = {0},
  year = {1971},
  month = {Nov},
  publisher = {American Physical Society},
  doi = {10.1103/PhysRevB.4.3174},
  url = {https://link.aps.org/doi/10.1103/PhysRevB.4.3174}
}

@book{Cardy96,
  author    = {Cardy, John},
  title     = {Scaling and Renormalization in Statistical Physics},
  publisher = {Cambridge University Press},
  address   = {Cambridge},
  year      = {1996},
  series    = {Cambridge Lecture Notes in Physics},
  volume    = {5},
  isbn      = {978-0521499590}
}

@article{Feynman82,
  author  = {Feynman, Richard P.},
  title   = {Simulating Physics with Computers},
  journal = {International Journal of Theoretical Physics},
  volume  = {21},
  number  = {6/7},
  pages   = {467--488},
  year    = {1982},
  doi     = {10.1007/BF02650179}
}

@article{Deutsch85,
  author  = {Deutsch, David},
  title   = {Quantum Theory, the Church--Turing Principle and the Universal Quantum Computer},
  journal = {Proceedings of the Royal Society A},
  volume  = {400},
  number  = {1818},
  pages   = {97--117},
  year    = {1985},
  doi     = {10.1098/rspa.1985.0070}
}

@article{Zurek91,
  author  = {Zurek, Wojciech H.},
  title   = {Decoherence and the Transition from Quantum to Classical},
  journal = {Physics Today},
  volume  = {44},
  number  = {10},
  pages   = {36--44},
  year    = {1991},
  doi     = {10.1063/1.881293}
}

@article{Fradkin78,
  author  = {Fradkin, Eduardo and Susskind, Leonard},
  title   = {Order and disorder in gauge systems and magnets},
  journal = {Physical Review D},
  volume  = {17},
  number  = {10},
  pages   = {2637--2658},
  year    = {1978},
  doi     = {10.1103/PhysRevD.17.2637}
}

@article{Anderson73,
title = {Resonating valence bonds: A new kind of insulator?},
journal = {Materials Research Bulletin},
volume = {8},
number = {2},
pages = {153-160},
year = {1973},
issn = {0025-5408},
doi = {https://doi.org/10.1016/0025-5408(73)90167-0},
url = {https://www.sciencedirect.com/science/article/pii/0025540873901670},
author = {P.W. Anderson}
}

@article{Elitzur75,
  title = {Impossibility of spontaneously breaking local symmetries},
  author = {Elitzur, S.},
  journal = {Phys. Rev. D},
  volume = {12},
  issue = {12},
  pages = {3978--3982},
  numpages = {0},
  year = {1975},
  month = {Dec},
  publisher = {American Physical Society},
  doi = {10.1103/PhysRevD.12.3978},
  url = {https://link.aps.org/doi/10.1103/PhysRevD.12.3978}
}

@article{Villain80,
  author  = {Villain, J. and Bidaux, R. and Carton, J.-P. and Conte, R.},
  title   = {Order as an effect of disorder},
  journal = {Journal de Physique},
  volume  = {41},
  pages   = {1263--1272},
  year    = {1980},
}

@book{Pathria96,
  author    = {Pathria, R. K.},
  title     = {Statistical Mechanics},
  edition   = {2},
  publisher = {Butterworth--Heinemann},
  address   = {Oxford},
  year      = {1996},
  isbn      = {978-0-7506-2469-5}
}

@article{Harris97,
  title = {Geometrical Frustration in the Ferromagnetic Pyrochlore ${\mathrm{Ho}}_{2}{\mathrm{Ti}}_{2}{O}_{7}$},
  author = {Harris, M. J. and Bramwell, S. T. and McMorrow, D. F. and Zeiske, T. and Godfrey, K. W.},
  journal = {Phys. Rev. Lett.},
  volume = {79},
  issue = {13},
  pages = {2554--2557},
  numpages = {0},
  year = {1997},
  month = {Sep},
  publisher = {American Physical Society},
  doi = {10.1103/PhysRevLett.79.2554},
  url = {https://link.aps.org/doi/10.1103/PhysRevLett.79.2554}
}

@article{Collins97,
   title={Review/Synthèse: Triangular antiferromagnets},
   volume={75},
   ISSN={1208-6045},
   url={http://dx.doi.org/10.1139/p97-007},
   DOI={10.1139/p97-007},
   number={9},
   journal={Canadian Journal of Physics},
   publisher={Canadian Science Publishing},
   author={Collins, M F and Petrenko, O A},
   year={1997},
   month=sep, pages={605–655} 
}

@article{Hastings00,
  title = {Dirac structure, RVB, and Goldstone modes in the kagom\'e antiferromagnet},
  author = {Hastings, M. B.},
  journal = {Phys. Rev. B},
  volume = {63},
  issue = {1},
  pages = {014413},
  numpages = {16},
  year = {2000},
  month = {Dec},
  publisher = {American Physical Society},
  doi = {10.1103/PhysRevB.63.014413},
  url = {https://link.aps.org/doi/10.1103/PhysRevB.63.014413}
}

@article{Bramwell01,
   title={Spin Ice State in Frustrated Magnetic Pyrochlore Materials},
   volume={294},
   ISSN={1095-9203},
   url={http://dx.doi.org/10.1126/science.1064761},
   DOI={10.1126/science.1064761},
   number={5546},
   journal={Science},
   publisher={American Association for the Advancement of Science (AAAS)},
   author={Bramwell, Steven T. and Gingras, Michel J. P.},
   year={2001},
   month=nov, pages={1495–1501}
}

@article{Nussinov05,
  title = {Discrete sliding symmetries, dualities, and self-dualities of quantum orbital compass models and $p+ip$ superconducting arrays},
  author = {Nussinov, Zohar and Fradkin, Eduardo},
  journal = {Phys. Rev. B},
  volume = {71},
  issue = {19},
  pages = {195120},
  numpages = {9},
  year = {2005},
  month = {May},
  publisher = {American Physical Society},
  doi = {10.1103/PhysRevB.71.195120},
  url = {https://link.aps.org/doi/10.1103/PhysRevB.71.195120}
}

@article{Racz83,
  title = {Ising transition into an order with extensive entropy at $T=0$: Potts antiferromagnets in a magnetic field},
  author = {R\'acz, Z. and Vicsek, T.},
  journal = {Phys. Rev. B},
  volume = {27},
  issue = {5},
  pages = {2992--3000},
  numpages = {0},
  year = {1983},
  month = {Mar},
  publisher = {American Physical Society},
  doi = {10.1103/PhysRevB.27.2992},
  url = {https://link.aps.org/doi/10.1103/PhysRevB.27.2992}
}

@article{Nussinov15,
  title = {Compass models: Theory and physical motivations},
  author = {Nussinov, Zohar and van den Brink, Jeroen},
  journal = {Rev. Mod. Phys.},
  volume = {87},
  issue = {1},
  pages = {1--59},
  numpages = {59},
  year = {2015},
  month = {Jan},
  publisher = {American Physical Society},
  doi = {10.1103/RevModPhys.87.1},
  url = {https://link.aps.org/doi/10.1103/RevModPhys.87.1}
}

@article{Moessner98,
  title = {Properties of a Classical Spin Liquid: The Heisenberg Pyrochlore Antiferromagnet},
  author = {Moessner, R. and Chalker, J. T.},
  journal = {Phys. Rev. Lett.},
  volume = {80},
  issue = {13},
  pages = {2929--2932},
  numpages = {0},
  year = {1998},
  month = {Mar},
  publisher = {American Physical Society},
  doi = {10.1103/PhysRevLett.80.2929},
  url = {https://link.aps.org/doi/10.1103/PhysRevLett.80.2929}
}

@book{Diep20,
author = {Diep, H.},
year = {2020},
month = {07},
pages = {},
title = {Frustrated Spin Systems},
isbn = {978-981-12-1413-4},
doi = {10.1142/11660}
}

@article{Binder86,
  title = {Spin glasses: Experimental facts, theoretical concepts, and open questions},
  author = {Binder, K. and Young, A. P.},
  journal = {Rev. Mod. Phys.},
  volume = {58},
  issue = {4},
  pages = {801--976},
  numpages = {0},
  year = {1986},
  month = {Oct},
  publisher = {American Physical Society},
  doi = {10.1103/RevModPhys.58.801},
  url = {https://link.aps.org/doi/10.1103/RevModPhys.58.801}
}

@article{Castelnovo08,
   title={Magnetic monopoles in spin ice},
   volume={451},
   ISSN={1476-4687},
   url={http://dx.doi.org/10.1038/nature06433},
   DOI={10.1038/nature06433},
   number={7174},
   journal={Nature},
   publisher={Springer Science and Business Media LLC},
   author={Castelnovo, C. and Moessner, R. and Sondhi, S. L.},
   year={2008},
   month=jan, pages={42–45} 
}

@article{Yan11,
   title={Spin-Liquid Ground State of the
            S
            = 1/2 Kagome Heisenberg Antiferromagnet},
   volume={332},
   ISSN={1095-9203},
   url={http://dx.doi.org/10.1126/science.1201080},
   DOI={10.1126/science.1201080},
   number={6034},
   journal={Science},
   publisher={American Association for the Advancement of Science (AAAS)},
   author={Yan, Simeng and Huse, David A. and White, Steven R.},
   year={2011},
   month=jun, pages={1173–1176} 
}

@book{Lacroix11,
title = "Introduction to Frustrated Magnetism",
author = "Lacroix C., Mendels P., Mila F.",
year = "2011",
series = "Springer Series in Solid-State Sciences",
publisher = "Springer"
}

@article{Khomskii03,
   title={Orbital ordering and frustrations},
   volume={36},
   ISSN={1361-6447},
   url={http://dx.doi.org/10.1088/0305-4470/36/35/307},
   DOI={10.1088/0305-4470/36/35/307},
   number={35},
   journal={Journal of Physics A: Mathematical and General},
   publisher={IOP Publishing},
   author={Khomskii, D I and Mostovoy, M V},
   year={2003},
   month=aug, pages={9197–9207} 
}

@book{Goldenfeld92,
  author    = {Goldenfeld, Nigel},
  title     = {Lectures on Phase Transitions and the Renormalization Group},
  publisher = {Addison-Wesley},
  address   = {Reading, Massachusetts},
  year      = {1992},
}

@article{Castelnovo07,
   title={Topological order and topological entropy in classical systems},
   volume={76},
   ISSN={1550-235X},
   url={http://dx.doi.org/10.1103/PhysRevB.76.174416},
   DOI={10.1103/physrevb.76.174416},
   number={17},
   journal={Physical Review B},
   publisher={American Physical Society (APS)},
   author={Castelnovo, Claudio and Chamon, Claudio},
   year={2007},
   month=nov }

@article{Lipowski97,
doi = {10.1088/0305-4470/30/21/012},
url = {https://doi.org/10.1088/0305-4470/30/21/012},
year = {1997},
month = {nov},
publisher = {},
volume = {30},
number = {21},
pages = {7365},
author = {Adam Lipowski},
title = {Glassy behaviour and semi-local invariance in Ising model with four-spin interaction},
journal = {Journal of Physics A: Mathematical and General}
}

@article{Park10,
  title = {Thermodynamic instability and first-order phase transition in an ideal Bose gas},
  author = {Park, Jeong-Hyuck and Kim, Sang-Woo},
  journal = {Phys. Rev. A},
  volume = {81},
  issue = {6},
  pages = {063636},
  numpages = {7},
  year = {2010},
  month = {Jun},
  publisher = {American Physical Society},
  doi = {10.1103/PhysRevA.81.063636},
  url = {https://link.aps.org/doi/10.1103/PhysRevA.81.063636}
}

@book{Gausterer92,
title = "Computational Methods in Field Theory",
author = "Gausterer H. and Lang C.B.",
year = "1992",
page = "59",
publisher = "Springer"
}

@article{Fernandez09,
  title = {Phase transition in the three dimensional Heisenberg spin glass: Finite-size scaling analysis},
  author = {Fernandez, L. A. and Martin-Mayor, V. and Perez-Gaviro, S. and Tarancon, A. and Young, A. P.},
  journal = {Phys. Rev. B},
  volume = {80},
  issue = {2},
  pages = {024422},
  numpages = {5},
  year = {2009},
  month = {Jul},
  publisher = {American Physical Society},
  doi = {10.1103/PhysRevB.80.024422},
  url = {https://link.aps.org/doi/10.1103/PhysRevB.80.024422}
}

@article{Gerlach15,
  title = {First-order directional ordering transition in the three-dimensional compass model},
  author = {Gerlach, Max H. and Janke, Wolfhard},
  journal = {Phys. Rev. B},
  volume = {91},
  issue = {4},
  pages = {045119},
  numpages = {8},
  year = {2015},
  month = {Jan},
  publisher = {American Physical Society},
  doi = {10.1103/PhysRevB.91.045119},
  url = {https://link.aps.org/doi/10.1103/PhysRevB.91.045119}
}

@article{Johnston14,
title = {Transmuted Finite-size Scaling at First-order Phase Transitions},
journal = {Physics Procedia},
volume = {57},
pages = {68-72},
year = {2014},
note = {Proceedings of the 27th Workshop on Computer Simulation Studies in Condensed Matter Physics (CSP2014)},
issn = {1875-3892},
doi = {https://doi.org/10.1016/j.phpro.2014.08.133},
url = {https://www.sciencedirect.com/science/article/pii/S1875389214002788},
author = {Marco Mueller and Wolfhard Janke and Desmond A. Johnston}
}

@article{Johnston17,
   title={Plaquette Ising models, degeneracy and scaling},
   volume={226},
   ISSN={1951-6401},
   url={http://dx.doi.org/10.1140/epjst/e2016-60329-4},
   DOI={10.1140/epjst/e2016-60329-4},
   number={4},
   journal={The European Physical Journal Special Topics},
   publisher={Springer Science and Business Media LLC},
   author={Johnston, Desmond A. and Mueller, Marco and Janke, Wolfhard},
   year={2017},
   month=apr, pages={749–764} 
}

@article{Binder87,
doi = {10.1088/0034-4885/50/7/001},
url = {https://doi.org/10.1088/0034-4885/50/7/001},
year = {1987},
month = {jul},
publisher = {},
volume = {50},
number = {7},
pages = {783},
author = {K Binder},
title = {Theory of first-order phase transitions},
journal = {Reports on Progress in Physics},
}

@article{Loison04,
   title={Canonical local algorithms for spin systems: heat bath and Hasting's methods},
   volume={41},
   ISSN={1434-6036},
   url={http://dx.doi.org/10.1140/epjb/e2004-00332-5},
   DOI={10.1140/epjb/e2004-00332-5},
   number={3},
   journal={The European Physical Journal B},
   publisher={Springer Science and Business Media LLC},
   author={Loison, D. and Qin, C. L. and Schotte, K. D. and Jin, X. F.},
   year={2004},
   month={Oct},
   pages={395–412}
}

@incollection{Janke08,
  author    = {Janke, Wolfhard},
  title     = {Monte Carlo Methods in Classical Statistical Physics},
  booktitle = {Computational Many-Particle Physics},
  editor    = {Fehske, H. and Schneider, R. and Wei{\ss}e, A.},
  publisher = {Springer},
  year      = {2008},
  pages     = {79--140},
  doi       = {10.1007/978-3-540-74686-7_4}
}

@article{Challa86,
  title = {Finite-size effects at temperature-driven first-order transitions},
  author = {Challa, Murty S. S. and Landau, D. P. and Binder, K.},
  journal = {Phys. Rev. B},
  volume = {34},
  issue = {3},
  pages = {1841--1852},
  numpages = {0},
  year = {1986},
  month = {Aug},
  publisher = {American Physical Society},
  doi = {10.1103/PhysRevB.34.1841},
  url = {https://link.aps.org/doi/10.1103/PhysRevB.34.1841}
}

@article{Borgs90,
  author       = {Borgs, Christian and Koteck{\'y}, Roman},
  title        = {A rigorous theory of finite-size scaling at first-order phase transitions},
  journal      = {Journal of Statistical Physics},
  volume       = {61},
  number       = {1/2},
  pages        = {79--119},
  year         = {1990},
  publisher    = {Springer},
  doi          = {10.1007/BF01013955},
  url          = {https://link.springer.com/article/10.1007/BF01013955}
}

@incollection{Janke03,
  author       = {Janke, Wolfhard},
  title        = {First-order Phase Transitions},
  booktitle    = {Computer Simulations in Condensed Matter Systems: From Materials to Chemical Biology Volume 1},
  editor       = {Binder, Kurt},
  publisher    = {Springer},
  year         = {2003},
  pages        = {111--135},
  series       = {Lecture Notes in Physics},
  volume       = {703},
  url          = {https://link.springer.com/chapter/10.1007/978-94-010-0173-1_6}
}

@article{Janke14,
   title={Nonstandard Finite-Size Scaling at First-Order Phase Transitions},
   volume={112},
   ISSN={1079-7114},
   url={http://dx.doi.org/10.1103/PhysRevLett.112.200601},
   DOI={10.1103/physrevlett.112.200601},
   number={20},
   journal={Physical Review Letters},
   publisher={American Physical Society (APS)},
   author={Mueller, Marco and Janke, Wolfhard and Johnston, Desmond A.},
   year={2014},
   month={May}
}

@book{Baxter82,
  author    = {Baxter, Rodney J.},
  title     = {Exactly Solved Models in Statistical Mechanics},
  publisher = {Academic Press},
  year      = {1982}
}

@book{Auerbach94,
  author    = {Auerbach, Assa},
  title     = {Interacting Electrons and Quantum Magnetism},
  publisher = {Springer},
  year      = {1994}
}

@article{Kosterlitz73,
  author  = {Kosterlitz, J. M. and Thouless, D. J.},
  title   = {Ordering, metastability and phase transitions in two-dimensional systems},
  journal = {Journal of Physics C},
  volume  = {6},
  pages   = {1181--1203},
  year    = {1973},
  doi     = {10.1088/0022-3719/6/7/010}
}

@article{Fisher89,
  author = {Fisher, Matthew P. A. and Weichman, Peter B. and Grinstein, Gershenzon and Fisher, Daniel S.},
  title   = {Boson localization and the superfluid--insulator transition},
  journal = {Physical Review B},
  volume  = {40},
  pages   = {546--570},
  year    = {1989},
  doi     = {10.1103/PhysRevB.40.546}
}

@article{Hubbard63,
  author  = {Hubbard, John},
  title   = {Electron correlations in narrow energy bands},
  journal = {Proceedings of the Royal Society A},
  volume  = {276},
  pages   = {238--257},
  year    = {1963},
  doi     = {10.1098/rspa.1963.0204}
}

@article{Greiner02,
  author  = {Greiner, Markus and Mandel, Olaf and Esslinger, Tilman and H{\"a}nsch, Theodor W. and Bloch, Immanuel},
  title   = {Quantum phase transition from a superfluid to a Mott insulator in a gas of ultracold atoms},
  journal = {Nature},
  volume  = {415},
  pages   = {39--44},
  year    = {2002},
  doi     = {10.1038/415039a}
}

@article{Bloch08,
  title = {Many-body physics with ultracold gases},
  author = {Bloch, Immanuel and Dalibard, Jean and Zwerger, Wilhelm},
  journal = {Rev. Mod. Phys.},
  volume = {80},
  issue = {3},
  pages = {885--964},
  numpages = {0},
  year = {2008},
  month = {Jul},
  publisher = {American Physical Society},
  doi = {10.1103/RevModPhys.80.885},
  url = {https://link.aps.org/doi/10.1103/RevModPhys.80.885}
}

@article{Manousakis91,
  title = {The spin-\textonehalf{} Heisenberg antiferromagnet on a square lattice and its application to the cuprous oxides},
  author = {Manousakis, Efstratios},
  journal = {Rev. Mod. Phys.},
  volume = {63},
  issue = {1},
  pages = {1--62},
  numpages = {0},
  year = {1991},
  month = {Jan},
  publisher = {American Physical Society},
  doi = {10.1103/RevModPhys.63.1},
  url = {https://link.aps.org/doi/10.1103/RevModPhys.63.1}
}

@article{Budrikis24,
  author       = {Budrikis, Zoë},
  title        = {100 Years of the Ising Model},
  journal      = {Nature Reviews Physics},
  year         = {2024},
  volume       = {6},
  pages        = {530--530},
  doi          = {10.1038/s42254-024-00760-x},
  publisher    = {Nature Publishing Group}
}

@book{Sachdev11,
  author       = {Sachdev, Subir},
  title        = {Quantum Phase Transitions},
  edition      = {2},
  publisher    = {Cambridge University Press},
  year         = {2011},
  isbn         = {9780521514682},
  doi          = {10.1017/CBO9780511973765},
  address      = {Cambridge, UK}
}

@article{Batista05,
   title={Generalized Elitzur’s theorem and dimensional reductions},
   volume={72},
   ISSN={1550-235X},
   url={http://dx.doi.org/10.1103/PhysRevB.72.045137},
   DOI={10.1103/physrevb.72.045137},
   number={4},
   journal={Physical Review B},
   publisher={American Physical Society (APS)},
   author={Batista, C. D. and Nussinov, Zohar},
   year={2005},
   month={Jul}
}

@article{Metropolis49,
  author  = {Metropolis, Nicholas and Ulam, Stanislaw},
  title   = {The {Monte Carlo} Method},
  journal = {Journal of the American Statistical Association},
  year    = {1949},
  volume  = {44},
  number  = {247},
  pages   = {335--341},
  doi     = {10.2307/2280232},
  url     = {https://www.jstor.org/stable/2280232}
}

@article{Metropolis53,
  author  = {Metropolis, Nicholas and Rosenbluth, Arianna W. and Rosenbluth, Marshall N. and Teller, Augusta H. and Teller, Edward},
  title   = {Equation of State Calculations by Fast Computing Machines},
  journal = {The Journal of Chemical Physics},
  year    = {1953},
  volume  = {21},
  number  = {6},
  pages   = {1087--1092},
  doi     = {10.1063/1.1699114},
  url     = {https://doi.org/10.1063/1.1699114}
}

@article{Brown87,
  title = {Overrelaxed heat-bath and Metropolis algorithms for accelerating pure gauge Monte Carlo calculations},
  author = {Brown, Frank R. and Woch, Thomas J.},
  journal = {Phys. Rev. Lett.},
  volume = {58},
  issue = {23},
  pages = {2394--2396},
  numpages = {0},
  year = {1987},
  month = {Jun},
  publisher = {American Physical Society},
  doi = {10.1103/PhysRevLett.58.2394},
  url = {https://link.aps.org/doi/10.1103/PhysRevLett.58.2394}
}

@article{Hukushima96,
   title={Exchange Monte Carlo Method and Application to Spin Glass Simulations},
   volume={65},
   ISSN={1347-4073},
   url={http://dx.doi.org/10.1143/JPSJ.65.1604},
   DOI={10.1143/jpsj.65.1604},
   number={6},
   journal={Journal of the Physical Society of Japan},
   publisher={Physical Society of Japan},
   author={Hukushima, Koji and Nemoto, Koji},
   year={1996},
   month={Jun},
   pages={1604–1608}
}

@article{Machta09,
   title={Strengths and weaknesses of parallel tempering},
   volume={80},
   ISSN={1550-2376},
   url={http://dx.doi.org/10.1103/PhysRevE.80.056706},
   DOI={10.1103/physreve.80.056706},
   number={5},
   journal={Physical Review E},
   publisher={American Physical Society (APS)},
   author={Machta, J.},
   year={2009},
   month={Nov}
}

@book{Bernd04,
  title={Markov Chain Monte Carlo Simulations and Their Statistical Analysis},
  author={Bernd A. Berg},
  year={2004}
}

@article{Janke93,
  title = {Accurate first-order transition points from finite-size data without power-law corrections},
  author = {Janke, W.},
  journal = {Phys. Rev. B},
  volume = {47},
  issue = {22},
  pages = {14757--14770},
  numpages = {0},
  year = {1993},
  month = {Jun},
  publisher = {American Physical Society},
  doi = {10.1103/PhysRevB.47.14757},
  url = {https://link.aps.org/doi/10.1103/PhysRevB.47.14757}
}

@article{Parisi98,
	doi = {10.1088/0305-4470/31/20/007},
	url = {https://doi.org/10.1088/0305-4470/31/20/007},
	year = 1998,
	month = {may},
	publisher = {{IOP} Publishing},
	volume = {31},
	number = {20},
	pages = {4657--4668},
	author = {Giorgio Parisi and Juan J Ruiz-Lorenzo and Daniel A Stariolo},
	title = {Crossovers in the two-dimensional Ising spin glass with ferromagnetic next-nearest-neighbour interactions},
	journal = {Journal of Physics A: Mathematical and General},
}

@incollection{Binder98,
title = {Finite Size Scaling Analysis of Ising Model Block Distribution Functions},
editor = {John L. CARDY},
series = {Current Physics–Sources and Comments},
publisher = {Elsevier},
volume = {2},
pages = {79-100},
year = {1988},
booktitle = {Finite-Size Scaling},
issn = {0922-503X},
doi = {https://doi.org/10.1016/B978-0-444-87109-1.50012-1},
url = {https://www.sciencedirect.com/science/article/pii/B9780444871091500121},
author = {K. Binder}
}

@Article{Nandkishore19,
  Title                    = {Fractons},
  Author                   = {Nandkishore, Rahul M. and Hermele, Michael},
  Journal                  = {Annual Review of Condensed Matter Physics},
  Year                     = {2019},
  Number                   = {1},
  Pages                    = {295-313},
  Volume                   = {10},
  Doi                      = {10.1146/annurev-conmatphys-031218-013604},
}

@article{Ma17,
  title = {Fracton topological order via coupled layers},
  author = {Ma, Han and Lake, Ethan and Chen, Xie and Hermele, Michael},
  journal = {Phys. Rev. B},
  volume = {95},
  issue = {24},
  pages = {245126},
  numpages = {18},
  year = {2017},
  month = {Jun},
  publisher = {American Physical Society},
  doi = {10.1103/PhysRevB.95.245126},
  url = {https://link.aps.org/doi/10.1103/PhysRevB.95.245126}
}

@article{Shirley19_2,
  title = {Foliated fracton order in the checkerboard model},
  author = {Shirley, Wilbur and Slagle, Kevin and Chen, Xie},
  journal = {Phys. Rev. B},
  volume = {99},
  issue = {11},
  pages = {115123},
  numpages = {8},
  year = {2019},
  month = {Mar},
  publisher = {American Physical Society},
  doi = {10.1103/PhysRevB.99.115123},
  url = {https://link.aps.org/doi/10.1103/PhysRevB.99.115123}
}

@article{Wang23,
  title = {Renormalization of Ising cage-net model and generalized foliation},
  author = {Wang, Zongyuan and Ma, Xiuqi and Stephen, David T. and Hermele, Michael and Chen, Xie},
  journal = {Phys. Rev. B},
  volume = {108},
  issue = {3},
  pages = {035148},
  numpages = {26},
  year = {2023},
  month = {Jul},
  publisher = {American Physical Society},
  doi = {10.1103/PhysRevB.108.035148},
  url = {https://link.aps.org/doi/10.1103/PhysRevB.108.035148}
}

@article{Tu21,
  title = {Non-Abelian fracton order from gauging a mixture of subsystem and global symmetries},
  author = {Tu, Yi-Ting and Chang, Po-Yao},
  journal = {Phys. Rev. Res.},
  volume = {3},
  issue = {4},
  pages = {043084},
  numpages = {24},
  year = {2021},
  month = {Oct},
  publisher = {American Physical Society},
  doi = {10.1103/PhysRevResearch.3.043084},
  url = {https://link.aps.org/doi/10.1103/PhysRevResearch.3.043084}
}

@article{Prem19_2,
  title = {Cage-Net Fracton Models},
  author = {Prem, Abhinav and Huang, Sheng-Jie and Song, Hao and Hermele, Michael},
  journal = {Phys. Rev. X},
  volume = {9},
  issue = {2},
  pages = {021010},
  numpages = {26},
  year = {2019},
  month = {Apr},
  publisher = {American Physical Society},
  doi = {10.1103/PhysRevX.9.021010},
  url = {https://link.aps.org/doi/10.1103/PhysRevX.9.021010}
}

@article{Prem19_3,
  title={Gauging permutation symmetries as a route to non-Abelian fractons},
  author={Abhinav Prem and Dominic J. Williamson},
  journal={SciPost Physics},
  year={2019},
  url={https://api.semanticscholar.org/CorpusID:155093155}
}

@article{Yoshida13,
  title = {Exotic topological order in fractal spin liquids},
  author = {Yoshida, Beni},
  journal = {Phys. Rev. B},
  volume = {88},
  issue = {12},
  pages = {125122},
  numpages = {17},
  year = {2013},
  month = {Sep},
  publisher = {American Physical Society},
  doi = {10.1103/PhysRevB.88.125122},
  url = {https://link.aps.org/doi/10.1103/PhysRevB.88.125122}
}

@article{Pretko17,
  title = {Subdimensional particle structure of higher rank $U(1)$ spin liquids},
  author = {Pretko, Michael},
  journal = {Phys. Rev. B},
  volume = {95},
  issue = {11},
  pages = {115139},
  numpages = {11},
  year = {2017},
  month = {Mar},
  publisher = {American Physical Society},
  doi = {10.1103/PhysRevB.95.115139},
  url = {https://link.aps.org/doi/10.1103/PhysRevB.95.115139}
}

@article{Pretko17_2,
  title = {Higher-spin Witten effect and two-dimensional fracton phases},
  author = {Pretko, Michael},
  journal = {Phys. Rev. B},
  volume = {96},
  issue = {12},
  pages = {125151},
  numpages = {15},
  year = {2017},
  month = {Sep},
  publisher = {American Physical Society},
  doi = {10.1103/PhysRevB.96.125151},
  url = {https://link.aps.org/doi/10.1103/PhysRevB.96.125151}
}

@article{You22,
  title = {Fracton critical point at a higher-order topological phase transition},
  author = {You, Yizhi and Bibo, Julian and Pollmann, Frank and Hughes, Taylor L.},
  journal = {Phys. Rev. B},
  volume = {106},
  issue = {23},
  pages = {235130},
  numpages = {5},
  year = {2022},
  month = {Dec},
  publisher = {American Physical Society},
  doi = {10.1103/PhysRevB.106.235130},
  url = {https://link.aps.org/doi/10.1103/PhysRevB.106.235130}
}

@article{Hirono24,
  title = {A Symmetry Principle for Gauge Theories with Fractons},
  author = {Hirono, Yuji and Hirono, Yoshimasa and Seiberg, Nathan and Shao, Shu-Heng and Wang, Kai-Yu},
  journal = {SciPost Physics},
  volume = {16},
  number = {2},
  pages = {050},
  year = {2024},
  doi = {10.21468/SciPostPhys.16.2.050},
  url = {https://scipost.org/10.21468/SciPostPhys.16.2.050}
}

@article{Nussinov09,
   title={A symmetry principle for topological quantum order},
   volume={324},
   ISSN={0003-4916},
   url={http://dx.doi.org/10.1016/j.aop.2008.11.002},
   DOI={10.1016/j.aop.2008.11.002},
   number={5},
   journal={Annals of Physics},
   publisher={Elsevier BV},
   author={Nussinov, Zohar and Ortiz, Gerardo},
   year={2009},
   month=may, pages={977–1057} 
}

@article{Seiberg20,
  title        = {Exotic $U(1)$ Symmetries, Duality, and Fractons in 3+1-Dimensional Quantum Field Theory},
  author       = {Seiberg, Nathan and Shao, Shu-Heng},
  journal      = {SciPost Physics},
  volume       = {9},
  number       = {4},
  pages        = {046},
  year         = {2020},
  doi          = {10.21468/SciPostPhys.9.4.046},
  url          = {https://scipost.org/10.21468/SciPostPhys.9.4.046}
}

@article{Seiberg21,
   title={Exotic symmetries, duality, and fractons in 2+1-dimensional quantum  field theory},
   volume={10},
   ISSN={2542-4653},
   url={http://dx.doi.org/10.21468/SciPostPhys.10.2.027},
   DOI={10.21468/scipostphys.10.2.027},
   number={2},
   journal={SciPost Physics},
   publisher={Stichting SciPost},
   author={Seiberg, Nathan and Shao, Shu-Heng},
   year={2021}
}

@article{Pretko18,
  title = {The fracton gauge principle},
  author = {Pretko, Michael},
  journal = {Phys. Rev. B},
  volume = {98},
  issue = {11},
  pages = {115134},
  numpages = {6},
  year = {2018},
  month = {Sep},
  publisher = {American Physical Society},
  doi = {10.1103/PhysRevB.98.115134},
  url = {https://link.aps.org/doi/10.1103/PhysRevB.98.115134}
}

@article{Gromov19,
  title = {Towards Classification of Fracton Phases: The Multipole Algebra},
  author = {Gromov, Andrey},
  journal = {Phys. Rev. X},
  volume = {9},
  issue = {3},
  pages = {031035},
  numpages = {19},
  year = {2019},
  month = {Aug},
  publisher = {American Physical Society},
  doi = {10.1103/PhysRevX.9.031035},
  url = {https://link.aps.org/doi/10.1103/PhysRevX.9.031035}
}

@article{Gromov20,
  title = {Fracton hydrodynamics},
  author = {Gromov, Andrey and Lucas, Andrew and Nandkishore, Rahul M.},
  journal = {Phys. Rev. Res.},
  volume = {2},
  issue = {3},
  pages = {033124},
  numpages = {11},
  year = {2020},
  month = {Jul},
  publisher = {American Physical Society},
  doi = {10.1103/PhysRevResearch.2.033124},
  url = {https://link.aps.org/doi/10.1103/PhysRevResearch.2.033124}
}

@article{Prem19,
  title = {Fractons from Vector Gauge Theory},
  author = {Prem, Abhinav and Pretko, Michael and Nandkishore, Rahul M.},
  journal = {Physical Review Letters},
  volume = {123},
  number = {13},
  pages = {136401},
  year = {2019},
  publisher = {American Physical Society},
  doi = {10.1103/PhysRevLett.123.136401}
}

@article{Pai19,
  title = {Localization in Fractonic Random Circuits},
  author = {Pai, Shriya and Pretko, Michael and Nandkishore, Rahul M.},
  journal = {Phys. Rev. X},
  volume = {9},
  issue = {2},
  pages = {021003},
  numpages = {21},
  year = {2019},
  month = {Apr},
  publisher = {American Physical Society},
  doi = {10.1103/PhysRevX.9.021003},
  url = {https://link.aps.org/doi/10.1103/PhysRevX.9.021003}
}

@article{Kalb74,
  title = {Classical direct interstring action},
  author = {Kalb, Michael and Ramond, P.},
  journal = {Phys. Rev. D},
  volume = {9},
  issue = {8},
  pages = {2273--2284},
  numpages = {0},
  year = {1974},
  month = {Apr},
  publisher = {American Physical Society},
  doi = {10.1103/PhysRevD.9.2273},
  url = {https://link.aps.org/doi/10.1103/PhysRevD.9.2273}
}

@article{Bulmash18,
  title = {Higgs mechanism in higher-rank symmetric U(1) gauge theories},
  author = {Bulmash, Daniel and Barkeshli, Maissam},
  journal = {Phys. Rev. B},
  volume = {97},
  issue = {23},
  pages = {235112},
  numpages = {25},
  year = {2018},
  month = {Jun},
  publisher = {American Physical Society},
  doi = {10.1103/PhysRevB.97.235112},
  url = {https://link.aps.org/doi/10.1103/PhysRevB.97.235112}
}

@article{Xu06,
  title = {Gapless bosonic excitation without symmetry breaking: An algebraic spin liquid with soft gravitons},
  author = {Xu, Cenke},
  journal = {Phys. Rev. B},
  volume = {74},
  issue = {22},
  pages = {224433},
  numpages = {10},
  year = {2006},
  month = {Dec},
  publisher = {American Physical Society},
  doi = {10.1103/PhysRevB.74.224433},
  url = {https://link.aps.org/doi/10.1103/PhysRevB.74.224433}
}

@article{Pankov07,
  title = {Resonating singlet valence plaquettes},
  author = {Pankov, S. and Moessner, R. and Sondhi, S. L.},
  journal = {Phys. Rev. B},
  volume = {76},
  issue = {10},
  pages = {104436},
  numpages = {10},
  year = {2007},
  month = {Sep},
  publisher = {American Physical Society},
  doi = {10.1103/PhysRevB.76.104436},
  url = {https://link.aps.org/doi/10.1103/PhysRevB.76.104436}
}

@article{Xu10,
  title = {Emergent gravity at a Lifshitz point from a Bose liquid on the lattice},
  author = {Xu, Cenke and Ho\ifmmode \check{r}\else \v{r}\fi{}ava, Petr},
  journal = {Phys. Rev. D},
  volume = {81},
  issue = {10},
  pages = {104033},
  numpages = {8},
  year = {2010},
  month = {May},
  publisher = {American Physical Society},
  doi = {10.1103/PhysRevD.81.104033},
  url = {https://link.aps.org/doi/10.1103/PhysRevD.81.104033}
}

@article{Prem18,
  title = {Emergent phases of fractonic matter},
  author = {Prem, Abhinav and Pretko, Michael and Nandkishore, Rahul M.},
  journal = {Phys. Rev. B},
  volume = {97},
  issue = {8},
  pages = {085116},
  numpages = {21},
  year = {2018},
  month = {Feb},
  publisher = {American Physical Society},
  doi = {10.1103/PhysRevB.97.085116},
  url = {https://link.aps.org/doi/10.1103/PhysRevB.97.085116}
}

@article{Jimenez02,
  doi = {10.1088/0953-8984/14/7/309},
  year = {2002},
  month = {feb},
  volume = {14},
  number = {7},
  pages = {1509},
  author = {M Jim\'{e}nez-Ruiz and A Criado and F J Bermejo and G J Cuello and F R Trouw and R Fern\'{a}ndez-Perea and H L\"{o}wen and C Cabrillo and H E Fischer},
  title = {Glassy dynamics of a kinetically constrained model: a direct comparison with experiment},
  journal = {J. Phys.: Condens. Matter}
}

@article{Elmatad09,
author = {Elmatad, Yael S. and Chandler, David and Garrahan, Juan P.},
title = {Corresponding States of Structural Glass Formers},
journal = {The Journal of Physical Chemistry B},
volume = {113},
number = {16},
pages = {5563-5567},
year = {2009},
doi = {10.1021/jp810362g},
    note ={PMID: 19254014}
}

@article{Biroli13,
  author    = {Giulio Biroli and Juan P. Garrahan},
  title     = {Perspective: The glass transition},
  journal   = {The Journal of Chemical Physics},
  volume    = {138},
  number    = {12},
  pages     = {12A301},
  year      = {2013},
  doi       = {10.1063/1.4795539},
  publisher = {American Institute of Physics}
}

@article{Hermele21,
title = {Fracton phases via exotic higher-form symmetry-breaking},
journal = {Annals of Physics},
volume = {424},
pages = {168360},
year = {2021},
issn = {0003-4916},
doi = {https://doi.org/10.1016/j.aop.2020.168360},
url = {https://www.sciencedirect.com/science/article/pii/S0003491620302943},
author = {Marvin Qi and Leo Radzihovsky and Michael Hermele}
}

@article{Ma18_2,
  title = {Fracton topological order from the Higgs and partial-confinement mechanisms of rank-two gauge theory},
  author = {Ma, Han and Hermele, Michael and Chen, Xie},
  journal = {Phys. Rev. B},
  volume = {98},
  issue = {3},
  pages = {035111},
  numpages = {15},
  year = {2018},
  month = {Jul},
  publisher = {American Physical Society},
  doi = {10.1103/PhysRevB.98.035111},
  url = {https://link.aps.org/doi/10.1103/PhysRevB.98.035111}
}

@article{Haah11,
  title = {Local stabilizer codes in three dimensions without string logical operators},
  author = {Haah, Jeongwan},
  journal = {Phys. Rev. A},
  volume = {83},
  issue = {4},
  pages = {042330},
  numpages = {16},
  year = {2011},
  month = {Apr},
  publisher = {American Physical Society},
  doi = {10.1103/PhysRevA.83.042330},
  url = {https://link.aps.org/doi/10.1103/PhysRevA.83.042330}
}

@article{Haah13,
  title = {Quantum Self-Correction in the 3D Cubic Code Model},
  author = {Bravyi, Sergey and Haah, Jeongwan},
  journal = {Phys. Rev. Lett.},
  volume = {111},
  issue = {20},
  pages = {200501},
  numpages = {5},
  year = {2013},
  month = {Nov},
  publisher = {American Physical Society},
  doi = {10.1103/PhysRevLett.111.200501},
  url = {https://link.aps.org/doi/10.1103/PhysRevLett.111.200501}
}

@article{Haah13_2,
   title={Commuting Pauli Hamiltonians as Maps between Free Modules},
   volume={324},
   ISSN={1432-0916},
   url={http://dx.doi.org/10.1007/s00220-013-1810-2},
   DOI={10.1007/s00220-013-1810-2},
   number={2},
   journal={Communications in Mathematical Physics},
   publisher={Springer Science and Business Media LLC},
   author={Haah, Jeongwan},
   year={2013},
   month=oct, pages={351–399} 
}

@Article{Vijai15,
  Title                    = {A new kind of topological quantum order: A dimensional hierarchy of quasiparticles built from stationary excitations},
  Author                   = {Vijay, Sagar and Haah, Jeongwan and Fu, Liang},
  Journal                  = {Phys. Rev. B},
  Year                     = {2015},
  Month                    = {Dec},
  Pages                    = {235136},
  Volume                   = {92},
  Doi                      = {10.1103/PhysRevB.92.235136},
  Issue                    = {23},
  Numpages                 = {11},
  Publisher                = {American Physical Society},
}

@Article{Vijai16,
  Title                    = {Fracton topological order, generalized lattice gauge theory, and duality},
  Author                   = {Vijay, Sagar and Haah, Jeongwan and Fu, Liang},
  Journal                  = {Phys. Rev. B},
  Year                     = {2016},
  Month                    = {Dec},
  Pages                    = {235157},
  Volume                   = {94},
  Doi                      = {10.1103/PhysRevB.94.235157},
  Issue                    = {23},
  Numpages                 = {9},
  Publisher                = {American Physical Society},
  Url                      = {https://link.aps.org/doi/10.1103/PhysRevB.94.235157}
}

@article{Slagle17,
  title = {Fracton topological order from nearest-neighbor two-spin interactions and dualities},
  author = {Slagle, Kevin and Kim, Yong Baek},
  journal = {Phys. Rev. B},
  volume = {96},
  issue = {16},
  pages = {165106},
  numpages = {22},
  year = {2017},
  month = {Oct},
  publisher = {American Physical Society},
  doi = {10.1103/PhysRevB.96.165106},
  url = {https://link.aps.org/doi/10.1103/PhysRevB.96.165106}
}

@article{Dua19,
  title = {Sorting topological stabilizer models in three dimensions},
  author = {Dua, Arpit and Kim, Isaac H. and Cheng, Meng and Williamson, Dominic J.},
  journal = {Phys. Rev. B},
  volume = {100},
  issue = {15},
  pages = {155137},
  numpages = {36},
  year = {2019},
  month = {Oct},
  publisher = {American Physical Society},
  doi = {10.1103/PhysRevB.100.155137},
  url = {https://link.aps.org/doi/10.1103/PhysRevB.100.155137}
}

@article{Song22,
  title = {Optimal Thresholds for Fracton Codes and Random Spin Models with Subsystem Symmetry},
  author = {Song, Hao and Sch\"onmeier-Kromer, Janik and Liu, Ke and Viyuela, Oscar and Pollet, Lode and Martin-Delgado, M. A.},
  journal = {Phys. Rev. Lett.},
  volume = {129},
  issue = {23},
  pages = {230502},
  numpages = {7},
  year = {2022},
  month = {Nov},
  publisher = {American Physical Society},
  doi = {10.1103/PhysRevLett.129.230502},
  url = {https://link.aps.org/doi/10.1103/PhysRevLett.129.230502}
}

@article{Toulouse83,
	author = {{Rammal, R.} and {Toulouse, G.}},
	title = {Random walks on fractal structures and percolation clusters},
	DOI= "10.1051/jphyslet:0198300440101300",
	url= "https://doi.org/10.1051/jphyslet:0198300440101300",
	journal = {J. Physique Lett.},
	year = 1983,
	volume = 44,
	number = 1,
	pages = "13-22",
}

@article{Kogut79,
  title = {An introduction to lattice gauge theory and spin systems},
  author = {Kogut, John B.},
  journal = {Rev. Mod. Phys.},
  volume = {51},
  issue = {4},
  pages = {659--713},
  numpages = {0},
  year = {1979},
  month = {Oct},
  publisher = {American Physical Society},
  doi = {10.1103/RevModPhys.51.659},
  url = {https://link.aps.org/doi/10.1103/RevModPhys.51.659}
}

@article{Kogut75,
  title = {Hamiltonian formulation of Wilson's lattice gauge theories},
  author = {Kogut, John and Susskind, Leonard},
  journal = {Phys. Rev. D},
  volume = {11},
  issue = {2},
  pages = {395--408},
  numpages = {0},
  year = {1975},
  month = {Jan},
  publisher = {American Physical Society},
  doi = {10.1103/PhysRevD.11.395},
  url = {https://link.aps.org/doi/10.1103/PhysRevD.11.395}
}

@article{Burrello15,
  title = {Formulation of lattice gauge theories for quantum simulations},
  author = {Zohar, Erez and Burrello, Michele},
  journal = {Phys. Rev. D},
  volume = {91},
  issue = {5},
  pages = {054506},
  numpages = {15},
  year = {2015},
  month = {Mar},
  publisher = {American Physical Society},
  doi = {10.1103/PhysRevD.91.054506},
  url = {https://link.aps.org/doi/10.1103/PhysRevD.91.054506}
}

@book{Schwartz14,
  author    = {Matthew D. Schwartz},
  title     = {Quantum Field Theory and the Standard Model},
  publisher = {Cambridge University Press},
  year      = {2014},
  address   = {Cambridge, UK},
  isbn      = {978-1-107-03473-0},
}

@article{Savary16,
  author  = {Lucile Savary and Leon Balents},
  title   = {Quantum spin liquids: a review},
  journal = {Reports on Progress in Physics},
  volume  = {80},
  number  = {1},
  pages   = {016502},
  year    = {2016},
  month   = {Nov},
  doi     = {10.1088/0034-4885/80/1/016502},
  publisher = {IOP Publishing}
}

@article{Capponi25,
  author  = {Sylvain Capponi},
  title   = {Classical and quantum spin liquids},
  journal = {arXiv preprint arXiv:2501.14433},
  year    = {2025},
  url     = {https://arxiv.org/abs/2501.14433}
}

@article{Wen17,
  title = {Colloquium: Zoo of quantum-topological phases of matter},
  author = {Wen, Xiao-Gang},
  journal = {Rev. Mod. Phys.},
  volume = {89},
  issue = {4},
  pages = {041004},
  numpages = {17},
  year = {2017},
  month = {Dec},
  publisher = {American Physical Society},
  doi = {10.1103/RevModPhys.89.041004},
  url = {https://link.aps.org/doi/10.1103/RevModPhys.89.041004}
}

@book{Wen04,
  author    = {Xiao-Gang Wen},
  title     = {Quantum Field Theory of Many-Body Systems: From the Origin of Sound to an Origin of Light and Electrons},
  year      = {2004},
  publisher = {Oxford University Press},
  address   = {Oxford, UK},
  isbn      = {9780199227259}
}

@article{Chamon05,
  title = {Quantum Glassiness in Strongly Correlated Clean Systems: An Example of Topological Overprotection},
  author = {Chamon, Claudio},
  journal = {Phys. Rev. Lett.},
  volume = {94},
  issue = {4},
  pages = {040402},
  numpages = {4},
  year = {2005},
  month = {Jan},
  publisher = {American Physical Society},
  doi = {10.1103/PhysRevLett.94.040402},
  url = {https://link.aps.org/doi/10.1103/PhysRevLett.94.040402}
}

@article{Slagle17_2,
  title = {Quantum field theory of X-cube fracton topological order and robust degeneracy from geometry},
  author = {Slagle, Kevin and Kim, Yong Baek},
  journal = {Phys. Rev. B},
  volume = {96},
  issue = {19},
  pages = {195139},
  numpages = {17},
  year = {2017},
  month = {Nov},
  publisher = {American Physical Society},
  doi = {10.1103/PhysRevB.96.195139},
  url = {https://link.aps.org/doi/10.1103/PhysRevB.96.195139}
}

@article{Albert17,
doi = {10.1088/1751-8121/aa9314},
url = {https://doi.org/10.1088/1751-8121/aa9314},
year = {2017},
month = {nov},
publisher = {IOP Publishing},
volume = {50},
number = {50},
pages = {504002},
author = {Albert, Victor V and Pascazio, Saverio and Devoret, Michel H},
title = {General phase spaces: from discrete variables to rotor and continuum limits},
journal = {Journal of Physics A: Mathematical and Theoretical}
}

@article{Castelnovo10,
  title = {Quantum mechanical and information theoretic view on classical glass transitions},
  author = {Castelnovo, Claudio and Chamon, Claudio and Sherrington, David},
  journal = {Phys. Rev. B},
  volume = {81},
  issue = {18},
  pages = {184303},
  numpages = {22},
  year = {2010},
  month = {May},
  publisher = {American Physical Society},
  doi = {10.1103/PhysRevB.81.184303},
  url = {https://link.aps.org/doi/10.1103/PhysRevB.81.184303}
}

@article{Keys11,
  title = {Excitations Are Localized and Relaxation Is Hierarchical in Glass-Forming Liquids},
  author = {Keys, Aaron S. and Hedges, Lester O. and Garrahan, Juan P. and Glotzer, Sharon C. and Chandler, David},
  journal = {Phys. Rev. X},
  volume = {1},
  issue = {2},
  pages = {021013},
  numpages = {15},
  year = {2011},
  month = {Nov},
  publisher = {American Physical Society},
  doi = {10.1103/PhysRevX.1.021013},
  url = {https://link.aps.org/doi/10.1103/PhysRevX.1.021013}
}

@article{Chleboun14,
doi = {10.1209/0295-5075/107/36002},
url = {https://doi.org/10.1209/0295-5075/107/36002},
year = {2014},
month = {jul},
publisher = {EDP Sciences, IOP Publishing and Società Italiana di Fisica},
volume = {107},
number = {3},
pages = {36002},
author = {Chleboun, P. and Faggionato, A. and Martinelli, F.},
title = {The influence of dimension on the relaxation process of East-like models: Rigorous results},
journal = {Europhysics Letters}
}

@article{Knill09,
  title = {Restrictions on Transversal Encoded Quantum Gate Sets},
  author = {Eastin, Bryan and Knill, Emanuel},
  journal = {Phys. Rev. Lett.},
  volume = {102},
  issue = {11},
  pages = {110502},
  numpages = {4},
  year = {2009},
  month = {Mar},
  publisher = {American Physical Society},
  doi = {10.1103/PhysRevLett.102.110502},
  url = {https://link.aps.org/doi/10.1103/PhysRevLett.102.110502}
}

@article{Fowler12,
  title = {Surface codes: Towards practical large-scale quantum computation},
  author = {Fowler, Austin G. and Mariantoni, Matteo and Martinis, John M. and Cleland, Andrew N.},
  journal = {Phys. Rev. A},
  volume = {86},
  issue = {3},
  pages = {032324},
  numpages = {48},
  year = {2012},
  month = {Sep},
  publisher = {American Physical Society},
  doi = {10.1103/PhysRevA.86.032324},
  url = {https://link.aps.org/doi/10.1103/PhysRevA.86.032324}
}

@article{Bravyi13,
  title = {Classification of Topologically Protected Gates for Local Stabilizer Codes},
  author = {Bravyi, Sergey and K\"onig, Robert},
  journal = {Phys. Rev. Lett.},
  volume = {110},
  issue = {17},
  pages = {170503},
  numpages = {5},
  year = {2013},
  month = {Apr},
  publisher = {American Physical Society},
  doi = {10.1103/PhysRevLett.110.170503},
  url = {https://link.aps.org/doi/10.1103/PhysRevLett.110.170503}
}

@article{Shirley18,
  title = {Fracton Models on General Three-Dimensional Manifolds},
  author = {Shirley, Wilbur and Slagle, Kevin and Wang, Zhenghan and Chen, Xie},
  journal = {Phys. Rev. X},
  volume = {8},
  issue = {3},
  pages = {031051},
  numpages = {13},
  year = {2018},
  month = {Aug},
  publisher = {American Physical Society},
  doi = {10.1103/PhysRevX.8.031051},
  url = {https://link.aps.org/doi/10.1103/PhysRevX.8.031051}
}

@article{Slagle18,
  title = {X-cube model on generic lattices: Fracton phases and geometric order},
  author = {Slagle, Kevin and Kim, Yong Baek},
  journal = {Phys. Rev. B},
  volume = {97},
  issue = {16},
  pages = {165106},
  numpages = {11},
  year = {2018},
  month = {Apr},
  publisher = {American Physical Society},
  doi = {10.1103/PhysRevB.97.165106},
  url = {https://link.aps.org/doi/10.1103/PhysRevB.97.165106}
}

@article{Shi18,
  title = {Deciphering the nonlocal entanglement entropy of fracton topological orders},
  author = {Shi, Bowen and Lu, Yuan-Ming},
  journal = {Phys. Rev. B},
  volume = {97},
  issue = {14},
  pages = {144106},
  numpages = {12},
  year = {2018},
  month = {Apr},
  publisher = {American Physical Society},
  doi = {10.1103/PhysRevB.97.144106},
  url = {https://link.aps.org/doi/10.1103/PhysRevB.97.144106}
}

@article{Ma18,
  title = {Topological entanglement entropy of fracton stabilizer codes},
  author = {Ma, Han and Schmitz, A. T. and Parameswaran, S. A. and Hermele, Michael and Nandkishore, Rahul M.},
  journal = {Phys. Rev. B},
  volume = {97},
  issue = {12},
  pages = {125101},
  numpages = {16},
  year = {2018},
  month = {Mar},
  publisher = {American Physical Society},
  doi = {10.1103/PhysRevB.97.125101},
  url = {https://link.aps.org/doi/10.1103/PhysRevB.97.125101}
}

@article{Schmitz18,
  title = {Recoverable information and emergent conservation laws in fracton stabilizer codes},
  author = {Schmitz, A. T. and Ma, Han and Nandkishore, Rahul M. and Parameswaran, S. A.},
  journal = {Phys. Rev. B},
  volume = {97},
  issue = {13},
  pages = {134426},
  numpages = {20},
  year = {2018},
  month = {Apr},
  publisher = {American Physical Society},
  doi = {10.1103/PhysRevB.97.134426},
  url = {https://link.aps.org/doi/10.1103/PhysRevB.97.134426}
}

@Article{Shirley19,
	title={{Foliated fracton order from gauging subsystem symmetries}},
	author={Wilbur Shirley and Kevin Slagle and Xie Chen},
	journal={SciPost Phys.},
	volume={6},
	pages={041},
	year={2019},
	publisher={SciPost},
	doi={10.21468/SciPostPhys.6.4.041},
	url={https://scipost.org/10.21468/SciPostPhys.6.4.041},
}

@article{Wegner71,
  author    = {Franz J. Wegner},
  title     = {Duality in Generalized Ising Models and Phase Transitions without Local Order Parameters},
  journal   = {Journal of Mathematical Physics},
  volume    = {12},
  number    = {10},
  pages     = {2259--2272},
  year      = {1971},
  doi       = {10.1063/1.1665530},
  publisher = {AIP Publishing},
  url       = {https://doi.org/10.1063/1.1665530}
}

@article{Brown20,
  title = {Parallelized quantum error correction with fracton topological codes},
  author = {Brown, Benjamin J. and Williamson, Dominic J.},
  journal = {Phys. Rev. Res.},
  volume = {2},
  issue = {1},
  pages = {013303},
  numpages = {15},
  year = {2020},
  month = {Mar},
  publisher = {American Physical Society},
  doi = {10.1103/PhysRevResearch.2.013303},
  url = {https://link.aps.org/doi/10.1103/PhysRevResearch.2.013303}
}

@article{Zhou22,
  title = {Evolution of dynamical signature in the X-cube fracton topological order},
  author = {Zhou, Chengkang and Li, Meng-Yuan and Yan, Zheng and Ye, Peng and Meng, Zi Yang},
  journal = {Phys. Rev. Res.},
  volume = {4},
  issue = {3},
  pages = {033111},
  numpages = {15},
  year = {2022},
  month = {Aug},
  publisher = {American Physical Society},
  doi = {10.1103/PhysRevResearch.4.033111},
  url = {https://link.aps.org/doi/10.1103/PhysRevResearch.4.033111}
}

@article{Wegner73,
title = {A transformation including the weak-graph theorem and the duality transformation},
journal = {Physica},
volume = {68},
number = {3},
pages = {570-578},
year = {1973},
issn = {0031-8914},
doi = {https://doi.org/10.1016/0031-8914(73)90381-9},
url = {https://www.sciencedirect.com/science/article/pii/0031891473903819},
author = {F.J. Wegner}
}

@misc{Wegner14,
      title={Duality in generalized Ising models}, 
      author={Franz J. Wegner},
      year={2014},
      eprint={1411.5815},
      archivePrefix={arXiv},
      primaryClass={hep-lat},
      url={https://arxiv.org/abs/1411.5815}, 
}

@inbook{Preskill98,
author = {John Preskill},
title = {Fault-Tolerant Quantum Computation},
booktitle = {Introduction to Quantum Computation and Information},
chapter = {},
pages = {213-269},
doi = {10.1142/9789812385253_0008},
URL = {https://www.worldscientific.com/doi/abs/10.1142/9789812385253_0008}
}

@book{Nielsen00,
  author = {Nielsen, Michael A. and Chuang, Isaac L.},
  publisher = {Cambridge University Press},
  title = {Quantum Computation and Quantum Information},
  year = 2000
}

@article{Delgado02,
  title = {Information and computation: Classical and quantum aspects},
  author = {Galindo, A. and Martin-Delgado, M. A.},
  journal = {Rev. Mod. Phys.},
  volume = {74},
  issue = {2},
  pages = {347--423},
  numpages = {0},
  year = {2002},
  month = {May},
  publisher = {American Physical Society},
  doi = {10.1103/RevModPhys.74.347},
  url = {https://link.aps.org/doi/10.1103/RevModPhys.74.347}
}

@article{Kitaev03,
title = {Fault-tolerant quantum computation by anyons},
journal = {Annals of Physics},
volume = {303},
number = {1},
pages = {2-30},
year = {2003},
issn = {0003-4916},
doi = {https://doi.org/10.1016/S0003-4916(02)00018-0},
url = {https://www.sciencedirect.com/science/article/pii/S0003491602000180},
author = {A.Yu. Kitaev}
}

@article{Shor95,
  title = {Scheme for reducing decoherence in quantum computer memory},
  author = {Shor, Peter W.},
  journal = {Phys. Rev. A},
  volume = {52},
  issue = {4},
  pages = {R2493--R2496},
  numpages = {0},
  year = {1995},
  month = {Oct},
  publisher = {American Physical Society},
  doi = {10.1103/PhysRevA.52.R2493},
  url = {https://link.aps.org/doi/10.1103/PhysRevA.52.R2493}
}

@article{Shor97,
  author  = {Shor, Peter W.},
  title   = {Polynomial-Time Algorithms for Prime Factorization and Discrete Logarithms on a Quantum Computer},
  journal = {SIAM Journal on Computing},
  volume  = {26},
  number  = {5},
  pages   = {1484--1509},
  year    = {1997},
  note    = {Originally presented in 1994},
  doi     = {10.1137/S0097539795293172}
}

@article{Shor96,
  author={Shor, P.W.},
  booktitle={Proceedings of 37th Conference on Foundations of Computer Science}, 
  title={Fault-tolerant quantum computation}, 
  year={1996},
  volume={},
  number={},
  pages={56-65},
  doi={10.1109/SFCS.1996.548464}
}

@article{Knill97,
  title = {Theory of quantum error-correcting codes},
  author = {Knill, Emanuel and Laflamme, Raymond},
  journal = {Phys. Rev. A},
  volume = {55},
  issue = {2},
  pages = {900--911},
  numpages = {0},
  year = {1997},
  month = {Feb},
  publisher = {American Physical Society},
  doi = {10.1103/PhysRevA.55.900},
  url = {https://link.aps.org/doi/10.1103/PhysRevA.55.900}
}

@article{Gottesman98,
  title = {Theory of fault-tolerant quantum computation},
  author = {Gottesman, Daniel},
  journal = {Phys. Rev. A},
  volume = {57},
  issue = {1},
  pages = {127--137},
  numpages = {0},
  year = {1998},
  month = {Jan},
  publisher = {American Physical Society},
  doi = {10.1103/PhysRevA.57.127},
  url = {https://link.aps.org/doi/10.1103/PhysRevA.57.127}
}

@article{Moses23,
  title = {A Race-Track Trapped-Ion Quantum Processor},
  author = {Moses, S. A. et al.},
  journal = {Phys. Rev. X},
  volume = {13},
  issue = {4},
  pages = {041052},
  numpages = {25},
  year = {2023},
  month = {Dec},
  publisher = {American Physical Society},
  doi = {10.1103/PhysRevX.13.041052},
  url = {https://link.aps.org/doi/10.1103/PhysRevX.13.041052}
}

@article{Bluvstein24,
  title = {Logical quantum processor based on reconfigurable atom arrays},
  author = {Lukin M.D.},
  journal = {Nature},
  issue = {626},
  pages = {58--65},
  year = {2024},
  publisher = {Nature},
  doi = {https://doi.org/10.1038/s41586-023-06927-3},
  url = {https://www.nature.com/articles/s41586-023-06927-3}
}

@article{Google25,
  title = {Quantum error correction below the surface code threshold},
  author = {Gogle Quantum AI and Collaborators},
  journal = {Nature},
  issue = {638},
  pages = {920--926},
  year = {2025},
  publisher = {Nature},
  doi = {https://doi.org/10.1038/s41586-024-08449-y},
  url = {https://www.nature.com/articles/s41586-024-08449-y}
}

@Article{Dennis02,
  Title                    = {Topological quantum memory},
  Author                   = {Dennis,Eric and Kitaev,Alexei and Landahl,Andrew and Preskill,John },
  Journal                  = {Journal of Mathematical Physics},
  Year                     = {2002},
  Number                   = {9},
  Pages                    = {4452-4505},
  Volume                   = {43},
  Doi                      = {10.1063/1.1499754}
}

@article{Wang03,
title = {Confinement-Higgs transition in a disordered gauge theory and the accuracy threshold for quantum memory},
journal = {Annals of Physics},
volume = {303},
number = {1},
pages = {31-58},
year = {2003},
issn = {0003-4916},
doi = {https://doi.org/10.1016/S0003-4916(02)00019-2},
url = {https://www.sciencedirect.com/science/article/pii/S0003491602000192},
author = {Chenyang Wang and Jim Harrington and John Preskill}
}

@article{Satzinger21,
author = {K. J. Satzinger, P. Roushan et al.},
title = {Realizing topologically ordered states on a quantum processor},
journal = {Science},
volume = {374},
number = {6572},
pages = {1237-1241},
year = {2021},
doi = {10.1126/science.abi8378},
URL = {https://www.science.org/doi/abs/10.1126/science.abi8378}
}

@article{Krinner22,
	author = {Krinner, Sebastian and Lacroix, Nathan and Remm, Ants and Di Paolo, Agustin and Genois, Elie and Leroux, Catherine and Hellings, Christoph and Lazar, Stefania and Swiadek, Francois and Herrmann, Johannes and Norris, Graham J. and Andersen, Christian Kraglund and M{\"u}ller, Markus and Blais, Alexandre and Eichler, Christopher and Wallraff, Andreas},
	doi = {10.1038/s41586-022-04566-8},
	id = {Krinner2022},
	isbn = {1476-4687},
	journal = {Nature},
	number = {7911},
	pages = {669--674},
	title = {Realizing repeated quantum error correction in a distance-three surface code},
	url = {https://doi.org/10.1038/s41586-022-04566-8},
	volume = {605},
	year = {2022}
}

@article{Zhao22,
  title = {Realization of an Error-Correcting Surface Code with Superconducting Qubits},
  author = {Zhao, Youwei and Ye, Yangsen and Huang, He-Liang and Zhang, Yiming and Wu, Dachao and Guan, Huijie and Zhu, Qingling and Wei, Zuolin and He, Tan and Cao, Sirui and Chen, Fusheng and Chung, Tung-Hsun and Deng, Hui and Fan, Daojin and Gong, Ming and Guo, Cheng and Guo, Shaojun and Han, Lianchen and Li, Na and Li, Shaowei and Li, Yuan and Liang, Futian and Lin, Jin and Qian, Haoran and Rong, Hao and Su, Hong and Sun, Lihua and Wang, Shiyu and Wu, Yulin and Xu, Yu and Ying, Chong and Yu, Jiale and Zha, Chen and Zhang, Kaili and Huo, Yong-Heng and Lu, Chao-Yang and Peng, Cheng-Zhi and Zhu, Xiaobo and Pan, Jian-Wei},
  journal = {Phys. Rev. Lett.},
  volume = {129},
  issue = {3},
  pages = {030501},
  numpages = {7},
  year = {2022},
  month = {Jul},
  publisher = {American Physical Society},
  doi = {10.1103/PhysRevLett.129.030501},
  url = {https://link.aps.org/doi/10.1103/PhysRevLett.129.030501}
}

@article{Bombin06,
  title = {Topological Quantum Distillation},
  author = {Bombin, H. and Martin-Delgado, M. A.},
  journal = {Phys. Rev. Lett.},
  volume = {97},
  issue = {18},
  pages = {180501},
  numpages = {4},
  year = {2006},
  month = {Oct},
  publisher = {American Physical Society},
  doi = {10.1103/PhysRevLett.97.180501},
  url = {https://link.aps.org/doi/10.1103/PhysRevLett.97.180501}
}

@Article{Katzgraber09,
  Title                    = {Error Threshold for Color Codes and Random Three-Body Ising Models},
  Author                   = {Katzgraber, Helmut G. and Bombin, H. and Martin-Delgado, M. A.},
  Journal                  = {Phys. Rev. Lett.},
  Year                     = {2009},
  Month                    = {Aug},
  Pages                    = {090501},
  Volume                   = {103},
  Doi                      = {10.1103/PhysRevLett.103.090501},
  Issue                    = {9},
  Numpages                 = {4},
  Publisher                = {American Physical Society},
  Url                      = {https://link.aps.org/doi/10.1103/PhysRevLett.103.090501}
}

@Article{Bombin12,
  Title                    = {Strong Resilience of Topological Codes to Depolarization},
  Author                   = {Bombin, H. and Andrist, Ruben S. and Ohzeki, Masayuki and Katzgraber, Helmut G. and Martin-Delgado, M. A.},
  Journal                  = {Phys. Rev. X},
  Year                     = {2012},
  Month                    = {Apr},
  Pages                    = {021004},
  Volume                   = {2},
  Doi                      = {10.1103/PhysRevX.2.021004},
  Issue                    = {2},
  Numpages                 = {10},
  Publisher                = {American Physical Society},
  Url                      = {https://link.aps.org/doi/10.1103/PhysRevX.2.021004}
}

@article{Breuckmann21,
  title = {Quantum Low-Density Parity-Check Codes},
  author = {Breuckmann, Nikolas P. and Eberhardt, Jens Niklas},
  journal = {PRX Quantum},
  volume = {2},
  issue = {4},
  pages = {040101},
  numpages = {19},
  year = {2021},
  month = {Oct},
  publisher = {American Physical Society},
  doi = {10.1103/PRXQuantum.2.040101},
  url = {https://link.aps.org/doi/10.1103/PRXQuantum.2.040101}
}

@article{Panteleev21,
  doi = {10.22331/q-2021-11-22-585},
  url = {https://doi.org/10.22331/q-2021-11-22-585},
  title = {Degenerate {Q}uantum {LDPC} {C}odes {W}ith {G}ood {F}inite {L}ength {P}erformance},
  author = {Panteleev, Pavel and Kalachev, Gleb},
  journal = {{Quantum}},
  issn = {2521-327X},
  publisher = {{Verein zur F{\"{o}}rderung des Open Access Publizierens in den Quantenwissenschaften}},
  volume = {5},
  pages = {585},
  month = nov,
  year = {2021}
}

@inproceedings{Panteleev22,
author = {Panteleev, Pavel and Kalachev, Gleb},
title = {Asymptotically good Quantum and locally testable classical LDPC codes},
year = {2022},
isbn = {9781450392648},
publisher = {Association for Computing Machinery},
address = {New York, NY, USA},
url = {https://doi.org/10.1145/3519935.3520017},
doi = {10.1145/3519935.3520017},
booktitle = {Proceedings of the 54th Annual ACM SIGACT Symposium on Theory of Computing},
pages = {375–388},
numpages = {14},
location = {Rome, Italy},
series = {STOC 2022}
}

@article{Calderbank96,
  title     = {Good quantum error-correcting codes exist},
  author    = {Calderbank, A. R. and Shor, P. W.},
  journal   = {Physical Review A},
  volume    = {54},
  number    = {2},
  pages     = {1098--1105},
  year      = {1996},
  publisher = {American Physical Society},
  doi       = {10.1103/PhysRevA.54.1098}
}

@inproceedings{Aharonov97,
  author    = {Aharonov, Dorit and Ben-Or, Michael},
  title     = {Fault-Tolerant Quantum Computation with Constant Error},
  booktitle = {Proceedings of the Twenty-Ninth Annual ACM Symposium on Theory of Computing},
  year      = {1997},
  pages     = {176--188},
  publisher = {ACM},
  address   = {New York, NY, USA},
  doi       = {10.1145/258533.258579},
}

@misc{Shtetl,
  author       = {Aaronson, Scott and Granade, Chris},
  title        = {Lecture 14: Skepticism of Quantum Computing},
  howpublished = {\textit{Shtetl-Optimized blog / PHYS771 Lecture}},
  year         = {2006},
  url          = {http://www.scottaaronson.com/democritus/lec14.html},
}

@article{Knill98,
  author    = {Emanuel Knill and
               Raymond Laflamme and
               Wojciech H. Zurek},
  title     = {Resilient quantum computation: Error models and thresholds},
  journal   = {Science},
  volume    = {279},
  number    = {5349},
  pages     = {342--345},
  year      = {1998},
  doi       = {10.1126/science.279.5349.342},
  url       = {https://doi.org/10.1126/science.279.5349.342}
}

@article{Li25,
  author    = {Yaodong Li and Nicholas O'Dea and Vedika Khemani},
  title     = {Perturbative Stability and Error--Correction Thresholds of Quantum Codes},
  journal   = {PRX Quantum},
  volume    = {6},
  page      = {010327},
  year      = {2025},
  doi       = {10.1103/PRXQuantum.6.010327},
  url       = {https://doi.org/10.1103/PRXQuantum.6.010327}
}

@article{Raussendorf07,
  title = {Fault-Tolerant Quantum Computation with High Threshold in Two Dimensions},
  author = {Raussendorf, Robert and Harrington, Jim},
  journal = {Phys. Rev. Lett.},
  volume = {98},
  issue = {19},
  pages = {190504},
  numpages = {4},
  year = {2007},
  month = {May},
  publisher = {American Physical Society},
  doi = {10.1103/PhysRevLett.98.190504},
  url = {https://link.aps.org/doi/10.1103/PhysRevLett.98.190504}
}

@article{Aharonov99,
author = {Aharonov, Dorit and Ben-Or, Michael},
title = {Fault-Tolerant Quantum Computation with Constant Error Rate},
journal = {SIAM Journal on Computing},
volume = {38},
number = {4},
pages = {1207-1282},
year = {2008},
doi = {10.1137/S0097539799359385},
URL = {https://doi.org/10.1137/S0097539799359385},
eprint = {https://doi.org/10.1137/S0097539799359385}
}

@article{Doussal88,
  title = {Location of the Ising Spin-Glass Multicritical Point on Nishimori's Line},
  author = {Le Doussal, Pierre and Harris, A. Brooks},
  journal = {Phys. Rev. Lett.},
  volume = {61},
  issue = {5},
  pages = {625--628},
  numpages = {0},
  year = {1988},
  month = {Aug},
  publisher = {American Physical Society},
  doi = {10.1103/PhysRevLett.61.625},
  url = {https://link.aps.org/doi/10.1103/PhysRevLett.61.625}
}

@article{Doussal89,
  title = {\ensuremath{\epsilon} expansion for the Nishimori multicritical point of spin glasses},
  author = {Le Doussal, Pierre and Harris, A. Brooks},
  journal = {Phys. Rev. B},
  volume = {40},
  issue = {13},
  pages = {9249--9252},
  numpages = {0},
  year = {1989},
  month = {Nov},
  publisher = {American Physical Society},
  doi = {10.1103/PhysRevB.40.9249},
  url = {https://link.aps.org/doi/10.1103/PhysRevB.40.9249}
}

@article{Iba99,
doi = {10.1088/0305-4470/32/21/302},
url = {https://dx.doi.org/10.1088/0305-4470/32/21/302},
year = {1999},
month = {may},
publisher = {},
volume = {32},
number = {21},
pages = {3875},
author = {Yukito Iba},
title = {The 
Nishimori line and Bayesian statistics},
journal = {Journal of Physics A: Mathematical and General},
}

@article{Chubb18,
title = {Statistical mechanical models for quantum codes with correlated noise},
journal = {Ann. Inst. Henri Poincaré Comb. Phys. Interact},
volume = {8},
number = {2},
pages = {269-321},
year = {2021},
doi = {10.4171/AIHPD/105},
url = {https://ems.press/journals/aihpd/articles/1146536},
}

@article{Harris74,
doi = {10.1088/0022-3719/7/9/009},
url = {https://doi.org/10.1088/0022-3719/7/9/009},
year = {1974},
month = {may},
publisher = {},
volume = {7},
number = {9},
pages = {1671},
author = {A B Harris},
title = {Effect of random defects on the critical behaviour of Ising models},
journal = {Journal of Physics C: Solid State Physics}
}

@article{Imry75,
  title = {Random-Field Instability of the Ordered State of Continuous Symmetry},
  author = {Imry, Yoseph and Ma, Shang-keng},
  journal = {Phys. Rev. Lett.},
  volume = {35},
  issue = {21},
  pages = {1399--1401},
  numpages = {0},
  year = {1975},
  month = {Nov},
  publisher = {American Physical Society},
  doi = {10.1103/PhysRevLett.35.1399},
  url = {https://link.aps.org/doi/10.1103/PhysRevLett.35.1399}
}

@article{Vojta06,
  author = {Thomas Vojta},
  title = {Rare region effects at classical, quantum and nonequilibrium phase transitions},
  journal = {J. Phys. A: Math. Gen.},
  volume = {39},
  pages = {R143--R205},
  year = {2006},
  doi = {10.1088/0305-4470/39/22/R01}
}

@Article{Nishimori07,
  Title                    = {Duality in Finite-Dimensional Spin Glasses},
  Author                   = {Nishimori, Hidetoshi},
  Journal                  = {Journal of Statistical Physics},
  Year                     = {2007},

  Month                    = {Mar},
  Number                   = {4},
  Pages                    = {977--986},
  Volume                   = {126},
  Day                      = {01},
  Doi                      = {10.1007/s10955-006-9156-1},
  ISSN                     = {1572-9613},
  Url                      = {https://doi.org/10.1007/s10955-006-9156-1}
}

@article{Takeda05,
doi = {10.1088/0305-4470/38/17/004},
url = {https://dx.doi.org/10.1088/0305-4470/38/17/004},
year = {2005},
month = {apr},
publisher = {},
volume = {38},
number = {17},
pages = {3751},
author = {Takeda, Koujin and Sasamoto, Tomohiro and Nishimori, Hidetoshi},
title = {Exact location of the multicritical point for finite-dimensional spin glasses: a conjecture},
journal = {Journal of Physics A: Mathematical and General}
}

@article{Mueller17,
title = {Exact solutions to plaquette Ising models with free and periodic boundaries},
journal = {Nuclear Physics B},
volume = {914},
pages = {388-404},
year = {2017},
issn = {0550-3213},
doi = {https://doi.org/10.1016/j.nuclphysb.2016.11.005},
url = {https://www.sciencedirect.com/science/article/pii/S0550321316303558},
author = {Marco Mueller and Desmond A. Johnston and Wolfhard Janke}
}

@article{Forney11,
  title={Partition Functions of Normal Factor Graphs},
  author={G. David Forney and Pascal O. Vontobel},
  journal={ArXiv},
  year={2011},
  volume={abs/1102.0316},
  url={https://api.semanticscholar.org/CorpusID:13978813}
}

@article{Kramers41a,
  title = {Statistics of the Two-Dimensional Ferromagnet. Part I},
  author = {Kramers, H. A. and Wannier, G. H.},
  journal = {Phys. Rev.},
  volume = {60},
  issue = {3},
  pages = {252--262},
  numpages = {0},
  year = {1941},
  month = {Aug},
  publisher = {American Physical Society},
  doi = {10.1103/PhysRev.60.252},
  url = {https://link.aps.org/doi/10.1103/PhysRev.60.252}
}

@article{Kramers41b,
  title = {Statistics of the Two-Dimensional Ferromagnet. Part II},
  author = {Kramers, H. A. and Wannier, G. H.},
  journal = {Phys. Rev.},
  volume = {60},
  issue = {3},
  pages = {263--276},
  numpages = {0},
  year = {1941},
  month = {Aug},
  publisher = {American Physical Society},
  doi = {10.1103/PhysRev.60.263},
  url = {https://link.aps.org/doi/10.1103/PhysRev.60.263}
}

@article{Savit80,
  title = {Duality in field theory and statistical systems},
  author = {Savit, Robert},
  journal = {Rev. Mod. Phys.},
  volume = {52},
  issue = {2},
  pages = {453--487},
  numpages = {0},
  year = {1980},
  month = {Apr},
  publisher = {American Physical Society},
  doi = {10.1103/RevModPhys.52.453},
  url = {https://link.aps.org/doi/10.1103/RevModPhys.52.453}
}

@incollection{Nishimori10,
    author = {Nishimori, Hidetoshi and Ortiz, Gerardo},
    isbn = {9780199577224},
    booktitle = {Elements of Phase Transitions and Critical Phenomena},
    publisher = {Oxford University Press},
    year = {2010},
    month = {12},
    doi = {10.1093/acprof:oso/9780199577224.003.0010},
    url = {https://doi.org/10.1093/acprof:oso/9780199577224.003.0010},
}

@article{Hinczewski05,
  title = {Multicritical point relations in three dual pairs of hierarchical-lattice Ising spin glasses},
  author = {Hinczewski, Michael and Berker, A. Nihat},
  journal = {Phys. Rev. B},
  volume = {72},
  issue = {14},
  pages = {144402},
  numpages = {6},
  year = {2005},
  month = {Oct},
  publisher = {American Physical Society},
  doi = {10.1103/PhysRevB.72.144402},
  url = {https://link.aps.org/doi/10.1103/PhysRevB.72.144402}
}

@article{Hasenbusch07,
  title = {Magnetic-glassy multicritical behavior of the three-dimensional $\ifmmode\pm\else\textpm\fi{}J$ Ising model},
  author = {Hasenbusch, Martin and Toldin, Francesco Parisen and Pelissetto, Andrea and Vicari, Ettore},
  journal = {Phys. Rev. B},
  volume = {76},
  issue = {18},
  pages = {184202},
  numpages = {7},
  year = {2007},
  month = {Nov},
  publisher = {American Physical Society},
  doi = {10.1103/PhysRevB.76.184202},
  url = {https://link.aps.org/doi/10.1103/PhysRevB.76.184202}
}

@article{Masayuki09,
  title = {Locations of multicritical points for spin glasses on regular lattices},
  author = {Ohzeki, Masayuki},
  journal = {Phys. Rev. E},
  volume = {79},
  issue = {2},
  pages = {021129},
  numpages = {14},
  year = {2009},
  month = {Feb},
  publisher = {American Physical Society},
  doi = {10.1103/PhysRevE.79.021129},
  url = {https://link.aps.org/doi/10.1103/PhysRevE.79.021129}
}

@Article{Kubica18,
  Title                    = {Three-Dimensional Color Code Thresholds via Statistical-Mechanical Mapping},
  Author                   = {Kubica, Aleksander and Beverland, Michael E. and Brand\~ao, Fernando and Preskill, John and Svore, Krysta M.},
  Journal                  = {Phys. Rev. Lett.},
  Year                     = {2018},

  Month                    = {May},
  Pages                    = {180501},
  Volume                   = {120},

  Doi                      = {10.1103/PhysRevLett.120.180501},
  Issue                    = {18},
  Numpages                 = {6},
  Publisher                = {American Physical Society},
  Url                      = {https://link.aps.org/doi/10.1103/PhysRevLett.120.180501}
}

@article{Gattringer18,
title = {Kramers–Wannier duality and worldline representation for the SU(2) principal chiral model},
journal = {Physics Letters B},
volume = {778},
pages = {435-441},
year = {2018},
issn = {0370-2693},
doi = {https://doi.org/10.1016/j.physletb.2018.01.065},
url = {https://www.sciencedirect.com/science/article/pii/S0370269318300832},
author = {Christof Gattringer and Daniel Göschl and Carlotta Marchis},
}

@article{Sachdev19,
doi = {10.1088/1361-6633/aae110},
url = {https://doi.org/10.1088/1361-6633/aae110},
year = {2018},
month = {nov},
publisher = {IOP Publishing},
volume = {82},
number = {1},
pages = {014001},
author = {Sachdev, Subir},
title = {Topological order, emergent gauge fields, and Fermi surface reconstruction},
journal = {Reports on Progress in Physics},
}

@article{Halime20,
  title = {Reliability of Lattice Gauge Theories},
  author = {Halimeh, Jad C. and Hauke, Philipp},
  journal = {Phys. Rev. Lett.},
  volume = {125},
  issue = {3},
  pages = {030503},
  numpages = {6},
  year = {2020},
  month = {Jul},
  publisher = {American Physical Society},
  doi = {10.1103/PhysRevLett.125.030503},
  url = {https://link.aps.org/doi/10.1103/PhysRevLett.125.030503}
}

@article{Homeier21,
  author = {Homeier, Lukas and Schweizer, Christian and Aidelsburger, Monika and Fedorov, Arkady and Grusdt, Fabian},
  title = {$\mathbb{Z}_2$ lattice gauge theories and Kitaev's toric code: A scheme for analog quantum simulation},
  journal = {Physical Review B},
  year = {2021},
}

@article{Lumia22,
  author = {Lumia, Luca and Torta, Pietro and Mbeng, Glen B. and et al.},
  title = {Two-Dimensional $\mathbb{Z}_2$ Lattice Gauge Theory on a Near-Term Quantum Simulator},
  journal = {PRX Quantum},
  year = {2022},
}

@article{Fradkin79,
  title = {Phase diagrams of lattice gauge theories with Higgs fields},
  author = {Fradkin, Eduardo and Shenker, Stephen H.},
  journal = {Phys. Rev. D},
  volume = {19},
  issue = {12},
  pages = {3682--3697},
  numpages = {0},
  year = {1979},
  month = {Jun},
  publisher = {American Physical Society},
  doi = {10.1103/PhysRevD.19.3682},
  url = {https://link.aps.org/doi/10.1103/PhysRevD.19.3682}
}

@article{Buchsbaum73,
title = {What makes a complex exact?},
journal = {Journal of Algebra},
volume = {25},
number = {2},
pages = {259-268},
year = {1973},
issn = {0021-8693},
doi = {https://doi.org/10.1016/0021-8693(73)90044-6},
url = {https://www.sciencedirect.com/science/article/pii/0021869373900446},
author = {David A Buchsbaum and David Eisenbud}
}

@ARTICLE{Tillich14,
  author={Tillich, Jean-Pierre and Zémor, Gilles},
  journal={IEEE Transactions on Information Theory}, 
  title={Quantum LDPC Codes With Positive Rate and Minimum Distance Proportional to the Square Root of the Blocklength}, 
  year={2014},
  volume={60},
  number={2},
  pages={1193-1202},
  doi={10.1109/TIT.2013.2292061}
}

@article{You18,
  title = {Subsystem symmetry protected topological order},
  author = {You, Yizhi and Devakul, Trithep and Burnell, F. J. and Sondhi, S. L.},
  journal = {Phys. Rev. B},
  volume = {98},
  issue = {3},
  pages = {035112},
  numpages = {18},
  year = {2018},
  month = {Jul},
  publisher = {American Physical Society},
  doi = {10.1103/PhysRevB.98.035112},
  url = {https://link.aps.org/doi/10.1103/PhysRevB.98.035112}
}

@article{Stephen20,
   title={Subsystem symmetry enriched topological order in three dimensions},
   volume={2},
   ISSN={2643-1564},
   url={http://dx.doi.org/10.1103/PhysRevResearch.2.033331},
   DOI={10.1103/physrevresearch.2.033331},
   number={3},
   journal={Physical Review Research},
   publisher={American Physical Society (APS)},
   author={Stephen, David T. and Garre-Rubio, José and Dua, Arpit and Williamson, Dominic J.},
   year={2020},
   month=aug }

@article{Yoshida16,
  title = {Topological phases with generalized global symmetries},
  author = {Yoshida, Beni},
  journal = {Phys. Rev. B},
  volume = {93},
  issue = {15},
  pages = {155131},
  numpages = {18},
  year = {2016},
  month = {Apr},
  publisher = {American Physical Society},
  doi = {10.1103/PhysRevB.93.155131},
  url = {https://link.aps.org/doi/10.1103/PhysRevB.93.155131}
}

@article{Devakul18,
  title = {Classification of subsystem symmetry-protected topological phases},
  author = {Devakul, Trithep and Williamson, Dominic J. and You, Yizhi},
  journal = {Phys. Rev. B},
  volume = {98},
  issue = {23},
  pages = {235121},
  numpages = {15},
  year = {2018},
  month = {Dec},
  publisher = {American Physical Society},
  doi = {10.1103/PhysRevB.98.235121},
  url = {https://link.aps.org/doi/10.1103/PhysRevB.98.235121}
}

@article{Xu04,
  title = {Strong-Weak Coupling Self-Duality in the Two-Dimensional Quantum Phase Transition of $p+ip$ Superconducting Arrays},
  author = {Xu, Cenke and Moore, J. E.},
  journal = {Phys. Rev. Lett.},
  volume = {93},
  issue = {4},
  pages = {047003},
  numpages = {4},
  year = {2004},
  month = {Jul},
  publisher = {American Physical Society},
  doi = {10.1103/PhysRevLett.93.047003},
  url = {https://link.aps.org/doi/10.1103/PhysRevLett.93.047003}
}

@article{Chen11,
  title = {Classification of gapped symmetric phases in one-dimensional spin systems},
  author = {Chen, Xie and Gu, Zheng-Cheng and Wen, Xiao-Gang},
  journal = {Phys. Rev. B},
  volume = {83},
  issue = {3},
  pages = {035107},
  numpages = {19},
  year = {2011},
  month = {Jan},
  publisher = {American Physical Society},
  doi = {10.1103/PhysRevB.83.035107},
  url = {https://link.aps.org/doi/10.1103/PhysRevB.83.035107}
}

@article{Aasen20,
   title={Topological defect networks for fractons of all types},
   volume={2},
   ISSN={2643-1564},
   url={http://dx.doi.org/10.1103/PhysRevResearch.2.043165},
   DOI={10.1103/physrevresearch.2.043165},
   number={4},
   journal={Physical Review Research},
   publisher={American Physical Society (APS)},
   author={Aasen, David and Bulmash, Daniel and Prem, Abhinav and Slagle, Kevin and Williamson, Dominic J.},
   year={2020},
   month=oct 
}

@article{Haah21,
   title={Classification of translation invariant topological Pauli stabilizer codes for prime dimensional qudits on two-dimensional lattices},
   volume={62},
   ISSN={1089-7658},
   url={http://dx.doi.org/10.1063/5.0021068},
   DOI={10.1063/5.0021068},
   number={1},
   journal={Journal of Mathematical Physics},
   publisher={AIP Publishing},
   author={Haah, Jeongwan},
   year={2021},
   month=jan 
}

@article{Song19,
  title = {Twisted fracton models in three dimensions},
  author = {Song, Hao and Prem, Abhinav and Huang, Sheng-Jie and Martin-Delgado, M. A.},
  journal = {Phys. Rev. B},
  volume = {99},
  issue = {15},
  pages = {155118},
  numpages = {53},
  year = {2019},
  month = {Apr},
  publisher = {American Physical Society},
  doi = {10.1103/PhysRevB.99.155118},
  url = {https://link.aps.org/doi/10.1103/PhysRevB.99.155118}
}

@article{You20,
   title={Symmetric fracton matter: Twisted and enriched},
   volume={416},
   ISSN={0003-4916},
   url={http://dx.doi.org/10.1016/j.aop.2020.168140},
   DOI={10.1016/j.aop.2020.168140},
   journal={Annals of Physics},
   publisher={Elsevier BV},
   author={You, Yizhi and Devakul, Trithep and Burnell, F.J. and Sondhi, S.L.},
   year={2020},
   month=may, pages={168140} 
}

@article{Bombin07_3,
   title={Homological error correction: Classical and quantum codes},
   volume={48},
   ISSN={1089-7658},
   url={http://dx.doi.org/10.1063/1.2731356},
   DOI={10.1063/1.2731356},
   number={5},
   journal={Journal of Mathematical Physics},
   publisher={AIP Publishing},
   author={Bombin, H. and Martin-Delgado, M. A.},
   year={2007},
   month=may }

@article{Bombin07,
  title = {Optimal resources for topological two-dimensional stabilizer codes: Comparative study},
  author = {Bombin, H. and Martin-Delgado, M. A.},
  journal = {Phys. Rev. A},
  volume = {76},
  issue = {1},
  pages = {012305},
  numpages = {6},
  year = {2007},
  month = {Jul},
  publisher = {American Physical Society},
  doi = {10.1103/PhysRevA.76.012305},
  url = {https://link.aps.org/doi/10.1103/PhysRevA.76.012305}
}

@article{Bombin07_2,
  title = {Exact topological quantum order in $D=3$ and beyond: Branyons and brane-net condensates},
  author = {Bombin, H. and Martin-Delgado, M. A.},
  journal = {Phys. Rev. B},
  volume = {75},
  issue = {7},
  pages = {075103},
  numpages = {18},
  year = {2007},
  month = {Feb},
  publisher = {American Physical Society},
  doi = {10.1103/PhysRevB.75.075103},
  url = {https://link.aps.org/doi/10.1103/PhysRevB.75.075103}
}

@article{Stephens14,
  title = {Fault-tolerant thresholds for quantum error correction with the surface code},
  author = {Stephens, Ashley M.},
  journal = {Phys. Rev. A},
  volume = {89},
  issue = {2},
  pages = {022321},
  numpages = {9},
  year = {2014},
  month = {Feb},
  publisher = {American Physical Society},
  doi = {10.1103/PhysRevA.89.022321},
  url = {https://link.aps.org/doi/10.1103/PhysRevA.89.022321}
}

@article{Pretko20,
   title={Fracton phases of matter},
   volume={35},
   ISSN={1793-656X},
   url={http://dx.doi.org/10.1142/S0217751X20300033},
   DOI={10.1142/s0217751x20300033},
   number={06},
   journal={International Journal of Modern Physics A},
   publisher={World Scientific Pub Co Pte Ltd},
   author={Pretko, Michael and Chen, Xie and You, Yizhi},
   year={2020},
   month=feb, pages={2030003} 
}

@article{Aitchison24,
   title={Boundaries and defects in the cubic code},
   volume={109},
   ISSN={2469-9969},
   url={http://dx.doi.org/10.1103/PhysRevB.109.205125},
   DOI={10.1103/physrevb.109.205125},
   number={20},
   journal={Physical Review B},
   publisher={American Physical Society (APS)},
   author={Aitchison, Cory T. and Bulmash, Daniel and Dua, Arpit and Doherty, Andrew C. and Williamson, Dominic J.},
   year={2024},
   month=may }

@article{Roberts20,
  title = {Symmetry-Protected Self-Correcting Quantum Memories},
  author = {Roberts, Sam and Bartlett, Stephen D.},
  journal = {Phys. Rev. X},
  volume = {10},
  issue = {3},
  pages = {031041},
  numpages = {36},
  year = {2020},
  month = {Aug},
  publisher = {American Physical Society},
  doi = {10.1103/PhysRevX.10.031041},
  url = {https://link.aps.org/doi/10.1103/PhysRevX.10.031041}
}

@article{Bombin13,
doi = {10.1088/1367-2630/15/5/055023},
url = {https://doi.org/10.1088/1367-2630/15/5/055023},
year = {2013},
month = {may},
publisher = {IOP Publishing},
volume = {15},
number = {5},
pages = {055023},
author = {Bombin, H and Chhajlany, R W and Horodecki, M and Martin-Delgado, M A},
title = {Self-correcting quantum computers},
journal = {New Journal of Physics}
}

@article{Nussinov08,
  title = {Autocorrelations and thermal fragility of anyonic loops in topologically quantum ordered systems},
  author = {Nussinov, Zohar and Ortiz, Gerardo},
  journal = {Phys. Rev. B},
  volume = {77},
  issue = {6},
  pages = {064302},
  numpages = {16},
  year = {2008},
  month = {Feb},
  publisher = {American Physical Society},
  doi = {10.1103/PhysRevB.77.064302},
  url = {https://link.aps.org/doi/10.1103/PhysRevB.77.064302}
}

@article{Haah12,
  title = {Logical-operator tradeoff for local quantum codes},
  author = {Haah, Jeongwan and Preskill, John},
  journal = {Phys. Rev. A},
  volume = {86},
  issue = {3},
  pages = {032308},
  numpages = {11},
  year = {2012},
  month = {Sep},
  publisher = {American Physical Society},
  doi = {10.1103/PhysRevA.86.032308},
  url = {https://link.aps.org/doi/10.1103/PhysRevA.86.032308}
}

@article{Bravyi09,
doi = {10.1088/1367-2630/11/4/043029},
url = {https://doi.org/10.1088/1367-2630/11/4/043029},
year = {2009},
month = {apr},
publisher = {},
volume = {11},
number = {4},
pages = {043029},
author = {Bravyi, Sergey and Terhal, Barbara},
title = {A no-go theorem for a two-dimensional self-correcting quantum memory based on stabilizer codes},
journal = {New Journal of Physics}
}

@article{Bravyi11,
  title = {Energy Landscape of 3D Spin Hamiltonians with Topological Order},
  author = {Bravyi, Sergey and Haah, Jeongwan},
  journal = {Phys. Rev. Lett.},
  volume = {107},
  issue = {15},
  pages = {150504},
  numpages = {4},
  year = {2011},
  month = {Oct},
  publisher = {American Physical Society},
  doi = {10.1103/PhysRevLett.107.150504},
  url = {https://link.aps.org/doi/10.1103/PhysRevLett.107.150504}
}

@article{Wen90,
author = {WEN, X. G.},
title = {TOPOLOGICAL ORDERS IN RIGID STATES},
journal = {International Journal of Modern Physics B},
volume = {04},
number = {02},
pages = {239-271},
year = {1990},
doi = {10.1142/S0217979290000139}
}

@article{Hamming50,
  author  = {Hamming, Richard W.},
  title   = {Error Detecting and Error Correcting Codes},
  journal = {Bell System Technical Journal},
  volume  = {29},
  number  = {2},
  pages   = {147--160},
  year    = {1950},
  doi     = {10.1002/j.1538-7305.1950.tb00463.x}
}

@article{Ozeki98,
doi = {10.1088/0305-4470/31/24/007},
url = {https://doi.org/10.1088/0305-4470/31/24/007},
year = {1998},
month = {jun},
publisher = {},
volume = {31},
number = {24},
pages = {5451},
author = {Yukiyasu Ozeki and Nobuyasu Ito},
title = {Multicritical dynamics for the ± J Ising Model},
journal = {Journal of Physics A: Mathematical and General}
}

@article{Ohno04,
  doi = {10.1016/j.nuclphysb.2004.07.003},
  url = {https://doi.org/10.1016/j.nuclphysb.2004.07.003},
  year = {2004},
  month = oct,
  publisher = {Elsevier {BV}},
  volume = {697},
  number = {3},
  pages = {462--480},
  author = {Takuya Ohno and Gaku Arakawa and Ikuo Ichinose and Tetsuo Matsui},
  title = {Phase structure of the random-plaquette gauge model: accuracy threshold for a toric quantum memory},
  journal = {Nucl. Phys. B.}
}

@article{Tan25,
   title={Fracton models from product codes},
   volume={7},
   ISSN={2643-1564},
   url={http://dx.doi.org/10.1103/f48m-rlh3},
   DOI={10.1103/f48m-rlh3},
   number={3},
   journal={Physical Review Research},
   publisher={American Physical Society (APS)},
   author={Tan, Yi and Roberts, Brenden and Tantivasadakarn, Nathanan and Yoshida, Beni and Yao, Norman Y.},
   year={2025},
   month=sep }

@article{Imry79,
  title = {Influence of quenched impurities on first-order phase transitions},
  author = {Imry, Yoseph and Wortis, Michael},
  journal = {Phys. Rev. B},
  volume = {19},
  issue = {7},
  pages = {3580--3585},
  numpages = {0},
  year = {1979},
  month = {Apr},
  publisher = {American Physical Society},
  doi = {10.1103/PhysRevB.19.3580},
  url = {https://link.aps.org/doi/10.1103/PhysRevB.19.3580}
}

@article{Aizenman89,
  title = {Rounding of first-order phase transitions in systems with quenched disorder},
  author = {Aizenman, Michael and Wehr, Jan},
  journal = {Phys. Rev. Lett.},
  volume = {62},
  issue = {21},
  pages = {2503--2506},
  numpages = {0},
  year = {1989},
  month = {May},
  publisher = {American Physical Society},
  doi = {10.1103/PhysRevLett.62.2503},
  url = {https://link.aps.org/doi/10.1103/PhysRevLett.62.2503}
}

@article{Cardy99,
  author = {Cardy, John},
  title = {Quenched randomness at first‐order transitions},
  journal = {Physica A},
  volume = {263},
  pages = {215--225},
  year = {1999}
}

@article{Chatelain01,
  TITLE = {{Evidence for softening of first-order transition in 3D by quenched disorder}},
  AUTHOR = {Chatelain, Christophe and Berche, Bertrand and Janke, Wolfhard and Berche, Pierre Emmanuel},
  URL = {https://hal.science/hal-00137974},
  NOTE = {LaTeX file with Revtex, 4 pages, 4 eps figures},
  JOURNAL = {{Physical Review E : Statistical, Nonlinear, and Soft Matter Physics [2001-2015]}},
  PUBLISHER = {{American Physical Society}},
  VOLUME = {64},
  PAGES = {036120},
  YEAR = {2001},
  HAL_ID = {hal-00137974},
  HAL_VERSION = {v1},
}

@article{Aharony96,
  title = {Absence of Self-Averaging and Universal Fluctuations in Random Systems near Critical Points},
  author = {Aharony, Amnon and Harris, A. Brooks},
  journal = {Phys. Rev. Lett.},
  volume = {77},
  issue = {18},
  pages = {3700--3703},
  numpages = {0},
  year = {1996},
  month = {Oct},
  publisher = {American Physical Society},
  doi = {10.1103/PhysRevLett.77.3700},
  url = {https://link.aps.org/doi/10.1103/PhysRevLett.77.3700}
}

@article{Katzgraber06,
doi = {10.1088/1742-5468/2006/03/P03018},
url = {https://doi.org/10.1088/1742-5468/2006/03/P03018},
year = {2006},
month = {mar},
publisher = {},
volume = {2006},
number = {03},
pages = {P03018},
author = {Katzgraber, Helmut G and Trebst, Simon and Huse, David A and Troyer, Matthias},
title = {Feedback-optimized parallel tempering Monte Carlo},
journal = {Journal of Statistical Mechanics: Theory and Experiment}
}

@misc{LRZ_Linux_Cluster,
  title = {Linux Cluster},
  howpublished = {\url{https://www.lrz.de/en/service-center/alle-lrz-services/high-performance-computing/linux-cluster}},
  note = {Accessed: 01.09.2025},
}

@article{Queiroz06,
  title = {Multicritical point of Ising spin glasses on triangular and honeycomb lattices},
  author = {de Queiroz, S. L. A.},
  journal = {Phys. Rev. B},
  volume = {73},
  issue = {6},
  pages = {064410},
  numpages = {7},
  year = {2006},
  month = {Feb},
  publisher = {American Physical Society},
  doi = {10.1103/PhysRevB.73.064410},
  url = {https://link.aps.org/doi/10.1103/PhysRevB.73.064410}
}
